\documentclass[aps,pra,twocolumn,superscriptaddress,groupedaddress,nofootinbib]{revtex4}

\usepackage{graphicx}
\usepackage{dcolumn}
\usepackage{bm}       
\usepackage{amssymb}  
\usepackage{epsfig}
\usepackage{epstopdf}
\usepackage{amsmath}
\usepackage[latin9]{inputenc}
\usepackage[title]{appendix}
\usepackage{graphics}
\usepackage{dsfont}
\usepackage{xcolor}
\usepackage{nccfoots}
\usepackage[flushleft]{threeparttable}
\usepackage{threeparttable}
\usepackage{graphicx}
\usepackage{float}
\usepackage{dsfont}
\usepackage{array}
\usepackage{multirow}
\usepackage{mathrsfs}

\begin{document}

\title{The $^{229}$Th isomer: prospects for a nuclear optical clock}

\author{Lars von der Wense}
\affiliation{Ludwig-Maximilians-Universit\"at M\"unchen, 85748 Garching, Germany.}
\author{Benedict Seiferle}
\affiliation{Ludwig-Maximilians-Universit\"at M\"unchen, 85748 Garching, Germany.}

\begin{abstract}
The proposal for the development of a nuclear optical clock has triggered a multitude of experimental and theoretical studies. In particular the prediction of an unprecedented systematic frequency uncertainty of about $10^{-19}$ has rendered a nuclear clock an interesting tool for many applications, potentially even for a re-definition of the second. The focus of the corresponding research is a nuclear transition of the $^{229}$Th nucleus, which possesses a uniquely low nuclear excitation energy of only $8.12\pm0.11$ eV ($152.7\pm2.1$~nm). This energy is sufficiently low to allow for nuclear laser spectroscopy, an inherent requirement for a nuclear clock. Recently, some significant progress toward the development of a nuclear frequency standard has been made and by today there is no doubt that a nuclear clock will become reality, most likely not even in the too far future. Here we present a comprehensive review of the current status of nuclear clock development with the objective of providing a rather complete list of literature related to the topic, which could serve as a reference for future investigations.
\end{abstract}
\maketitle

\tableofcontents

\setcounter{page}{1}

\section{Introduction}
Since ancient human history it has been important to subdivide the continuous flow of time into certain repeating intervals or cycles \cite{McCarthy2018}. This is of particular practical use as it allows defining absolute points in time (e.g., for harvesting or scheduling other events). In a natural way, the earth's rotation around its own axis and around the sun, as well as the rotation of the moon around the earth, do provide us with such repeating time intervals, and it is a remarkable fact that the rotation of the earth around its own axis was used for the definition of the SI second until 1960, followed by the ephemeris second, which was based on the rotation of the earth around the sun and was used until 1967 \cite{Cesium}. In early historic times, the earth's rotation was measured with the help of well positioned stones or holes through which the sun could shine. Famous examples are Stonehenge and the Nebra Sky Disc \cite{Herten2018}. Later, more advanced sundials were used, while the underlying principle remained unchanged \cite{Higgins2004}. Nevertheless, there is a disadvantage of using only the earth's rotation for time measurement: It requires the observation of celestial bodies like the sun or the stars that may not always be visible. For this reason, other notions of time measurement were developed, which are independent of the earth's rotation. These were, for example, water clocks, which make use of the continuous flow of water, or candle clocks in European monasteries. Here the term ``clock" refers to any instrument that was used to measure time intervals. In the late 13th century mechanical clocks arose, which made use of a weight, e.g. a stone, that was mechanically hindered from free fall \cite{McCarthy2018}. Such clocks showed large inaccuracies of about 15 minutes per day when compared to the earth's rotation so that they had to be reset on a daily basis \cite{Usher}. A mechanical clock that can be considered as accurate and independent of the earth's rotation was the pendulum clock developed and built by Christian Huygens in 1657 \cite{Sorge}. Early pendulum clocks achieved inaccuracies of less than 1 minute per day which were later improved to less than 15 seconds per day \cite{Bennet}. For centuries, pendulum clocks remained the most accurate clocks and were constantly developed. In 1921, the Shortt-Synchronome free pendulum clock was built, which achieved an inaccuracy of a few milliseconds per day \cite{Jackson1928}. The techniques used for time measurement improved quickly within the following decades. By 1927, the quartz clock was developed, which uses the piezo-electric effect to introduce oscillations into the quartz crystal \cite{Horton1928,Marrison}. The inaccuracies achieved by the very early quartz clocks were 1~s in 4 months and therefore larger than what was achievable with the pendulum clocks at the time. However, as no macroscopic mechanical motion is involved in their process of operation, these clocks soon surpassed the mechanical clocks, approaching an inaccuracy of $10^{-9}$, corresponding to 1~s in 32 years \cite{Marrison}. The quartz clocks were more accurate than the earth's rotation itself and allowed to measure the variations of the earth's rotation in 1935 \cite{Scheibe1936}.\\[0.2cm]
A revolution in time measurement started in the late 1930s in the group of Isidor Rabi, who developed concepts for using the magnetic interaction between the nucleus and the valence electron in the atomic shell for time measurement \cite{Ramsey}. This led to the development of an atomic clock based on ammonia in 1949 by Harold Lyons at the National Bureau of Standards in the US \cite{Lyons}. The original idea to use atomic particles for metrology dates back to the 19th century to Kelvin and Maxwell. In 1955 Luis Essen and Jack Parry built an atomic clock providing an inaccuracy of $10^{-9}$ at the National Physics Laboratory (NPL) in the UK \cite{Essen}. In the following years the accuracies of atomic clocks improved significantly, approaching inaccuracies of $10^{-13}$ in the 1960s, corresponding to 1~s in 300,000 years \cite{Ramsey}. These extraordinary accuracies led to the re-definition of the second in 1967 \cite{Cesium}, when the second was defined as the time elapsed after 9,192,631,770 cycles of a microwave stabilized to the hyperfine splitting of the $^{133}$Cs $^2$S$_{1/2}$ electronic level \cite{McCarthy2018}. Cesium was chosen as it possesses only one natural isotope, so that no reduction of signal-to-noise ratio occurs as would be the case if different underlying hyperfine-structure splittings were present. The accuracy of the Cs-clock was further improved in the following decades. Important use was made of laser cooling, leading to the development of the Cs-fountain clocks in the early 1990s \cite{Clairon1991} that achieved inaccuracies in the $10^{-16}$ range around 2005 \cite{Wynands}.\\[0.2cm]
The next leap in technology occurred with the development of the frequency comb in 1998 \cite{Udem,Hall2006,Haensch2006}. While previously, optical frequency measurements required complex frequency divider chains (see, e.g., \cite{Schnatz1996}), the frequency comb allowed in a more practical and precise way to directly count the oscillations of visible light in the optical range, which possesses five orders of magnitude larger frequencies compared to electromagnetic waves in the microwave region. As the frequency of the atomic transition significantly affects the accuracy of time measurement, the possibility to count the oscillations of laser light had an immediate impact on clock technology, leading to the development of optical atomic clocks \cite{Poli,Ludlow}. Today, optical atomic clocks are the most accurate time-keeping devices, with an inaccuracy below $10^{-18}$ \cite{Brewer}, corresponding to 1 second in 30 billion years, significantly longer than the age of the universe. Some of the most accurate optical atomic clocks operational today are listed in Sec.~\ref{NOC.ACC}. Clearly, this 100-fold improved accuracy compared to the Cs fountain clocks will lead to a re-definition of the second in the near future. Reasons why the second was not re-defined in 2019 together with many other SI units \cite{CGPM2018} are that the field of time measurement continues to be fast developing and many criteria need to be matched before the decision for a new definition of the second can be made \cite{Gill2011,Riehle2015,Gill2016,Riehle2018}.\\[0.2cm]
Obviously, given the accuracies of clocks, their applications have also changed. Important use of atomic clocks is made in satellite-based navigation, where a position is determined via the time it takes for electromagnetic pulses to travel certain distances. With a future generation of optical atomic clocks the precision of navigation may improve up to a point where, for example, small strains of the earth's crust can be measured and applied for earthquake prediction \cite{Gill2008,Ludlow}. Importantly, however, optical atomic clocks of the highest accuracies have even led to the creation of the novel field of chronometric geodesy \cite{Delva2019}. At the achieved accuracies, general relativistic effects enter the game: time dilation caused by the local gravitational field becomes observable. Close to the surface of the earth, a height difference of 1~cm corresponds to a relative frequency shift of $10^{-18}$, which is measurable with the most accurate optical atomic clocks today \cite{McGrew}. Due to earth's tides, local weather and seismic activities, local gravitational fields are subject to frequent changes, which become observable. For this reason, we can think of the most accurate optical atomic clocks on earth as gravity sensors \cite{Mehlstaeubler}. Further, optical atomic clocks currently provide the most stringent tests for potential time variations of fundamental constants \cite{Rosenband,Godun} and are considered as instruments for dark matter searches \cite{Derevianko,Roberts2017}.\\[0.2cm]
Here, a new type of clock is discussed, the so called ``nuclear clock", which uses a nuclear transition instead of an atomic shell transition for time measurement \cite{Peik}. The nuclear clock has the potential for unprecedented performance \cite{Campbell1} and may ultimately lead to the introduction of a new primary time standard. In addition, the potential for a solid-state nuclear clock has been discussed \cite{Rellergert,Kazakov1,Wense2019c}, which may have advantages in terms of stability, compactness and power consumption. In the focus of the corresponding research is a metastable nuclear excited state of the $^{229}$Th nucleus (usually denoted as $^{229\text{m}}$Th and called the ``thorium isomer"). With an energy difference of only about 8~eV to the nuclear ground state, $^{229\text{m}}$Th is the nuclear excited state of lowest know excitation energy, thereby offering the possibility for laser excitation. While an imprecise knowledge of the transition energy has so far hindered the development of a nuclear clock, recently, three new measurements have led to a significant increase in confidence about the isomer's excitation energy \cite{Seiferle2019b,Yamaguchi2019,Sikorsky2020}. This new knowledge is likely to lead to a phase transition in the $^{229\text{m}}$Th-related research, away from previous ``high-energy" nuclear-physics dominated research and more towards ``low-energy" precision laser spectroscopy. This has motivated us to review the history of experiments up until this point and to provide a complete discussion of existing literature in this field. The review is addressed to the reader from inside or outside the field who is interested in a comprehensive overview over the topic. The reader who prefers a shorter discussion is referred to \cite{Peik4,Wense2018,Thirolf2019b}.
\section{A nuclear optical clock for time measurement}\label{NOC}
It might be a natural new step to use a nuclear transition instead of an atomic shell transition for time measurement \cite{Peik}. Such a ``nuclear clock" is expected to approach an accuracy of 10$^{-19}$, corresponding to 1~s of inaccuracy in 300 billion years \cite{Campbell1}. The central reason for the expected improved accuracy of a nuclear clock compared to existing atomic clocks is that the nucleus is several orders of magnitude smaller than the atomic shell, resulting in a smaller coupling to electric and magnetic fields. Such reduced coupling is expected to improve the achievable accuracy of time measurement \cite{Peik4}. Alternative approaches being investigated are atomic clocks based on highly charged ions \cite{Schmoeger,Kozlov2018,Micke2020}, superradiant optical clocks \cite{Bohnet2012,Norcia2016}, neutral atoms in tweezer arrays \cite{Norcia2019,Madjarov2019} and multiple ion clocks \cite{Herschbach2012,Pyka2014,Keller2019,Keller2019b}. A detailed discussion of the nuclear clock concept will be given in the following.
\subsection{The general principle of clock operation}\label{NOC.GEN}
Although different clocks significantly differ from each other in their technical realization, the fundamental underlying principle is always the same. What all clocks have in common is something that changes with time in a predictive manner. This might be the rotation of the earth, the flow of water, or a pendulum \cite{McCarthy2018}. Further, these changes have to be measured. In case of the earth's rotation the detection of changes can be either done by comparing the position of the sun with a stick or, in a more modern way, by monitoring the stars with the help of a telescope. In a pendulum clock, the number of cycles of a pendulum are mechanically counted with the help of a clockwork. While pendulum clocks are macroscopic mechanical devices and therefore subject to differences between individual items, in atomic clocks the frequency of a particular atomic transition is used as an oscillator. As atoms of the same type are fundamentally identical, in this way, the time intervals can be measured free from mechanical artifacts in a universal manner. In modern optical atomic clocks the laser light (i.e., the oscillations of the electromagnetic field) tuned to a particular atomic transition is used as an oscillator, transferred to the microwave region with the help of a frequency comb and counted, as will be detailed in the following. \\[0.2cm] 
\subsection{Operational principle of optical atomic clocks}\label{NOC.OAC}
\label{01_opticalclocks}
\begin{figure}[t]
 \begin{center}
 \includegraphics[width=8cm]{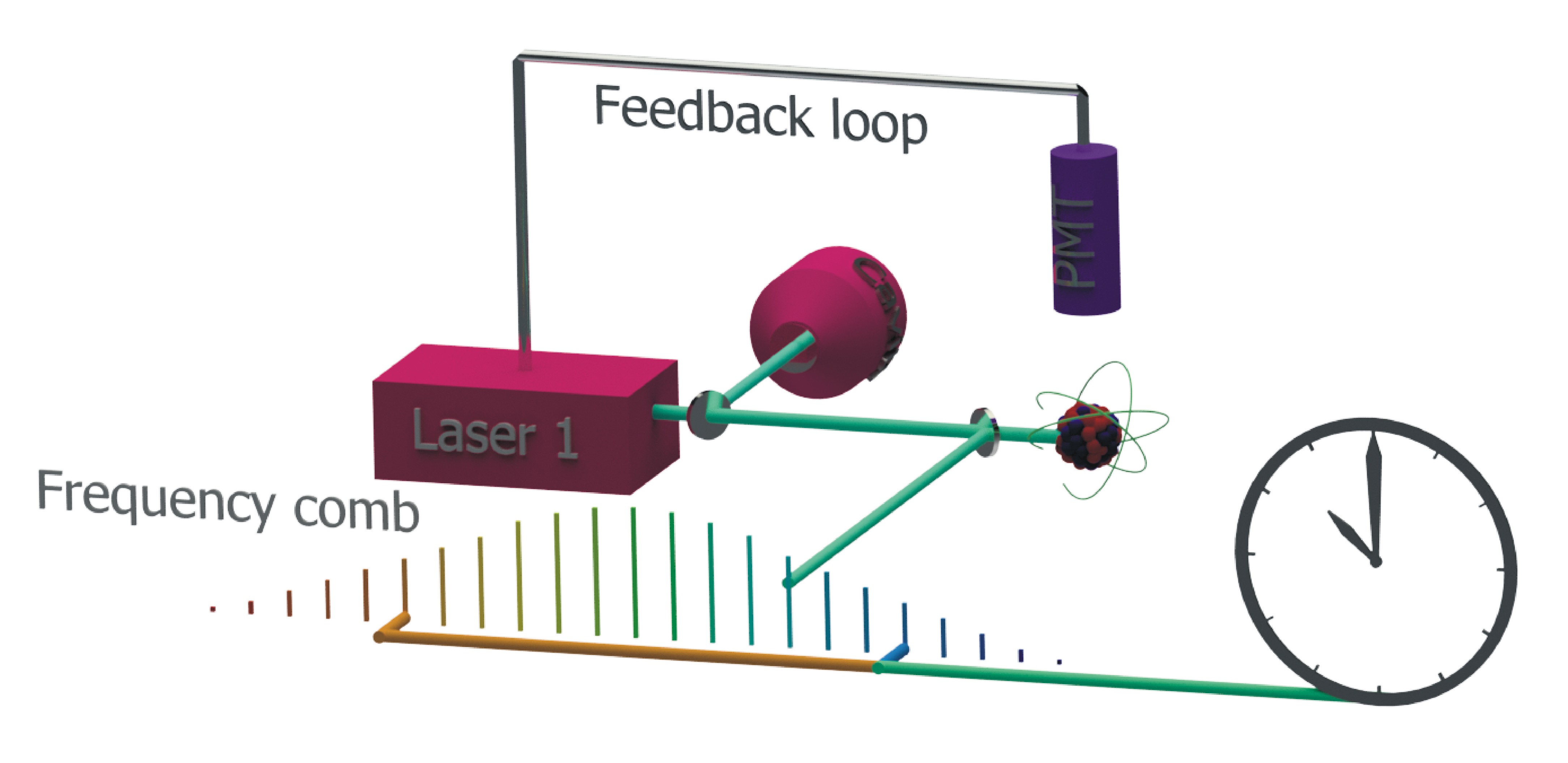}
  \caption{\footnotesize Schematic concept of an optical atomic clock. A spectroscopy laser is tuned to an atomic resonance line. Stabilization of the laser to the resonance is achieved via the detection of fluorescence light emitted from the atomic resonance in combination with a feedback loop. When stabilized to the atomic transition, the laser frequency can be considered as constant and is counted with the help of a frequency comb. After a certain (pre-defined) number of periods of the light wave generated by the spectroscopy laser, one second has elapsed \cite{Gill2011,Riehle2017}. The operational principle of a nuclear optical clock is identical, however with the atomic transition replaced by a nuclear transition \cite{Peik,Campbell1}.}
 \label{optical_clock_principle}
 \end{center}
\end{figure}
Modern optical atomic clocks are today the most accurate frequency standards \cite{Poli,Ludlow}. Their underlying principle of operation is based on the fact that the energy corresponding to an atomic transition is the same for all identical atoms and remains constant (independent of space and time). If laser light is used to optically excite an atomic transition, the frequency of the laser light has to match the corresponding energy of the atomic transition. For this reason, if a laser is stabilized to an atomic transition, the laser frequency will remain constant. In a practical implementation of an optical atomic clock, the laser light is locked to a highly-stable optical cavity to achieve short-term stabilization \cite{Kessler,Matei2017}. Stabilization to the atomic transition is then required to correct for long-term drifts of the cavity resonator. This stabilization can be achieved, for example, using a detector to monitor the excitation of the electronic transition and a feedback loop, which corrects the laser tuning via some servo electronics. While at earlier times, the frequency of the light wave could be counted with the help of complex frequency divider chains \cite{Schnatz1996}, the Nobel-prize winning technology of the frequency comb has led to a drastic simplification of optical frequency measurements \cite{Udem,Hall2006,Haensch2006}. In this technique, the frequency comb acts as a transmission gearing between optical and microwave frequencies. The latter ones can be electronically counted. This allows one to measure time by counting the number of oscillations of the laser wave that has been stabilized to a particular atomic transition. A conceptual sketch of the principle of operation of an optical atomic clock is shown in Fig.~\ref{optical_clock_principle}.\\[0.2cm]
Currently, two different types of optical atomic clocks dominate the field: single ion clocks and optical lattice clocks. In single ion optical atomic clocks the laser is stabilized to an atomic transition of a single, laser-cooled ion in a Paul trap. Opposed to that, in optical lattice clocks a cloud of atoms stored in an optical lattice is used for laser stabilization.\\[0.2cm]
The currently most accurate clocks in the world are: an optical atomic clock based on a single $^{27}$Al$^+$ ion \cite{Brewer}, an optical lattice clock based on $^{171}$Yb \cite{Schioppo,McGrew}, an $^{171}$Yb$^+$ single-ion clock \cite{Huntemann,Sanner2019} and a $^{87}$Sr optical lattice clock \cite{Nicholson,Bothwell2019}. Each of these clocks achieves an accuracy around $10^{-18}$, corresponding to 1~s of inaccuracy in 30 billion years (see Sec.~\ref{NOC.ACC}).
\subsection{Accuracy and stability}\label{NOC.ACC}
\begin{table*}[t]
\begin{center}
\caption{\footnotesize Accuracies and stabilities of several existing optical atomic clocks in comparison to the predicted performance of a $^{229}$Th single-ion nuclear clock \cite{Campbell1}.}
\begin{footnotesize}
\begin{tabular}{ccccc}
\noalign{\smallskip}\hline\noalign{\smallskip}
 Isotope  & Type & \parbox[pt][2em][c]{1.5cm}{\centering{Accuracy}}  & \parbox[pt][2em][c]{2.5cm}{\centering{Stability ($\tau$ in s)}} & \parbox[pt][2em][c]{1.5cm}{\centering{Reference}} \\
\noalign{\smallskip}\hline\noalign{\smallskip}
\parbox[0pt][1em][c]{2cm}{$^{87}$Sr} & \parbox[pt][1em][c]{2.5cm}{\centering{optical lattice}} & \parbox[pt][1em][c]{1.5cm}{\centering{$2.0\cdot10^{-18}$}} & \parbox[pt][1em][c]{2.5cm}{\centering{$2.2\cdot10^{-16}/\sqrt{\tau}$}} & \parbox[pt][1em][c]{3cm}{\cite{Bothwell2019} }\\
\parbox[0pt][1em][c]{2cm}{$^{87}$Sr} & \parbox[pt][1em][c]{2.5cm}{\centering{optical lattice}} & \parbox[pt][1em][c]{1.5cm}{\centering{}} & \parbox[pt][1em][c]{2.5cm}{\centering{$4.8\cdot10^{-17}/\sqrt{\tau}$}} & \parbox[pt][1em][c]{3cm}{\cite{Oelker2019} }\\
\parbox[0pt][1em][c]{2cm}{$^{171}$Yb$^+$ (E3)} &  \parbox[pt][1em][c]{1.5cm}{\centering{single ion}} & \parbox[pt][1em][c]{1.5cm}{\centering{$2.8\cdot10^{-18}$}} & \parbox[pt][1em][c]{2.5cm}{\centering{$1.4\cdot10^{-15}/\sqrt{\tau}$}} & \parbox[pt][1em][c]{3cm}{\cite{Sanner2019}} \\
\parbox[0pt][1em][c]{2cm}{$^{171}$Yb} & \parbox[pt][1em][c]{2.5cm}{\centering{optical lattice}} & \parbox[pt][1em][c]{1.5cm}{\centering{$1.4\cdot10^{-18}$}} &  \parbox[pt][1em][c]{2.5cm}{\centering{$1.5\cdot10^{-16}/\sqrt{\tau}$}} & \parbox[pt][1em][c]{3cm}{\cite{McGrew}}\\
\parbox[0pt][1em][c]{2cm}{$^{171}$Yb} & \parbox[pt][1em][c]{2.5cm}{\centering{optical lattice}} & \parbox[pt][1em][c]{1.5cm}{\centering{}} &  \parbox[pt][1em][c]{2.5cm}{\centering{$6\cdot10^{-17}/\sqrt{\tau}$}} & \parbox[pt][1em][c]{3cm}{\cite{Schioppo}}\\
\parbox[0pt][1em][c]{2cm}{$^{27}$Al$^+$} & \parbox[pt][1em][c]{2.5cm}{\centering{single ion}} & \parbox[pt][1em][c]{1.5cm}{\centering{$9.4\cdot10^{-19}$}} & \parbox[pt][1em][c]{2.5cm}{\centering{$1.2\cdot10^{-15}/\sqrt{\tau}$}}& \parbox[pt][1em][c]{3cm}{\cite{Brewer}}\\
\noalign{\smallskip}\hline\noalign{\smallskip}
\parbox[0pt][1em][c]{2cm}{$^{229}$Th$^{3+}$} & \parbox[pt][1em][c]{3cm}{\centering{single ion nuclear}} & \parbox[pt][1em][c]{2cm}{\centering{$\sim1\cdot10^{-19}\ ^{*}$}} & \parbox[pt][1em][c]{2.5cm}{\centering{$\sim5\cdot10^{-16}/\sqrt{\tau}\ ^{*}$}}& \parbox[pt][1em][c]{3cm}{\cite{Campbell1}}\\
\noalign{\smallskip}\hline\noalign{\smallskip}
\end{tabular}
\end{footnotesize}
\begin{tablenotes}
\footnotesize $^{*}$Estimated values.
\end{tablenotes}
\label{clockshifts}
\end{center}
\end{table*}
Two fundamental quantities determine the quality of a clock: (in)accuracy and (in)stability \cite{Riehle2006,Riehle2017}. Quite generally, the inaccuracy describes how much a measured value systematically differs from the correct value. Therefore, it includes all systematic uncertainties of the measurement. Opposed to that, the instability can be associated with the statistical measurement uncertainty.\\[0.2cm]
The notion of time measurement is closely related to frequency measurement: if the frequency of an unperturbed oscillator $\omega_0$ is known, a time interval can be measured by counting the number of oscillations. Usually, however, the measured frequency will be subject to perturbations. Here a systematic frequency shift $y_\text{sys}$ in fractional units as well as time-dependent statistical fluctuations $y_\text{stat}(t)$ are distinguished, leading to a measured frequency of $\omega_\text{meas}(t)=\omega_0(1+y_\text{sys}+y_\text{stat}(t))$. In agreement with this notation, one can define the fractional frequency offset as $y(t)=y_\text{sys}+y_\text{stat}(t)=(\omega_\text{meas}(t)-\omega_0)/\omega_0$. The systematic frequency shift $y_\text{sys}$ is generally considered to be known and can be corrected for, up to its uncertainty. The remaining ``systematic frequency uncertainty" is an important parameter for the quality of a clock and is usually referred to as the clock's (in)accuracy. The reason is, that this remaining uncertainty will lead to an unavoidable deviation of the frequency used for time measurement to the unperturbed frequency and thus to a continuously growing error in time measurement, which equals the clock's (in)accuracy under the assumption that the unperturbed frequency of the oscillator would be used to define the second.\\[0.2cm]
The stability of a clock is a measure for the statistical fluctuations (introduced by $y_\text{stat}(t)$) that occur during the frequency measurement. Here $y_\text{stat}(t)$ is called the frequency noise and the corresponding Allan variance allows to estimate the achievable stability of the clock. The Allan variance $\sigma^2_y(\tau)$ is defined as the comparison of two successive frequency deviations for a given averaging time $\tau$. More exactly, we have $\sigma^2_y(\tau)=1/2\langle\left(\bar{y}_{n+1}-\bar{y}_n\right)^2 \rangle$, with $\bar{y}_n$ as the $n$th fractional frequency average over the time $\tau$. For current atomic clocks, Allan deviations $\sigma_y(\tau)$ are often limited by the quantum projection noise (QPN) and given by \cite{Itano,Ludlow}
\begin{equation}
\label{Allan1}
\sigma_y(\tau)\approx \frac{1}{\omega}\sqrt{\frac{1}{NT\tau}}.
\end{equation}
Here $\omega$ is the angular frequency of the resonance, $T$ the coherence time, $N$ the number of irradiated atoms and $\tau$ the averaging time. High clock performance can only be realized if the systematic frequency uncertainty is small and if the statistical uncertainty can be brought to a value comparable to the systematic uncertainty on a realistic time scale. The stabilities and accuracies achieved by some optical atomic clocks are compared to the expected performance of a single ion nuclear clock in Tab.~\ref{clockshifts}.
\subsection{The idea of a nuclear optical clock}\label{NOC.IDEA}
Although the accuracy that is achieved by optical atomic clocks today is already stunning, surpassing $10^{-18}$ corresponding to an error of 1 second after $3\cdot10^{10}$~years, significantly longer than the age of the universe \cite{Brewer}, it is reasonable to ask if it is possible to push the limits further. A conceptual idea that has been proposed is to use a nuclear transition instead of an atomic shell transition for time measurement \cite{Tkalya1,Peik,Campbell1,Safronova2018a}. The principle of operation of this ``nuclear optical clock" remains unchanged compared to that of optical atomic clocks discussed in Sec.~\ref{01_opticalclocks}, except that a nuclear transition instead of an atomic shell transition is used for laser stabilization.\\[0.2cm]
Historically, the term ``nuclear clock" was sometimes used in the context of radiocarbon dating (see, e.g., \cite{radiocarbon}). Also, when in 1960 Mössbauer spectroscopy of the ultra-narrow $\gamma$-ray resonance at 93.3~keV of $^{67}$Zn was achieved, it was referred to as a ``nuclear clock" \cite{Nagle1960}. With a half-life of $9$~$\mu$s, to date this state provides the Mössbauer line with the smallest relative linewidth of $\Delta E/E=5.5\cdot10^{-16}$. At that time, however, the term ``nuclear clock" was used for a precise measurement of a relative frequency shift, rather than for an absolute frequency determination.\\[0.2cm]
Early discussions about the potential use of $^{229\text{m}}$Th for metrology date back to 1995, when it was proposed to investigate the isomeric properties for the ``{\it development of a nuclear source of light with reference narrow line}" \cite{Tkalya1995}. In 1996 the ``{\it development of a high stability nuclear source of light for metrology}" was discussed as an application for $^{229\text{m}}$Th \cite{Tkalya1}.\\[0.2cm]
A detailed concept and analysis of a nuclear clock was published in 2003 by E. Peik and C. Tamm \cite{Peik}. Conceptually, nuclear transitions provide three advantages compared to atomic transitions (see also \cite{Peik2}): 1. The atomic nucleus is about 5 orders of magnitude smaller than the atomic shell, which leads to significantly reduced magnetic dipole and electric quadrupole moments and therefore to a higher stability against external influences, resulting in an expected improved accuracy and stability of the clock. 2. Nuclear transition energies are typically larger than those in atoms. This leads to higher frequencies which allow for small instabilities. 3. The nucleus is largely unaffected by the atomic shell, for this reason it is intriguing to develop a solid-state nuclear clock based on Mössbauer spectroscopy. Such a solid-state clock could contain a large number of nuclei of about $10^{14}$, thus leading to improved statistical uncertainties when compared to atomic lattice clocks typically providing about $10^4$ atoms.\\[0.2cm]
\subsection{Nuclear transition requirements}\label{NOC.REQ}
\begin{figure}[t]
 \begin{center}
 \includegraphics[width=8cm]{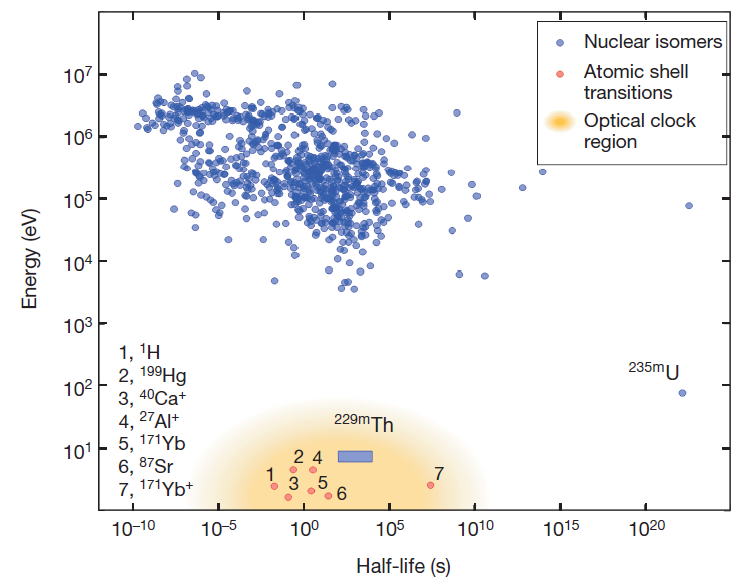}
  \caption{\footnotesize Energy-half-life diagram of nuclear isomeric states with lifetimes longer than 1~ns (blue circles) together with selected atomic shell transitions used for frequency metrology. $^{229\text{m}}$Th (expected region shown as a blue box) exhibits a special position located in the optical clock region rather than in the region of typical nuclear excited states. Reprinted from \cite{Wense2} with kind permission of Springer Nature.}
 \label{energy_halflife}
 \end{center}
\end{figure}
From the principle of clock operation, it is evident that a nuclear optical clock requires narrow-bandwidth laser excitation of a nucleus. As narrow-bandwidth laser technology with significant intensity is only available up to photon energies below 100~eV, the central and most important nuclear transition requirement is that its energy must be sufficiently low to permit laser excitation. This requirement poses a strong constraint on the nuclear transitions useful for clock operation as, from the more than 176,000 known nuclear excited states, only 2 exhibit an energy below 100~eV. These are $^{229\text{m}}$Th, a metastable excited state of the $^{229}$Th nucleus with an excitation energy of only about 8~eV \cite{Beck1,Beck2,Seiferle2019b,Yamaguchi2019,Sikorsky2020}, making it the nuclear state of lowest known excitation energy, and $^{235\text{m}}$U, a metastable state of $^{235}$U with an excitation energy of $\approx76.7$~eV \cite{Ponce2018a}. The existence of a potential third nuclear excited state below 100~eV excitation energy, $^{229\text{m}}$Pa, is still under investigation \cite{Ahmad2015}.\\[0.2cm]
Besides the low excitation energy, the radiative lifetime should be sufficiently long to result in a narrow natural linewidth of the transition. In this way the quality factor $Q=\omega/\Delta\omega$ of the resonance is large, leading to a high potential stability. On the other hand, a linewidth deep in the sub-Hz range is not favorable in terms of clock performance, as in this case the stability will not be limited by the transition itself, but instead by the coherence time of the laser light used for irradiation. Moreover, a narrower linewidth of the transition leads to a larger required laser power, which might be technologically more challenging and results in larger laser-induced frequency shifts of the transition. For these reasons the transition linewidth should be sufficiently narrow to lead to a high stability of the clock, while still remaining sufficiently broad to allow for laser excitation with convenient laser power. $^{229\text{m}}$Th possesses a calculated radiative lifetime between $10^3$ and $10^4$~s (see App. \ref{THEO.DECEXC.GAMMADEC}), corresponding to a linewidth of $10^{-3}$ to $10^{-4}$~Hz, which can be considered as nearly ideal for nuclear clock development. $^{235\text{m}}$U, however, has an extraordinary long radiative lifetime of $\approx10^{24}$~s\footnote{The reason for the actually observed lifetime of 37.5 minutes is a large internal conversion coefficient of up to $10^{21}$ \cite{Zon}.}, leading to no significant probability for direct laser excitation \cite{Wense2019d}. However, excitation via an electronic-bridge mechanism (see App.~\ref{THEO.HIGHEXC.IEB}) might be possible \cite{Berengut2018}. The special position of $^{229\text{m}}$Th compared to other nuclear isomeric states is shown in the energy-half-life diagram Fig.~\ref{energy_halflife}. The thorium isomer possesses an energy that is orders of magnitude below the usual nuclear energy scale and is instead located in a region typical of optical atomic clocks.\\[0.2cm]
There are some further requirements concerning the nucleus in its ground-state. It should be sufficiently available and long-lived to be able to obtain and handle moderate quantities of the material. In case that the nucleus was short-lived, it would have to be continuously generated via a nuclear decay or even a nuclear fusion process, which would make it very impractical or even impossible to use the transition for time measurement. Fortunately, the $^{229}$Th ground state possesses a long half-life of $\approx7917$ years \cite{Varga2014}, which allows for relatively easy handling. Further, it is a daughter product of $^{233}$U, which is available in large quantities ($^{233}$U is a fissile material that was produced in large amounts during the Cold War). Opposed to that, $^{229\text{m}}$Pa, if it exists, will be significantly harder to obtain, as the lifetime of $^{229}$Pa is only 1.5 days, and it has to be produced in nuclear fusion reactions at accelerator facilities \cite{Ahmad2015}.
\subsection{The special properties of $^{229\text{m}}$Th}\label{NOC.PROP}
\begin{table}[t]
\begin{center}
\caption{\footnotesize $^{229\text{m}}$Th energy constraints found in literature.}
\begin{footnotesize}
\begin{tabular}{ccc}
\noalign{\smallskip}\hline\noalign{\smallskip}
 Energy & \parbox[pt][2em][c]{2.4cm}{\centering{Wavelength}}  & \parbox[pt][2em][c]{2.5cm}{\centering{Reference}} \\
\noalign{\smallskip}\hline\noalign{\smallskip}
\parbox[0pt][1.5em][c]{2.5cm}{\centering{$<100$ eV}} &  \parbox[pt][1em][c]{2cm}{\centering{$>12.4$ nm}} & \parbox[pt][1em][c]{2.5cm}{\cite{Kroger_Reich}} \\
\parbox[0pt][1.5em][c]{2.5cm}{\centering{$<10$ eV$^{(1)}$}} &  \parbox[pt][1em][c]{2cm}{\centering{$>124$ nm}} & \parbox[pt][1em][c]{2.5cm}{\cite{Akavoli}} \\
\parbox[0pt][1.5em][c]{2.5cm}{\centering{$(-)1\pm 4$ eV}} &  \parbox[pt][1em][c]{2cm}{\centering{$>250$ nm}} & \parbox[pt][1em][c]{2.5cm}{\cite{Reich_Helmer}} \\
\parbox[0pt][1.5em][c]{2.5cm}{\centering{$4.5\pm 1$ eV}} &  \parbox[pt][1em][c]{2cm}{\centering{$275_{-50}^{+75}$ nm}} & \parbox[pt][1em][c]{2.5cm}{\cite{Reich_Helmer2}} \\
\parbox[0pt][1.5em][c]{2.5cm}{\centering{$3.5\pm 1$ eV$^{(2)}$}} &  \parbox[pt][1em][c]{2cm}{\centering{$354_{-75}^{+150}$ nm}} & \parbox[pt][1em][c]{2.5cm}{\cite{Helmer_Reich}} \\
\parbox[0pt][1.5em][c]{2.5cm}{\centering{$5.5\pm 1$ eV}} &  \parbox[pt][1em][c]{2cm}{\centering{$225_{-35}^{+50}$ nm}} & \parbox[pt][1em][c]{2.5cm}{\cite{Filho}} \\
\parbox[0pt][1.5em][c]{2.5cm}{\centering{$7.6\pm 0.5$ eV}} &  \parbox[pt][1em][c]{2cm}{\centering{$163.1^{+11.5}_{-10.0}$ nm}} & \parbox[pt][1em][c]{2.5cm}{\cite{Beck1}} \\
\parbox[0pt][1.5em][c]{2.5cm}{\centering{$7.8\pm 0.5$ eV}} &  \parbox[pt][1em][c]{2cm}{\centering{$159.0_{-9.6}^{+10.8}$ nm}} & \parbox[pt][1em][c]{2.5cm}{\cite{Beck2}} \\
\parbox[0pt][1.5em][c]{2.6cm}{\centering{$6.3<E<18.3$ eV}} &  \parbox[pt][1em][c]{2.5cm}{\centering{$68<\lambda<197$ nm}} & \parbox[pt][1em][c]{2.5cm}{\cite{Wense2}} \\
\parbox[0pt][1.5em][c]{2.5cm}{\centering{$8.28\pm 0.17$ eV}} &  \parbox[pt][1em][c]{2cm}{\centering{$149.7^{+3.2}_{-3.0}$ nm}} & \parbox[pt][1em][c]{2.5cm}{\cite{Seiferle2019b}} \\
\parbox[0pt][1.5em][c]{2.6cm}{\centering{$2.5<E<8.9$ eV}} &  \parbox[pt][1em][c]{2.6cm}{\centering{$139<\lambda<496$ nm}} & \parbox[pt][1em][c]{2.5cm}{\cite{Masuda2019}} \\
\parbox[0pt][1.5em][c]{2.5cm}{\centering{$8.30\pm 0.92$ eV}} &  \parbox[pt][1em][c]{2cm}{\centering{$149.4^{+18.6}_{-14.9}$ nm}} & \parbox[pt][1em][c]{2.6cm}{\cite{Yamaguchi2019}} \\
\parbox[0pt][1.5em][c]{2.5cm}{\centering{$8.1\pm 0.7$ eV$^{(3)}$}} &  \parbox[pt][1em][c]{2cm}{\centering{$153.1_{-12.2}^{+14.4}$ nm}} & \parbox[pt][1em][c]{2.6cm}{\cite{Yamaguchi2019}} \\
\parbox[0pt][1.5em][c]{2.5cm}{\centering{$7.84\pm 0.29$ eV$^{(4)}$}} &  \parbox[pt][1em][c]{2cm}{\centering{$158.1^{+6.1}_{-5.6}$ nm}} & \parbox[pt][1em][c]{2.5cm}{\cite{Sikorsky2020}} \\
\parbox[0pt][1.5em][c]{2.5cm}{\centering{$8.10\pm 0.17$ eV$^{(5)}$}} &  \parbox[pt][1em][c]{2cm}{\centering{$153.1^{+3.2}_{-3.2}$ nm}} & \parbox[pt][1em][c]{2.5cm}{\cite{Sikorsky2020}} \\
\parbox[0pt][1.5em][c]{2.5cm}{\centering{$8.1\pm 1.3$ eV$^{(6)}$}} &  \parbox[pt][1em][c]{2cm}{\centering{$153.1^{+29.2}_{-21.2}$ nm}} & \parbox[pt][1em][c]{2.5cm}{\cite{Sikorsky2020}} \\
\parbox[0pt][1.5em][c]{2.5cm}{\centering{$7.8\pm 0.8$ eV$^{(7)}$}} &  \parbox[pt][1em][c]{2cm}{\centering{$159.0^{+18.1}_{-14.8}$ nm}} & \parbox[pt][1em][c]{2.5cm}{\cite{Sikorsky2020}} \\
\noalign{\smallskip}\hline\noalign{\smallskip}
\end{tabular}
\end{footnotesize}
\begin{tablenotes}
\footnotesize $^{(1)}$This value is based on private communication with Helmer and Reich. $^{(2)}$The same value was already published earlier in \cite{Helmer1993}.  $^{(3)}$Re-evaluation of the measurement of \cite{Beck1,Beck2} using the branching ratio obtained in \cite{Masuda2019}. $^{(4)}$Based on the asymmetry of the 29.19 keV doublet. $^{(5)}$Based on the double-difference technique like in \cite{Beck1,Beck2}. $^{(6)}$Based on the difference between three $\gamma$ lines. $^{(7)}$Based on the difference to the 29.19~keV excitation energy like in \cite{Yamaguchi2019}.
\end{tablenotes}
\label{energyconstraints}
\end{center}
\end{table}
\begin{figure}[t]
 \begin{center}
 \includegraphics[width=9cm]{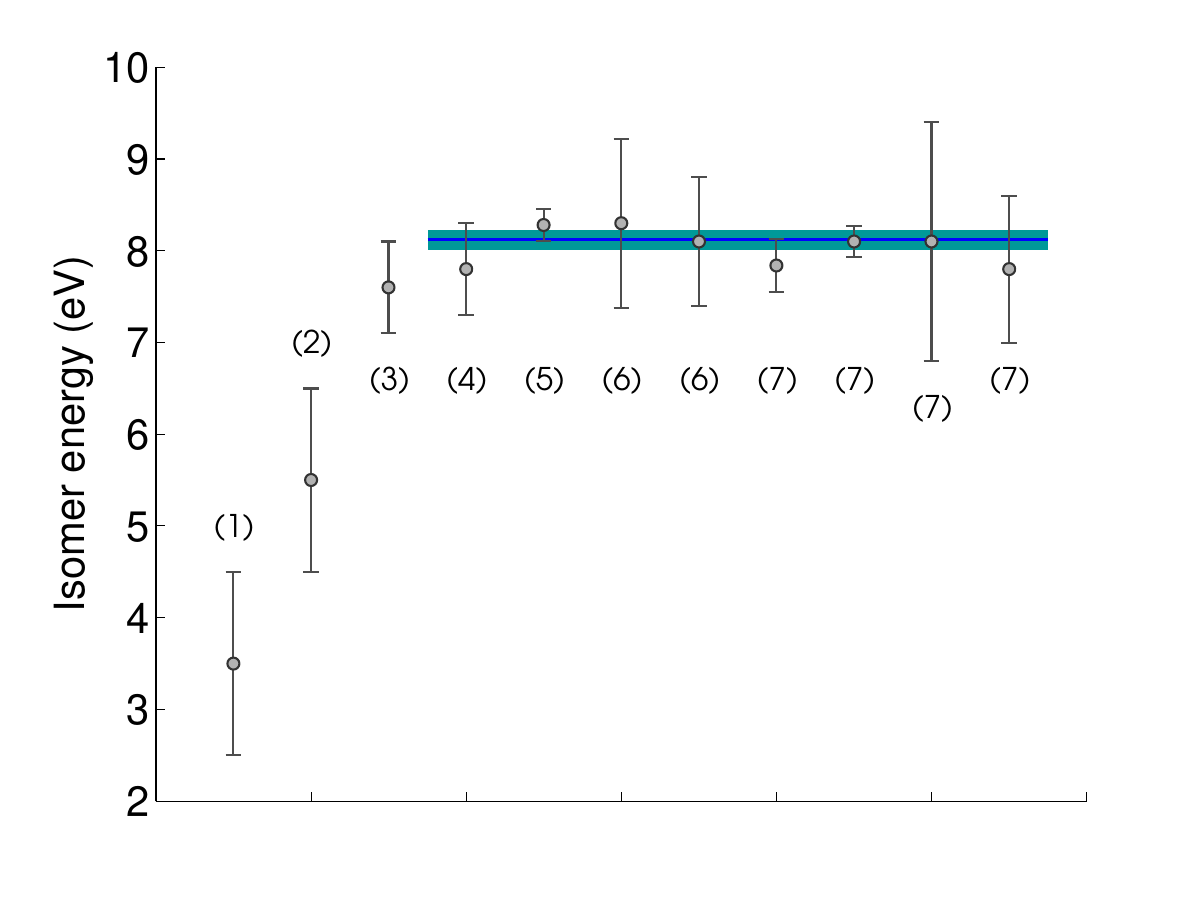}
  \caption{\footnotesize Selected isomeric energy measurements together with the weighted mean (blue, $8.12\pm0.11$~eV) of the eight most recently published values. The references are as follows: (1):\cite{Helmer_Reich}, (2):\cite{Filho}, (3):\cite{Beck1}, (4):\cite{Beck2}, (5):\cite{Seiferle2019b}, (6):\cite{Yamaguchi2019}, (7): \cite{Sikorsky2020}.}
 \label{energyplot}
 \end{center}
\end{figure}
Based on the above considerations, it is evident that $^{229\text{m}}$Th is the most promising candidate for the development of a nuclear clock using existing laser technology. In the following, a short review of $^{229\text{m}}$Th properties is given. A detailed experimental history of $^{229\text{m}}$Th is presented in Sec.~\ref{EXP}.\\[0.2cm]
$^{229}$Th was considered to possess a nuclear excited state below 100~eV in the early 1970s \cite{Kroger,Kroger_Reich}, when its existence was inferred from particular features of the $^{229}$Th level scheme as populated in the $\alpha$ decay of $^{233}$U, that remained otherwise unexplained. The excited state was inferred to possess spin, parity and Nilsson quantum numbers $3/2^+[631]$, whereas the nuclear ground-state has quantum numbers $5/2^+[633]$ (For the nuclear structure model, the interested reader is referred to \cite{Nilsson,Wense3}). The isomer-to-ground-state transition is thus of multipolarity {\it M1}. Based on $\gamma$-ray spectroscopy of nuclear states of higher energies, the $^{229\text{m}}$Th energy was constrained to be below 10~eV in 1990 \cite{Reich_Helmer} and an energy value of $3.5\pm1.0$~eV was determined in 1994 \cite{Helmer_Reich}. Interestingly, the existence of a nuclear excited state of such low energy seems to be a coincidence and there is currently no conclusive theoretical calculation that allows to predict nuclear levels to this precision. However, low-energy nuclear first excited states appear to be a peculiarity of the isotones with neutron number $139$ \cite{Kotthaus}.\\[0.2cm]
At the time of the nuclear clock proposal the parameters of $^{229\text{m}}$Th, in particular its energy, were not known to sufficient precision to allow for nuclear laser spectroscopy of individual thorium ions and thus the development of a nuclear clock. This fact triggered a multitude of experimental efforts to experimentally determine the excited state's energy and half-life. The detection of light emitted in the direct decay of $^{229\text{m}}$Th would significantly help to determine its energy to higher accuracy. However, to date all efforts have failed to observe an unambiguous signal of light emitted in the decay of the isomeric state (see Sec.~\ref{EXP.DIRECT} for a review). In 2007, an improved measurement led to a shift of the literature value for the isomer's energy from 3.5~eV to $7.6\pm0.5$~eV \cite{Beck1} (slightly corrected to $7.8\pm0.5$~eV in 2009 \cite{Beck2}). This partly explains the failure of earlier experiments to observe any direct $^{229\text{m}}$Th decay signal by an incorrect search range. However, all recent experiments have also failed to observe any unambiguous signal of light emitted in the isomer's direct decay, potentially pointing towards a strong non-radiative decay channel \cite{Jeet,Yamaguchi,Wense3,Stellmer2018a,Knize2019}. In 2012 \cite{Zhao} and again in 2018 \cite{Borisyuk2018b} the detection of light emitted in the decay of $^{229\text{m}}$Th was reported, but the observed signals are subject to controversial discussions within the community \cite{Peik3,Thirolf2019b}.\\[0.2cm]
The direct detection of electrons emitted in the isomer's internal conversion decay channel was achieved in 2016 \cite{Wense2}. In the internal conversion (IC) decay, the nucleus couples to the electronic shell, transferring its energy to a shell electron, which is subsequently ejected (see, e.g. \cite{Cohen1971} and App. \ref{THEO.HIGHDEC.IC}). Generally, for all nuclear excited states, IC poses a competing decay channel to the ordinary radiative decay. The ratio of IC compared to radiative decay defines the internal conversion coefficient, denoted by $\alpha_{ic}$, which is usually large for nuclear excited states of low energies. In 2019, spectroscopy of the IC electrons emitted in the isomer's direct decay constrained the transition energy to $8.28\pm0.17$~eV \cite{Seiferle2019b}.\\[0.2cm]
In the same year, the population of the isomer from the nuclear ground-state was achieved via a Lambda-excitation scheme involving the second nuclear excited state of $^{229}$Th at about 29.19~keV \cite{Masuda2019}. This allowed for a precision determination of the excitation energy of the 29.19~keV state. Based on this measurement, in combination with precision $\gamma$-ray spectroscopy, the isomeric energy was constrained to $8.30\pm0.92$~eV in \cite{Yamaguchi2019}. Although being of lower precision, this measurement provides an important consistency check. In \cite{Masuda2019}, the $\gamma$-decay branching ratio of the 29.19~keV state to the ground-state was re-measured and obtained as $1/(9.4\pm2.4)$. A re-evaluation of the earlier Beck et al. data from 2009 \cite{Beck1,Beck2} based on this new branching ratio was carried out in \cite{Yamaguchi2019}, resulting in an energy value of $8.1\pm0.7$ eV.\\[0.2cm]
In 2020 results of a $\gamma$-ray spectroscopy measurement performed at the Kirchhoff Institute of Physics in Heidelberg were published \cite{Geist2020,Sikorsky2020}. The experimental concept is comparable to the earlier measurements carried out by Beck et al. \cite{Beck1,Beck2}, however with three-fold improved detector resolution. An energy of $8.10\pm0.17$ eV was obtained \cite{Sikorsky2020}, which is of same precision like the value obtained from IC spectroscopy \cite{Seiferle2019b} and the values agree within a $2\sigma$ uncertainty interval (a preliminary value of $8.09^{+0.14}_{-0.19}$ eV was published in \cite{Geist2020}). In the same study, three further methods for the isomer's energy determination were applied: Due to the high detector resolution, the expected asymmetry of the 29.19~keV $\gamma$ line became observable. This was used for determining the energy to $7.84\pm0.29$ eV. By analysis of a further gamma-ray cascade, a value of $8.1\pm1.3$ eV was obtained. Finally, based on the same method already applied by Yamaguchi et al. \cite{Yamaguchi2019}, an energy of $7.8\pm0.8$ eV was determined. $^{229\text{m}}$Th energy values that can be found in literature are listed in Tab.~\ref{energyconstraints} and selected values are shown in Fig.~\ref{energyplot}. The weighted mean of the eight most recent measurements is obtained as $8.12\pm0.11$ eV and shown as a blue line. This recent progress determines the laser technology required for nuclear laser excitation. However, for precision laser spectroscopy of individual $^{229}$Th ions, a further energy constraint still appears to be advantageous (see Sec.~\ref{STEPS}).\\[0.2cm]
\begin{figure}[t]
 \begin{center}
 \includegraphics[width=6cm]{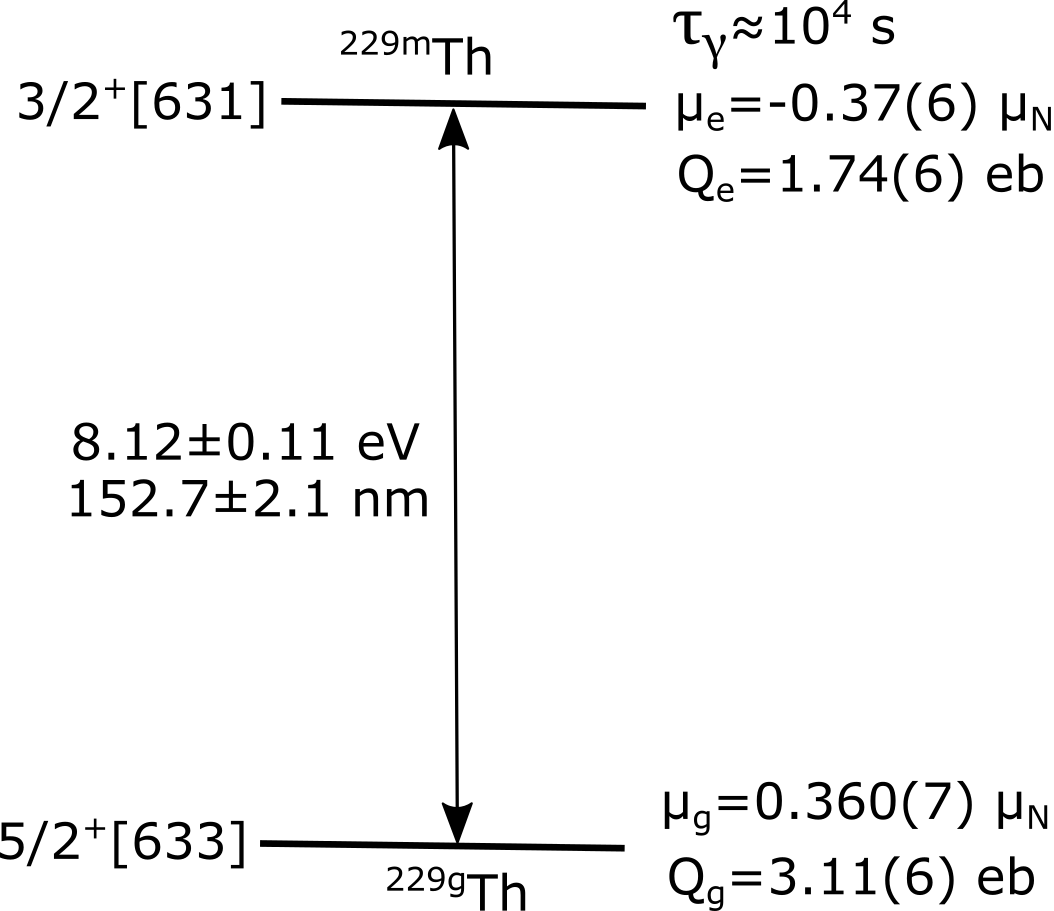}
 \vspace{-0.0cm}
  \caption{\footnotesize $^{229}$Th doublet consisting of ground and excited state together with the most important properties. The references for the parameters are listed in Tab.~\ref{stateparameters}.}
 \label{twolevel}
 \end{center}
\end{figure}
\begin{table}[t]
\begin{center}
\caption{\footnotesize Important parameters of the $^{229}$Th nuclear transition. All parameters except for the radiative lifetime have been experimentally determined.}
\begin{footnotesize}
\begin{tabular}{cccc}
\noalign{\smallskip}\hline\noalign{\smallskip}
 Parameter  & \parbox[pt][2em][c]{1cm}{\centering{Variable}}  & \parbox[pt][2em][c]{1cm}{\centering{Value}} & \parbox[pt][2em][c]{2.5cm}{\centering{Reference}}\\
\noalign{\smallskip}\hline\noalign{\smallskip}
\parbox[0pt][1.5em][c]{2cm}{Energy} &  \parbox[pt][1em][c]{1cm}{\centering{$E$}} & \parbox[pt][1em][c]{2cm}{\centering{$8.12\pm0.11$ eV}} & \parbox[pt][1em][c]{2.5cm}{\centering{weighted mean}}\\
\parbox[0pt][1.5em][c]{2cm}{Rad. Lifetime} &  \parbox[pt][1em][c]{1cm}{\centering{$\tau_\gamma$}} & \parbox[pt][1em][c]{2cm}{\centering{$\approx10^4$ s}} & \parbox[pt][1em][c]{2.5cm}{\centering{\cite{Minkov2019}}}\\
\parbox[0pt][1.5em][c]{2cm}{\centering{Exc.Magn.Dip.}} &  \parbox[pt][1em][c]{1cm}{\centering{$\mu_e$}} & \parbox[pt][1em][c]{2cm}{\centering{$-0.37(6)$ $\mu_N$}} & \parbox[pt][1em][c]{2.5cm}{\centering{\cite{Thielking2018}}}\\
\parbox[0pt][1.5em][c]{2cm}{\centering{Exc.Elec.Quad.}} &  \parbox[pt][1em][c]{1cm}{\centering{$Q_e$}} & \parbox[pt][1em][c]{2cm}{\centering{$1.74(6)$ eb}} & \parbox[pt][1em][c]{2.5cm}{\centering{\cite{Thielking2018}}}\\
\parbox[0pt][1.5em][c]{2cm}{\centering{Gnd.Magn.Dip.}} &  \parbox[pt][1em][c]{1cm}{\centering{$\mu_g$}} & \parbox[pt][1em][c]{2cm}{\centering{$0.360(7)$ $\mu_N$}} & \parbox[pt][1em][c]{2.5cm}{\centering{\cite{Safronova2}}}\\
\parbox[0pt][1.5em][c]{2cm}{\centering{Gnd.Elec.Quad.}} &  \parbox[pt][1em][c]{1cm}{\centering{$Q_g$}} & \parbox[pt][1em][c]{2cm}{\centering{$3.11(6)$ eb}} & \parbox[pt][1em][c]{2.5cm}{\centering{\cite{Safronova2}}}\\
\parbox[0pt][1.5em][c]{2cm}{\centering{IC coefficient}} &  \parbox[pt][1em][c]{1cm}{\centering{$\alpha_{ic}$}} & \parbox[pt][1em][c]{2cm}{\centering{$\approx10^9$$^*$ }} & \parbox[pt][1em][c]{2.5cm}{\centering{\cite{Seiferle3}}}\\
\noalign{\smallskip}\hline\noalign{\smallskip}
\end{tabular}
\begin{tablenotes}\footnotesize 
 $^*$For neutral surface-bound $^{229}$Th atoms.
\end{tablenotes}
\label{stateparameters}
\end{footnotesize}
\end{center}
\end{table}
\noindent As a peculiarity of the extremely low excitation energy, the lifetime of $^{229\text{m}}$Th strongly depends on the electronic environment of the nucleus (see, e.g., \cite{Karpeshin6}). In $^{229}$Th ions, the internal conversion decay channel is energetically forbidden, as the isomeric energy is below the energy that is required for further ionization of Th$^+$ ($11.9$~eV). This leads to a lifetime that may approach the radiative lifetime of $^{229\text{m}}$Th, for which no measurement exists, but which was predicted to be $1.2\cdot10^4$~s (assuming an energy of 8.1~eV) \cite{Minkov2019}. A detailed discussion of the isomer's radiative lifetime can be found in App.~\ref{THEO.DECEXC.GAMMADEC}. Experimentally, for $^{229\text{m}}$Th$^{2+}$ ions, an isomeric lifetime of longer than 1 minute was found \cite{Wense2}. Opposed to that, in neutral $^{229}$Th atoms the internal conversion decay channel is energetically allowed. The internal conversion coefficient for $^{229\text{m}}$Th was theoretically predicted to be $\alpha_{ic}\approx10^9$ \cite{Strizhov,Karpeshin6,Tkalya99}, leading to an isomeric lifetime which is reduced by 9 orders of magnitude to about 10 microseconds. A half-life of $7\pm1$~$\mu$s (corresponding to about $10$~$\mu$s lifetime) was indeed confirmed in 2017 for neutral, surface bound $^{229\text{m}}$Th atoms \cite{Seiferle3}.\\[0.2cm]
In 2018 a laser-spectroscopic characterization of the nuclear properties of $^{229\text{m}}$Th was performed \cite{Thielking2018}. In this experiment, laser spectroscopy of the $^{229}$Th atomic shell was conducted using a $^{229}$Th$^{2+}$ ion cloud with 2\% of the ions in the nuclear excited state. This allowed to probe for the hyperfine shift induced by the different nuclear spin states of the ground and the isomeric state. In this way, experimental values for the magnetic dipole and the electric quadrupole moment of $^{229\text{m}}$Th$^{2+}$ could be inferred, which were found to be $1.74(6)$~eb (spectroscopic quadrupole moment) and $-0.37(6)$~$\mu_N$ (magnetic dipole moment). In particular, the magnetic dipole moment differed by a factor of five from the previous theoretical prediction. A detailed analysis of the theoretically predicted and experimentally determined hyperfine structure confirmed this discrepancy \cite{Mueller2018}, which was finally resolved by an improved nuclear structure model including Coriolis mixing \cite{Minkov2019}. Further, the difference in the mean-square charge radii of the ground and excited state was inferred to 0.012(2) fm$^2$ \cite{Thielking2018}. This value is of high importance for the determination of the sensitivity of a nuclear clock for time variations of fundamental constants \cite{Berengut}. For this reason the experimental data were analyzed in more detail in \cite{Safronova2018b}, resulting in a value of 0.0105(13) fm$^2$. Direct laser cooling and trapping of $^{229}$Th$^{3+}$ ions, as required for the development of a single-ion nuclear clock, has already been accomplished in 2011 \cite{Campbell3}. This experiment also determined the electric quadrupole moment of the $^{229}$Th nuclear ground-state to be $3.11(16)$ eb and the magnetic dipole moment to 0.36~$\mu_N$ (see also \cite{Safronova2}). The main important properties of the $^{229\text{m}}$Th nuclear transition are sketched in Fig.~\ref{twolevel} and listed in Tab.~\ref{stateparameters}. 
\subsection{Different $^{229\text{m}}$Th-based nuclear optical clock concepts}\label{NOC.CON}
\begin{figure*}[t]
 \begin{center}
 \includegraphics[width=14cm]{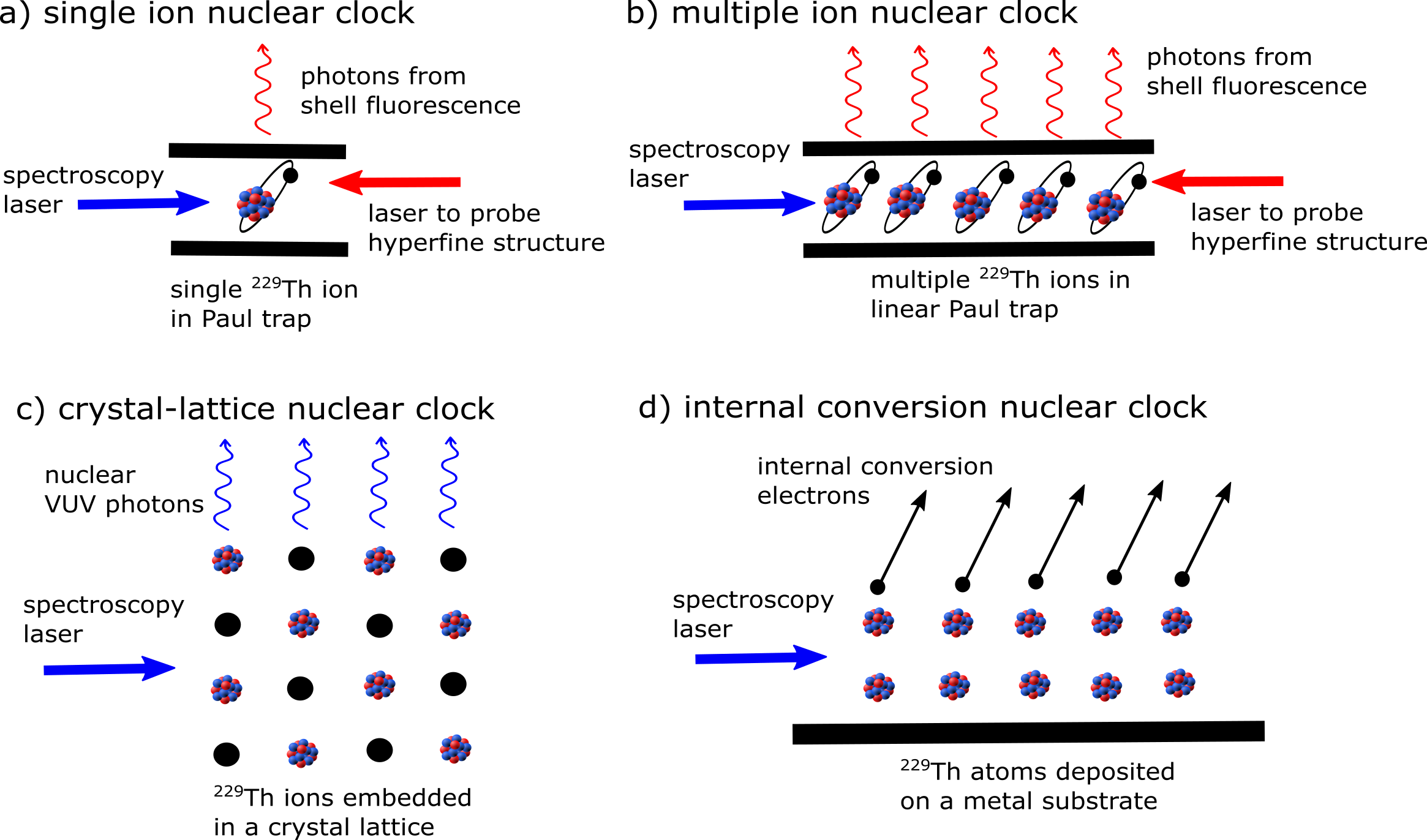}
  \caption{\footnotesize Different nuclear optical clock concepts: a) single-ion nuclear clock, b) multiple ion nuclear clock, c) crystal-lattice nuclear clock and d) internal-conversion nuclear clock. The expected accuracy and stability performances of the individual clock types are listed in Tab.~\ref{clockconcepts1}.}
 \label{clockconcepts}
 \end{center}
\end{figure*}
Four types of nuclear clock concepts can be distinguished as shown in Fig.~\ref{clockconcepts}. These are:\\[0.2cm]
\noindent \textbf{a) A single ion nuclear clock,} which uses a single trapped and laser-cooled ion for time measurement (Fig. \ref{clockconcepts} a). As the environmental conditions of individual trapped ions can be well controlled, such a clock is expected to provide the highest accuracy \cite{Peik,Campbell1}.\\[0.2cm]
\noindent \textbf{b) A multiple ion nuclear clock} based on multiple trapped $^{229}$Th ions (Fig. \ref{clockconcepts} b). Due to higher read-out statistics a multiple-ion nuclear clock would provide a better stability performance than the single ion clock.\\[0.2cm]
\noindent \textbf{c) A crystal lattice nuclear clock} that makes use of laser-based Mössbauer spectroscopy in a large band-gap material (Fig. \ref{clockconcepts} c). In this approach about $10^{14}$ $^{229}$Th$^{4+}$ ions, embedded in a crystal-lattice environment are irradiated in parallel leading to a superior stability performance. This, however, is expected to come at the cost of achievable accuracy, which is assumed to be limited by line-broadening and temperature uncertainty \cite{Peik,Rellergert,Kazakov1}.\\[0.2cm]
\noindent \textbf{d) An internal-conversion-based nuclear clock,} which is conceptually similar to the crystal-lattice nuclear clock, with the difference that the nuclear excitation is probed in the internal conversion (IC) instead of the radiative decay channel (Fig. \ref{clockconcepts} d). The performance was shown to be comparable to the crystal-lattice clock approach \cite{Wense2019c}.\\[0.2cm]
\noindent The different clock concepts are compared in Tab.~\ref{clockconcepts1}, where the most important parameters are listed. In the following, a detailed discussion is provided.
\begin{table*}[t]
\begin{center}
\caption{\footnotesize Comparison of different nuclear optical clock concepts. All values are order-of-magnitude estimates only. See text for explanation.}
\begin{footnotesize}
\begin{tabular}{ccccc}
\noalign{\smallskip}\hline\noalign{\smallskip}
 Concept  & \parbox[pt][2em][c]{2.5cm}{\centering{Accuracy}}  & \parbox[pt][2.5em][c]{2.5cm}{\centering{Stability ($\tau$ in s)}}  & \parbox[pt][2em][c]{2.5cm}{\centering{$\tau_\text{min}^*$}} & \parbox[pt][2em][c]{1.5cm}{\centering{References}} \\
\noalign{\smallskip}\hline\noalign{\smallskip}
\parbox[0pt][1em][c]{3cm}{Single ion} & \parbox[pt][1em][c]{2.5cm}{\centering{$\sim1\cdot10^{-19}$}} & \parbox[pt][1em][c]{3cm}{\centering{$\sim5\cdot10^{-16}/\sqrt{\tau}$}} & \parbox[pt][1em][c]{2.5cm}{\centering{ $\sim290$ days }} & \parbox[pt][1em][c]{2.5cm}{\cite{Peik} }\\
\parbox[0pt][1em][c]{3cm}{} & \parbox[pt][1em][c]{2.5cm}{\centering{}} & \parbox[pt][1em][c]{3cm}{\centering{}} & \parbox[pt][1em][c]{2.5cm}{\centering{  }} & \parbox[pt][1em][c]{2.5cm}{\cite{Campbell1} }\\
\parbox[0pt][1em][c]{3cm}{Multiple ion} & \parbox[pt][1em][c]{2.5cm}{\centering{$\sim1\cdot10^{-19}$ }} & \parbox[pt][1em][c]{2.5cm}{\centering{$\sim5\cdot10^{-17}/\sqrt{\tau}$}} &  \parbox[pt][1em][c]{2.5cm}{\centering{$\sim2.9$ days}} & \parbox[pt][1em][c]{2.5cm}{}\\
\parbox[0pt][1em][c]{3cm}{} & \parbox[pt][1em][c]{2.5cm}{\centering{}} & \parbox[pt][1em][c]{3cm}{\centering{}} & \parbox[pt][1em][c]{2.5cm}{\centering{  }} & \parbox[pt][1em][c]{2.5cm}{}\\
\parbox[0pt][1em][c]{3cm}{Crystal lattice} & \parbox[pt][1em][c]{2.5cm}{\centering{$\sim2\cdot10^{-16}$}} & \parbox[pt][1em][c]{3cm}{\centering{$\sim4.5\cdot10^{-13}/\sqrt{\tau^{3}}\ ^{**}$}} & \parbox[pt][1em][c]{2.5cm}{\centering{$\sim170$ s}}& \parbox[pt][1em][c]{2.7cm}{\cite{Rellergert}}\\
\parbox[0pt][1em][c]{3cm}{} & \parbox[pt][1em][c]{2.5cm}{\centering{}} & \parbox[pt][1em][c]{3cm}{\centering{}} & \parbox[pt][1em][c]{2.5cm}{\centering{ }} & \parbox[pt][1em][c]{2.5cm}{\cite{Kazakov1}}\\
\parbox[0pt][1em][c]{3cm}{Internal conversion} & \parbox[pt][1em][c]{2.5cm}{\centering{$\sim2\cdot10^{-16}$}} & \parbox[pt][1em][c]{3cm}{\centering{$\sim7\cdot10^{-15}/\sqrt{\tau}\ ^{***}$}} & \parbox[pt][1em][c]{2.5cm}{\centering{$\sim1200$ s}}& \parbox[pt][1em][c]{2.5cm}{\cite{Wense2019c}}\\
\noalign{\smallskip}\hline\noalign{\smallskip}
\end{tabular}
 \end{footnotesize}
\begin{tablenotes}\footnotesize 
$^{*}$Minimum averaging time required for the stability to surpass the accuracy. $^{**}$ Based on Eq.~(\ref{stability_crystal}) for $T_\text{int}=\tau$, as the interrogation time $T_\text{int}$ cannot exceed the averaging time $\tau$. $^{***}$ Based on Eq.~(\ref{stability_crystal}) for $T_\text{int}=10^{-5}$ s, corresponding to the IC-shortened isomeric lifetime.
\end{tablenotes}
\label{clockconcepts1}
\end{center}
\end{table*}
\subsubsection{The single (or multiple) ion nuclear clock}
\begin{figure}[t]
 \begin{center}
 \includegraphics[width=7cm]{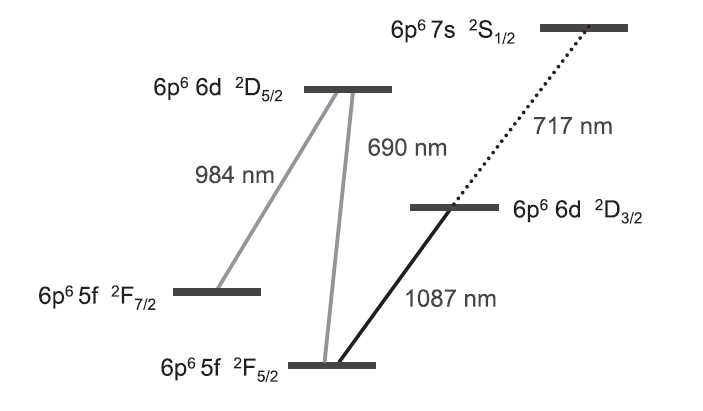}
  \caption{\footnotesize Electronic level configuration of Th$^{3+}$. A closed two-level as well as a closed three-level $\Lambda$ system are available for doppler- and low-temperture laser-cooling, respectively. It was proposed to excite the long-lived $7s^2S_{1/2}$ electronic shell state for nuclear clock operation in order to achieve a high accuracy of the clock. Reprinted from \cite{Peik} with kind permission of EDP Sciences.}
 \label{Peikconcept}
 \end{center}zjr 
\end{figure}

\begin{figure}[t]
 \begin{center}
 \includegraphics[width=7cm]{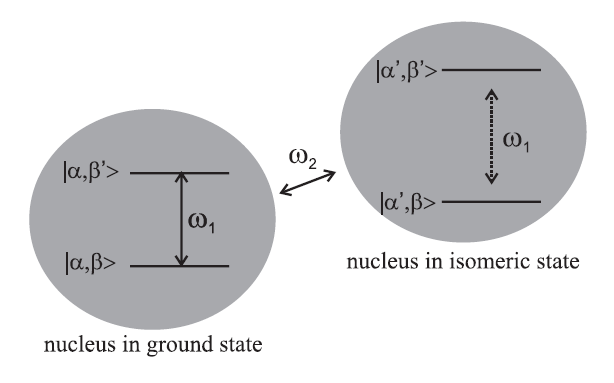}
  \caption{\footnotesize Conceptual sketch of the double-resonance method proposed for the secure identification of the excitation of the metastable nuclear state. The different spin states of nuclear ground ($I_g=5/2$) and excited ($I_e=3/2$) state lead to different hyperfinesplittings of the electronic shell states. A laser with angular frequency $\omega_1$ is tuned to an electronic resonance. Excitation of the nuclear state with a second laser ($\omega_2$) will cause $\omega_1$ to be out of resonance. Reprinted from \cite{Peik} with kind permission of EDP Sciences.}
 \label{doubleresonance}
 \end{center}
\end{figure}

\begin{figure}[t]
 \begin{center}
 \includegraphics[width=7cm]{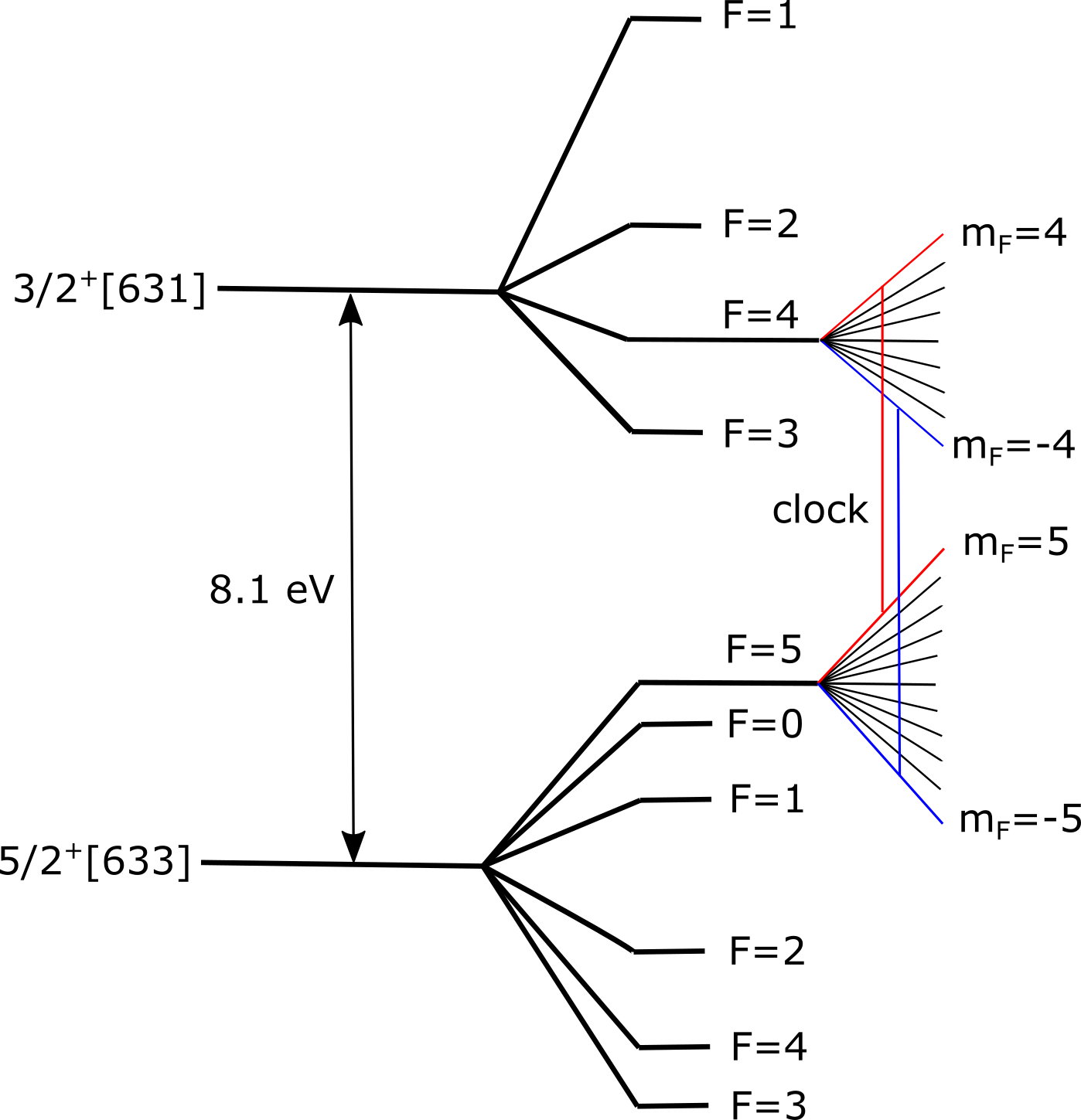}
  \caption{\footnotesize Stretched pair of nuclear hyperfine states proposed for the $^{229}$Th$^{3+}$ single-ion nuclear clock by Campbell et al. \cite{Campbell1}. Importantly, the electronic shell remains in the $5F_{5/2}$ electronic ground level. The line splitting of the isomeric state was corrected based on the experimentally determined magnetic dipole and electric quadrupole moment of $^{229\text{m}}$Th \cite{Thielking2018}. Reprinted from \cite{Campbell1} with kind permission of the American Physical Society.}
 \label{ClocklinesCampbell}
 \end{center}
\end{figure}

\paragraph*{The single ion nuclear clock}\ The first detailed proposal for the development of a nuclear optical clock based on individual $^{229}$Th ions was published by E. Peik and C. Tamm in 2003 \cite{Peik}. In their work, Peik and Tamm proposed to perform nuclear laser spectroscopy of $^{229}$Th$^{3+}$. The $3+$ charge state was chosen, as it possesses a simple electronic configuration, corresponding to a closed shell plus one valence electron, in this way allowing for direct laser cooling. Laser cooling of Th$^{3+}$ can be achieved in a two step process (see Fig.~\ref{Peikconcept}): In a first step the closed two-level system between the $5$F$_{5/2}$ electronic ground-state and the $6$D$_{3/2}$ excited state is used for Doppler cooling. The required laser wavelength is 1087~nm. In a second step, cooling to crystallization is achieved with a closed three-level $\Lambda$ system consisting of the $5$F$_{5/2}$, the $6$D$_{5/2}$ and the $5$F$_{7/2}$ states. The required wavelengths are 690~nm and 984~nm, respectively. Direct laser cooling of $^{229}$Th$^{3+}$ ions was experimentally achieved in 2011 \cite{Campbell3} (see also Sec. \ref{STEPS.COOL.DIRECT}).
\noindent An important advantage of a nuclear clock compared to usual optical atomic clocks is that the electronic shell can be chosen such that the combined quantum numbers of the system of shell plus nucleus are ``good" in a sense that the entire system provides lowest sensitivity to external perturbations \cite{Peik2}. In addition, due to the small nuclear moments\footnote{The magnetic dipole moments of the ground and excited state of $^{229}$Th were measured to be $\mu_g=0.360(7)\cdot\mu_N$ \cite{Safronova2} and $\mu_m\approx -0.37(6)\cdot\mu_N$ \cite{Thielking2018}, where $\mu_N$ is the nuclear magneton $\mu_N=5.05\cdot10^{-27}$ J/T, while typical atomic magnetic moments are in the order of the Bohr magneton $\mu_B=9.27\cdot10^{-24}$ J/T. The $^{229}$Th nuclear electric quadrupole moments of ground and excited state are $Q_g=3.11(6)$~eb \cite{Safronova2} and $Q_m=1.74(6)$~eb \cite{Thielking2018}, where the conventional unit of electron-barn is used (1~b = $10^{-28}$~m$^2$). Compared to that, the quadrupole moments of electronic states are in the range of up to hundreds of ea$_0^2$, where a$_0$ is the Bohr radius (1 a$_0^2$ = $2.80\cdot10^{-21}$~m$^2$).}, direct coupling of external electric fields to the nucleus is negligible. However, shell-nucleus coupling via the hyperfine interaction still has to be considered as a potential source of perturbations.\\[0.2cm]
\noindent One of the central ideas of the 2003 nuclear-clock concept is to populate an excited state of the Th$^{3+}$ electronic shell during clock operation. It was shown that the metastable $7s^2S_{1/2}$ shell state in Th$^{3+}$ with 1~s lifetime would be an appropriate choice. In this way, perturbing effects due to the linear Zeeman effect, the tensor part of the quadratic Stark effect and atomic quadrupole interactions could be avoided. Further, as no shifts can play a role which are entirely dependent on the electronic quantum numbers, no shifts from static electric fields, electromagnetic radiation or collisions have to be considered, leading to the proposal of a highly accurate nuclear clock. Only a small fractional black-body radiation shift of $10^{-19}$ at room temperature is expected due to the hyperfine Stark shift. The nuclear excitation with laser light was proposed to be probed using the hyperfine shift of the $^{229}$Th shell, as induced by the change of nuclear spin and magnetic moment during the excitation of the nuclear isomeric state. This method is known as the ``double-resonance method" (see Fig.~\ref{doubleresonance}) \cite{Peik}.
\noindent This pioneering nuclear clock concept, besides promising to lead to a nuclear optical clock of extraordinary high accuracy, has two disadvantages \cite{Campbell1}: One problem is that the second-order differential Zeeman shift is large ($\sim70$ kHz/mT$^2$). Further, the required choice of an electronic level leaves us with the metastable $7s^2S_{1/2}$ electronic state in $^{229}$Th$^{3+}$ as the only appropriate choice. This state, however, has an expected lifetime of only 1~s (compared to the up to $10^4$~s expected nuclear isomeric lifetime), which reduces the  quality-factor of the entire system.\\[0.2cm]
In 2012, an alternative approach for a nuclear clock based on $^{229\mathrm{m}}$Th was proposed by Campbell at al. \cite{Campbell1}. This proposal aims at a solution of the above mentioned problems of the earlier clock approach \cite{Peik}. In their work, Campbell et al. propose to use a pair of stretched nuclear hyperfine states for the clock transition, while $^{229}$Th$^{3+}$ remains in its $5F_{5/2}$ electronic ground state (see Fig.~\ref{ClocklinesCampbell}).
\begin{table}[t]
\begin{center}
\caption{\footnotesize Expected systematic shifts and uncertainties of a $^{229\mathrm{m}}$Th$^{3+}$ single-ion nuclear clock. Reprinted from \cite{Campbell1} with kind permission of the American Physical Society.}
\begin{footnotesize}
\begin{tabular}{ccc}
\noalign{\smallskip}\hline\noalign{\smallskip}
 Type of shift  & \parbox[pt][2em][c]{2cm}{\centering{Shift ($\times 10^{-20}$)}}  & \parbox[pt][2em][c]{2cm}{\centering{Uncertainty ($\times 10^{-20}$)}} \\
\noalign{\smallskip}\hline\noalign{\smallskip}
\parbox[0pt][1em][c]{3cm}{Excess micromotion} &  \parbox[pt][1em][c]{2cm}{\centering{10}} & \parbox[pt][1em][c]{2cm}{\centering{10}} \\
\parbox[0pt][1em][c]{3cm}{Gravitational} & \parbox[pt][1em][c]{2cm}{\centering{0}} & \parbox[pt][1em][c]{2cm}{\centering{10}} \\
\parbox[0pt][1em][c]{3cm}{Cooling laser Stark} & \parbox[pt][1em][c]{2cm}{\centering{0}} & \parbox[pt][1em][c]{2cm}{\centering{5}} \\
\parbox[0pt][1em][c]{3cm}{Electric quadrupole} & \parbox[pt][1em][c]{2cm}{\centering{3}} & \parbox[pt][1em][c]{2cm}{\centering{3}}  \\
\parbox[0pt][1em][c]{3cm}{Secular motion} & \parbox[pt][1em][c]{2cm}{\centering{5}} & \parbox[pt][1em][c]{2cm}{\centering{1}} \\
\parbox[0pt][1em][c]{3cm}{Linear Doppler} & \parbox[pt][1em][c]{2cm}{\centering{0}} & \parbox[pt][1em][c]{2cm}{\centering{1}} \\
\parbox[0pt][1em][c]{3cm}{Linear Zeeman} & \parbox[pt][1em][c]{2cm}{\centering{0}} & \parbox[pt][1em][c]{2cm}{\centering{1}} \\
\parbox[0pt][1em][c]{3cm}{Collisions} & \parbox[pt][1em][c]{2cm}{\centering{0}} &  \parbox[pt][1em][c]{2cm}{\centering{1}}\\
\parbox[0pt][1em][c]{3cm}{Blackbody radiation} & \parbox[pt][1em][c]{2cm}{\centering{0.013}} & \parbox[pt][1em][c]{2cm}{\centering{0.013}} \\
\parbox[0pt][1em][c]{3cm}{Clock laser Stark} & \parbox[pt][1em][c]{2cm}{\centering{0}} & \parbox[pt][1em][c]{2cm}{\centering{$\ll$ 0.01}} \\
\parbox[0pt][1em][c]{3cm}{Trapping field Stark} & \parbox[pt][1em][c]{2cm}{\centering{0}} & \parbox[pt][1em][c]{2cm}{\centering{ $\ll$ 0.01}} \\
\parbox[0pt][1em][c]{3cm}{Quadratic Zeeman} & \parbox[pt][1em][c]{2cm}{\centering{0}} & \parbox[pt][1em][c]{2cm}{\centering{0}} \\
\parbox[0pt][2em][c]{3cm}{Total} & \parbox[pt][2em][c]{2cm}{\centering{18}} & \parbox[pt][2em][c]{2cm}{\centering{15}} \\
\noalign{\smallskip}\hline\noalign{\smallskip}
\end{tabular}
\label{nuclclockshifts}
\end{footnotesize}
\end{center}
\end{table}
\noindent By a detailed analysis, partly based on numerical simulations, they were able to show that such a nuclear clock has the potential to approach a systematic frequency uncertainty of $\sim1\cdot10^{-19}$, superior to existing atomic clock technology. The expected systematic error budget of this clock is shown in Tab.~\ref{nuclclockshifts}, as taken from \cite{Campbell1}. A very detailed discussion of nuclear clock performance as well as a quantitative evaluation of clock shifts can be found in \cite{Campbell4}.\\[0.2cm]
\noindent There is, however, a remaining challenge of the single-ion nuclear clock approach: the clock stability will be limited by the coherence time of the laser light used for irradiation. In the deep VUV at about 150~nm one cannot expect to achieve a laser bandwidth significantly below 1~Hz in the near future. For this reason the quality factor of the clock has to be expected to be reduced to $Q=\omega/\Delta\omega_L\approx 2\cdot10^{15}$, resulting in a quantum-projection-noise (QPN) limited stability performance calculated based on Eq.~(\ref{Allan1}) of $\sigma_y=5\cdot10^{-16}/\sqrt{\tau/\text{s}}$ ($\omega=1.3\cdot10^{16}$, $T=1/(2\pi)$~s, $N=1$), comparable to those of single-ion optical atomic clocks (see Tab.~\ref{clockshifts}). In order to make use of the low expected systematic frequency uncertainty, the stability has to be brought to a level corresponding to the accuracy, which will require long averaging times on the order of 290 days.\\[0.2cm]
\paragraph*{The multiple ion nuclear clock}\ A complementary approach, which could lead to a better stability performance, would be to use multiple $^{229}$Th$^{3+}$ ions, like a linear Coulomb crystal \cite{Campbell3}. This concept is currently also being investigated for atomic clocks \cite{Herschbach2012,Pyka2014,Keller2019}. Importantly, systematic frequency uncertainties in the $10^{-19}$ range for a linear Coulomb crystal appear to be feasible \cite{Keller2019b}. If $100$ laser-cooled $^{229}$Th$^{3+}$ ions would be irradiated in parallel, the clock stability would improve by a factor of $10$, resulting in $\sigma_y\approx5\cdot10^{-17}/\sqrt{\tau/\text{s}}$ ($\omega=1.3\cdot10^{16}$, $T=1/(2\pi)$~s, $N=100$). In this case, the stability would approach the systematic frequency uncertainty after the relatively short averaging time of 2.9 days. An even more radical approach would be to embed a huge amount ($\approx10^{14}$ per mm$^3$) of $^{229}$Th ions into a crystal lattice structure \cite{Peik,Rellergert,Kazakov1} or to deposit a thin layer of $^{229}$Th atoms on a surface \cite{Wense2019c}. This concept is known as the ``solid-state nuclear clock" and will be discussed in the following section.\\[0.2cm]
While there is no conceptual hindrance for the development of a single or multiple ion nuclear optical clock, the reason that it has not yet been built is that the $^{229}$Th isomeric energy value has not been constrained to sufficient precision to allow for narrow-band laser excitation of the nucleus in a Paul trap. For this reason, there are ongoing worldwide efforts to further constrain the $^{229\text{m}}$Th energy value. A detailed review is given in Sec.~\ref{EXP}. Recent reviews on the experimental status of the nuclear-clock development can also be found in \cite{Peik4,Wense2018,Thirolf2019b}.
\subsubsection{The solid-state nuclear clock}\label{NOC.CON.SOLID}
The idea of using the nucleus instead of the atomic shell for time measurement makes it conceptually possible to embed a large number of $^{229}$Th nuclei with a high density into a solid-state environment and laser-spectroscopically excite the nuclear transition. This concept is known as the solid-state nuclear clock. Although predicted to achieve a systematic frequency uncertainty of $2\cdot10^{-16}$ \cite{Rellergert}, inferior compared to optical atomic clocks, such a solid-state optical clock is expected to have a wide range of practical applications, due to expected advantages in terms of stability, mechanical robustness and power consumption.\\[0.2cm]
Two different approaches for a solid-state nuclear clock can be distinguished: the first approach, known as the crystal-lattice clock, makes use of a suppression of the isomer's IC decay channel in large band-gap material \cite{Tkalya2b}. This leads to a lifetime prolongation and a corresponding narrowing of the nuclear transition linewidth, while in parallel allowing for an improved stability performance due to the large number of irradiated nuclei \cite{Peik,Rellergert,Kazakov1}. The nuclear excitation is probed via the direct detection of photons emitted in the nuclear deexcitation.\\[0.2cm]
In the second approach, a large number of neutral $^{229}$Th atoms is deposited as a thin layer on a surface and nuclear excitation is monitored in the IC decay channel. The concept was introduced as the internal-conversion (IC)-based nuclear clock. Although in this case the isomeric lifetime is significantly shortened, both approaches were shown to offer the potential for comparable performance \cite{Wense2019c}. In the following, both concepts will be evaluated.\\[0.2cm]
The idea for a solid-state nuclear clock was reported together with the single-ion nuclear clock in 2003 \cite{Peik}. The proposal makes important use of an earlier idea to directly excite the nucleus by laser light in a dielectric medium \cite{Tkalya2b}. A detailed analysis of the concept was carried out in 2010 by Rellergert et al. \cite{Rellergert}. A long-term accuracy of $2\cdot10^{-16}$ was predicted, limited by line-broadening effects that arise due to spin fluctuations, and temperature-induced frequency shifts. A 1~mK temperature difference results in a fractional shift of the frequency by $6\cdot10^{-15}$. State of the art thermometer accuracies are on the order of $0.04$~mK, resulting in a systematic frequency uncertainty of $2.4\cdot10^{-16}$ \cite{Jeet2018}.\\[0.2cm]
An analysis of the expected stability performance of a solid-state nuclear clock was presented by G. Kazakov et al. in 2012 \cite{Kazakov1}. Due to the radically different approach compared to that of trap-based optical clocks, the stability will not be limited by the quantum-projection noise and Eq.~(\ref{Allan1}) cannot be used for the stability estimate. Instead it was shown, that the stability will be Shot-noise limited and can be estimated under the assumptions of short interrogation times ($T_\text{int}\le\Gamma^{-1}$, with $\Gamma$ being the total nuclear decay rate) and low laser intensities by \cite{Kazakov1}\footnote{The equation is obtained by inserting Eq.~(37) of \cite{Kazakov1} into Eq.~(48) of the same paper and using that by definition $t=4T_\text{int}$, where $t$ denotes the time for a complete interrogation cycle used in \cite{Kazakov1}.}
\begin{equation}
\label{stability_crystal}
\sigma_y\approx\frac{1}{\omega T_\text{int}}\frac{\tilde{\Gamma}}{\Gamma}\frac{1}{\sqrt{\Gamma_\text{exc}N_\text{eff}\tau}}.
\end{equation}
Here $T_\text{int}$ denotes the interrogation time (not the coherence time, as the clock does not operate in a coherent regime) and $\tilde{\Gamma}$ denotes the relaxation rate of the optical coherences in the solid-state environment. $\Gamma=(1+\alpha_{ic})\Gamma_\gamma$ (with the internal conversion coefficient $\alpha_{ic}$ and the radiative decay rate $\Gamma_\gamma$) corresponds to the total decay rate of the nuclear excitation, including potential non-radiative decay channels. $\Gamma_\text{exc}$ is the nuclear excitation rate during laser irradiation and $N_\text{eff}$ denotes the effective number of irradiated nuclei defined as $N_\text{eff}\approx\Gamma_\gamma/\Gamma \cdot k\Omega/4\pi\cdot N$, with $k$ the quantum efficiency of the detector and $\Omega$ the effective solid angle covered by the detector. All other variables are defined as previously. Note, that based on Eq.~(\ref{stability_crystal}) in order to approach a small instability, one should use the maximum permitted interrogation time $T_\text{int}$, however the averaging time $\tau$ can never be chosen shorter than $T_\text{int}$. Therefore, in combination with the previous condition $T_\text{int}\le\Gamma^{-1}$, one obtains for $T_\text{int}$ the condition $\text{max}(T_\text{int})=\text{min}(\Gamma^{-1},\tau)$. The nuclear excitation rate can be estimated as \cite{Wense2019d}
\begin{equation}
\Gamma_\text{exc}\approx\frac{\pi c^2 I \Gamma_\gamma}{\hbar\omega^3\tilde{\Gamma}},
\end{equation}
with $I$ being the laser intensity used for nuclear excitation. In \cite{Kazakov1} it was estimated, that the decay rate of the coherences will be on the order of $\tilde{\Gamma}\approx 1\cdot10^3$~1/s due to coupling to the solid-state environment.\\[0.2cm]
\paragraph*{Stability performance of a crystal-lattice clock}
\ The crystal-lattice nuclear clock approach makes important use of a suppression of the internal-conversion decay of the isomeric state by choosing large-band-gap crystals like, e.g., MgF$_2$, CaF$_2$ or LiSrAlF$_6$ as host materials for the $^{229}$Th ions, which possess a $4+$ charge state due to their position in the crystal lattice \cite{Jackson,Rellergert2,Hehlen,Dessovic,Ellis,Pimon}. In this way, the radiative decay of $^{229\text{m}}$Th will become observable and no shortening of the isomer's lifetime due to internal conversion occurs, leading to a reduction of the transition linewidth and thus to a higher sensitivity for any laser detuning. For this reason the total decay rate equals the radiative decay rate of the transition $\Gamma=\Gamma_\gamma=10^{-4}$~Hz. An enhancement of the nuclear transition rate in a crystal-lattice environment proportional to $n^3$, with $n$ being the refractive index of the host material, has been predicted \cite{Tkalya8,Jeet}. For a typical value of $n\approx1.5$, a lifetime shortening by a factor of $3.4$ can be expected. As in the following, only order-of-magnitude estimates are made, this factor will be neglected. Under the assumption that a single tooth of a 150~nm VUV frequency comb with 5 nW of power per comb tooth and less than 1~kHz bandwidth is focused to a spot size of 300~$\mu$m diameter for nuclear excitation, the laser intensity amounts to $I=7\cdot10^{-6}$ W/cm$^2$. This leads to a nuclear excitation rate of  $\Gamma_\text{exc}\approx1\cdot10^{-5}$~1/s. Note, that about 5~nW of out-coupled laser power per comb tooth in the XUV around 97~nm was already achieved via cavity-enhanced high-harmonic generation (HHG) \cite{Zhang2020}. Obtaining a bandwidth of less than 1~kHz is still subject to investigation, but appears to be doable \cite{Benko,Benko2016}. Importantly, for the crystal-lattice clock approach, no narrowing of the bandwidth to significantly below 1~kHz appears to be advantageous, as the nuclear linewidth is assumed to be broadened to about this value. A $^{229}$Th-doped crystal with a density of $4.1\cdot10^{14}$ per mm$^3$ was already grown \cite{Jeet}. Considering the laser spot of 0.3 mm diameter and a crystal depth of 1~mm, about $N=2.9\cdot10^{13}$ nuclei are irradiated. If a detector that covers one tenth of the solid angle $\Omega=4\pi/10$ and has a quantum efficiency of $k=0.1$ is used for photon detection, based on Eq.~(\ref{stability_crystal}) the stability of the crystal lattice nuclear clock is estimated to be $\sigma_y=4.5\cdot10^{-13}/\sqrt{(\tau/\text{s})^3}$, where $T_\text{int}=\tau$ was used, as the interrogation time cannot exceed the averaging time. Thus, after about $\tau=170$~s of averaging the instability approaches the expected systematic frequency uncertainty of $2\cdot10^{-16}$.\\[0.2cm]
\paragraph*{Stability performance of an IC-based nuclear clock}
\ The second type of solid-state nuclear clock would make use of the internal-conversion (IC) as opposed to the radiative decay channel of $^{229\text{m}}$Th \cite{Wense2019c}. Here a thin layer of $^{229}$Th atoms on a surface is laser excited and the IC electrons emitted in the isomeric decay are probed. At first glance, this approach seems to be disadvantageous compared to the crystal-lattice clock, as the linewidth of the nuclear transition is broadened to about 16~kHz due to the short IC lifetime of about 10~$\mu$s \cite{Seiferle3}. However, as the IC decay occurs on a considerably shorter time-scale than the radiative decay, the read-out times would be shorter, which partly compensates for the line-broadening. The expected stability performance can again be estimated based on Eq.~(\ref{stability_crystal}). For this purpose it is assumed, that the optical decoherence rate is dominated by the short IC lifetime of $^{229\text{m}}$Th, such that $\tilde{\Gamma}=\Gamma/2\approx5\cdot10^4$~Hz holds. Here $\Gamma$ denotes the total decay rate of the nuclear excited state, including non-radiative decay channels: $\Gamma=\Gamma_\gamma+\Gamma_\text{ic}$. As $\Gamma_\text{ic}\gg\Gamma_\gamma$, one has $\Gamma\approx\Gamma_\text{ic}\approx 1\cdot10^5$~Hz. The electrons emitted in the IC decay can be guided towards the detector by electric fields leading to the detection of most of the emitted electrons. The effective number of nuclei can be estimated as $N_\text{eff}=k\cdot N$, where $k$ has to take account for the detection efficiency and electron losses in the target surface. Assuming the irradiation of a round $^{229}$ThO$_2$ surface of 0.3~mm diameter and 10 nm thickness, about $1.6\cdot10^{13}$ nuclei are irradiated. For an interrogation time of $T_\text{int}=10$~$\mu$s an instability of $\sigma_y\approx7\cdot10^{-15}/\sqrt{\tau/\text{s}}$ is obtained. Here, as previously $I=7\cdot10^{-6}$~W/cm$^2$ and $k=0.1$ were chosen as input parameters.\\[0.2cm]
While obviously showing an inferior stability performance compared to the crystal-lattice nuclear clock approach, the instability is sufficiently small to approach a level below the expected systematic frequency uncertainty of $2\cdot10^{-16}$ after an averaging time of $\tau\approx1200$~s. There is no reason for reducing the instability to a value significantly below the systematic frequency uncertainty, as this will not improve the clock performance. It might be seen as an advantage of the IC-based solid-state clock approach that no suppression of the non-radiative decay channels is required.
\section{Steps towards a nuclear clock}\label{STEPS}
\begin{table*}[t]
\begin{center}
\caption{\footnotesize Comparison of different experiments aiming for a precise $^{229\text{m}}$Th energy determination.}
\begin{footnotesize}
\begin{tabular}{cccccc}
\noalign{\smallskip}\hline\noalign{\smallskip}
 Concept  & \parbox[pt][2em][c]{2.5cm}{\centering{Exc. channel}}  & \parbox[pt][2.5em][c]{2.5cm}{\centering{Dec. channel}}  & \parbox[pt][2em][c]{2.5cm}{\centering{Achievable uncertainty }} & \parbox[pt][2em][c]{1.5cm}{\centering{Reference}} & \parbox[pt][2em][c]{1.5cm}{\centering{Section}} \\
\noalign{\smallskip}\hline\noalign{\smallskip}
\parbox[0pt][1em][c]{3cm}{$\gamma$-ray spectroscopy} & \parbox[pt][1em][c]{2.5cm}{\centering{$^{233}$U $\alpha$ decay}} & \parbox[pt][1em][c]{1.5cm}{\centering{indirect}} & \parbox[pt][1em][c]{2.5cm}{ $\approx40$ meV } & \parbox[pt][1em][c]{2.5cm}{\cite{Kazakov3} } & \parbox[pt][1em][c]{2.5cm}{\centering{Sec. 5.1.4}}\\
\parbox[0pt][1em][c]{3cm}{$^{233}$U in crystal} & \parbox[pt][1em][c]{2.5cm}{\centering{$^{233}$U $\alpha$ decay}} & \parbox[pt][1em][c]{1.5cm}{\centering{rad. decay}} & \parbox[pt][1em][c]{2.5cm}{$\approx 5$ meV$^{**}$}& \parbox[pt][1em][c]{2.5cm}{\cite{Stellmer4}}& \parbox[pt][1em][c]{2.5cm}{\centering{Sec. 5.2.1} }\\
\parbox[0pt][1em][c]{3cm}{$^{229}$Ac in crystal} & \parbox[pt][1em][c]{2.5cm}{\centering{$^{229}$Ac $\beta$ decay}} & \parbox[pt][1em][c]{1.5cm}{\centering{rad. decay}} & \parbox[pt][1em][c]{2.5cm}{$\approx 5$ meV$^{**}$}& \parbox[pt][1em][c]{2.5cm}{\cite{Verlinde2019}}& \parbox[pt][1em][c]{2.5cm}{\centering{Sec. 5.2.1} }\\
\parbox[0pt][1em][c]{3cm}{Laser in crystal} & \parbox[pt][1em][c]{2.5cm}{\centering{direct photon}} & \parbox[pt][1em][c]{1.5cm}{\centering{rad. decay}} & \parbox[pt][1em][c]{2.5cm}{$\approx 40$ $\mu$eV$^{*}$}& \parbox[pt][1em][c]{2.5cm}{\cite{Jeet2018}}& \parbox[pt][1em][c]{2.5cm}{\centering{Sec. 5.2.2} }\\
\parbox[0pt][1em][c]{3cm}{Electronic bridge} & \parbox[pt][1em][c]{2.5cm}{\centering{electronic bridge}} & \parbox[pt][1em][c]{2.5cm}{\centering{rad. \& indirect}} &  \parbox[pt][1em][c]{2.5cm}{$\approx40$ $\mu$eV$^{*}$} & \parbox[pt][1em][c]{2.5cm}{\cite{Porsev3}} & \parbox[pt][1em][c]{2.5cm}{\centering{Sec. 5.2.3 \& 5.4.3} }\\
\parbox[0pt][1em][c]{3cm}{$^{229}$Th in SiO$_2$} & \parbox[pt][1em][c]{2.5cm}{\centering{plasma excitation}} & \parbox[pt][1em][c]{1.5cm}{\centering{rad. decay}} & \parbox[pt][1em][c]{2.5cm}{$\approx5$ meV$^{**}$}& \parbox[pt][1em][c]{2.5cm}{\cite{Borisyuk2018a}}& \parbox[pt][1em][c]{2.5cm}{\centering{Sec. 5.2.4} }\\
\parbox[0pt][1em][c]{3cm}{Storage ring} & \parbox[pt][1em][c]{2.5cm}{\centering{NEEC}} & \parbox[pt][1em][c]{1.5cm}{\centering{indirect}} & \parbox[pt][1em][c]{2.5cm}{$\approx 1$ meV}& \parbox[pt][1em][c]{2.5cm}{\cite{Ma}} & \parbox[pt][1em][c]{2.5cm}{\centering{Sec. 5.2.4} }\\
\parbox[0pt][1em][c]{3cm}{via 29~keV state} & \parbox[pt][1em][c]{2.5cm}{\centering{$\Lambda$ excitation}} & \parbox[pt][1em][c]{1.5cm}{\centering{rad. decay}} & \parbox[pt][1em][c]{2.5cm}{$\approx 5$ meV$^{**}$}& \parbox[pt][1em][c]{2.5cm}{\cite{Masuda2019}}& \parbox[pt][1em][c]{2.5cm}{\centering{Sec. 5.2.5} }\\
\parbox[0pt][1em][c]{3cm}{Tunnel Junction} & \parbox[pt][1em][c]{2.5cm}{\centering{$^{233}$U $\alpha$ decay}} & \parbox[pt][1em][c]{1.5cm}{\centering{IC decay}} & \parbox[pt][1em][c]{2.5cm}{$\approx10$ meV}& \parbox[pt][1em][c]{2.5cm}{\cite{Ponce2018b}} & \parbox[pt][1em][c]{2.5cm}{\centering{Sec. 5.3.4} }\\
\parbox[0pt][1em][c]{3cm}{IC from exc. el. states} & \parbox[pt][1em][c]{2.5cm}{\centering{$^{233}$U $\alpha$ decay}} & \parbox[pt][1em][c]{1.5cm}{\centering{IC decay}} & \parbox[pt][1em][c]{2.5cm}{$\approx10$ $\mu$eV$^{*}$}& \parbox[pt][1em][c]{2.5cm}{\cite{Bilous2}} & \parbox[pt][1em][c]{2.5cm}{\centering{Sec. 5.3.4} }\\
\parbox[0pt][1em][c]{3cm}{Laser on surface} & \parbox[pt][1em][c]{2.5cm}{\centering{direct photon}} & \parbox[pt][1em][c]{1.5cm}{\centering{IC decay}} & \parbox[pt][1em][c]{2.5cm}{$\approx40$ $\mu$eV$^{*}$}& \parbox[pt][1em][c]{2.5cm}{\cite{Wense6}} & \parbox[pt][1em][c]{2.5cm}{\centering{Sec. 5.3.4} }\\
\parbox[0pt][1em][c]{3cm}{Highly charged ion} & \parbox[pt][1em][c]{2.5cm}{\centering{electronic bridge}} & \parbox[pt][1em][c]{1.5cm}{\centering{IC decay}} & \parbox[pt][1em][c]{2.5cm}{$\approx40$ $\mu$eV$^{*}$}& \parbox[pt][1em][c]{2.5cm}{\cite{Bilous2020}} & \parbox[pt][1em][c]{2.5cm}{\centering{Sec. 5.3.4} }\\
\noalign{\smallskip}\hline\noalign{\smallskip}
\end{tabular}
 \end{footnotesize}
\begin{tablenotes}\footnotesize 
$^{*}$assuming 10~GHz laser bandwidth, $^{**}$assuming 0.1~nm spectrometer resolution.
\end{tablenotes}
\label{energyconcepts}
\end{center}
\end{table*}
\noindent Several steps have to be taken on the road towards a nuclear clock. Here it is important to distinguish between the different types of nuclear clocks as introduced in Sec.~\ref{NOC.CON}, as each clock has its own requirements. In the following, we will focus on the steps that are required for the development of a single (or multiple) ion nuclear clock. The requirements for the development of a solid-state nuclear clock (crystal lattice or internal conversion) will be addressed in Sec. \ref{STEPS.SOLID}.\\[0.2cm]
Due to its high expected accuracy, the development of a single or multiple ion nuclear clock can be considered as the ultimate goal. However, it is also the most experimentally demanding, as it will require driving nuclear Rabi oscillations of individual laser-cooled $^{229}$Th$^{3+}$ ions in a Paul trap. Such Rabi oscillations are necessary for the Ramsey interrogation scheme for clock operation, which improves the clock's stability to the QPN limit \cite{Ludlow}.\\[0.2cm]
Driving nuclear Rabi oscillations requires that the nuclear transition energy is pinned down into the Hz-range, which would allow the laser used for excitation exactly to be tuned onto the nuclear resonance. However, the currently available 1-$\sigma$ energy uncertainty interval spans 0.34~eV, which corresponds to about 100~THz. The precision of the nuclear transition energy has therefore to be improved by $14$ orders of magnitude before a single or multiple ion nuclear clock will become reality.\\[0.2cm]
Certainly, a $14$ orders of magnitude improvement in energy precision will not be achieved in one single step. Instead, several steps making use of different technological tools will have to be employed. A VUV frequency comb would allow bridging the 10 orders of magnitude energy difference between its total optical bandwidth (about 10~THz) and the bandwidth of an individual comb mode (below 1~kHz). During the search for the nuclear transition, only the comb mode spacing of about 100~MHz would have to be bridged, which would require $\approx10^5$ scan steps (assuming 1~kHz mode bandwidth). For 100 irradiated $^{229}$Th$^{3+}$ ions an irradiation time of 10~s per scan step might be sufficient for a secure detection of the isomer's excitation \cite{Wense2019c}. This results in a reasonably short scanning time of about $10^6$~s. As soon as the energy has been constrained to about 1~kHz, a narrow-band laser could be used to further pin down the energy into the Hz range. The same laser could then also be used to drive the transition in the clock concept.\\[0.2cm]
There is, however, a central challenge that has to be overcome before the above searching method can start: the isomeric energy has to be constrained to an uncertainty that corresponds to the total bandwidth of the frequency comb used for excitation (10~THz corresponds to about 40~meV). This is an improvement of about a factor of ten compared to the currently available energy uncertainty of 0.34~eV \cite{Seiferle2019b,Sikorsky2020}. Importantly, any improvement of the isomer's energy uncertainty to below the total bandwidth of the frequency comb will not shorten the required times for laser-based scanning, as the mode-spacing of the comb will always have to be bridged when searching for the nuclear excitation with all comb-modes in parallel. Constraining the isomer's transition energy by 14 orders of magnitude can therefore be achieved in three steps:\\[0.2cm]
{\bf Step 1}: Reduce the energy uncertainty by a factor of about 10 to allow for step 2. Various approaches that are under investigation to achieve this goal are listed in Tab.~\ref{energyconcepts}.\\[0.2cm]
{\bf Step 2}: Reduce the energy uncertainty by further ten orders of magnitude with direct frequency-comb spectroscopy.\\[0.2cm]
{\bf Step 3}: Use a narrow-band laser to ultimately pin down the energy into the Hz range. The same laser could also be used as a driving laser in the nuclear clock.\\[0.2cm] 
Many experiments are currently aiming to achieve the first goal, namely the improvement of our knowledge about the isomer's energy value by a factor of 10. These experiments will be detailed in Sec. \ref{EXP}. A tabulated overview of the main important approaches is shown in Tab.~\ref{energyconcepts}. The following parts of this section will focus on other important steps: several ways to generate $^{229}$Th$^{3+}$ ions are discussed in Sec.~\ref{STEPS.IONGEN}. Paul trapping of $^{229}$Th will be discussed in Sec.~\ref{STEPS.TRAP}. The laser cooling of $^{229}$Th$^{3+}$ ions in a Paul trap has already been achieved and will be presented in Sec.~\ref{STEPS.COOL}. Different concepts for laser systems that could be used as clock lasers will be presented in Sec.~\ref{STEPS.LASER}. Finally, in Sec.~\ref{STEPS.SOLID}, the requirements for the development of a solid-state nuclear clock will be discussed.
\subsection{Generation of $^{229}$Th$^{3+}$ ions} \label{STEPS.IONGEN}
A central requirement for the development of a single or multiple ion nuclear clock is the generation of $^{229}$Th ions, preferably in the $3+$ charge state. Several different ways for $^{229}$Th$^{3+}$ ion generation have been experimentally investigated. In the main approach thorium ions are generated by laser ablation from a solid $^{229}$Th target. The method was successfully applied for loading of thorium ($^{229}$Th or $^{232}$Th) ions into Paul traps at the Georgia Institute of Technology \cite{Steele,Campbell2,Churchill2010,Campbell3,Campbell4,Radnaev,Radnaev2}, at PTB \cite{Zimmermann,Zimmermann1,Herrera2,Thielking2018,Meier2019}, at the Moscow Engineering Physics Institute (MePhI) \cite{Troyan,Borisyuk4,Borisyuk2017e} at a collaboration of different institutes in Mainz \cite{Groot-Berning2019,Stopp2019} and most recently at Griffith University in Brisbane \cite{Piotrowski2020}.\\[0.2cm]
Alternative approaches that have been investigated are inductively coupled plasma techniques \cite{Troyan}, electron-beam evaporation \cite{Troyan1,Troyan3} and the extraction of ions from a $^{233}$U source \cite{Wense4,Wense5,Pohjalainen2019}. The production of thorium ions by laser ablation has the advantages that it is simple and does not require a helium buffer-gas environment. This comes at the cost of a limited time of use and a radioactive contamination of the vacuum chamber used for laser ablation. Opposed to that, the extraction from $^{233}$U requires a more complex setup but can deliver a continuous and pure $^{229}$Th ion beam of high intensity. Interestingly, 2\% of the $^{229}$Th nuclei are even in the isomeric state. The disadvantage is that it requires a buffer-gas environment and therefore a differential pumping scheme. A concept for the extraction of $^{229}$Th from a $^{233}$U source without the requirement of buffergas is currently under investigation \cite{Haas2019}.
\subsection{Paul trapping of $^{229}$Th ions}\label{STEPS.TRAP}
Laser cooling of $^{229}$Th and the development of a single-ion nuclear clock requires Paul trapping of individual thorium ions. Paul trapping of $^{229}$Th ions has been achieved at PTB \cite{Zimmermann,Herrera2,Thielking2018,Meier2019} at the Georgia Institute of Technology \cite{Campbell3,Campbell4,Radnaev,Radnaev2} and at LMU \cite{Wense2,Seiferle2019b}. Several further experiments have reported preparatory trapping of $^{232}$Th ions, which has the advantage of a lower radioactivity \cite{Troyan1,Troyan2,Troyan3,Borisyuk4,Borisyuk5,Borisyuk2017e,Groot-Berning2019,Stopp2019,Piotrowski2020}. Importantly, thorium is a very reactive element. For example, a reaction rate of $6.0\cdot10^{-10}$~cm$^3$/s for the Th$^{+}+$O$_2$ molecule formation is listed in \cite{Kudryavtsev, Johnsen}, which is a factor of 10 larger than typical ion reaction rates. Therefore long $^{229}$Th trapping times do require ultra-high vacuum conditions.\\[0.2cm]
Ideally for a nuclear clock, the ions have to be confined in the Lamb-Dicke regime, in which a photon absorption does not lead to the excitation of vibrational modes of the trapped ion. This requires a steep trapping potential for a linear Paul trap, so that the Lamb-Dicke parameter \cite{Stenholm1985}
$$
\eta=\frac{2\pi}{\lambda}\sqrt{\frac{\hbar}{2M\omega_\text{trap}}}
$$
is much smaller than $1$. Here $\lambda$ denotes the wavelength of the nuclear transition, $M$ is the $^{229}$Th atomic mass and $\omega_\text{trap}$ the secular trap frequency. Trap parameters that would be suitable for trapping and laser cooling of $^{229}$Th$^{3+}$ as given in \cite{Bussmann} are listed in Tab.~\ref{trapparameters}. For these parameters one has $\eta\approx0.09\ll 1$.
\begin{table}[t]
\begin{center}
\caption{\footnotesize Paul trap parameters proposed in \cite{Bussmann} for confinement of $^{229}$Th$^{3+}$ ions in the Lamb-Dicke regime.}
\begin{footnotesize}
\begin{tabular}{ccc}
\noalign{\smallskip}\hline\noalign{\smallskip}
 Parameter  & \parbox[pt][2em][c]{2cm}{\centering{Variable}}  & \parbox[pt][2em][c]{2cm}{\centering{Value}} \\
\noalign{\smallskip}\hline\noalign{\smallskip}
\parbox[0pt][1em][c]{3cm}{Aperture radius} &  \parbox[pt][1em][c]{2cm}{\centering{$r_0$}} & \parbox[pt][1em][c]{2cm}{\centering{400 $\mu$m}} \\
\parbox[0pt][1em][c]{3cm}{Trapping voltage} & \parbox[pt][1em][c]{2cm}{\centering{$U_{RF}$}} & \parbox[pt][1em][c]{2cm}{\centering{850~V}} \\
\parbox[0pt][1em][c]{3cm}{Trapping frequ.} & \parbox[pt][1em][c]{2cm}{\centering{$\Omega_{RF}$}} & \parbox[pt][1em][c]{2cm}{\centering{$2\pi\cdot29$~MHz}} \\
\parbox[0pt][1em][c]{3cm}{Secular frequ.} & \parbox[pt][1em][c]{2cm}{\centering{$\omega_\text{trap}$}} & \parbox[pt][1em][c]{2cm}{\centering{$2\pi\cdot4.1$~MHz}}\\
\parbox[0pt][1em][c]{3cm}{Stability param.} & \parbox[pt][1em][c]{2cm}{\centering{$q$}} & \parbox[pt][1em][c]{2cm}{\centering{$0.41$}} \\
\noalign{\smallskip}\hline\noalign{\smallskip}
\end{tabular}
\label{trapparameters}
\end{footnotesize}
\end{center}
\end{table}
\subsection{Laser cooling of trapped $^{229}$Th ions} \label{STEPS.COOL}
In preparation for precision laser-spectroscopy experiments of $^{229}$Th ions in a Paul trap, laser cooling is required. Two different concepts for $^{229}$Th laser cooling have been investigated: direct laser cooling of $^{229}$Th$^{3+}$ ions and sympathetic cooling.\\[0.2cm]
\subsubsection{Direct laser cooling}\label{STEPS.COOL.DIRECT}
\begin{figure}[t]
 \begin{center}
 \includegraphics[width=7cm]{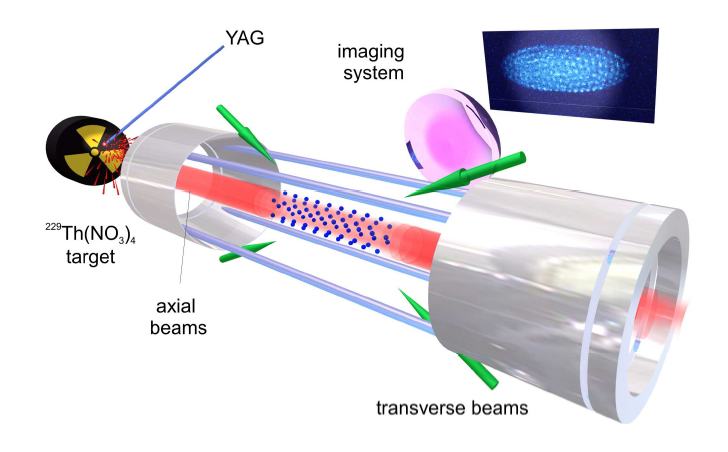}
  \caption{\footnotesize Drawing of the experimental setup used for direct laser cooling of $^{229}$Th$^{3+}$ \cite{Campbell3}. The ions are produced by laser-ablation from a radioactive source. Doppler cooling is achieved with four lasers from the sides (green) and further cooling with collinear laser beams of two different wavelengths (red). Reprinted from \cite{Campbell3} with kind permission of the American Physical Society.}
 \label{campbelltrap}
 \end{center}
\end{figure}

\begin{figure}[t]
 \begin{center}
 \includegraphics[width=7cm]{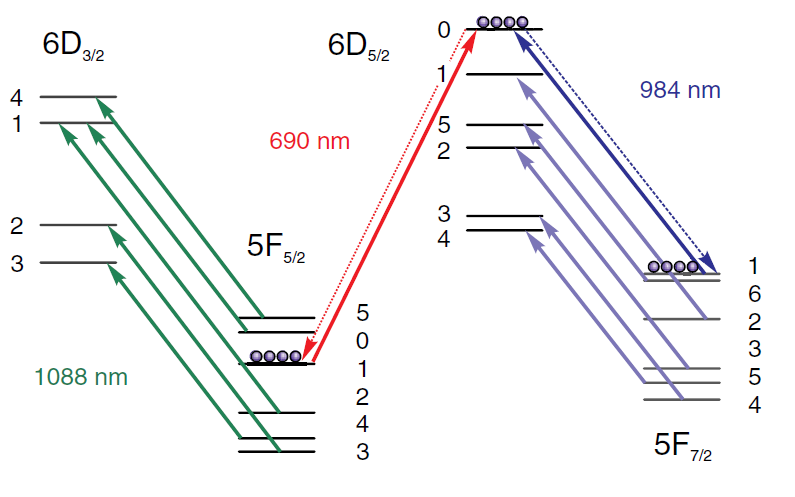}
  \caption{\footnotesize Level scheme of different atomic shell states of $^{229}$Th$^{3+}$ that are involved in the laser cooling scheme \cite{Campbell3}. The nuclear spin state of $I_g=5/2$ adds significantly to the complexity. Reprinted from \cite{Campbell3} with kind permission of the American Physical Society.}
 \label{campbelllines}
 \end{center}
\end{figure}

\begin{figure}[t]
 \begin{center}
 \includegraphics[width=7cm]{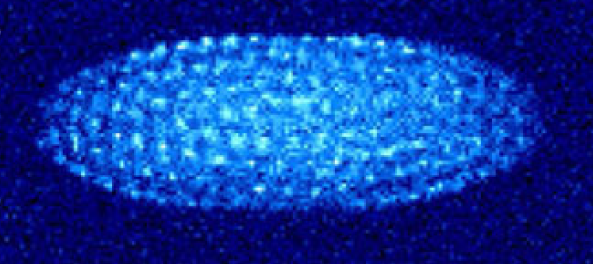}
  \caption{\footnotesize Coulomb crystal of $^{229}$Th$^{3+}$ as observed after direct laser cooling \cite{Campbell3}. Reprinted from \cite{Campbell3} with kind permission of the American Physical Society.}
 \label{campbellcrystal}
 \end{center}
\end{figure}

As a predecessor for $^{229}$Th$^{3+}$, $^{232}$Th$^{3+}$ ions were laser cooled at the Georgia Institute of Technology for reasons of better availability of the material, lower radioactivity and an easier laser-cooling scheme due to the lack of hyperfine structure caused by a nuclear spin equal to zero \cite{Steele,Campbell2,Churchill2010,Churchill2011}. Direct laser cooling of $^{229}$Th$^{3+}$ ions was experimentally achieved in 2011 at the same institute \cite{Campbell3,Campbell4,Radnaev2011,Radnaev2}. For the experiment, $^{229}$Th$^{3+}$ ions were laser ablated from a radioactive $^{229}$Th target and trapped in a linear Paul trap (Fig.~\ref{campbelltrap}).
The laser cooling was then performed in a two-step process as proposed in 2003 \cite{Peik}. In the first step, Doppler cooling was performed using the 1087~nm transition between the $5$F$_{5/2}$ ground state and the $6$D$_{3/2}$ state. In the second step, additional cooling was applied using the three-level $\Lambda$ system consisting of $5$F$_{5/2}$, $6$D$_{5/2}$ and $5$F$_{7/2}$ states. Wavelengths of 690~nm and 984~nm are required for this purpose. Direct laser cooling of $^{229}$Th$^{3+}$ is slightly complicated by the hyperfine structure induced by the nuclear ground-state spin of $I_g=5/2$. The complex electronic level structure, including individual hyperfine levels, is shown in Fig.~\ref{campbelllines} (a coarse electronic level diagram for $^{229}$Th$^{3+}$ and $^{232}$Th$^{3+}$ is provided in Fig.~\ref{Peikconcept}). To overcome this challenge experimentally, the laser light of the cooling lasers was fed into electro-optical modulators (EOMs) in order to introduce sidebands onto the laser frequencies, which allowed for laser pumping of the HFS \cite{Campbell3,Campbell4}. A photograph of the resulting $^{229}$Th$^{3+}$ Coulomb crystal is shown in Fig.~\ref{campbellcrystal}.
\noindent Direct laser cooling of $^{229}$Th$^{3+}$ is also in preparation at MEPhI to provide the basis for nuclear laser spectroscopy in a Paul trap and the development of a nuclear clock \cite{Troyan1,Troyan2,Troyan3,Borisyuk4,Borisyuk5,Borisyuk2017c,Borisyuk2017e}.\\[0.2cm]
\subsubsection{Sympathetic cooling}\label{STEPS.COOL.SYMP}
As direct $^{229}$Th$^{3+}$ laser cooling requires a significant technological effort, sympathetic cooling of thorium ions has also been investigated. A sympathetic laser-cooling experiment of thorium ions was reported in 2019 from the University of Mainz in collaboration with other institutes \cite{Groot-Berning2019,Stopp2019}. In this experiment, $^{232}$Th$^+$ ions were generated via laser ablation and sympathetically cooled with $^{40}$Ca$^+$. It is planned to investigate $^{229}$Th ions of different charge states and also $^{229\text{m}}$Th in the future. For this purpose, a $^{229}$Th recoil source is under development, which will populate the isomeric state in the $\alpha$ decay of $^{233}$U without the requirement of a buffer-gas environment \cite{Haas2019}.\\[0.2cm]
Most experiments are aiming for sympathetic cooling with $^{88}$Sr$^+$. This has the advantage that the mass-to-charge ratio is closer to that of $^{229}$Th$^{3+}$, making the ions easier to trap in parallel. Experiments along this line are prepared at PTB, LMU and NIST. Importantly, for the nuclear clock, direct laser cooling would be advantageous, as it allows for better cooling also of the transverse modes of motion.
\subsection{Laser systems for nuclear clock development}\label{STEPS.LASER}
\begin{figure*}[t]
 \begin{center}
 \includegraphics[width=17cm]{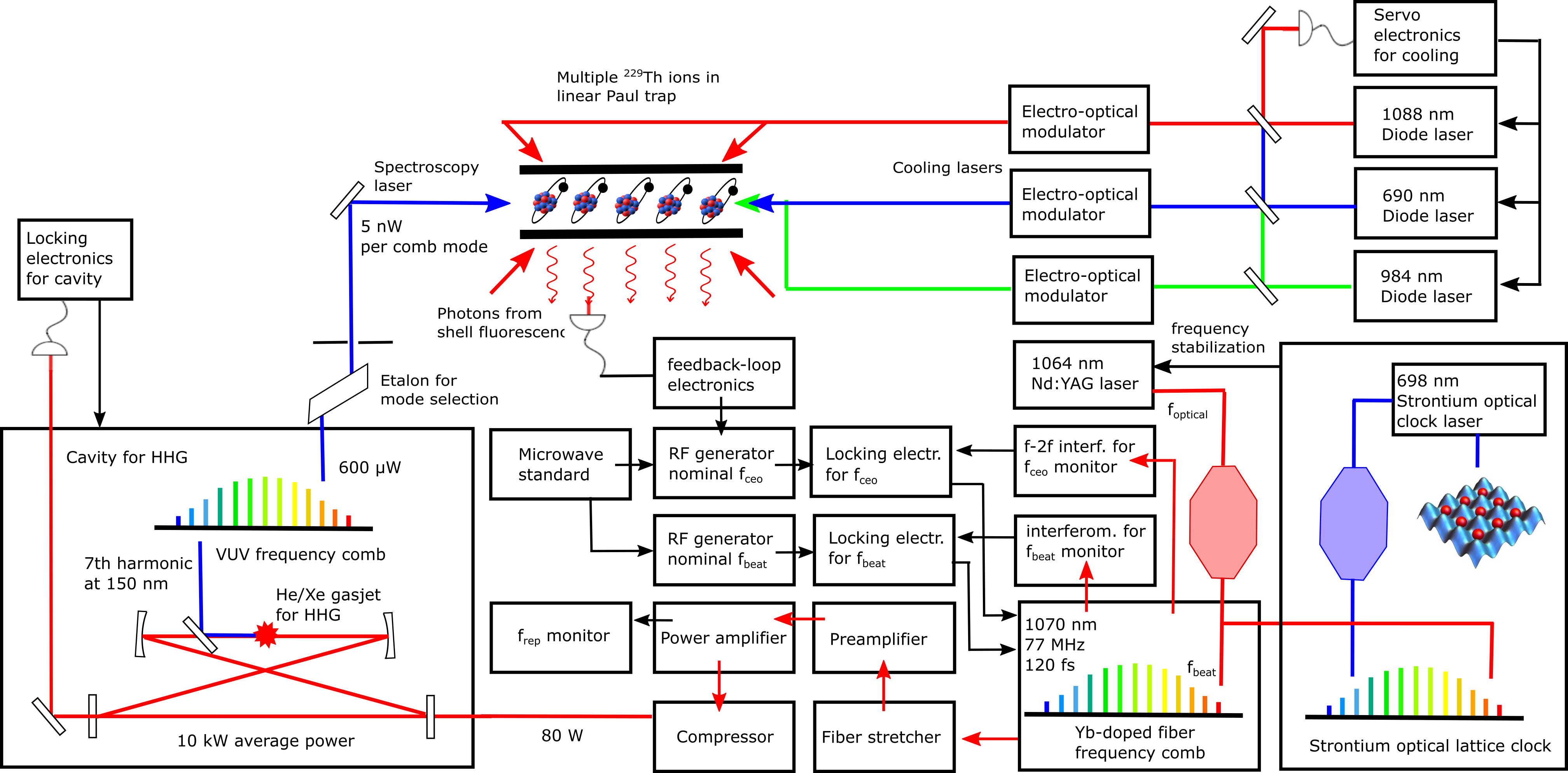}
  \caption{\footnotesize Detailed sketch of an experimental setup that could be used for high-precision frequency determination of $^{229\text{m}}$Th. The same setup could also serve as a nuclear frequency standard and be used for a precision search for potential time variations of fundamental constants. An IR frequency comb based on an Yb-doped fiber laser system at about 1070~nm wavelength is transferred into the VUV wavelength range via intra-cavity HHG \cite{Ozawa2013,Porat2018,Zhang2020}. Irradiation of multiple laser-cooled $^{229}$Th$^{3+}$ ions in a Paul trap is performed with the 7th harmonic at around 150~nm providing a power of about 5~nW per comb mode. The excitation of the nuclei is probed via the different hyperfine splittings of the electronic shell. Fluorescence light from one of the cooling lasers is used as a feedback signal to actively stabilize the beat-frequency $f_\text{beat}$ of a narrow-bandwidth Nd:YAG laser with the comb, thereby acting on the repetition frequency $f_\text{rep}$. $f_\text{ceo}$ is monitored via the $f-2f$ interferometry of the frequency comb and stabilized by comparison to a microwave standard. If the absolute frequency $f_\text{optical}$ of the Nd:YAG laser is known to high precision from active stabilization to an optical frequency standard (Sr-optical lattice clock), one can obtain $f_\text{rep}$ with high precision from $f_\text{optical}=f_\text{ceo}+N_\text{optical}\cdot f_\text{rep}+f_\text{beat}$ and the nuclear transition frequency is obtained as $f_\text{nuclear}=f_\text{ceo}+N_\text{nuclear}\cdot f_\text{rep}$. Here $N_\text{optical}$ and $N_\text{nuclear}$ denote the comb-mode numbers related to the laser light and the nuclear transition, respectively. They can be obtained by performing several measurements with different repetition rates. We acknowledge discussions with Chuankun Zhang, Christian Sanner and Jun Ye.}
 \label{Yb7th}
 \end{center}
\end{figure*}
A central requirement and probably one of the biggest challenges is the development of a laser system suitable for driving the nuclear transition in the clock concept. Ideally, the laser light should be extremely stable, cw and ultra-narrow-band with a bandwidth very well in the mHz-range \cite{Kessler,Matei2017}. Unfortunately no such laser at about 150~nm wavelength is available today. This may, however, be subject to change in the future. The non-linear crystal currently used for second-harmonic generation (and therefore also for the generation of cw laser light at the shortest wavelength) is KBe$_2$BO$_3$F$_2$ (KBBF) \cite{Chen2009}. While this crystal offers a cut-off wavelength of 147~nm \cite{Chen2009}, unfortunately the minimum wavelength that still allows for second-harmonic generation (SHG) phase-matching is about 164~nm \cite{Halasyamani2017}. The generation of pulsed laser light below 150~nm via sum-frequency mixing using KBBF was experimentally demonstrated \cite{Nakazato2016}. However, applying the same scheme for the generation of cw laser light would require a high intensity cw source of light below 190~nm, which is not available today. There are a few other materials known that exhibit non-linear optical behavior and possess a cut-off wavelength shorter than 150~nm, for example BPO$_4$ \cite{Zhang2011} and BaMgF$_4$ (BMF) \cite{Shimamura2005}. Both crystals, however, do not offer the possibility of phase-matching in the deep VUV range and quasi phase-matching (QPM) schemes are still under investigation \cite{Zhang2011}. Several new nonlinear optical materials were proposed based on theoretical calculations to exhibit optical transparency as well as phase matching below 150~nm \cite{Kang2018}. The most promising are AlCO$_3$F, PB$_3$O$_6$F$_2$ and SiCO$_3$F$_2$. To our knowledge, these materials have not yet been experimentally investigated. A potential future narrow-band laser system that would be suitable to operate a nuclear optical clock could potentially be based on SHG generation of 300~nm light in one of these crystals.\\[0.2cm]
As long as no cw laser source for nuclear clock operation is available, direct frequency comb spectroscopy appears to be the only alternative. Frequency combs in the visible and infrared have been converted into the VUV region via high-harmonic generation (HHG) \cite{Gohle2005,Jones2005,Ozawa2008,Yost2009,Bernhardt2012,Cingoez,Pupeza,Benko,Ozawa2015,Holzberger2015,Carstens,Porat2018,Seres2019,Saule2019,Zhang2020}. Also, direct frequency-comb spectroscopy of electronic shell transitions has already been performed \cite{Cingoez,Ozawa2013,Ozawa2017}. Using existing technology, there are two promising paths to generate a high-power frequency comb at around 150~nm. One way is via 7th harmonic generation of an Yb-doped fiber frequency comb \cite{Wense2019c}, the second way is via 5th harmonic generation of a Ti:Sapphire system \cite{Seres2019}. Direct frequency-comb spectroscopy of Xe using the 7th harmonic of an Yb-doped fiber system has already been performed, and the same concept could also be used for nuclear laser spectroscopy of $^{229\text{m}}$Th \cite{Ozawa2013}. A conceptual sketch of an experimental setup that could be used for precision spectroscopy of $^{229\text{m}}$Th and as a nuclear optical clock is shown in Fig~\ref{Yb7th}. A frequency comb is generated in the IR at around 1070~nm using an Yb-doped fiber laser system. The comb is amplified and coupled into an enhancement cavity. High harmonics are generated in a noble-gas jet, with the 7th harmonic matching the expected energy of the nuclear transition. An out-coupled power of 5~nW per comb mode in the 7th harmonic at about 150~nm has already been generated in this way \cite{Zhang2020}\footnote{About 600~$\mu$W total out-coupled harmonic power in the 11th harmonic were shown. The 7th harmonic is known to provide roughly the same power and the number of comb modes is $1.2\cdot10^5$, resulting in 5~nW power per mode.}. Laser excitation rates are discussed in App.~\ref{THEO.DECEXC.GAMMAEXC} and the nuclear population density can be calculated based on Eq.~(\ref{Torrey}).
\subsection{Steps towards a solid-state nuclear clock}\label{STEPS.SOLID}
The original idea of a solid-state nuclear clock based on the crystal-lattice approach makes important use of the detection of photons emitted in the isomer's decay. These photons allow for probing the excitation of $^{229\text{m}}$Th in the solid-state environment, similar to the double-resonance method in the single-ion clock approach. For this reason, several experiments search for photons emitted in the isomer's direct decay in a solid-state environment (see Tab.~\ref{energyconcepts}). If observed, this would not only lay the foundation for a crystal-lattice nuclear clock, but it would also pave the way for a high-precision measurement of the nuclear excitation energy via photon spectroscopy.\\[0.2cm]
As an important step, decay channels other than the isomer's direct radiative decay have to be suppressed in a solid-state environment. Such decay channels are internal conversion (IC, see App. \ref{THEO.HIGHDEC.IC}) and the electronic bridge decay (EB, see App. \ref{THEO.HIGHDEC.EB}). Theoretically it is expected that both competing decay channels are suppressed, if materials with a band-gap larger than the isomer's energy are chosen as host materials \cite{Tkalya2b,Tkalya9}. Materials under investigation are LiSrAlF$_6$ \cite{Rellergert,Jeet} and CaF$_2$ \cite{Stellmer2,Stellmer2018a}. The verification, that non-radiative decay channels of the isomeric state can be successfully suppressed in a solid-state environment is still outstanding and an active field of continued research.\\[0.2cm]
As an alternative, the concept of an IC-based nuclear clock makes use of IC-electron detection for monitoring the nuclear excitation \cite{Wense2019c}. IC electrons emitted in the isomeric decay have already been observed and the laser excitation of the $^{229\text{m}}$Th from the ground-state is currently being investigated. It would be straightforward to develop an IC-based nuclear clock as soon as frequency-comb spectroscopy of surface-bound $^{229\text{m}}$Th atoms has been achieved. If successful, the isomer's energy would also be sufficiently well constrained to perform laser spectroscopy of $^{229}$Th ions.
\section{Potential applications}\label{APPL}
Any improvement in time measurement has led to technological advances. One famous historic example is the development of the marine chronometer by John Harrison around 1735, which solved the longitude problem and allowed for reliable navigation at sea \cite{Gould}. Today's cesium atomic clocks are used for satellite-based navigation \cite{Hofmann} and the development of optical atomic clocks with their corresponding improvement in accuracy by a factor of 100 will certainly lead to a new generation of satellite-based clocks \cite{Gill2008}. Tests along this line have already been performed \cite{Lezius}.\\[0.2cm]
A potential use of a nuclear optical clock for satellite-based navigation is frequently considered. In this context, either the lower expected inaccuracy of an ion-based nuclear clock is brought up as an advantage for the precision of navigation, or the solid-state nuclear clock is advertised due to the expected advantages in terms of stability, robustness and power consumption. Both statements have to be brought into perspective: At the time of the nuclear clock proposal in 2003 \cite{Peik}, optical atomic clocks were a young research field and the nuclear clock was expected to offer a significant advantage compared to the most accurate clocks at that time. However, the recent progress in atomic clock technology has shown, that also atomic clocks offer the potential for inaccuracies in the $10^{-19}$ range \cite{Brewer} and no fundamental limit has yet been approached. For this reason it is likely that both, nuclear and atomic frequency standards, will approach comparable accuracy limits in the future. In this case the nuclear clock will have the disadvantage that it requires a more complicated laser technology. It can be considered as unlikely, that this disadvantage will be compensated by the solid-state nuclear clock approach.\\[0.2cm]
A different practical application that is compelling to consider is the use of a single- or multiple ion nuclear clock as a primary standard and maybe for a re-definition of the second. Also this consideration requires a critical discussion, which goes beyond the scope of this review and will only be touched briefly: Even in the case that a future nuclear optical clock will be the frequency standard of smallest systematic frequency uncertainty, the practical use of a $10^{-19}$ frequency uncertainty on the surface of the earth is unclear, as solid earth tides will lead to frequency changes on a significantly larger scale due to gravitational time dilation. Ideally, the definition of the second should therefore be based on a network of space-based clocks. However, considering that strict criteria for a re-definition of the second have already been put forward and it is expected that atomic clocks will fulfill these criteria before 2030 \cite{Riehle2018}, it is unlikely that a re-definition of the second will be postponed to a time when the nuclear frequency standard may eventually fulfill the same criteria.\\[0.2cm]
The most promising potential applications for a nuclear frequency standard can therefore be found in fundamental physics. More precisely, a high sensitivity of a nuclear clock for potential time variations of fundamental constants, e.g., the fine structure constant was proposed \cite{Flambaum1,Berengut2}. Based on the same effect, it was also considered to use a nuclear clock as a probe for topological dark matter \cite{Derevianko}. Only recently, a sensitivity factor of $-(0.9\pm0.3)\cdot10^4$ for time variation in $\alpha$ was estimated, significantly larger than for any atomic frequency standard \cite{Fadeev}. If confirmed, this would offer a great potential for a nuclear clock to put further constraints on time variations of fundamental constants. In the meanwhile, experimental and theoretical work toward the development of a nuclear clock are continued, as the realization of one or even more of the above mentioned applications justify these efforts.\\[0.2cm]
In the following, several applications will be considered in more detail. A thorough discussion of potential applications for a nuclear clock was provided in \cite{Thirolf2019}.
\subsection{Search for temporal variations of fundamental constants}\label{APPL.FUND}
A particularly high sensitivity of a nuclear clock for potential time variations of fundamental constants, e.g., the fine-structure constant, was highlighted \cite{Peik,Flambaum1,Rellergert}. Such temporal variations are predicted by some theories beyond the standard model \cite{Uzan} and have already been investigated for atomic clock transitions \cite{Rosenband,Godun,Safronova2019}.\\[0.2cm]
Already in their 2003 work, Peik and Tamm proposed that nuclear laser spectroscopy with $^{229}$Th would open new possibilities for probing for potential temporal variation of fundamental constants \cite{Peik}. A quantitative analysis was performed by V.V.~Flambaum in 2006 \cite{Flambaum1}. The result predicts an extremely high sensitivity for potential temporal changes of the fine structure constant $\alpha$ as well as the dimensionless strong interaction scale parameter $m_q/\Lambda_\mathrm{QCD}$. It was discussed in \cite{Flambaum1} that the sensitivity to these variations may be below $10^{-20}$ per year and thus at least three orders of magnitude more sensitive than existing constraints based on atomic-shell transitions (which currently pose the most stringent limits on such variations) \cite{Rosenband,Godun}.\\[0.2cm]
The reason for the predicted large sensitivity is that, from nuclear structure theory, the energies of the nuclear ground and excited states are proposed to be dominated by two individual high energy (MeV) terms which, by coincidence, cancel each other down to the eV range when subtracted in the very special case of the $^{229}$Th nucleus. The first term is the Coulomb-energy term, which is responsible for the sensitivity to variations in the fine structure constant $\alpha$, the second term results from various contributions of the strong interaction and thus leads to a sensitivity with respect to the strong interaction scale parameter $m_q/\Lambda_\mathrm{QCD}$. The ratio between the typical nuclear energy scale and the energy scale of $^{229\mathrm{m}}$Th can then be directly transferred to the sensitivity for changes in fundamental constants. The estimate for the variation of the $^{229\mathrm{m}}$Th transition frequency as a function of the variation of $\alpha$ and $\Lambda_\mathrm{QCD}$ is given as follows \cite{Flambaum1,Flambaum2}:
\begin{equation}
\frac{\delta\omega}{\omega}\approx 10^5\left(2\frac{\delta\alpha}{\alpha}+0.5\frac{\delta X_q}{X_q}-5\frac{\delta X_s}{X_s}\right)\frac{8\ \mathrm{eV}}{\omega},
\end{equation}
with $X_q=m_q/\Lambda_\mathrm{QCD}$ and $X_s=m_s/\Lambda_\mathrm{QCD}$, where $m_q=(m_u+m_d)/2$ and $m_s$ denote the light quark mass ($\approx5$ MeV) and the strange quark mass ($\approx120$ MeV), respectively. This expression has already been corrected for the new energy value of about 8~eV and contains an expected enhancement in sensitivity by five orders of magnitude for changes in the two fundamental constants (based on the Walecka model).\\[0.2cm]
Opposed to that, Hayes et al. came to the result that no significant sensitivity for potential temporal variations in fundamental constants could be achieved from a nuclear frequency standard \cite{Hayes1,Hayes2}. Based on the Feynman-Hellmann Theorem, which is a fundamental theorem of quantum-field theory, they derived the simple formula
\begin{equation}
\frac{\delta\omega}{\omega}=k\cdot\frac{\delta\alpha}{\alpha},\ \text{with}\ k=\frac{\Delta V_C}{\omega}
\end{equation}
for the $\alpha$-dependence of the relative nuclear frequency shift \cite{Hayes1}. Here $k$ is the sensitivity factor between a variation in $\alpha$ and the corresponding frequency variation and $\Delta V_C$ denotes the Coulomb energy difference between the ground and the excited nuclear state. Taking the Nilsson model as the basis for nuclear energy calculations \cite{Nilsson}, there is no Coulomb interaction included, leading to the prediction of $\Delta V_C=0$ and thus no expected sensitivity for a potential $\dot{\alpha}$.\\[0.2cm]
Given these inconsistent results, in the following years multiple theoretical investigations were made, however, no conclusive results for the sensitivity factor $k$ could be drawn \cite{He2,Xiao,He3,Flambaum3,Flambaum4}. It was concluded in \cite{Litvinova} that the calculated sensitivity of the $^{229\mathrm{m}}$Th nuclear transition to temporal variations in $\alpha$ heavily depends on the applied nuclear model and therefore no reliable prediction can be made (see also \cite{Feldmeier}).\\[0.2cm]
As a consequence, \cite{Berengut} proposed measuring the nuclear charge radii as well as the electric quadrupole moments of $^{229}$Th and $^{229\mathrm{m}}$Th, respectively, by laser spectroscopy of the atomic shells of both nuclear states. A technique is presented for deducing $V_C$ for both states from these parameters and thus determining the expected sensitivity factor for potential variations in $\alpha$. As this method is directly based on experimental data, it is independent of any particular choice of the nuclear structure model. In 2018 an experimental determination of the nuclear charge radius and electric quadrupole moments of $^{229\text{m}}$Th was achieved \cite{Thielking2018}. Although the precision of this measurement did not allow for an immediate determination of the sensitivity to variations in $\alpha$ \cite{Safronova2018b}, recently a value for the sensitivity factor of $k=-(0.9\pm0.3)\cdot10^4$ was obtained under the additional assumption of an identical charge density of the ground and the isomeric state \cite{Fadeev}.\\[0.2cm]
In 2010, a review article on the search for potential temporal variations of fundamental constants with $^{229}$Th was published by Berengut and Flambaum \cite{Berengut2}. Rellergert at al. \cite{Rellergert} highlighted the potentials of a solid-state nuclear clock with respect to searches for fundamental constant variation. In 2012, Berengut and Flambaum proposed that, besides temporal variations, spatial variations of fundamental constants could also be probed \cite{Berengut3}. In 2016 it was proposed that effects of Lorentz invariance and equivalence principle violation could be probed with the help of $^{229\mathrm{m}}$Th \cite{Flambaum6}. Comparable tests with optical atomic clocks have already been performed \cite{Sanner2019}.
\subsection{Chronometric geodesy}
The most accurate optical atomic clocks operational today have already opened a new field of application, namely chronometric geodesy \cite{Delva2019,Mehlstaeubler}. The underlying principle is that, following Einstein's theory of general relativity, time dilation occurs: a clock in a gravitational potential will slow down with respect to a clock in empty space. Therefore, two clocks located at different heights in the earth's potential will propagate differently. At the surface of the earth, 1~cm of height difference corresponds to a relative change in clock speed of $10^{-18}$. Consequently, the most accurate optical atomic clocks can measure a difference in height at the 1~cm level \cite{McGrew}.\\[0.2cm]
One may hope, that a more accurate clock would push this limit even further and that at the envisaged accuracy of $10^{-19}$ for a nuclear optical clock, geodesy at the 1~mm would be achievable \cite{Piotrowski,Piotrowski2020}. Even practical applications were proposed, for example, in the fields of earthquake prediction and the search for natural resources \cite{Safronova}. Also more precise tests of general relativity were discussed in literature \cite{Delva}, as well as clock-based gravitational wave detection \cite{Kolkowitz}.\\[0.2cm]
It is, however, important to note that this field is not specific to a nuclear clock and that the same advantages would be obtained from any atomic clock of comparable systematic frequency uncertainty. Further it can be expected that a potential improvement is limited by other effects, e.g., the stability of frequency transfer as well as height deviations of the earth due to solid earth tides, which amount to tens of cm per lunar day, thereby diluting any mm-scale height measurement. Therefore, it remains subject to speculation whether a nuclear clock will be able to advance this field. 
\subsection{Dark matter detection}
One of the big open questions of modern physics is the nature and origin of dark matter. Although predicted almost a century ago by indirect evidence in the solar system through its gravitational interaction, there has still not been a solution to the dark matter puzzle \cite{Safronovareview}. The idea of using a nuclear clock for dark matter detection is closely related to the previously discussed field of the search for temporal variations of fundamental constants. A particular class of dark matter, the so called ``topological dark matter", can be considered as large fields (on the 1000 km scale or even larger) that move through the universe interacting gravitationally. It has been proposed that such a field may lead to dark-matter induced time-variations of fundamental constants \cite{Derevianko}. Therefore, if a topological dark matter object were to propagate through the earth, it could induce frequency shifts of atomic and nuclear transitions, which would result in a relative time difference when compared to clocks that are not affected by the dark-matter field. Searches for signals potentially originating from transient topological dark matter using existing atomic clock data were already performed \cite{Roberts2017}. However, the obvious advantage of a nuclear clock would be the expected higher sensitivity to variations of fundamental constants if the sensitivity enhancement turns out to be present.\\[0.2cm]
In 2019, the potential for axionic dark matter detection using $^{229\text{m}}$Th was also discussed \cite{Flambaum2019}. Axionic dark matter is predicted to generate an oscillating Schiff moment in the nucleus, which is significantly enhanced in $^{229}$Th due to the strong octupole deformation. An oscillating Schiff moment could potentially be determined via nuclear laser spectroscopy of $^{229\text{m}}$Th. A high sensitivity for time (T), parity (P), and CP-violations in $^{229}$ThO molecules was also predicted \cite{Flambaum2019}.
\subsection{A $^{229\mathrm{m}}$Th-based nuclear laser}
A nuclear laser based on $^{229\mathrm{m}}$Th was conceptually proposed by Oganessian and Karamian in a publication from 1995, in which the thorium isomer was discussed in an individual section within the more general framework of nuclear $\gamma$-ray lasers \cite{Oganessian}.\\[0.2cm]
The working principle of a nuclear laser would be the same as for atomic-shell based lasers, but for using nuclear transitions instead. While there are significant problems to overcome when developing a high energy $\gamma$-ray laser based on nuclear transitions (see for example \cite{Rivlin}), the isomeric state in $^{229}$Th could allow the development a proof-of-principle device \cite{Oganessian}. The main issue is to achieve population inversion between the ground and the isomeric first excited state. Oganessian and Karamian proposed to achieve this inversion by heating to temperatures of $10^4$~K. However, this was at a time when the isomeric energy was still assumed to be about 3.5~eV and thus the internal-conversion decay channel of the isomer was expected to be suppressed in neutral $^{229}$Th. An alternative way to achieve population inversion was proposed by Karpeshin et al. via nuclear excitation by electron transition (NEET) \cite{Karpeshin2}.\\[0.2cm]
A quantitative analysis of the possibility of a $^{229\mathrm{m}}$Th-based nuclear laser was performed by Tkalya in 2011 \cite{Tkalya3}. In this work, a nuclear laser based on $^{229}$Th-doped solid-state crystals is investigated. The population inversion is proposed to be achieved in a two-step approach: First, the isomeric state is populated via direct laser excitation. As the nuclear ground and isomeric state provide only a two-level system, no population inversion can be achieved in this way. However, it would still be possible to excite a significant amount of nuclei. In a second step, it is proposed to apply a strong magnetic field (up to 100~T), to achieve a Zeeman splitting of the nuclear ground and excited state into corresponding sub-levels. By further cooling the crystal into the temperature region of about 0.01~K, the nuclei are expected to predominantly populate the energetically lowest Zeeman sub-levels of each nuclear state. In this way, a population inversion is achieved between the lower Zeeman sub-levels of the excited state and the upper Zeeman sub-levels of the ground state. The transition between these sub-levels would allow for light amplification by stimulated emission in the nucleus.\\[0.2cm]
Alternatively, line splitting into nuclear sub-levels could also be achieved via electric quadrupole splitting and is also discussed in \cite{Tkalya3}. In 2013, an alternative to the cooling method in order to achieve population inversion was proposed by Tkalya and Yatsenko \cite{Tkalya3b}. In this approach, a narrow-band laser is used to drive individual transitions in the Zeeman-split nuclear multiplet. A nuclear laser based on $^{229\mathrm{m}}$Th could be a proof-of-principle device for the development of high-energy nuclear $\gamma$-ray lasers, however, its practical realization still appears to be challenging. 
\begin{figure*}
 \begin{center}
 \includegraphics[width=13.5cm]{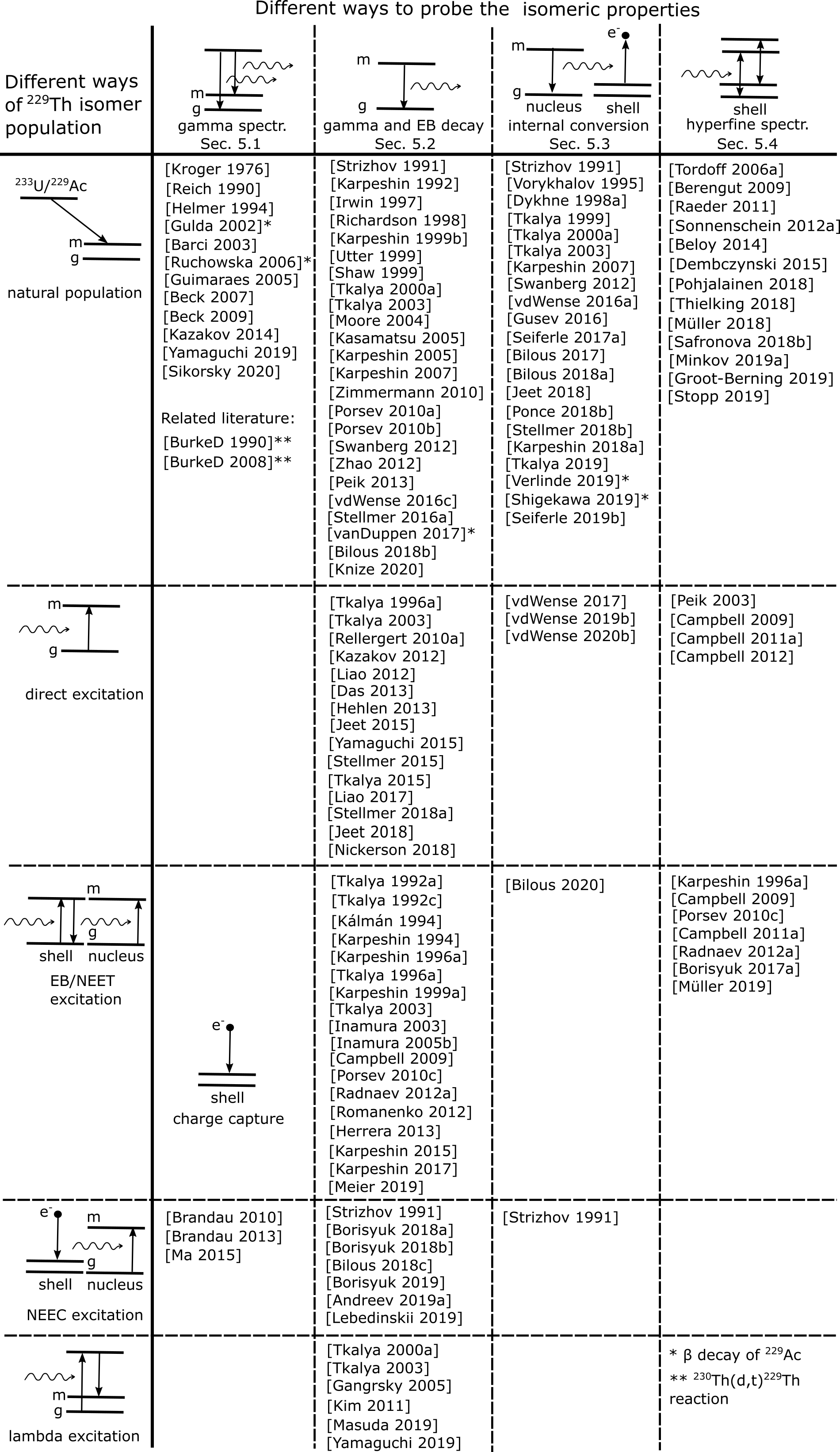}
  \caption{\footnotesize Overview of different ways of probing $^{229\text{m}}$Th properties and different ways of populating the isomeric state together with selected experiment-related literature relevant for each sub-field. A few experiments and proposals cannot be presented within this scheme and are included at an appropriate point in the text. Short notations: EB dec.: electronic bridge decay, EB exc.: electronic bridge excitation, NEET: nuclear excitation by electron transition, NEEC: nuclear excitation by electron capture.}
 \label{energy_overview}
 \end{center}
\end{figure*}
\section{$^{229\text{m}}$Th: past and present research}\label{EXP}
Since the prediction of its existence, constraining the isomer's energy has played the most important role in $^{229\text{m}}$Th-related research, and these investigations continue to the present. Different $^{229\text{m}}$Th energies that are reported in literature were listed in Tab.~\ref{energyconstraints}, and continued experimental efforts were tabulated in Tab.~\ref{energyconcepts}. In the following, past and continued $^{229\text{m}}$Th-related efforts will be discussed in detail.\\[0.2cm]
This section is structured according to the different ways to investigate the isomeric properties as shown in Fig.~\ref{energy_overview}. In Sec. \ref{EXP.GAMMA} $\gamma$ spectroscopy measurements will be described, which showed the existence of $^{229\text{m}}$Th in the early 1970s and have since then been developed for precision spectroscopy experiments. The search for photons emitted in the ground-state decay of $^{229\text{m}}$Th will be presented in Sec.~\ref{EXP.DIRECT}. These photons can either be emitted in the isomer's direct decay or in the decay via the electronic bridge (EB) mechanism (see App. \ref{THEO.HIGHDEC.EB}). In Sec. \ref{EXP.IC}, the investigation of the isomer's internal conversion (IC) decay channel will be discussed (see App. \ref{THEO.HIGHDEC.IC}). Finally, experiments dealing with the determination of the isomeric properties based on hyperfine spectroscopy of the $^{229}$Th shell will be presented in Sec. \ref{EXP.HYPER}.\\[0.2cm]
One further path of experimental investigation, namely the $^{229\text{m}}$Th $\alpha$ decay \cite{Dykhne2}, is not discussed. Corresponding experiments were carried out before 2010 \cite{Mitsugashira,Inamura1,Kikunaga1,Inamura2,Inamura3,Kikunaga3,Kikunaga2,Inamura4}, however, cannot be considered as realistic given today's knowledge, as they require a long isomeric lifetime in the neutral $^{229}$Th atom. A similar approach, however based on the observation of $\gamma$ rays was investigated in \cite{Browne}. A critical discussion of these and other experiments was provided in \cite{Sakharov}.
\subsection{$^{229}$Th $\gamma$-ray spectroscopy}\label{EXP.GAMMA}
The existence of a nuclear excited state of low energy in $^{229}$Th was inferred by Kroger and Reich in the early 1970ies from a precise study of the low-energy $\gamma$-ray spectrum of $^{229}$Th, as produced in the $\alpha$ decay of $^{233}$U \cite{Kroger,Kroger_Reich}. An upper limit of 100~eV for the transition energy was estimated, solely based on the non-observation of the excited states' direct decay. In 1990, further evidence for $^{229\mathrm{m}}$Th via the $^{230}$Th(d,t)$^{229}$Th nuclear reaction was given by Burke et al. \cite{Burke1} and an improved version of essentially the same measurement was published in 2008 \cite{Burke2}.
These early experiments securely established the existence of a low energy nuclear excited state in $^{229}$Th. For a detailed revision, the interested reader is referred to \cite{Wense3}. Investigations of the $^{229}$Th level scheme based on the $\alpha$ decay of $^{233}$U were also performed by Canty \cite{Canty} and later by Barci \cite{Barci}. Further investigations of the $^{229}$Th level structure were performed by Coulomb excitation \cite{Bemis} and by $\beta^-$ decay of $^{229}$Ac \cite{Chaya,Gulda,Ruchowska}.\\[0.2cm]
\subsubsection{First energy constraints}
\begin{figure}[t]
 \begin{center}
 \includegraphics[width=8cm]{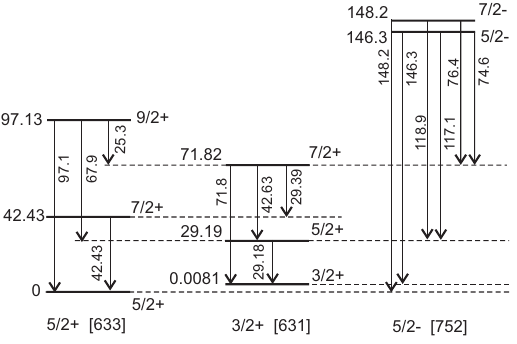}
 \vspace{-0.0cm}
  \caption{\footnotesize $^{229}$Th level scheme containing only levels that were used in the isomer's energy determination of Helmer and Reich \cite{Reich_Helmer,Helmer_Reich}. The spin and parity values as well as Nilsson quantum numbers \cite{Nilsson} for each band head are assigned below the bands. Each state is given together with its energy value in keV as well as its spin and parity assignments. The transition energies are also given. The $^{229}$Th ground-state rotational band is shown on the left side together with the 5/2+ [633] ground state. The isomeric state is the 3/2+ [631] band head of the second rotational band. For this state, the energy gap to the ground state is not shown to scale to improve clarity. Reprinted from \cite{Wense3} with kind permission of Springer Nature.}
  \label{thorium_levels1994}
 \end{center}
\end{figure}
As soon as the existence of a low-energy state in $^{229}$Th was established, the determination of its energy became an important experimental objective. In 1990, after more than a decade of effort, Reich and Helmer published a study in which they constrained the excited state's energy to be $(-1\pm4)$~eV \cite{Reich_Helmer}\footnote{Initial preparatory measurements had already been published in 1984 \cite{Reich_Helmer1}. Also, a value less than 10~eV, which was based on private communication with Helmer and Reich, had already been published in 1989 \cite{Akavoli}.} (see also \cite{Reich3}). The value was inferred based on the differences of nuclear levels of higher energies, populating the ground and the low-energy state, respectively. A sketch of the low-energy nuclear level structure of $^{229}$Th used in this study is shown in Fig.~\ref{thorium_levels1994}. Three different energy combinations were used, these are:
\begin{equation}
\begin{aligned}
\Delta_1&=E(97.1)-E(71.8)-E(25.3),\\
\Delta_2&=E(97.1)-E(67.9)-E(29.1),\\
\Delta_3&=[E(148.1)-E(146.3)]-[E(118.9)-E(117.1)].
\end{aligned}
\label{energydifferences}
\end{equation}
From this study, the authors concluded that the energy separation between the ground state and the first excited state of $^{229}$Th is smaller than the precision of the presented measurement and ``almost certainly less than 10~eV". Also a half-life estimate for a tentative 1~eV $M1$ transition between the excited state and the ground state was given as about 7 hours, placing the excitation to be a relatively long-lived isomer.\\[0.2cm]
The work of Reich and Helmer provided evidence for the existence of a nuclear transition with an extremely low energy of only a few eV. While Reich and Helmer themselves did not propose any applications for this nuclear state in their publication, their work was the basis for an increasing interest leading to the proposal of many interesting applications in the subsequent years.
\subsubsection{Continued experimental investigations}
Driven by the increasing interest in the newly discovered low-energy excited nuclear state, Helmer and Reich aimed for an improved energy determination of this excitation. This led to a publication in 1994, in which an energy value of $3.5\pm1.0$~eV was presented\footnote{A preliminary value of $(4.5\pm1.0)$~eV is given in \cite{Reich_Helmer2}} \cite{Helmer_Reich} (see also \cite{Helmer1993}). The techniques applied in this study were in principle the same as used in their earlier work, this time, however, with improved statistics and more accurate values of $\gamma$-ray energies of higher lying levels. For their analysis they used the same three $\gamma$-ray transition energy differences as before (Eq.~\ref{energydifferences}) together with one further difference:
\begin{equation}
\Delta_4= [E(148.1)-E(146.3)]-[E(76.4)-E(74.6)].
\end{equation}
The presented energy value of $3.5\pm1.0$~eV was the most accepted one until 2007. This value is below the ionization potential of thorium of 6.3~eV. For this reason, internal conversion, as a potential decay channel, was expected to be energetically forbidden, leading to an enhanced radiative decay and an increased half-life of 20 to 120 hours (assuming no coupling to the electronic environment) \cite{Helmer_Reich}. These assumptions had to be corrected following further energy investigations. Helmer and Reich assumed already in their 1994 work that no ``unique half-life" for $^{229\mathrm{m}}$Th might exist, as this will depend on the electronic environment of the sample.\\[0.2cm]
\noindent In 2005, motivated by an improved understanding of the $^{229}$Th level-scheme branching ratios \cite{Barci,Gulda} and the non-observation of the direct $^{229\mathrm{m}}$Th $\gamma$ ray, Guimar\~{a}es-Filho and Helene published a re-analysis of the spectroscopy data obtained by Helmer and Reich \cite{Filho}. The central technique applied in this work was the same as in the 1994 analysis of Helmer and Reich; however, this time an improved matrix formalism was used, including many reference lines to obtain better statistics. Improved branching ratios were also applied for the 29.18~keV and the 71.8~keV lines. These were assumed to decay by 100\% branching ratio into the 3/2+ isomeric state in the 1994 work of Helmer and Reich. However, more recent work proposed that the decay of these states might populate the ground state by branching ratios of 25\% and 40\%, respectively. In this re-analysis, the $\gamma$-ray transition energies were also corrected for recoil effects, leading to the different value of $5.5\pm1.0$~eV for the isomeric energy.\\[0.2cm]
\subsubsection{A corrected energy value}
\begin{figure}[t]
 \begin{center}
 \includegraphics[width=7cm]{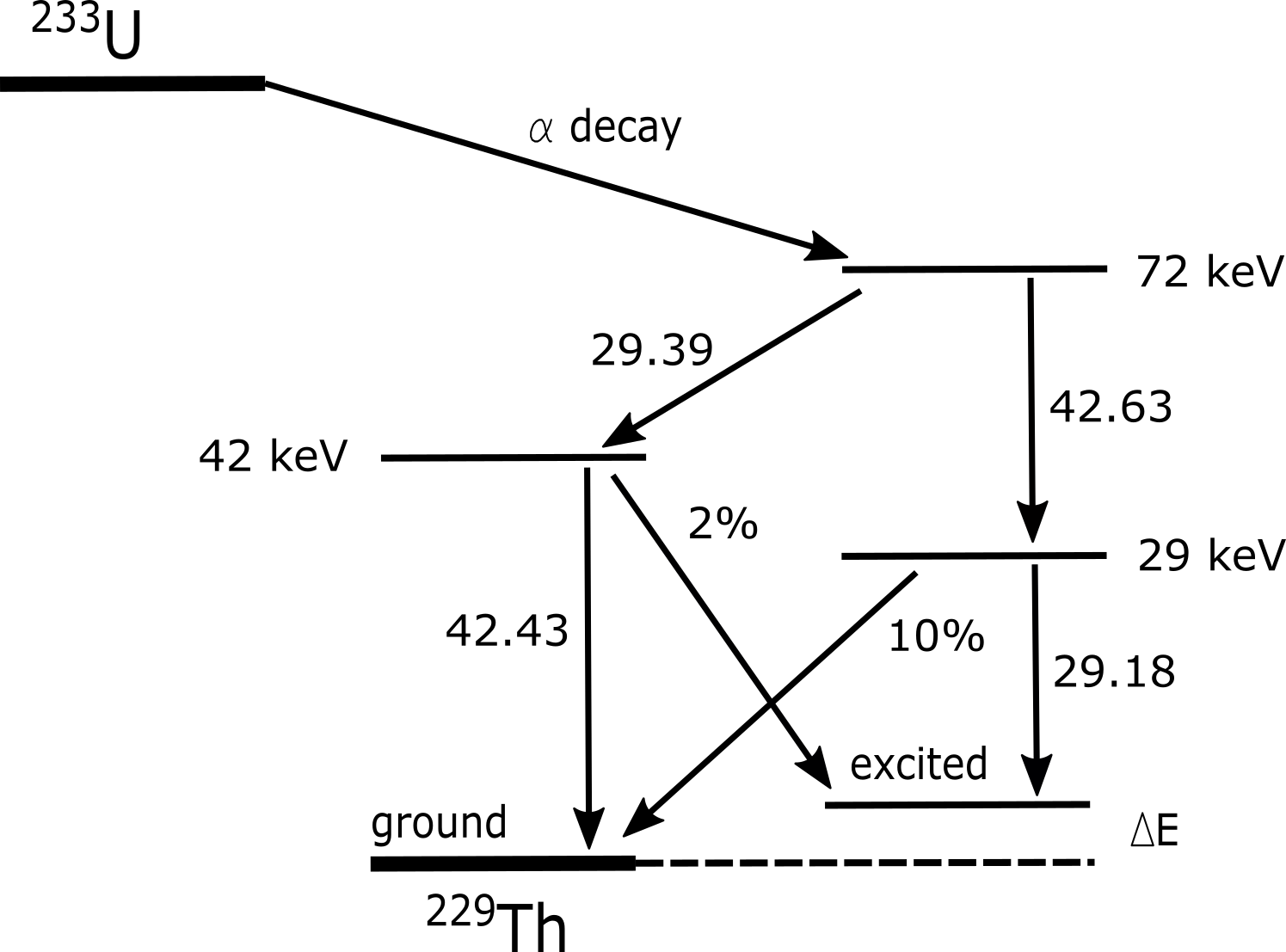}
  \caption{\footnotesize Partial $^{229}$Th level scheme used for the improved energy determination by Beck et al. in 2007 \cite{Beck1}. The high resolution of the NASA X-ray micro-calorimeter spectrometer of 26~eV (FWHM) allowed the closely spaced doublets at about 29 keV and 42 keV to be resolved (see Fig.~\ref{beckdata}).}
 \label{Beck2007}
 \end{center}
\end{figure}
\begin{figure*}[t]
 \begin{center}
 \includegraphics[width=14cm]{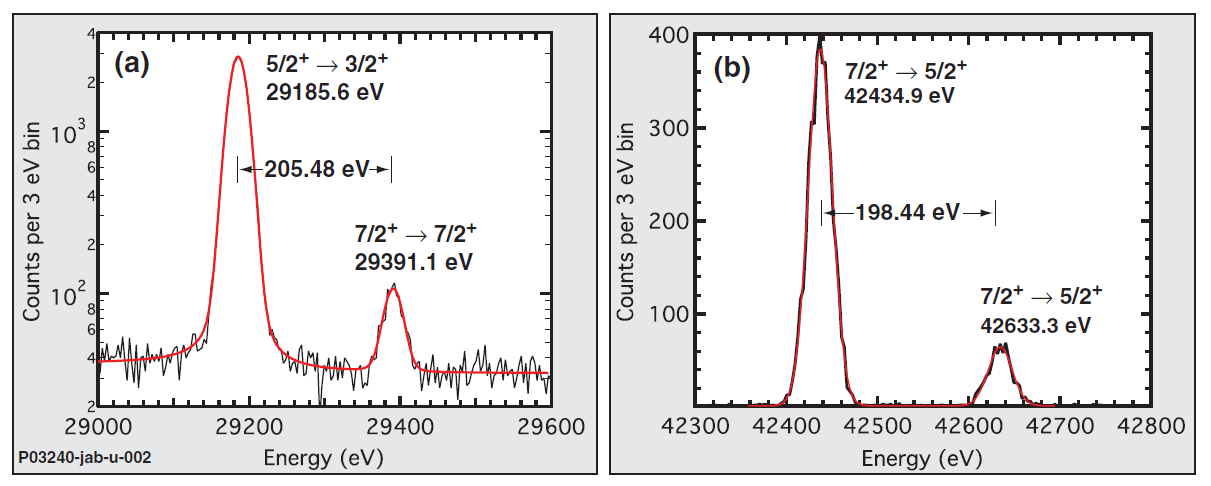}
 \vspace{-0.0cm}
  \caption{\footnotesize Spectroscopic resolution of the closely spaced $\gamma$-ray doublets of $^{229}$Th at around 29~keV and 42~keV \cite{Beck1}. The NASA micro-calorimeter spectrometer system XRS, with a measured spectral resolution of 26~eV, was used for this measurement. Based on this data the isomeric energy was constrained to $7.6\pm0.5$~eV \cite{Beck1}. The value was later slightly corrected to $7.8\pm0.5$~eV, when the 2\% inter-band mixing of the 42~keV state to the ground state was taken into consideration \cite{Beck2}. Reprinted from \cite{Beck1} with kind permission of the American Physical Society.}
 \label{beckdata}
 \end{center}
\end{figure*}
The value of 5.5~eV, as obtained in 2005 \cite{Filho}, was still below the threshold of the first ionization potential of thorium and an internal-conversion decay of the isomeric state was therefore expected to be suppressed. In 2007, however, a new measurement was published by Beck et al., which obtained the significantly larger energy value of $7.6\pm0.5$~eV \cite{Beck1}. This measurement made use of a different detection technique, using a cryogenically cooled micro-calorimeter spectrometer with a resolution of about 26~eV. By applying this significantly improved resolution, it was possible to resolve the closely spaced $\gamma$-ray lines of 29.18~keV and 29.36~keV as well as 42.43~keV and 42.63~keV (see Figs.~\ref{Beck2007} and \ref{beckdata}). This in turn allowed a new transition energy difference to be used for the energy determination of the isomeric state (Fig.~\ref{Beck2007}):
\begin{equation}
\Delta_5= [E(29.39)-E(29.18)]-[E(42.63)-E(42.43)].
\end{equation}
A further correction for the branching ratio of the 29.19~keV to ground-state decay was included, which was estimated to be 1/13 (as opposed to 1/4, assumed in \cite{Filho}).\\[0.2cm]
The value of $7.6\pm0.5$~eV, which was deduced in this way, poses a significant change in technology required for the direct detection of the isomeric decay. As the transition energy is placed above the ionization potential of neutral thorium of about 6.3~eV, internal conversion is allowed as an isomeric decay channel. Therefore, any significant chance to detect a photonic decay is only given for charged $^{229}$Th. In this case, the isomeric half-life was suggested to be about 5 hours by Beck et al.\\[0.2cm]
\noindent A minor correction to this value was introduced in 2009 by the same group \cite{Beck2}. While a possible non-zero branching ratio for the 29.19~keV to ground-state transition was already included in their previous publication, this time a non-zero branching ratio for the 42.43~keV to $^{229\mathrm{m}}$Th inter-band transition was also introduced. The estimated branching ratio is 2\%, leading only to a small correction for the isomeric energy to $7.8\pm0.5$~eV. This value was the most precise between 2009 and 2019. In 2017, a detailed re-evaluation of the Beck et al. data was performed, essentially resulting in a confirmation of the earlier result \cite{Kazakov4}. In 2019 a new branching ratio of the 29~keV state to the ground state of $1/(9.4\pm2.4)$ was obtained \cite{Masuda2019}. Considering this new branching ratio, a re-evaluation of the Beck et al. measurement leads to an energy of $8.1\pm0.7$~eV \cite{Yamaguchi2019}.
\subsubsection{Improved detector resolution}
\begin{figure}[t]
 \begin{center}
 \includegraphics[width=7cm]{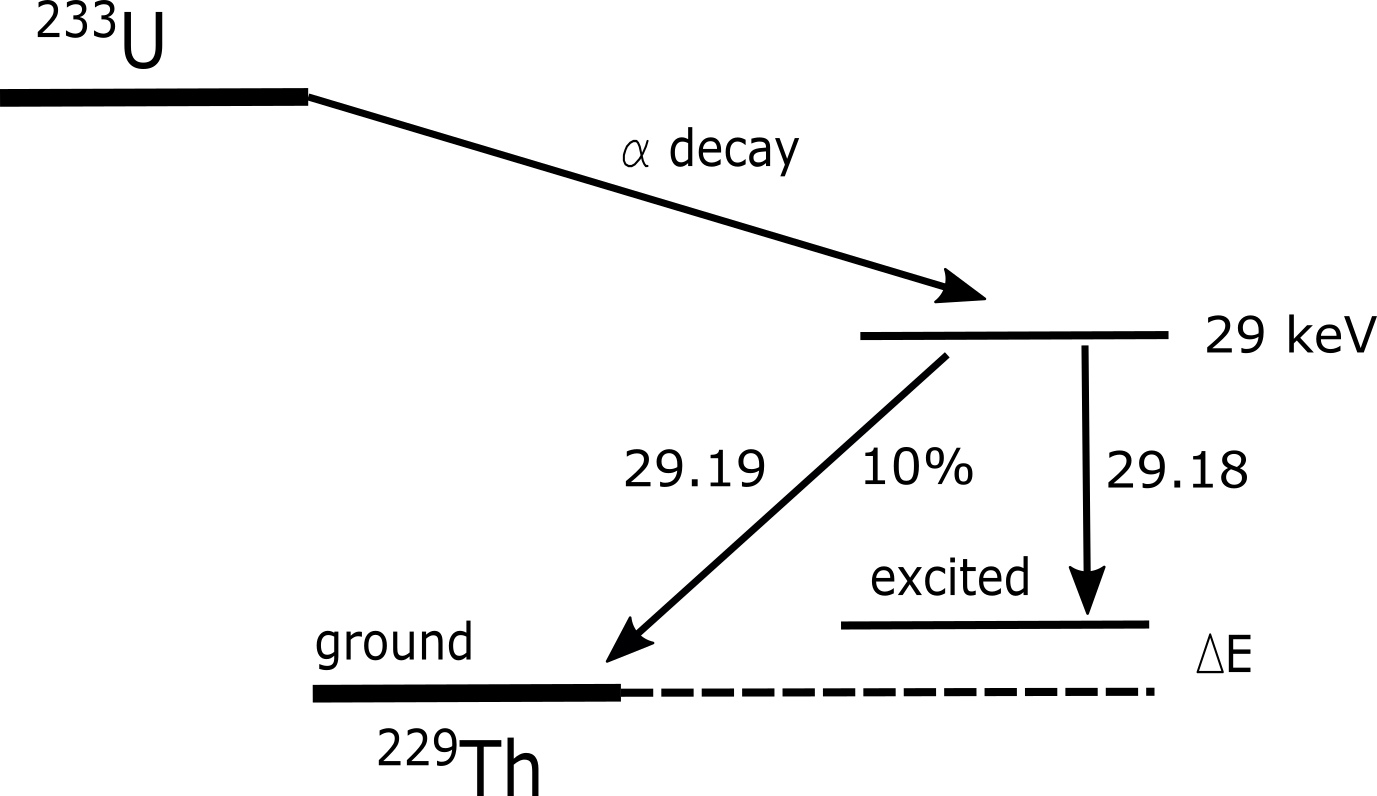}
  \caption{\footnotesize Low energy nuclear excited states of $^{229}$Th. The second excited state at 29~keV decays mostly into the metastable excited state, but with a 10\% probability also to the ground-state. This leads to a closely-spaced doublet of $\gamma$-rays that is visible as a shoulder of the 29.18~keV main peak in Fig.~\ref{geistdoublet} \cite{Sikorsky2020} (see also \cite{Geist2020}). Resolving this doublet has the potential to determine the $^{229\mathrm{m}}$Th energy with about 40~meV uncertainty \cite{Kazakov3}.}
 \label{29kevdoublet}
 \end{center}
\end{figure}
\begin{figure}[t]
 \begin{center}
 \includegraphics[width=8cm]{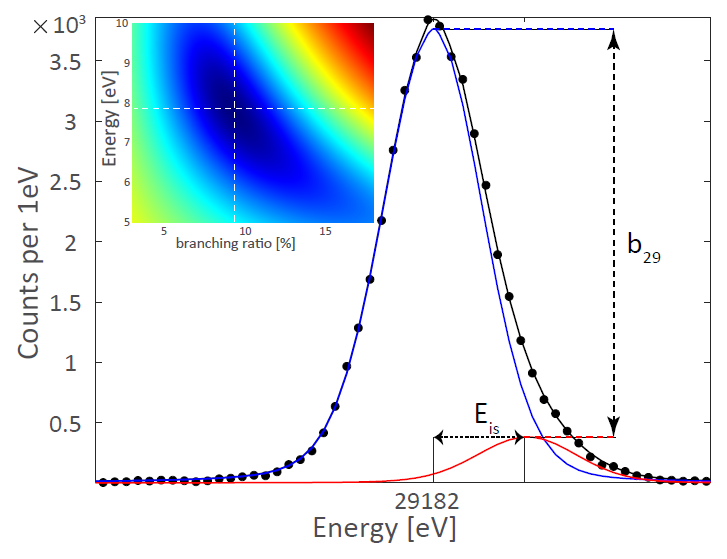}
  \caption{\footnotesize The closely spaced doublet of $\gamma$ lines at about 29.18 keV. The measured data is shown as black dots, The underlying gaussians of the 29.18~eV and 29.19~eV $\gamma$ lines are shown in blue and red, respectively. The sum of the two lines is shown as a black line. The asymmetry of the doublet is visible. The inset shows the most likely range of the branching ratio as a function of isomer energy. Reprinted from \cite{Sikorsky2020} with kind permission of the American Physical Society.}
 \label{geistdoublet}
 \end{center}
\end{figure}
With an energy resolution of 26 eV FWHM at several 10 keV absolute energy, the micro-calorimeter spectrometer system used in the Beck et al. measurements was one of the most accurate available systems at that time. However, between 2009 and 2020, a further improvement in technology arose, and metallic-magnetic micro-calorimeters have achieved resolutions of about 10~eV at 30~keV absolute energy \cite{Geist2020,Sikorsky2020,Muramatsu}. At 6~keV absolute energy, a resolution of even 2.7~eV FWHM was reported \cite{Fleischmann2009}. It is therefore intriguing to repeat the measurement with such improved detector resolution. A corresponding experiment was conducted at the Kirchhoff-Institute for Physics in Heidelberg (see \cite{Schneider}) and an energy of $8.10\pm0.17$ eV was reported in 2020 \cite{Sikorsky2020} (a preliminary value of $8.09^{+0.14}_{-0.19}$ eV can be found in \cite{Geist2020}). This is the reported value of smallest error margin obtained from $\gamma$ spectroscopy and consistent with other measurements \cite{Beck2,Seiferle2019b,Yamaguchi2019}.\\[0.2cm]
With a factor of about three further improvement of detector resolution, it would even be possible to resolve the closely spaced doublet of $\gamma$ rays originating from the second nuclear excited state of $^{229}$Th at about 29~keV. This state decays with a 90\% probability to $^{229\mathrm{m}}$Th and with 10\% probability to the nuclear ground-state \cite{Masuda2019}. A corresponding experiment was proposed in 2014 by Kazakov et al. and has the potential to determine the isomeric energy to about 40~meV uncertainty, if the 29~keV doublet were to be fully resolved \cite{Kazakov3}. The concept is visualized in Fig.~\ref{29kevdoublet}. Importantly, the asymmetry of the 29.19 keV doublet has already been made visible with 10~eV detector resolution as shown in Fig.~\ref{geistdoublet} \cite{Geist2020,Sikorsky2020}. Based on this asymmetry, the isomer's energy was constrained to be $7.84\pm0.29$~eV \cite{Sikorsky2020}.\\[0.2cm]
A very accurate measurement of the excitation energy of the 29 keV state of $^{229}$Th from the ground state was achieved in \cite{Masuda2019} (see also Sec. \ref{EXP.DIRECT.LAMBDA}). This opened up a new path for the determination of the isomer's energy via the comparison of the excitation energy from the ground-state to the decay energy to $^{229\text{m}}$Th. An energy value based on such an evaluation was published in 2019 as $8.30\pm0.92$ eV \cite{Yamaguchi2019}. In \cite{Sikorsky2020} the same technique was also applied, resulting in an isomeric energy value of $7.8\pm0.8$ eV. Although less precise than the energies obtained by other methods, these measurements provide an important consistency check. All recently published energies are reasonably consistent within their error margins.
\subsection{Search for $^{229\text{m}}$Th via direct radiative and EB decay}\label{EXP.DIRECT}
The uniquely low nuclear excitation energy of $^{229\text{m}}$Th and its potential applications have triggered a significant amount of experimental efforts aiming at the detection of photons emitted during the isomeric decay. Photons can be emitted in two processes: either in the direct ground-state decay of the isomeric state (see also App. \ref{THEO.DECEXC.GAMMADEC}) or in a process that is called ``electronic bridge" (EB) \cite{Strizhov}. During the EB decay the nucleus transfers its energy to the electronic shell, leading to the excitation of a (virtual) electronic shell state. Subsequently, this excited shell state decays under the emission of photons (see also App. \ref{THEO.HIGHDEC.EB}). Importantly, photons emitted during the EB decay can be of lower energy than the actual isomeric energy, as the electronic decay will in general proceed in several steps. The emission of photons, ideally observed from the isomer's direct decay to the nuclear ground state, would offer great potentials for a precise $^{229\text{m}}$Th energy determination. However, until today all efforts have failed to observe a secure signal of photons emitted during the isomeric decay. A sufficient suppression of non-radiative decay channels, most importantly internal conversion, is a central challenge for all experiments that are searching for light emitted during the isomeric decay.\\[0.2cm]
This subsection is structured according to the different ways used for populating the isomeric state as indicated in Fig.~\ref{energy_overview}. These can be: (1) the population via a natural nuclear decay branch (e.g., the $\alpha$ decay of $^{233}$U or the $\beta$ decay of $^{229}$Ac) discussed in Sec. \ref{EXP.DIRECT.NATURAL}; (2) the direct excitation from the ground state with light, presented in Sec. \ref{EXP.DIRECT.DIRECT}; (3) excitation making use of the electronic bridge (EB) mechanism or nuclear excitation by electron transition (NEET), discussed in Sec. \ref{EXP.DIRECT.IEB}; (4) experiments making use of the process of nuclear excitation by electron capture (NEEC), which is the reverse of the internal conversion decay, discussed in Sec. \ref{EXP.DIRECT.NEEC}; and (5) the excitation of $^{229\text{m}}$Th from the ground state making use of a $\Lambda$ excitation scheme using a nuclear excited state of higher energy, presented in Sec. \ref{EXP.DIRECT.LAMBDA}.
\subsubsection{Population via a natural decay branch} \label{EXP.DIRECT.NATURAL}
\begin{figure}[t]
 \begin{center}
 \includegraphics[width=6cm]{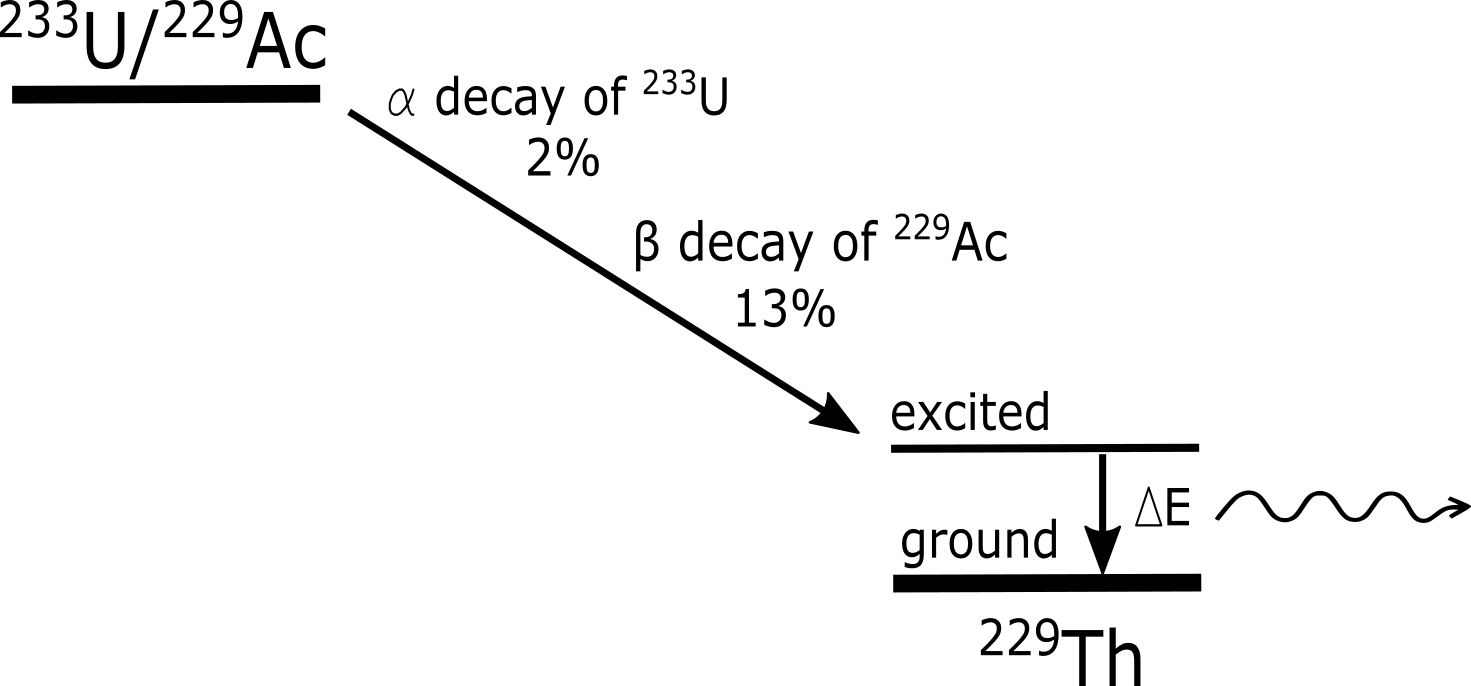}
  \caption{\footnotesize Population of $^{229\text{m}}$Th via the 2\% decay branch in the $\alpha$ decay of $^{233}$U, or via the 13\% decay branch in the $\beta$ decay of $^{229}$Ac, followed by photon emission.}
 \label{naturalpop}
 \end{center}
\end{figure}
The $\gamma$ spectroscopic measurements discussed in Sec.~\ref{EXP.GAMMA} make use of $^{229\text{m}}$Th populated naturally in nuclear decay. Most experiments populate the isomeric state in the $\alpha$ decay of $^{233}$U for the practical reason that the $^{233}$U half-life is long ($1.59\cdot10^5$ years), and it is therefore a continuous source of $^{229}$Th nuclei. However, the branching ratio of the $^{233}$U $\alpha$ decay to $^{229\text{m}}$Th is rather small, about 2\%. Opposed to that, the $\beta$ decay of $^{229}$Ac has a significantly larger branching ratio to $^{229\text{m}}$Th of more than 13\%. This advantage comes at the cost of a short $^{229}$Ac half-life of only 62.7 minutes. Therefore any experiment with $^{229}$Ac requires frequent replenishment via nuclear reactions at accelerator facilities. The concept is visualized in Fig.~\ref{naturalpop}.\\[0.2cm]
\paragraph*{Early experiments searching for a radiative decay}
\ Before 2007, the internal conversion (IC) decay channel of $^{229\text{m}}$Th was assumed to be energetically forbidden, as the isomeric energy was expected to be 3.5 eV and therefore lower than the ionization potential of the neutral thorium atom of 6.3~eV \cite{Trautmann1994}. For this reason, searches for a radiative $^{229\text{m}}$Th decay were performed with neutral $^{229}$Th. Research along this line was published in \cite{Irwin} by Irwin and Kim, where the detection of light emitted by the direct decay of the $^{229}$Th isomeric state was claimed. Some theoretical work was published as a consequence of this observation \cite{Dykhne3,Karpeshin9,Karpeshin10,Kalman2}.\\[0.2cm]
The same spectral features were re-observed by Richardson et al. \cite{Richardson}, where a liquid $^{233}$U source was used and comparisons between $^{233}$U and $^{232}$U were performed. In contrast to the earlier work of Irwin and Kim, Richardson et al. obtained better spectroscopic resolution and a substructure of the 3.5~eV line became visible. The assignment of this line structure to the $^{229}$Th isomeric decay was considered to be possible, but not unambiguous \cite{Richardson}.\\[0.2cm]
In 1999, light emitted from $^{233}$U samples was unambiguously shown to originate from $\alpha$-particle induced fluorescence of nitrogen \cite{Utter,Shaw} and from luminescence of the uranyl ion \cite{Young}. Therefore the direct detection of $^{229\text{m}}$Th remained an open question \cite{Irwin2}. A review of these early attempts of $^{229}$Th isomer detection can be found in \cite{Kim}.\\[0.2cm]
The search for light emitted during the $^{229\text{m}}$Th decay was continued in the following years in liquid solutions \cite{Moore,Zimmermann}, however with a negative result. Given today's knowledge about the isomer's energy, these early attempts did not have any chance to detect the isomeric decay as the water solutions used are not transparent to VUV photons. Studies with solid $^{233}$U samples were also carried out, but no photons emitted in the isomeric decay were observed \cite{Kasamatsu}. Here the isomeric decay will occur on a short timescale via IC under emission of an electron.\\[0.2cm]
\paragraph*{$^{229\text{m}}$Th in large band-gap materials}
\ In 2007, a new and improved value of $7.6$ eV (later corrected to $7.8$ eV) for the isomeric energy was determined \cite{Beck1,Beck2}. This energy is larger than the ionization potential of neutral $^{229}$Th, resulting in IC being the dominant isomeric decay mode in the neutral $^{229}$Th atom and leading to a lifetime reduction by a factor of $10^9$ \cite{Karpeshin6}. This discovery led to a new class of experiments, where $^{229}$Th, produced in the $\alpha$ decay of $^{233}$U, was embedded into materials with a large band-gap. In a solid-state environment, the material's band-gap replaces the ionization potential. If it is larger than the isomer's energy, IC is expected to be energetically forbidden \cite{Tkalya2b,Tkalya9}. In this case, the radiative decay (direct or EB) would again become the dominant decay channel, and the lifetime would be prolonged. Materials with a band-gap larger than 7.8~eV include CaF$_2$ and MgF$_2$.\\[0.2cm]
For these experiments, typically a thin layer of $^{233}$U with a large surface area is used for the production of $^{229}$Th. A significant amount of the $^{229}$Th $\alpha$-recoil isotopes, produced in the $^{233}$U decay, can leave the source material and is implanted into a large band-gap absorber plate. Photons emitted from the absorber plate are detected, e.g., with a photo-multiplier-tube (PMT). Spectral filters can be used to obtain information about the emitted photon energies. Experiments along this line were carried out since 2007 at various institutes around the world \cite{Zimmermann,JBurke,Swanberg,Zhao,Yasuda2017,Barker2018,Knize2019}, however with negative results. A direct detection of the isomeric decay by this method was reported in 2012 \cite{Zhao}. This result is, however, controversial \cite{Swanberg,Peik3,Bilous} and has so far not been reproduced by any other group.\\[0.2cm]
A similar experiment was carried out at the Maier-Leibnitz-Laboratory (MLL) in Garching, Germany \cite{Thirolf1,Thirolf2008,Wense1,Thirolf2,Seiferle2,Seiferle,Wense3}. Here a $^{229}$Th$^{3+}$ ion beam was generated from $\alpha$ recoil isotopes emitted by a $^{233}$U source. The ions were implanted into a MgF$_2$-coated micro-electrode of 50~$\mu$m diameter to obtain a point-like light source. Light, emitted in the isomeric decay was focused onto a CsI-coated micro-channel plate (MCP) detector with two parabolic mirrors, offering a high signal-to-background ratio of up to $10,000:1$. Despite a high sensitivity to photons emitted in the isomeric decay, no signal was detected, pointing toward a significant non-radiative decay branch \cite{Wense3}.\\[0.2cm]
An alternative method making use of a large-band-gap material was proposed in \cite{Hehlen} and is currently under investigation at TU Vienna. In this concept, $^{233}$U-doped CaF$_2$ crystals are grown, and the search for light emitted in the isomeric decay is performed. In \cite{Stellmer4} it was shown that, despite the background radio-luminescence originating from the $^{233}$U decay chain, there is a chance of detecting the isomeric decay. In case of a non-observation, an upper limit for the isomer's radiative decay branch could be obtained. So far no result has been reported.\\[0.2cm]
Investigation of the isomer following its excitation in the $^{229}$Ac $\beta^-$ decay and implantation into a crystal lattice was proposed in \cite{Duppen} (see also \cite{Kraemer}). The possibility of annealing the crystal to remove lattice defects after the $^{229}$Ac implantation, but before the $^{229}$Ac $\beta$ decay, could be advantageous in the search for a radiative decay channel in this approach. Importantly, the efficiency of populating $^{229\text{m}}$Th in the $^{229}$Ac $\beta$ decay (13\%) is significantly larger than that in the $^{233}$U $\alpha$ decay (2\%). $^{229}$Ac has a half-life of 62.7 minutes. For this reason, it has to be continuously produced by nuclear fusion processes at the ISOLDE Facility at CERN. Observation of the isomer's radiative decay would provide a spectroscopic determination of its energy. First experimental results were reported \cite{Verlinde2019}.
\subsubsection{Population via direct excitation} \label{EXP.DIRECT.DIRECT}
\begin{figure}[t]
 \begin{center}
 \includegraphics[width=6cm]{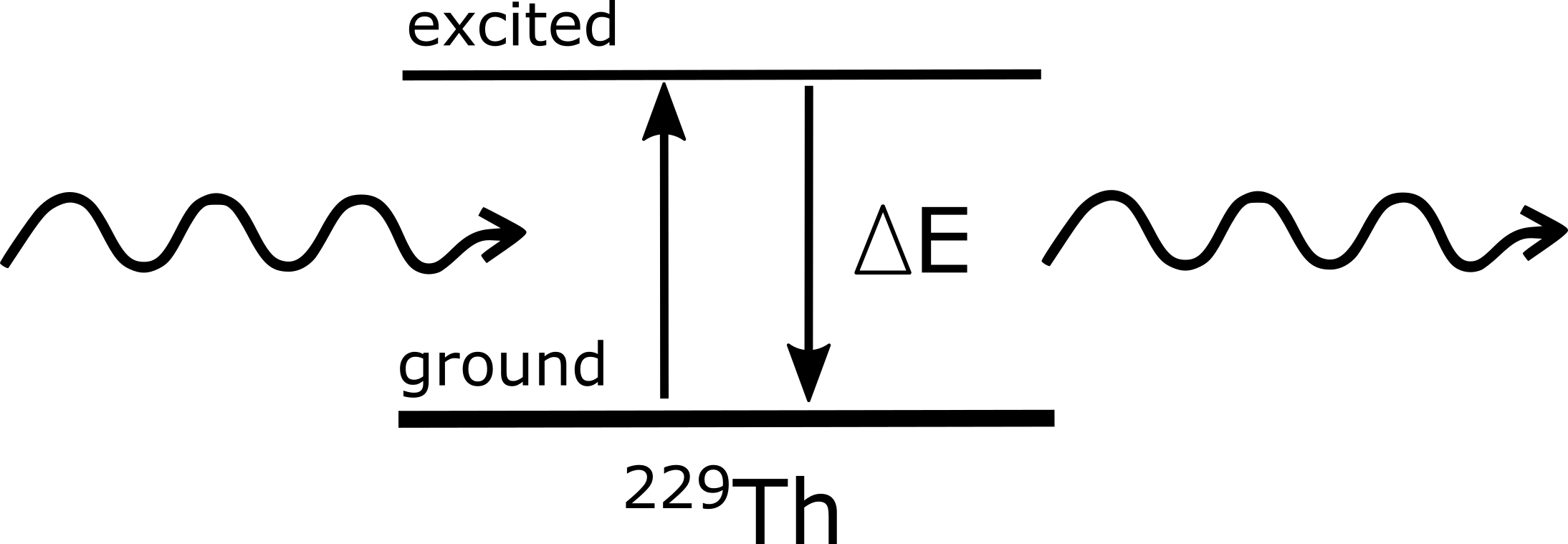}
  \caption{\footnotesize Concept of direct photon excitation of $^{229\text{m}}$Th from the nuclear ground state, followed by the observation of light emitted in the isomer's radiative decay. Excitation and decay will occur time separated with the time scale of the isomeric lifetime.}
 \label{direct_concept}
 \end{center}
\end{figure}
A different path of investigation of the isomeric properties is via the direct population of the isomeric state from the nuclear ground state, followed by the observation of photons emitted in the isomeric decay as sketched in Fig. \ref{direct_concept}. The probability of direct absorption of light was discussed in theoretical work in the 1990s (see App. \ref{THEO.DECEXC.GAMMAEXC}) \cite{Tkalya1,Tkalya9}.\\[0.2cm]
\paragraph*{Search for $^{229\text{m}}$Th with synchrotron light}
\ An experiment that makes use of the isomer's direct excitation was proposed in \cite{Rellergert}. For this purpose, a large band-gap material (which suppresses non-radiative decay channels) is doped with a high concentration of $^{229}$Th. These crystals, with a $^{229}$Th doping concentration of up to $4.1\cdot10^{17}$ cm$^{-3}$, are irradiated with broad-band VUV light as, for example, provided by synchrotrons or D$_2$ lamps, to excite the isomeric state. The successful excitation is inferred from photons emitted during the isomeric decay. The development of $^{229}$Th-doped crystals is driven by two groups, located at the University of California (UCLA), where LiSrAlF$_6$ is investigated as a host material, and at the Technical University (TU) Vienna, where CaF$_2$ is considered. These crystals are expected to also provide the basis for solid-state nuclear frequency standards \cite{Kazakov1}. Significant progress has been made in crystal development and theoretical understanding \cite{Jackson,Rellergert2,Hehlen,Kazakov2,Dessovic,Ellis,Stellmer2,Stellmer3,Schreitl2016,Barker2018,Gouder2019,Pimon}. Two experimental results using synchrotron radiation have been reported, which were, however, negative: the first one performed by UCLA, using the Advanced Light Source (ALS) of the Law\-rence Berkeley National Laboratory \cite{Jeet,Jeet2018} the second one performed by the Technical University of Vienna, using the Metrology Light Source (MLS) in Berlin, Germany \cite{Stellmer2018a}. A theoretical investigation that supports the experiments was published in \cite{Tkalya99}.\\[0.2cm]
Results of a slightly different experiment were published in 2015 \cite{Yamaguchi}. Here $^{229}$Th was adsorbed onto a CaF$_2$ surface and irradiated with undulator radiation. No photons in the expected wavelength range could be observed, which can be explained by the chemical structure of thorium on the CaF$_2$ surface \cite{Borisyuk}.\\[0.2cm]
\paragraph*{Search for $^{229\text{m}}$Th with laser light}
\begin{figure*}[t]
 \begin{center}
 \includegraphics[width=14cm]{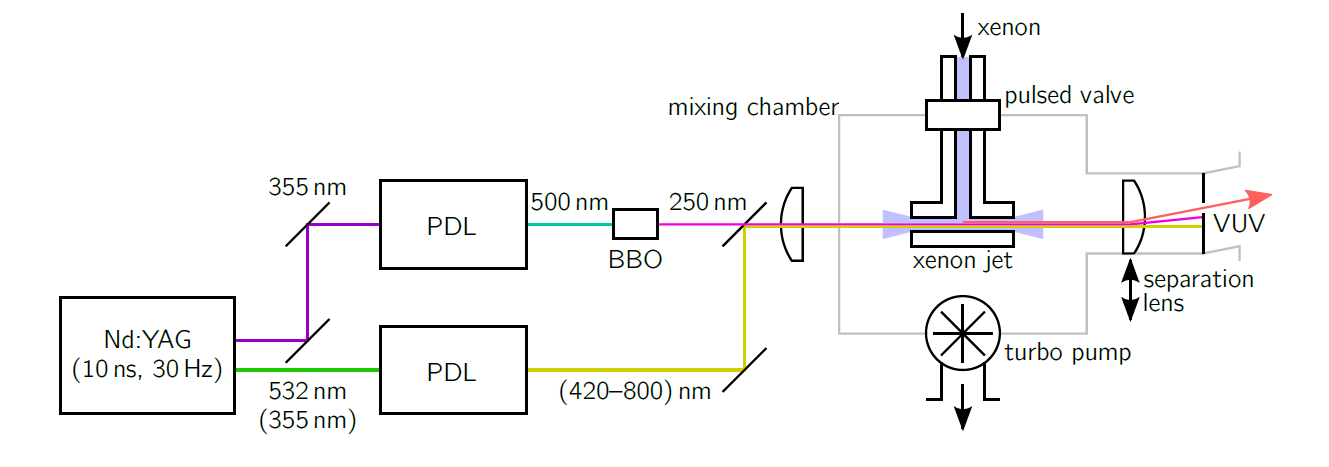}
  \caption{\footnotesize Conceptual sketch of the laser system used at UCLA in the search for the nuclear excitation. Two pulsed dye lasers (PDLs) are pumped by a Nd:YAG laser and are mixed in xenon gas via four-wave mixing in order to generate tunable VUV light \cite{Jeet2018}. Reprinted from \cite{Jeet2018} with kind permission of J. Jeet.}
 \label{UCLA_lasersystem}
 \end{center}
\end{figure*}
\ An alternative approach, which is currently being investigated at UCLA, is to excite $^{229\text{m}}$Th in a crystal lattice environment from the nuclear ground-state via direct laser excitation using a broad-band VUV laser source, generated by four-wave mixing in a noble-gas \cite{Jeet2018}. The laser system consists of two pulsed dye lasers (PDLs), which are mixed via four-wave mixing in a xenon environment to generate VUV light as shown in Fig.~\ref{UCLA_lasersystem}. The generated VUV light is tunable between 148 and 179~nm. While the population of $^{229\text{m}}$Th could be probed via the isomer's radiative decay, a precise energy determination would be achieved in the excitation channel. The achievable precision would be determined by the bandwidth of the laser light used for excitation, which is $\approx10$~GHz  ($40$~$\mu$eV). An alternative approach is based on a VUV frequency comb generated from the 5th harmonic of a Ti:Sapphire laser and is under development at TU Vienna \cite{Winkler,Seres2019}. A system that operates with the 7th harmonic of an Yb:doped fiber laser is located at JILA, Boulder, CO \cite{Wense2019c,Zhang2020}.
\subsubsection{Population via EB excitation} \label{EXP.DIRECT.IEB}
\begin{figure}[t]
 \begin{center}
 \includegraphics[width=8cm]{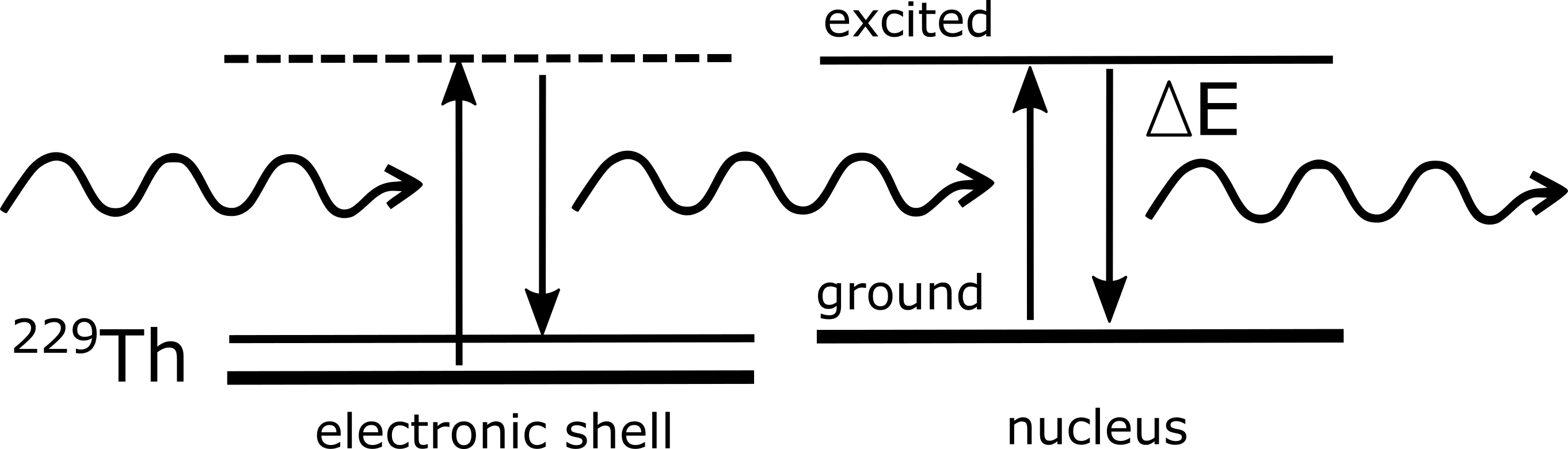}
  \caption{\footnotesize Concept of the isomer's excitation via electronic bridge (EB). The electronic shell is excited and the energy is transferred to the nucleus, which subsequently decays via emission of a photon. Note, that the electronic excited level could either be a real state (in the case of NEET) or a virtual level (in the case of EB excitation).}
 \label{electronicbridge}
 \end{center}
\end{figure}
In the electronic bridge (EB) excitation mechanism an electronic shell level is excited, and the energy is subsequently transferred to the nucleus (see App. \ref{THEO.HIGHEXC.IEB} and Fig.~\ref{electronicbridge}). The process is closely related to nuclear excitation via electron transition (NEET, see App. \ref{THEO.HIGHEXC.NEET}) and the expressions are often used as synonyms. Importantly, the EB excitation, although a high-order process, is expected to possess a larger efficiency than the isomer's direct excitation. It has therefore been the subject of significant theoretical investigations (important publications along this line are \cite{Tkalya5,Tkalya2,Kalman,Karpeshin1994,Karpeshin2,Tkalya1,Karpeshin3,Tkalya9,Porsev3,Karpeshin7,Karpeshin12}). Further, it is a relevant excitation channel for many experimental concepts.\\[0.2cm]
An early experimental proposal that makes use of the EB excitation mechanism can be found in \cite{Inamura1}. Here it was the approach to excite the isomeric state in a solid $^{229}$Th sample with the help of a hollow-cathode discharge lamp. Delayed photons as emitted either in the isomer's direct decay or in an electronic bridge channel were proposed to be used for identification of the isomer. In later work, also the potential for $\alpha$ decay was considered \cite{Inamura2,Inamura3}. Results were reported in \cite{Inamura4} purely based on the observation of $\alpha$ decay.\\[0.2cm]
In more modern experimental proposals, a two-step EB scheme is used for the excitation of $^{229\text{m}}$Th \cite{Campbell2,Porsev3,Campbell3,Radnaev,Karpeshin7,Karpeshin11}. In the two-step EB process, an electronic shell state is first excited from the atomic ground state and the EB excitation of the nucleus is achieved in a second step, starting from the excited electronic state (see App. \ref{THEO.HIGHEXC.IEB}). The two-step electronic-bridge excitation has the advantage that the laser light used for spectroscopy is of longer wavelength and therefore available with larger intensities. The idea is to load $^{229}$Th ions into a Paul trap and excite the electronic shell via laser irradiation. The nuclear excitation in the EB process can then either be probed via delayed photons emitted in the isomer's direct decay, via the EB decay channel or by laser-spectroscopy of the atomic shell's hyperfine structure. Here we focus on experimental efforts aimed at probing the nuclear excitation by emitted photons, which has been under investigation at PTB in Germany. For experiments aimed at probing the hyperfine structure, the reader is referred to Sec. \ref{EXP.HYPER.IEB}.\\[0.2cm]
Detailed theoretical investigations of the thorium electronic shell were performed to lay the foundations for the experimental studies of the EB process \cite{Dzuba,Safronova3}. At PTB the thorium ions are produced by laser ablation \cite{Zimmermann1} and stored in a linear Paul trap \cite{Zimmermann}. A pulsed laser system has been developed to excite the $^{229}$Th ion shell. Subsequently emitted photons are detected by a PMT \cite{Herrera1,Herrera3,Herrera2}. Two Paul traps are in operation in parallel, one for $^{229}$Th and one for $^{232}$Th, as an isomeric signal can only be identified by performing comparisons.\\[0.2cm]
Reported experimental results include the detection of 43 previously unknown energy levels in $^{232}$Th$^{1+}$ \cite{Herrera2} and the observation of an unexpected negative isotope shift in $^{229}$Th$^{1+}$ \cite{Okhapkin}. In 2019 the observation of 166 previously unknown levels between 7.8 and 9.8~eV in $^{229}$Th$^{1+}$ was reported, offering the opportunity to investigate the EB mechanism also at higher energy levels \cite{Meier2019}.\\[0.2cm]
\subsubsection{Population via nuclear excitation by electron capture} \label{EXP.DIRECT.NEEC}
\begin{figure}[t]
 \begin{center}
 \includegraphics[width=7cm]{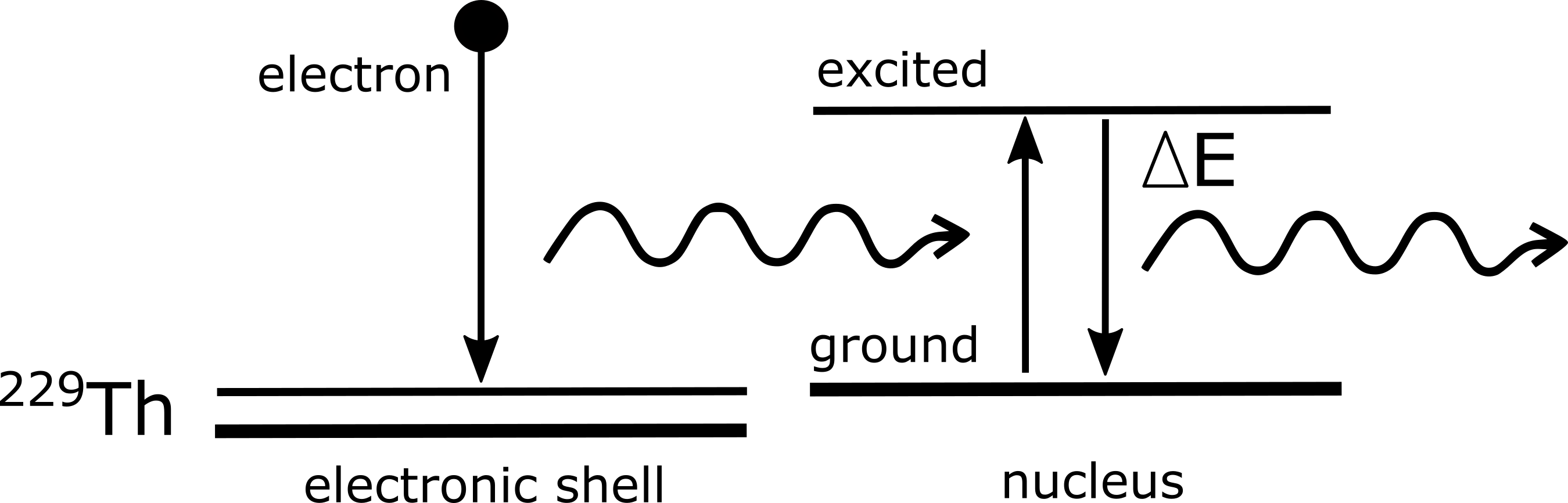}
  \caption{\footnotesize Concept of the isomer's excitation via NEEC. A $^{229}$Th ion captures an electron from the continuum. The released energy is transferred to the nucleus, which can subsequently decay by photon emission.}
 \label{NEEC}
 \end{center}
\end{figure}
Nuclear excitation by electron capture (NEEC, see App. \ref{THEO.HIGHEXC.NEEC}) is a process in which an electron recombines with an ion and part of the recombination energy is transferred to the nucleus, thereby exciting a nuclear state as shown in Fig.~\ref{NEEC}. NEEC is the reverse of the isomer's decay via internal conversion. Due to the large expected NEEC efficiency, it was already proposed to be used for the isomer's excitation in 1991, at the time under the original name ``reverse internal electron conversion" (RIEC) \cite{Strizhov}. Two different lines of experimental studies where proposed, making use of the NEEC process for the excitation of $^{229\text{m}}$Th. These are: (1) experiments with highly charged ions at storage rings, (2) experiments making use of a $^{229}$Th plasma.\\[0.2cm]
\paragraph*{NEEC at storage rings}
\ For an experimental investigation of the NEEC process, $^{229}$Th ions are stored in a high-energy storage ring (e.g., the ESR at GSI in Germany or the CSR at the IMP in Lanzhou, China). When these ions catch electrons that fulfill the resonance condition, namely that the electron's kinetic energy plus their binding energy after recombination equals the energy of the isomeric state, there is an enhanced probability for exciting $^{229}$Th into its isomeric state \cite{Palffy2,Palffy3}. By tuning the energy of an electron beam and monitoring the number of recombinations by detecting the ions' charge states, it is possible to find the resonance and thus to determine the isomer's energy. Corresponding proposals for the excitation of $^{229\text{m}}$Th can be found in \cite{Brandau1,Brandau2,Ma}. For shell processes, the method is known as dielectronic recombination (DR) \cite{Brandau1}. In addition to $^{229}$Th, other low-energy nuclear states such as $^{235\mathrm{m}}$U$^{89+}$ have also been considered with this approach \cite{Brandau2}.\\[0.2cm]
\paragraph*{NEEC in a plasma}
\ Excitation of $^{229\text{m}}$Th via NEEC in a laser-generated plasma is studied at MEPhI (Moscow Engineering Physics Institute) \cite{Borisyuk2018a,Lebedinskii2019}. A $^{229}$Th plasma is generated by laser ablation from a solid thorium target. Excitation of the isomeric state is predominantly achieved via NEEC in the plasma. Subsequently, the ions are implanted into a SiO$_2$ surface, which is expected to provide a sufficiently large band-gap to suppress the isomer's internal conversion decay channel. Thorium-oxide coatings on a SiO$_2$ surface were investigated as a substrate for observing the isomer's radiative decay and for the development of a solid-state nuclear clock in \cite{Borisyuk1,Borisyuk2,Borisyuk3,Troyan5,Troyan2015c,Troyan4,Borisyuk2017d}. In 2018, laser-plasma excitation of the isomeric state was reported, together with a determination of the isomer's excitation energy to $7.1\pm0.1$~eV and a value for the isomer's radiative half-life of $1880\pm170$~s \cite{Borisyuk2018b} (see also \cite{Borisyuk2018c,Lebedinskii2019}). The observations are, however, subject to controversial discussion within the community \cite{Thirolf2019b}.\\[0.2cm]
\subsubsection{Lambda excitation of $^{229\text{m}}$Th}\label{EXP.DIRECT.LAMBDA}
\begin{figure}[t]
 \begin{center}
 \includegraphics[width=6cm]{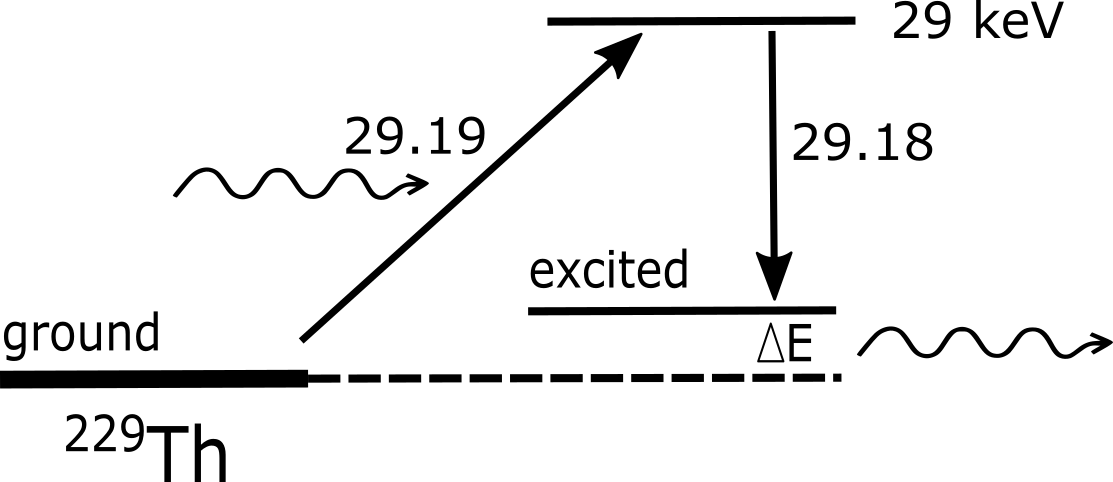}
  \caption{\footnotesize Population of $^{229\text{m}}$Th by means of a Lambda-excitation scheme via the 29~keV second nuclear excited state of $^{229}$Th. In a VUV transparent crystal, the isomeric state is expected to decay via emission of a photon. In case of observation of these photons, the isomeric energy could be spectroscopically determined \cite{Masuda2019}.}
 \label{29keVphoton}
 \end{center}
\end{figure}
\begin{figure*}[t]
 \begin{center}
 \includegraphics[width=14cm]{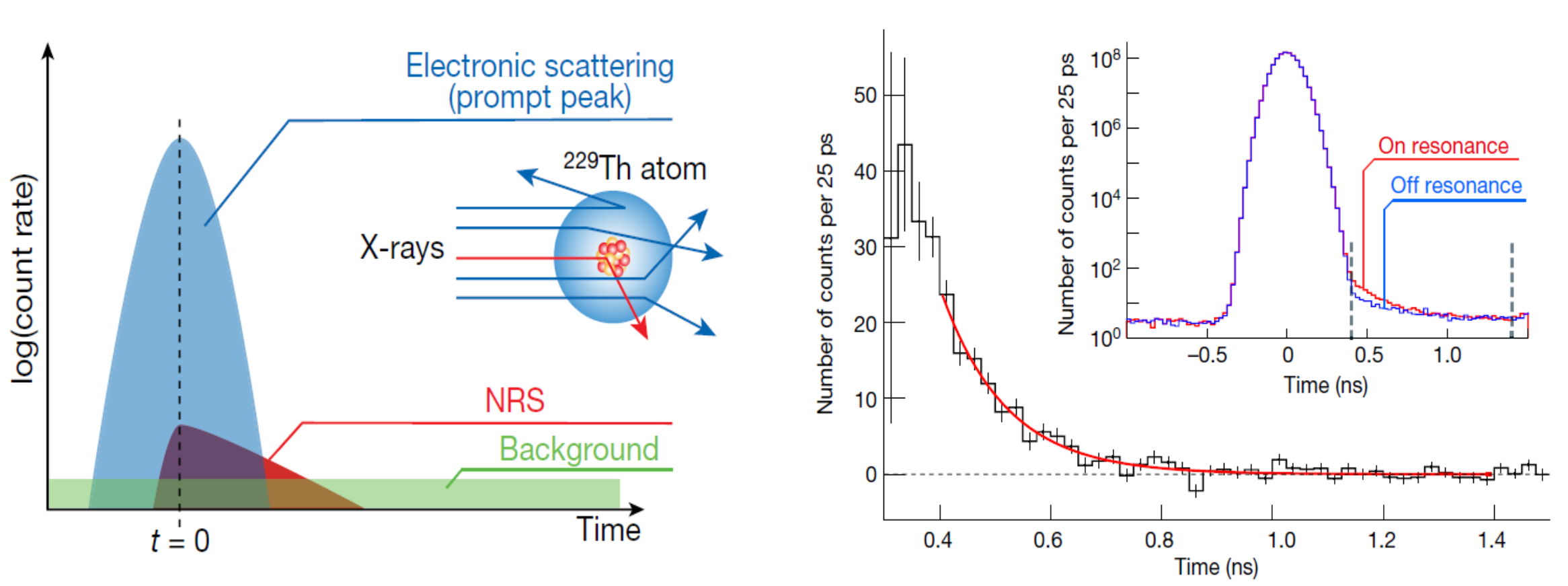}
  \caption{\footnotesize Left: experimental concept used by Masuda et al. for the excitation of the second nuclear excited state of $^{229}$Th at 29.19~keV energy \cite{Masuda2019}. A $^{229}$Th target is irradiated with pulsed synchrotron light. The excitation of the nuclear state is probed via delayed X-ray photons emitted during the nuclear deexcitation. Right: experimentally obtained X-ray signal. In its decay, the 29~keV nuclear excited state populates the isomeric state with a probability of about 90\%, achieving a secure experimental excitation of $^{229\text{m}}$Th from the ground-state. Reprinted from \cite{Masuda2019} (Springer Nature) with kind permission of the European Physical Journal and K. Yoshimura.}
 \label{Masudagraphics}
 \end{center}
\end{figure*}
In the Lambda excitation scheme of $^{229\text{m}}$Th, synchrotron light is used for excitation of a higher lying nuclear state of $^{229}$Th that decays with a certain probability to the isomeric state. A sketch of the experimental concept is shown in Fig. \ref{29keVphoton}. The population scheme was proposed in \cite{Tkalya2b} and later also discussed in \cite{Tkalya9}. The nuclear states at 29 keV and 97 keV excitation energy were considered as good candidates. An early experimental study along this line of research was published in \cite{Gangrsky}. A similar interesting proposal is the excitation of the 97 keV state via Mössbauer spectroscopy, as discussed in \cite{Kim}.\\[0.2cm]
In 2019, excitation of the nuclear state at 29~keV via synchrotron radiation was reported \cite{Masuda2019} (see also \cite{Yoshimi2018}). A $^{229}$Th oxide target of 1.8~kBq activity and 0.4~mm diameter was placed in the focus of a 29.19~keV X-ray beam generated via undulators at the high-brilliance X-ray beamline of the SPring-8 synchrotron facility in Japan. The 29.19~keV state of $^{229}$Th decays either to the isomeric excited state or to the ground-state, mainly by internal conversion with a short half-life of about 80~ps (the shortest half-life that has ever been measured with this technique). The IC decay of the state is accompanied by the emission of many specific X-rays, of which the L-shell lines were used to probe the state's excitation. The incident light was pulsed with 40 ps pulse duration and X-rays emitted after the synchrotron pulse were detected with an array of silicon avalanche photodiode (Si-APD) sensors from Hamamatsu that were specifically developed for this experiment. In parallel, the energy of the incident X-rays was determined to an accuracy of 0.07 eV via measurement of the pair of Bragg angles generated in a Si(440) reference crystal. A scheme of the concept is shown in Fig.~\ref{Masudagraphics} together with experimental data obtained from the nuclear resonance scattering (NRS) of the nuclear excited state at 29.19~keV \cite{Masuda2019}.\\[0.2cm]
In this experiment a secure population of the isomer from the ground state is achieved, opening a new path for the exploration of the isomer's radiative decay. The 29~keV state decays with a probability of about 90\% to $^{229\text{m}}$Th, which can be used to populate the isomeric state in a crystal-lattice environment. If the radiative decay of $^{229\text{m}}$Th in the crystal is observed, the isomeric energy could be spectroscopically determined with high precision.\\[0.2cm] \cite{Masuda2019} already constrains the isomeric energy to between 2.5 and 8.9~eV based on the comparison between the energy required to excite the 29.19~keV state and the energy of the $\gamma$ radiation emitted in its decay to the first excited state that was determined in previous studies. As the excitation energy of the 29~keV state could be determined to less than a tenth of an eV accuracy, the precision of this measurement is limited by the accuracy of the $\gamma$ decay and the branching ratios of the decay to the ground and first excited state. The branching ratio to the ground-state was determined in \cite{Masuda2019} to be $1/(9.4\pm2.4)$, which is slightly different than the previously assumed value of $1/(13\pm1)$ \cite{Beck1}.\\[0.2cm]
Based on this new information in combination with additional $\gamma$ spectroscopy data, a value for the isomer's energy of $8.30\pm0.92$~eV was obtained in \cite{Yamaguchi2019}. A re-evaluation of the measurement of \cite{Beck1,Beck2} was also carried out considering the new branching ratio, from which an energy of $8.1\pm0.7$~eV is inferred \cite{Yamaguchi2019}. A further comparison of the decay energy with the 29.19~keV excitation energy was performed in \cite{Sikorsky2020}, resulting in an energy value of $7.8\pm0.8$~eV.
\subsection{Search for $^{229\mathrm{m}}$Th via internal conversion}\label{EXP.IC}
\begin{figure}[t]
 \begin{center}
 \includegraphics[width=8cm]{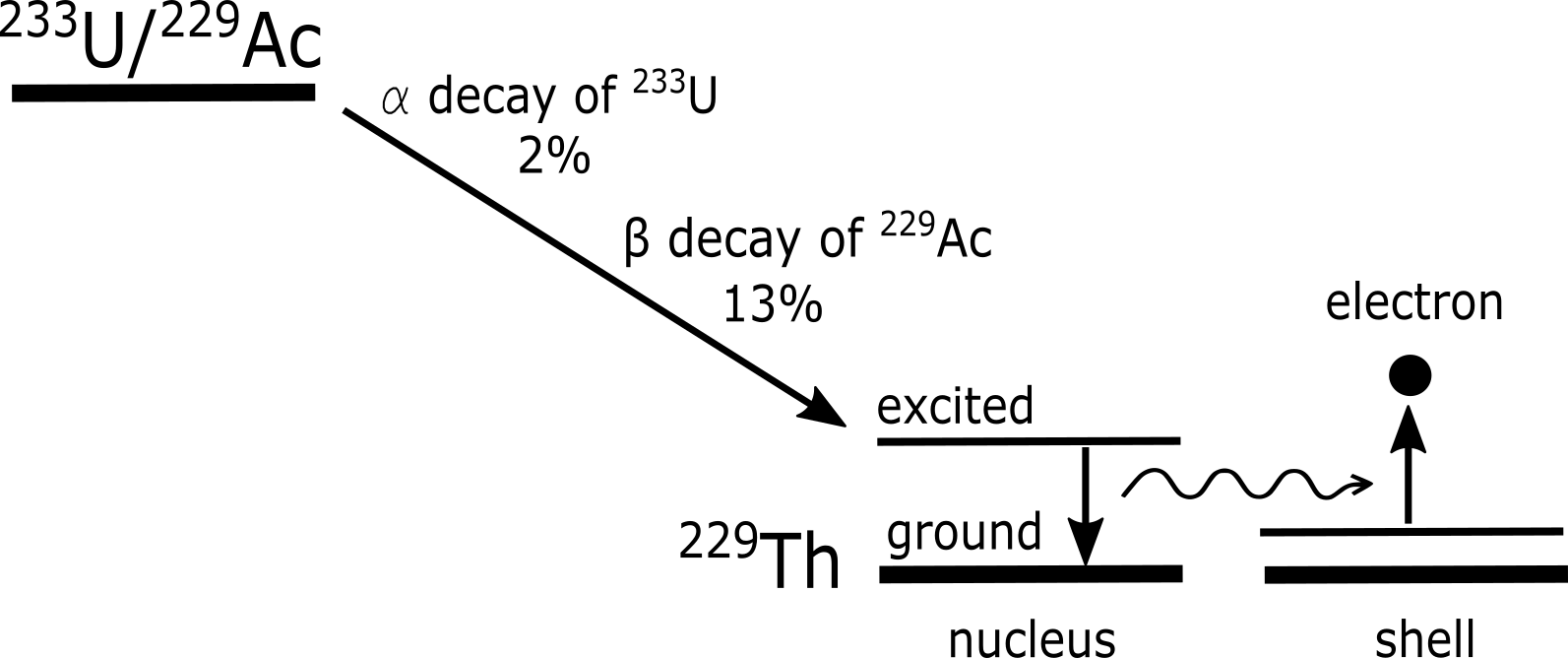}
  \caption{\footnotesize Detection of the isomer via the observation of IC electrons. The isomeric state is populated either in the $\alpha$ decay of $^{233}$U or in the $\beta$ decay of $^{229}$Ac. It then transfers its energy to the electronic shell, where an electron is ejected.}
 \label{ICdecay}
 \end{center}
\end{figure}
With the revision of the isomeric energy value to 7.6~eV in 2007 \cite{Beck1}, it became evident that, in the neutral thorium atom, the isomeric state will predominantly decay by internal conversion (IC). In IC, the nucleus couples to the electronic shell, transferring its excitation energy to a shell electron, which is subsequently ejected (see Fig.~\ref{ICdecay} and App. \ref{THEO.HIGHDEC.IC}). For $^{229\text{m}}$Th, an IC branching ratio of about $10^9$ was theoretically predicted \cite{Karpeshin6}. It is an advantage in the search for an IC decay channel that experimental conditions under which IC will dominate can easily be prepared. Therefore one can ignore other decay branches such as radiative decay. Early considerations of an IC decay of $^{229\mathrm{m}}$Th were already made in 1991 by Strizhov and Tkalya \cite{Strizhov}, and experimental investigations followed in 1995 \cite{Vorykhalov}. These experiments, however, assumed an IC-decay half-life in the range of hours, and there was no chance of observing the isomeric decay. In the following years IC was not further investigated, as it was thought to be energetically forbidden due to the expected isomeric energy of only 3.5~eV \cite{Helmer_Reich}. After 2007, experiments searching for an IC decay channel were carried out at the Law\-rence Livermore National Laboratory (LLNL), Livermore, USA, however with a negative result \cite{Swanberg,Wakeling}.\\[0.2cm]
In the following, a discussion of the detection of the isomer's IC decay channel \cite{Wense2} will be provided, followed by the determination of the isomer's IC lifetime \cite{Seiferle3} and improved constraints of the $^{229\text{m}}$Th energy value \cite{Seiferle2019b}. Finally, continued experimental efforts will be detailed. 
\begin{figure*}[t]
 \begin{center}
 \includegraphics[width=16cm]{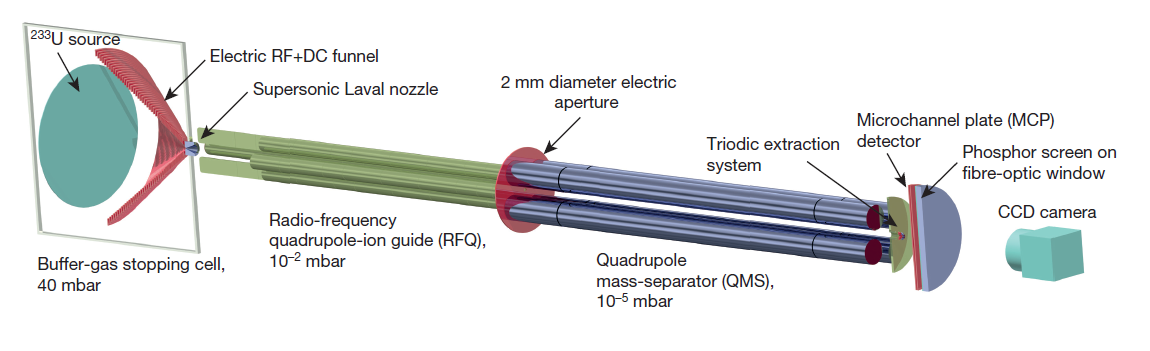}
  \caption{\footnotesize Experimental setup used for the detection of the internal conversion (IC) electrons emitted during the isomeric deexcitation. $^{229\text{m}}$Th is populated via the 2\% decay branch in the $\alpha$ decay of $^{233}$U. The $^{233}$U source is placed in a buffer-gas stopping cell for thermalization of the fast $^{229}$Th recoil ions that are leaving the source. During thermalization, charge capture occurs, leading to the formation of $^{229}$Th$^{2+/3+}$ ions. An isotopically pure ion beam is formed with the help of an RF funnel system, a radio-frequency quadrupole (RFQ) and a quadrupole mass separator (QMS). The ions are implanted into the surface of a micro-channel plate (MCP) detector, where they are neutralized, resulting in the isomer's decay via IC. The low-energy IC electrons are detected by the MCP detector in combination with a phosphor screen and a CCD camera. Reprinted from \cite{Wense2} with kind permission of Springer Nature.}
 \label{munich_setup}
 \end{center}
\end{figure*}  
\subsubsection{Detection of the $^{229\text{m}}$Th IC decay}\label{EXP.IC.DET}
\begin{figure}[t]
 \begin{center}
 \includegraphics[width=7cm]{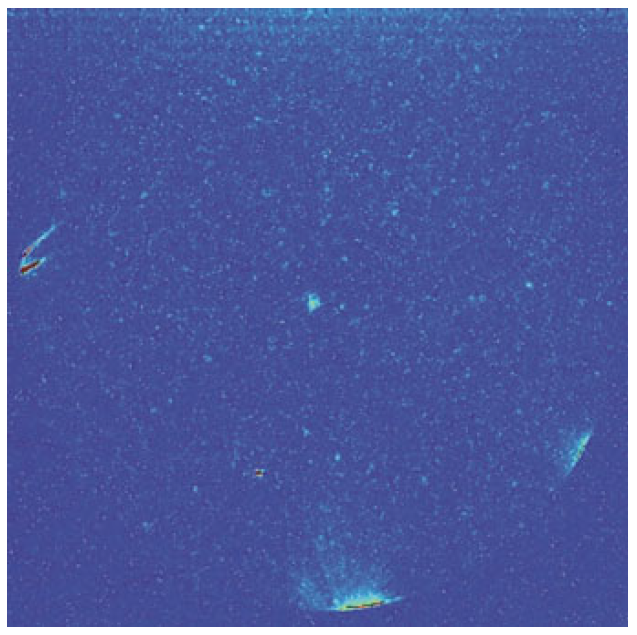}
  \caption{\footnotesize Indication of the decay of $^{229\text{m}}$Th (central spot) as observed in the raw data on October 15, 2014. The picture shows a CCD camera image of the phosphor screen connected to the MCP detector used for IC electron detection. The image was taken with a field-of-view of about 75 $\times$ 75 mm$^2$ and 400~s exposure time. Reprinted from \cite{Wense3} with kind permission of Springer Nature.}
 \label{firstsignal}
 \end{center}
\end{figure}

The direct detection of the internal conversion (IC) decay of $^{229\mathrm{m}}$Th was published in 2016 \cite{Wense2} (see also \cite{Wense3, Thirolf2017, Wense2019a}). The experimental setup used for IC electron detection is shown in Fig.~\ref{munich_setup}. For the detection of the IC decay channel of neutral $^{229\mathrm{m}}$Th, the isomeric state was populated via the 2\% decay branch in the $\alpha$ decay of $^{233}$U. A large-area (90~mm diameter), thin ($\approx7$~nm) $^{233}$U source of 290~kBq activity was used and placed in a buffer-gas stopping cell \cite{Neumayr2006}. $^{229}$Th nuclei produced in the $\alpha$ decay of $^{233}$U possess a kinetic energy of about 84~keV, sufficiently large to allow them to propagate through the thin $^{233}$U source material and to enter the stopping volume as highly charged ions. This process is fast, and the isomer has no time to decay when propagating through the source material. As the buffer-gas stopping cell is filled with 30 to 40 mbar of ultra-pure helium, the highly-charged $^{229}$Th ions are stopped by collisions with the He atoms. During the stopping process, charge exchange occurs, forming $^{229}$Th ions predominantly in the $2+$ and $3+$ charge states. In both charge states, the IC decay channel of $^{229\mathrm{m}}$Th is energetically forbidden, leading to a long isomer lifetime, which may approach the radiative lifetime of up to $10^4$~s. Electric guiding fields and a quadrupole-mass-separator (QMS) form an isotopically pure $^{229}$Th ion beam (of either 2+ or 3+ charge state), with 2\% of the ions in the nuclear excited state. This low-energy, isotopically pure $^{229}$Th ion beam provided the starting point for all further experimental investigations \cite{Wense4,Wense5}.\\[0.2cm]
For the detection of the IC electrons, the low-energy ion beam was directly accumulated on the surface of a CsI-coated MCP detector. Charge capture on the MCP surface leads to neutralization of the ions, thus triggering the IC decay of the isomeric state. The electrons produced in this process were detected by the MCP detector, which is connected to a CCD camera monitored phosphor screen. A CCD camera image with the isomeric decay signal is shown in Fig.~\ref{firstsignal}. Several exclusion measurements, most importantly a comparison to $^{230}$Th, were required to prove that the detected signal originates from the isomeric decay of $^{229\mathrm{m}}$Th \cite{Wense2}. The decay properties of $^{229\text{m}}$Th on a CsI-surface were discussed in \cite{Meyer2018,Borisyuk2018c}. An experimental proposal, based on the same detection scheme, was published by Gusev et al. \cite{Gusev}.
\subsubsection{Measuring the isomeric lifetime}\label{EXP.IC.LIFETIME}
\begin{figure*}[t]
 \begin{center}
 \includegraphics[width=14cm]{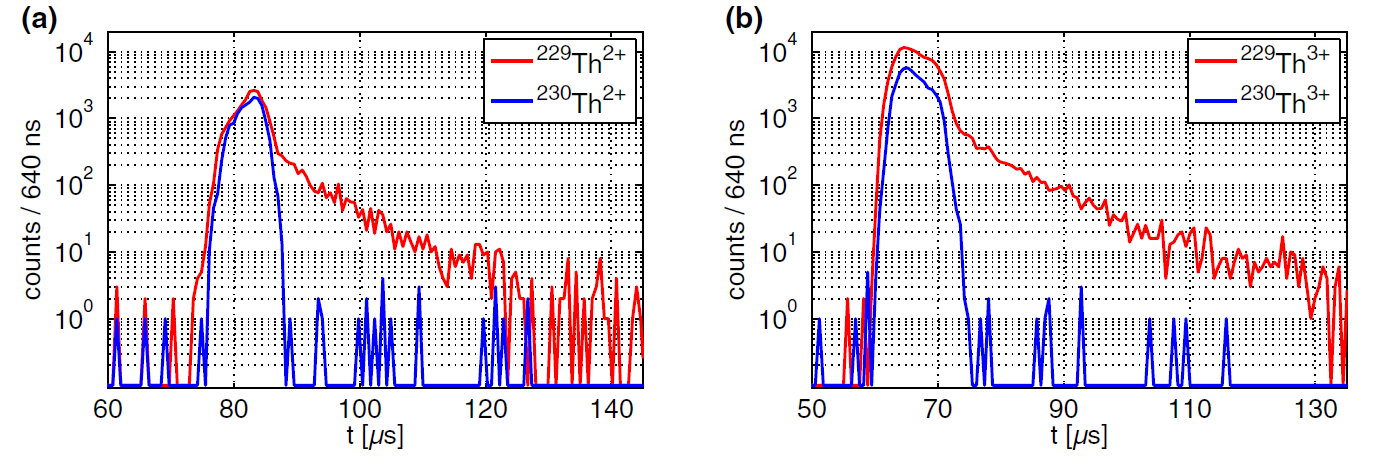}
  \caption{\footnotesize Lifetime determination of the IC decay channel for neutral $^{229\text{m}}$Th atoms deposited on an Inconel surface \cite{Seiferle3}. The lifetime was measured for atoms generated from the $2+$ and $3+$ ionic species. As expected, no difference in lifetime is observed. The isomeric half-life was determined to be $7\pm1$~$\mu$s. Comparative measurements with $^{230}$Th are also shown, where no isomeric species is present. Reprinted from \cite{Seiferle3} with kind permission of the American Physical Society.}
 \label{lifetime}
 \end{center}
\end{figure*}  
It has long been known from theory that, due to the extraordinarily low excitation energy of $^{229\mathrm{m}}$Th, no single isomeric lifetime exists. Instead, the lifetime is heavily dependent on the electronic environment of the $^{229}$Th nuclei \cite{Reich3,Strizhov,Karpeshin6,Tkalya99,Tkalya100,Kasamatsu2018}. If the internal conversion decay channel is energetically allowed, the isomeric state will decay on the timescale of a few microseconds via internal conversion and emission of an electron. This is the case for neutral $^{229}$Th atoms because the thorium ionization energy is about 6.3~eV \cite{Trautmann1994}, significantly below the isomer's energy. It is also the case for $^{229}$Th ions if they are in a sufficiently excited electronic state \cite{Bilous2}. For $^{229}$Th ions in their electronic ground-state, the IC decay channel should always be energetically forbidden. Nevertheless, bound internal conversion or electronic bridge channels may still be present, and their efficiencies would heavily depend on the exact electronic configuration (see App.~\ref{THEO.HIGHDEC.EB}). If the electronic density of states is low, the isomer's lifetime might approach its maximum value of the purely radiative lifetime, which has been theoretically predicted to be on the order of $10^3$ to $10^4$~s (see App. \ref{THEO.DECEXC.GAMMADEC} and Tab.~\ref{halflives}). A long radiative lifetime was the basis for the nuclear clock proposal, as it leads to long coherence times and a high quality factor.\\[0.2cm]
In a solid-state environment the same energy considerations hold, albeit with the ionization energy replaced by the band-gap. If the material's band-gap is smaller than the isomeric energy, electrons can be transferred from the valence band to the conduction band, which will lead to a short lifetime. On the other hand, if the band-gap is large, the radiative decay channel should be dominant, and the lifetime might be prolonged up to the radiative lifetime. This is used in the crystal-lattice nuclear clock approach, where only large band-gap materials are applied.\\[0.2cm]
Experimentally, the detection of the IC decay channel enabled measuring the isomeric lifetime in neutral, surface bound thorium atoms, as well as for the different extracted charge states. To probe the IC lifetime of neutral atoms, a bunched ion beam was generated from the setup described in Sec.~\ref{EXP.IC.DET}. As previously, the ions were collected and neutralized directly on a micro-channel plate (MCP) detector. The neutralization leads to the isomeric decay via electron emission. By triggering the IC electron detection with the release of the ion bunch, the exponential isomer decay becomes detectable, and the isomeric half-life was determined to be $7\pm1$~$\mu$s (corresponding to about 10~$\mu$s lifetime) \cite{Seiferle3,Seiferle2019c}. The experimentally obtained IC decay signals are shown in Fig.~\ref{lifetime}.\\[0.2cm]
The isomer's lifetime in $^{229}$Th$^{2+}$ was determined by storing an ion cloud in a linear Paul trap and waiting for the isomer to decay. After 1 minute of storage time, no measurable isomeric decay was observed, leading to the conclusion that the lifetime must be significantly longer than one minute \cite{Wense2}. The lifetime bound was limited by the achieved storage time of $^{229}$Th$^{2+}$ in the Paul trap. As thorium is a highly reactive element, strong signal degradation occurs for insufficient vacuum conditions. For $^{229}$Th$^{3+}$ only a few seconds of storage time was achieved and not surprisingly no fall-off of isomeric decay signal could be observed on this time scale. Achieving sufficiently long storage times to measure the isomeric lifetimes for $^{229}$Th$^{2+/3+}$ may be possible with a cryogenic linear Paul trap \cite{Schmoeger,Thirolf2020}.\\[0.2cm]
The isomeric lifetime in $^{229}$Th$^{+}$ exhibits a unique position: no IC decay has so far been observed, pointing towards a lifetime significantly shorter than $10$~ms \cite{Seiferle3} (see also \cite{Seiferle2019c}), which is the time required for the ions to be extracted from the buffer-gas stopping cell. As the IC decay channel is energetically forbidden for singly-charged ions, this lifetime is unexpectedly short. The reason might be a coincidental resonance with an electronic level, leading to a fast deexcitation via bound-internal conversion or electronic bridge decay. The probability for such an effect might be enhanced by the particularly dense electronic spectrum in Th$^{+}$ and line-broadening effects in the helium buffer-gas environment \cite{Karpeshin2018}. The actual reason is currently subject to speculation and will require further experimental investigations.\\[0.2cm]
An indication for lifetime variation dependent on the chemical environment was published in \cite{Seiferle2019c}. It was found that the isomeric lifetime for nuclei closer to the substrate surface decreases compared to the isomeric decay that occurs deeper in the material. The half-life varies by almost a factor of 2 between 3 and 6~$\mu$s. It would be interesting to investigate the isomer's lifetime for different surface materials. In particular, large band-gap materials are of interest in this context, as they could offer the potential for a significant lifetime prolongation, which is important for the development of a solid-state nuclear clock. Frozen noble gases offer an especially high potential for lifetime prolongation and have not yet been experimentally investigated \cite{Tkalya1}.
\subsubsection{An IC-based energy determination}
\begin{figure*}[t]
 \begin{center}
 \includegraphics[width=16cm]{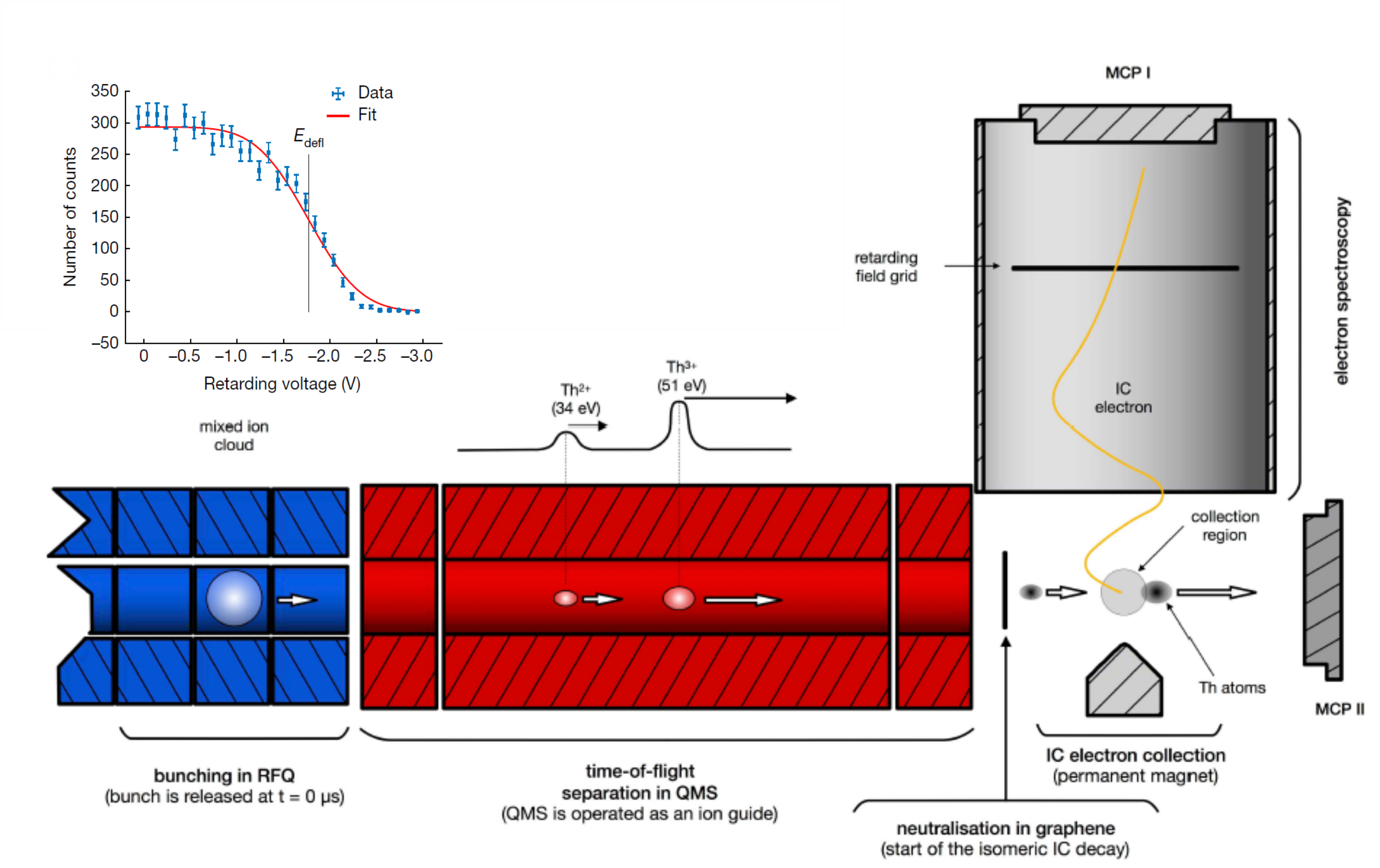}
  \caption{\footnotesize Experimental setup used for the determination of the isomeric energy based on IC electron spectroscopy \cite{Seiferle2019b}. The isomeric state is populated via the $\alpha$ decay of $^{233}$U as shown in Fig.~\ref{munich_setup}. $^{229}$Th$^{2+/3+}$ ion bunches are formed in the last segment of the radio-frequency quadrupole (RFQ). The two charge species are separated via time-of-flight separation in the QMS. The ion bunches are neutralized by charge capture in a graphene foil, which triggers the IC decay. A fraction of the IC decay will occur in the gray-shaded region above a permanent magnet marked as ``collection region". IC electrons that are emitted in this region will be collected and guided by magnetic fields towards the micro-channel plate detector MCP-1 located at the end of the electron spectrometer. A tunable negative voltage is applied to a retarding field grid placed between the collection region and the MCP detector to determine the kinetic energy of the IC electrons. The resulting electron spectrum is shown as an inset. The isomeric energy was inferred to be $8.28\pm0.17$~eV. Reprinted from \cite{Seiferle2019b} with kind permission of Springer Nature.}
 \label{setupenergy}
 \end{center}
\end{figure*}
A precise energy determination of $^{229\mathrm{m}}$Th is the most important requirement for the development of a nuclear optical clock (see Sec.~\ref{STEPS}). Preliminary energy constraints following IC observation were obtained in 2016 \cite{Wense2}: based on the observed long isomeric lifetime in $^{229}$Th$^{2+}$ ions, it was concluded that the IC decay channel must be energetically forbidden. Therefore the isomeric energy must be below 18.3~eV, which is the energy required for ionization of Th$^{2+}$ to Th$^{3+}$. Further, as the IC decay was observed for neutral $^{229}$Th atoms, it was concluded that the energy must be above the Th ionization threshold of $6.3$~eV. However, it has been argued that observation of the IC electrons took place on a surface, which could lead to a reduction of the lower energy constraint \cite{Meyer2018}.\\[0.2cm]
A precise energy determination based on IC electron spectroscopy was achieved in 2019 \cite{Seiferle2019b,Seiferle2019c,Seiferle2019a}. The experimental setup used for this detection is shown in Fig.~\ref{setupenergy} together with the observed electron spectrum. $^{229}$Th$^{2+/3+}$ ions, were extracted from a $^{233}$U source as described in Sec. \ref{EXP.IC.DET}, bunched and accelerated towards a graphene foil. At a kinetic energy of about 1~keV the ions traversed the graphene foil, thereby catching electrons and neutralizing. In this way a bunched $^{229}$Th atom beam was formed, with 2\% of the nuclei in the excited isomeric state. In neutral atoms, the isomer decays via internal conversion under emission of an electron with about 10~$\mu$s lifetime \cite{Seiferle3} (corresponding to $\approx30$~cm travel length at 1~keV thorium kinetic energy). In order to measure the electron's kinetic energy, the $^{229}$Th atom beam was injected into an electron spectrometer. A magnetic-bottle type retarding field spectrometer was chosen for this purpose, as it provides a high acceptance and therefore leads to a reasonably high detection rate for the IC electrons \cite{Seiferle4,Seiferle2019a}.\\[0.2cm]
In the magnetic-bottle type retarding field spectrometer, IC electrons emitted from the beam axis are vertically collected with the help of a strong magnetic field gradient (generated by a permanent magnet and a solenoid coil). Here the magnetic gradient field fulfills two purposes: (1) It acts as a magnetic mirror, collecting nearly all electrons that are emitted in the volume just above the permanent magnet. (2) It leads to a collimation of the electrons as soon as they enter the low-field region in the solenoid coil. The collimation of the electrons is required to use retarding electric fields to repel electrons below a certain energy threshold. Only the electron's velocity component perpendicular to the retarding grid is affected, which leads to the requirement of good electron collimation. Behind the retarding grid, the electrons are post-accelerated towards an MCP detector, where they are counted as a function of the applied retarding voltage, which results in an integrated electron spectrum \cite{Seiferle2019a}.\\[0.2cm]
The isomeric energy is constrained by the sum of the electron's kinetic energy and the binding energy of the electron. The energy evaluation is slightly complicated by the fact that not all $^{229}$Th atoms are in the electronic ground-state after neutralization in the graphene foil. Instead, multiple electronic states are excited, from which an electron can be ejected in the IC process. The $^{229}$Th$^+$ ion, generated in the IC process, will also not always be in its electronic ground-state configuration. This leads to the observation of a complex IC electron energy spectrum, which dominates the uncertainty of the isomer's energy constrained in this way to $8.28\pm0.17$~eV \cite{Seiferle2019b}. A review of the most recent progress in the determination of the isomeric properties was provided in \cite{Thirolf2019b} (see also \cite{Burke2019}). An alternative method for the isomer's energy determination via IC electron spectroscopy was proposed in \cite{Gusev2019}.
\subsubsection{Continued efforts for IC detection}
Three further experiments have reported results of the search for an IC decay channel of $^{229\text{m}}$Th. In the first approach, $^{233}$U $\alpha$-recoil ions were directly implanted into a 4~nm thick gold layer. IC electrons were expected to leave the gold layer and were monitored with an MCP detector in coincidence with the $^{233}$U $\alpha$ decay.
No isomeric decay signal could be observed, most likely due to background electrons accompanying the $\alpha$ decay \cite{Stellmer2018b}. The approach is comparable to an experiment carried out by Swanberg et al. \cite{Swanberg}.\\[0.2cm]
Early results of an experiment carried out at ISOLDE (CERN) with the goal of populating the isomer via the $\approx13$\% branching ratio in the $\beta$ decay of $^{229}$Ac were reported in \cite{Verlinde2019}. While ultimately aiming for the detection of photons emitted during the isomeric decay, a search for the internal conversion electrons was also performed. However, no IC electrons have so far been observed.\\[0.2cm]
In \cite{Shigekawa2019} an observation of IC electrons from $^{229}$Th populated in the $\beta$ decay of $^{229}$Ac was reported. Here low-energy electrons were detected in coincidence with high-energy electrons mainly produced in the $\beta$ decay of $^{229}$Ac. It is planned to use this technique to investigate the isomeric lifetime in different chemical environments. In the following, several new approaches for the isomer's precise energy determination based on the observation of IC electrons will be discussed.\\[0.2cm]
\paragraph*{IC from excited electronic states}
\ A short discussion of the potential to observe IC electrons from excited electronic shell states can be found in \cite{Strizhov}. The concept was considered in more detail in \cite{Dykhne3}. Here, an interesting scheme for laser-triggered isomeric decay detection was proposed. At that time, the isomeric lifetime in the neutral $^{229}$Th atom was assumed to be long, and a significant amount of nuclei in the isomeric state could have accumulated during $\alpha$ decay in $^{233}$U material. If the atomic shell is excited by pulsed laser light, IC from excited electronic shell states would be allowed, leading to a fast depopulation of the isomers under the emission of electrons, which could be detected.\\[0.2cm]
The idea was transferred to $^{229}$Th ions together with a detailed theoretical investigation of the expected IC decay rates for different excited states in \cite{Bilous2} (see also \cite{Bilous2018a}). It is proposed to extract $^{229}$Th$^{1+/2+}$ ions produced either in the $\alpha$ decay of $^{233}$U or in the $\beta$ decay of $^{229}$Ac from a buffer-gas stopping cell located at the IGISOL facility in Jyväskylä, Finland. Several ways of populating particular excited electronic shell states are discussed. It is proposed to observe the IC from the excited electronic state via the change of the ion's charge state during isomeric deexcitation. The technique can be used to constrain the isomeric energy by investigating the IC energy threshold. In 2019 the special case of IC decay from electronic Rydberg states was also theoretically discussed \cite{Tkalya2019}.\\[0.2cm]
\paragraph*{Detection of $^{229\text{m}}$Th with microcalorimetric techniques} 
\ Perhaps one of the most straightforward ways to detect IC decay and determine the isomeric energy would be via a microcalorimetric or transition-edge detection technique. In this method, the total energy deposited in the detector during the isomeric decay would be determined. In \cite{Ponce,Ponce2018b} it was proposed to use a superconducting tunnel junction (STJ) detector for this purpose, and a determination of the $^{235\text{m}}$U energy has already been reported \cite{Friedrich,Ponce,Ponce2018a}. The challenge that one encounters with $^{229\text{m}}$Th, however, is that the detectors are usually slower than the isomer's IC decay time. For this reason one has to expect an energy offset originating from the isomer's implantation into the detector. Nevertheless, it was estimated that an energy uncertainty of 10~meV could potentially be achieved \cite{Ponce2018b}.\\[0.2cm]
An alternative would be to use a superconducting-nanowire single-photon detector (SNSPD) \cite{Natarajan2012}, which allows for fast signal detection in the nanosecond range \cite{Jeet2018}. These detectors are based on a transition-edge detection technique: A nanowire is cooled down to superconductivity and biased with a small current. If energy is deposited into the nanowire, the superconductivity is locally broken, leading to an increase of the bias voltage, which can be detected. An energy value can be obtained by considering, that for a fixed energy, the detection efficiency strongly depends on the applied bias voltage. For this reason, measuring the count rate as a function of bias voltage would constrain the isomer's excitation energy. However, the expected achievable energy uncertainty of 100 to 200~meV is significantly larger than for the STJ detectors.\\[0.2cm]
\paragraph*{Direct laser excitation}
\begin{figure}[t]
 \begin{center}
 \includegraphics[width=7cm]{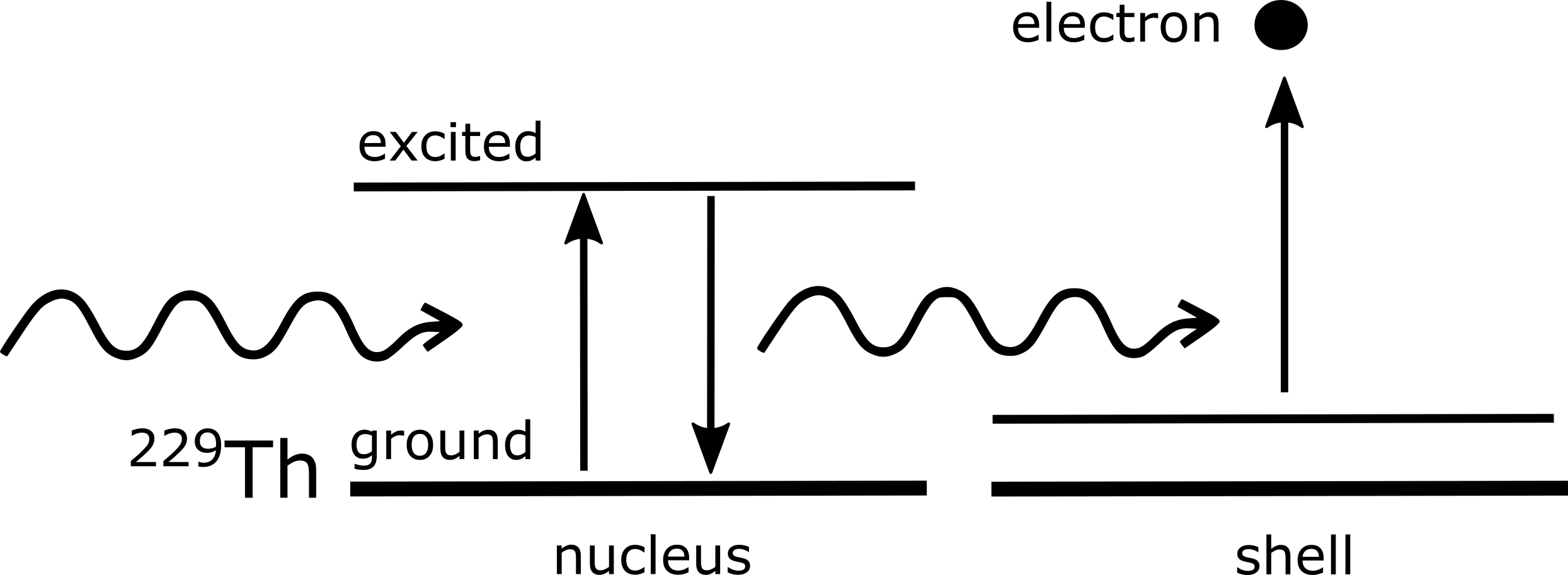}
  \caption{\footnotesize Conceptual sketch of direct laser excitation of $^{229\text{m}}$Th followed by IC electron emission \cite{Wense6}. The technique could measure the isomeric energy to a precision determined by the bandwidth of the laser light used for nuclear excitation. The laser system operated at UCLA and shown in Fig.~\ref{UCLA_lasersystem} could be used for this purpose. A VUV frequency comb based on an Yb-doped fiber laser system as shown in Fig.~\ref{Yb7th} could provide an alternative \cite{Wense2019c}.}
 \label{laserIC}
 \end{center}
\end{figure}
\ Another way to precisely determine the isomer's energy could be via direct laser excitation of $^{229}$Th atoms on a surface, followed by the observation of IC electrons \cite{Wense6,Wense2019b,Wense2019c}. While direct laser spectroscopy of individual $^{229}$Th ions is not yet possible (see Sec.~\ref{STEPS}), during the simultaneous irradiation of a large number (on the order of $10^{13}$) of atoms on a surface, a significant absolute number of nuclei could be excited into the isomeric state. These nuclei would decay via IC and the emission of an electron within the time scale of 10~$\mu$s. The experimental concept is visualized in Fig.~\ref{laserIC}.\\[0.2cm]
The same laser system already in operation in the search for the nuclear resonance in the crystal-lattice environment at UCLA and shown in Fig.~\ref{UCLA_lasersystem} could be used for this purpose \cite{Jeet2018}. The laser system is based on four-wave mixing in a noble gas and delivers tunable, pulsed VUV light around 150~nm. A system with comparable parameters, however, based on the 5th harmonic of a Ti:Sapphire laser is currently in preparation at the University of Hannover. If successful, the isomeric energy could be determined to about 40~$\mu$eV precision, corresponding to the 10~GHz bandwidth of the laser light used for excitation.\\[0.2cm]
In \cite{Wense2019c} it was proposed to perform narrow-band direct frequency comb spectroscopy of $^{229\text{m}}$Th using the 7th harmonic of an Yb-doped fiber system. This concept would enable a higher precision energy determination (ultimately limited by the natural IC-broadened isomeric linewidth of about 16~kHz) and could be used to realize an IC-based solid-state nuclear clock when a single comb-mode is stabilized to the nuclear transition (see Sec.~\ref{NOC.CON.SOLID}).\\[0.2cm]
\paragraph*{Electronic bridge in highly charged ions}
\ In 2020 it was proposed to investigate the EB excitation in highly charged $^{229}$Th ions \cite{Bilous2020}. For this purpose it is planned to generate $^{229}$Th$^{35+}$ in an electron-beam ion trap (EBIT). Laser excitation of a virtual electronic shell level is performed, and excitation of the isomeric state in the EB process is detected via the emission of electrons after implantation of the thorium ions into an MCP detector. Knowledge of the energies of the initial and final state of the electronic shell in combination with the observation of the nuclear resonance would constrain the isomer's excitation energy. 
\subsection{Probing the hyperfine structure of $^{229\text{m}}$Th}\label{EXP.HYPER}
Observation of $^{229\text{m}}$Th by means of the hyperfine shift introduced into the electronic shell due to the different spins of the ground ($5/3+$) and excited ($3/2+$) nuclear states has long been discussed in literature. An early consideration of this method for $^{229}$Th can be found in \cite{Karpeshin2}. Probing the isomer's hyperfine structure does not only provide further independent evidence of the isomer's existence and an alternative detection technique, but is also an important step towards a nuclear clock. The single-ion nuclear clock concept makes use of probing the isomer's excitation by the double-resonance method, which requires knowledge of the isomer's hyperfine structure \cite{Peik}. Further, it allows to determine the $^{229\text{m}}$Th nuclear magnetic dipole and electric quadrupole moments as well as the nuclear charge radius, which is a requirement for a quantitative estimate of the sensitivity-enhancement factor for time variations of fundamental constants \cite{Berengut}. In 2014 it was emphasized that an improved isomeric energy value could even be inferred by measuring the hyperfine structure of $^{229}$Th$^{3+}$ \cite{Beloy}. A measurement of the isomer's hyperfine structure was reported in \cite{Thielking2018}.\\[0.2cm]
Different experiments that make use of the atomic shell's hyperfine structure for isomer identification can be distinguished by the method used for populating $^{229\text{m}}$Th. This can either be a natural route (e.g., via the $\alpha$ decay of $^{233}$U); by direct excitation from the ground-state; or an excitation via the electronic bridge process. All concepts will be discussed individually in the following.
\subsubsection{Population via a natural decay branch}\label{EXP.HYPER.NAT}
\begin{figure*}[t]
 \begin{center}
 \includegraphics[width=16cm]{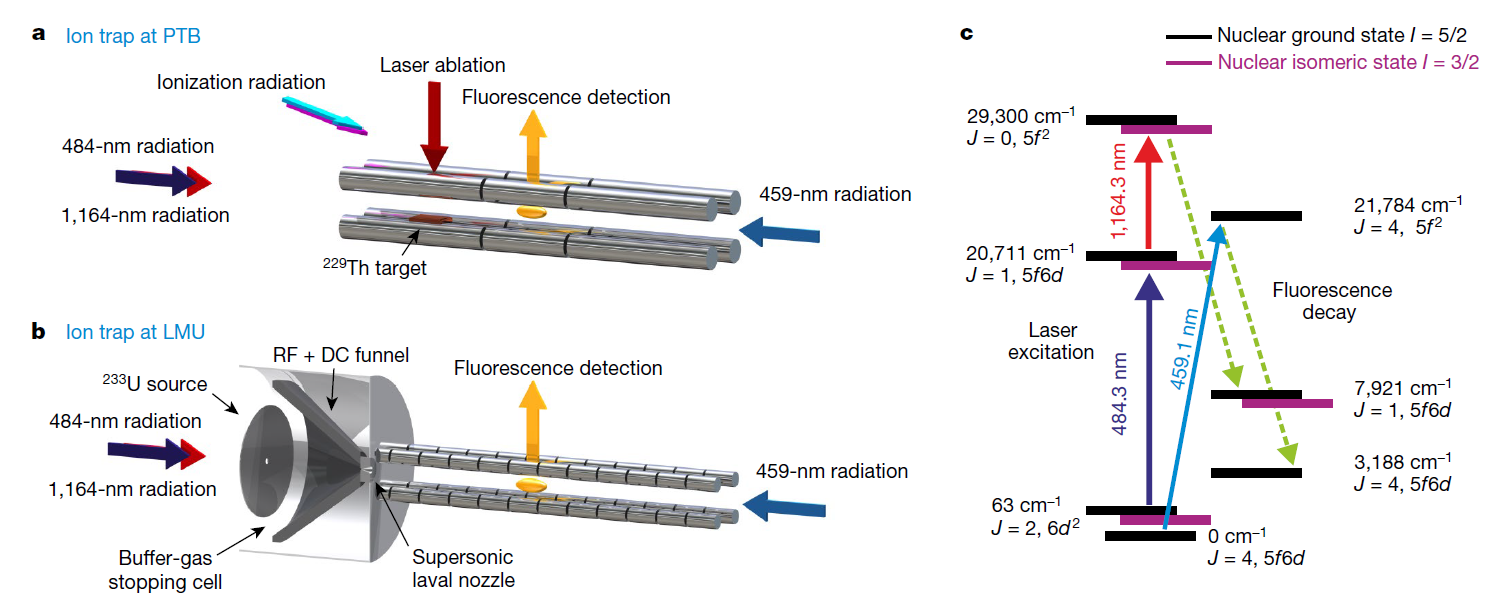}
  \caption{\footnotesize Experimental concept used for the investigation of the $^{229\text{m}}$Th atomic shell hyperfine structure \cite{Thielking2018}. Doppler-free collinear laser spectroscopy was performed in $^{229}$Th$^{2+}$ ions with 2\% in the isomeric state, as populated in the $\alpha$ decay of $^{233}$U (LMU trap). Laser light of 484~nm and 1164~nm wavelengths was used to excite the electronic shell in a two-step excitation scheme. A third laser at 459~nm was used to probe the relative number of trapped particles in parallel. Fluorescence light was probed at a different wavelength to reduce background. For comparison, the same spectroscopy was performed for $^{229}$Th$^{2+}$ ions in the nuclear ground-state obtained from laser ablation of a $^{229}$Th source (PTB trap). In this way, lines originating from the isomeric state were clearly distinguished. Reprinted from \cite{Thielking2018} with kind permission of Springer Nature.}
 \label{thielking}
 \end{center}
\end{figure*}
\begin{figure}[t]
 \begin{center}
 \includegraphics[width=8.5cm]{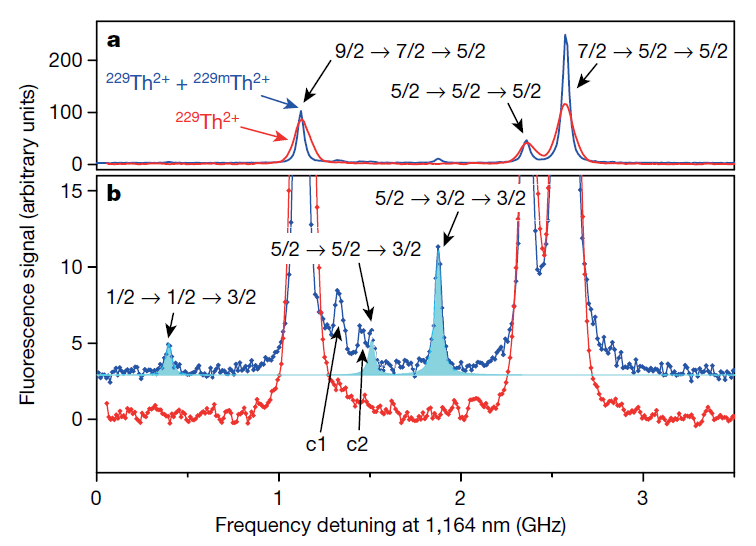}
  \caption{\footnotesize Typical hyperfine spectroscopy data obtained during Doppler-free collinear laser spectroscopy of $^{229\text{m}}$Th \cite{Thielking2018}. Lines originating from $^{229\text{m}}$Th are shaded blue. In total, seven out of eight hyperfine lines were experimentally observed. Reprinted from \cite{Thielking2018} with kind permission of Springer Nature.}
 \label{thielkingdata}
 \end{center}
\end{figure}   
It was proposed by Tordoff et al. in 2006 to probe the $^{229\text{m}}$Th hyperfine structure via collinear laser spectroscopy \cite{Tordoff}. Preparatory work has been completed by a collaboration of the University of Jyväskylä and the University of Mainz. In this approach, $^{229}$Th ions are extracted from a $^{233}$U source by a buffer-gas stopping-cell to form an ion beam \cite{Tordoff2,Sonnenschein1,Pohjalainen2018}. Detailed comparative studies of $^{233}$U sources for $^{229}$Th production are reported in \cite{Pohjalainen2018,Pohjalainen2019}. Experimental results include the detection of 20 previously unknown states in the $^{232}$Th level scheme, as well as numerous auto-ionizing states \cite{Raeder} and the measurement of the ground-state hyperfine structure in neutral $^{229}$Th \cite{Sonnenschein2}. Experimental overviews can be found in \cite{Sonnenschein3,Pohjalainen2018}. Another proposal along this line was made in 2015. Here it was proposed to probe the hyperfine structure of Th$^{+}$ ions \cite{Dembczynski}.\\[0.2cm]
The observation of the $^{229\text{m}}$Th hyperfine structure was achieved in a joint experiment of PTB, LMU and University of Mainz in 2018 \cite{Thielking2018}. The experimental concept is visualized in Fig.~\ref{thielking}, and spectroscopy data are shown in Fig.~\ref{thielkingdata}. In this experiment collinear laser spectroscopy of the electronic shell of $^{229\text{m}}$Th$^{2+}$ ions was performed. The ions were produced in the $\alpha$-decay of $^{233}$U and trapped in a linear Paul trap using the same setup as described in Sec.~\ref{EXP.IC.DET}. Due to the production process, 2\% of the $^{229}$Th ions were in the isomeric state. The spectroscopy scheme is shown in Fig.~\ref{thielking} c: in total, three lasers were used for spectroscopy. The first laser, with a wavelength of 484.3~nm, was used to excite the 5f6d state from the 6d$^2$ state, which is thermally populated from the 5f6d ground state due to collisions with the background gas. The second laser (1164.3~nm) excited the 5f$^2$ state from the previously populated 5f6d state as required for Doppler-free collinear laser spectroscopy. The states were chosen in a way to provide high excitation probabilities with diode lasers and low angular momentum quantum numbers to reduce the number of lines involved in the hyperfine splitting. A third laser at 459.1~nm was used to probe the number of trapped $^{229}$Th$^{2+}$ ions via excitation of the 5f$^2$ transition from the electronic ground-state in parallel to the collinear spectroscopy. The excitation of the states was measured via background-free detection of fluorescence at 467.7~nm and 537.8~nm.\\[0.2cm]
With this technique, seven out of eight spectral lines corresponding to the isomeric hyperfine structure could be observed \cite{Thielking2018}. Only the last line could not be detected due to its low intensity. This experiment determined the magnetic dipole and electric quadrupole moments of $^{229\text{m}}$Th to values of $-0.37(6)$~$\mu_N$ and $1.74(6)$~eb, respectively. The observed magnetic dipole moment was a factor of five larger than the previously predicted value of $-0.076$ $\mu_N$ \cite{Dykhne}. This discrepancy was confirmed by a detailed comparison of experimental data with theoretical predictions of the hyperfine structure \cite{Mueller2018}. It was resolved with an improved nuclear structure model including Coriolis mixing \cite{Minkov2019} (see also \cite{Minkov2018,Minkov2019b}). The difference of the mean-square charge radii of the ground and isomeric states was measrued to be $0.012(2)$ fm$^2$ \cite{Thielking2018}. Furthermore, the sensitivity to time variation of the fine-structure constant $\alpha$ was investigated as proposed in \cite{Berengut}. However, at that time, only an upper limit of $10^4$ for the sensitivity factor could be obtained (see Sec.~\ref{APPL.FUND} for details), as the uncertainties of the hyperfine constants were still too large. The mean-square charge-radii of the ground- and isomeric states was investigated in finer detail in \cite{Safronova2018b}, leading to a value of $0.0105(13)$ fm$^2$. Using the additional assumption of identical nuclear charge density for the ground and isomeric state, a sensitivity factor of $-(0.9\pm0.3)\cdot10^4$ was obtained in \cite{Fadeev}.
\subsubsection{Population by direct laser excitation}\label{EXP.HYPER.DIRECT}
Performing direct nuclear laser spectroscopy of $^{229\text{m}}$Th followed by hyperfine spectroscopy of the electronic shell to monitor excitation of the isomeric state provides the basis of the nuclear clock proposal \cite{Peik}. It is the objective of all groups that are planning to develop a nuclear clock based on individual $^{229}$Th ions stored in a Paul trap. Corresponding discussions can also be found in \cite{Campbell2,Campbell3,Campbell4,Campbell1}. However, the isomeric energy has not yet been constrained to sufficient precision to allow for direct nuclear laser spectroscopy of individual trapped $^{229}$Th ions. For this reason, alternative ways of populating the isomeric state from the nuclear ground state are being investigated. The most important is the electronic bridge (EB) scheme as will be discussed in the following section.
\subsubsection{Population via EB excitation}\label{EXP.HYPER.IEB}
The population of $^{229\text{m}}$Th via the electronic bridge (EB) mechanism (see App. \ref{THEO.HIGHEXC.IEB}) was already discussed in the context of the isomer's radiative decay in Sec. \ref{EXP.DIRECT.IEB}. Here the EB excitation is revisited in the context of the observation of $^{229\text{m}}$Th by hyperfine spectroscopy.\\[0.2cm]
An early proposal for excitation via EB followed by hyperfine spectroscopy to probe the isomer's population can be found in \cite{Karpeshin2} (EB exc. and NEET were used as synonyms at that time). Later, the concept was considered in more detail in \cite{Campbell2,Porsev3,Mueller2019}. In all recent proposals, a two-step EB excitation process is considered, which is technologically easier to realize. The isomer's excitation energy can be inferred from the laser energy used for excitation of the atomic shell in combination with the resonance condition of the EB process. The concept requires storing $^{229}$Th ions in a Paul trap and performing hyperfine spectroscopy of the electronic shell, which has been achieved at the Georgia Institute of Technology \cite{Campbell2,Campbell3} and at PTB in Germany \cite{Thielking2018}.
\section{Concluding remarks}
Within the past decades, significant progress has been made toward the realization of a nuclear clock. Important steps include a correction of the $^{229\text{m}}$Th energy to a value above the ionization threshold of neutral thorium \cite{Beck1}, direct laser cooling of triply ionized $^{229}$Th in a Paul trap \cite{Campbell3}, direct detection of the isomer's internal conversion (IC) decay channel \cite{Wense2}, the determination of the isomer's IC lifetime \cite{Seiferle3}, investigation of the $^{229\text{m}}$Th hyperfine structure \cite{Thielking2018}, population of the isomer from the nuclear ground state via a Lambda-excitation scheme \cite{Masuda2019} and a more precise energy determination \cite{Seiferle2019b,Sikorsky2020}.\\[0.2cm]
The parameters investigated so far are promising for the development of a nuclear frequency standard of unprecedented accuracy. With an energy of about 8.1~eV the transition will enable nuclear laser spectroscopy, and the isomeric lifetime was found to be larger than 1 minute for thorium ions of charge state equal to or larger than $2+$, resulting in a narrow resonance linewidth.\\[0.2cm]
Nevertheless, challenges remain: most importantly a determination of the isomeric energy to sufficient precision for narrow-band laser spectroscopy of individual thorium ions in a Paul trap (about $\pm20$~meV). Multiple experiments listed in Tab.~\ref{energyconcepts} are aiming for this goal. As soon as it is achieved, direct frequency comb spectroscopy will constrain the isomeric energy down to the Hz-range, e.g., with an experimental setup comparable to the one presented in Fig.~\ref{Yb7th}. A $^{229}$Th-based nuclear optical clock of high accuracy will then be realized.\\[0.2cm]
A nuclear-based frequency standard is expected to offer various applications as detailed in Sec.~\ref{APPL}. In particular, a high sensitivity to potential time variations of fundamental constants was highlighted \cite{Flambaum1}. Ideally, superior accuracy may lead to a re-definition of the second based on the unique $^{229}$Th nuclear transition. $^{229\text{m}}$Th undoubtedly offers exciting prospects for many new experimental avenues.\\[0.5cm]
\begin{figure*}
 \begin{center}
 \includegraphics[width=17cm]{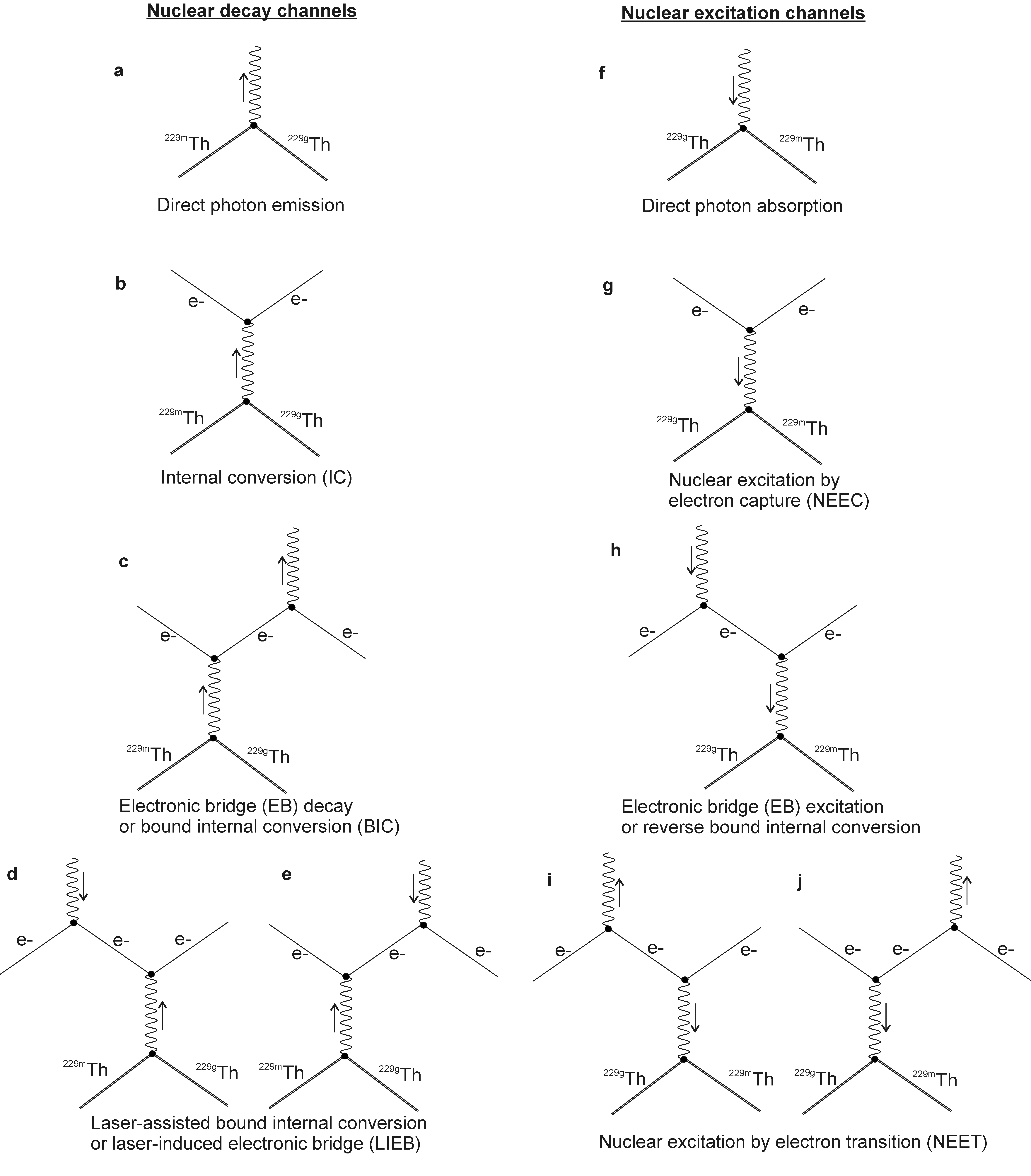}
  \caption{\footnotesize Fundamental processes of isomeric decay and excitation. \textbf{a} The nuclear deexcitation via direct photon emission. \textbf{b} The second-order process corresponding to internal conversion (IC). \textbf{c} Electronic bridge (EB) decay, a third order process in which a virtual electronic shell state is excited and subsequently decays by emission of a photon. Bound internal conversion (BIC) can be considered as a synonym \cite{Karpeshin2018b}. \textbf{d,e} Both diagrams correspond to the process of laser-assisted BIC. \textbf{f} Direct photon excitation of the nucleus. \textbf{g} Nuclear excitation by electron capture (NEEC), which is the reverse of the IC decay. \textbf{h} Electronic bridge (EB) excitation, which is the reverse of the EB decay. Reverse BIC is a frequently used synonym. \textbf{i,j} Nuclear excitation by electron transition (NEET). In this process the nucleus is excited during deexcitation of a real electronic shell state. See text for details.}
 \label{excdecay}
 \end{center}
\end{figure*}
\begin{appendices}
\section{The theory of $^{229\text{m}}$Th excitation and decay} \label{THEO}
A steep increase of theoretical interest in $^{229\text{m}}$Th occurred after the isomer's energy had been constrained to be below 10~eV in 1990 \cite{Reich_Helmer}. A nuclear excited state of such a low energy is unique in the nuclear landscape. In particular, the close proximity of the nuclear energy to the energy range of electronic-shell-based processes opens up some interesting opportunities to study the interplay between the electronic shell and the nucleus, most importantly the electronic bridge (EB) processes. The interested reader is also referred to the review articles \cite{Matinyan,Tkalya9}.\\[0.2cm]
Theoretical investigation mainly focused on the different excitation and decay channels of the isomeric state. An overview of the most important Feynman diagrams of the different processes is provided in Fig.~\ref{excdecay}. The decay channels can be divided into four categories: (1) direct radiative decay (Fig. \ref{excdecay}~a), (2) internal conversion (IC, Fig. \ref{excdecay}~b), (3) electronic bridge (EB) decay or bound internal conversion (BIC), which can be considered as synonyms (Fig. \ref{excdecay}~c) and (4) laser-assisted bound internal conversion (Fig. \ref{excdecay}~d,e). The processes can be distinguished by the number of vertices and thus the order of the process. Direct radiative decay corresponds to the emission of a photon in the nuclear deexcitation and is a first-order process. In the second-order internal conversion (IC) process, the nucleus transfers its energy to the electronic shell via a virtual photon. If the transferred energy is large enough, an electron is subsequently ejected from the electronic shell. In the third-order electronic-bridge (EB) process, a virtual electronic shell state is excited and immediately decays via the emission of a photon. In laser-assisted BIC, which is also of third-order, the missing energy between the nuclear transition and a particular electronic shell transition is provided by an external photon.\\[0.2cm]
Excitation of the isomeric state can in principle occur via the reverse of the four decay processes. These are: (1) direct photon excitation (Fig. \ref{excdecay}~f), which is a first-order process, (2) nuclear excitation by electron capture (NEEC), a second-order process in which an electron is captured by the shell, and the energy is transferred to the nucleus (the reverse of the IC decay, Fig. \ref{excdecay}~g), (3) electronic bridge (EB) excitation or reverse bound internal conversion shown in Fig. \ref{excdecay}~h (the reverse of the third-order EB decay process, sometimes also referred to as inverse electronic bridge, IEB).  And finally, (4) nuclear excitation by electron transition (NEET), which is not consistently defined in literature, but which we here consider to be a process of nuclear excitation during the decay of a real electronic shell state as shown in Fig. \ref{excdecay}~i,j. NEET is also a third-order process. In the following, each process will be discussed individually.
\subsection{Gamma decay and direct photon excitation}\label{THEO.DECEXC}
\subsubsection{$^{229\text{m}}$Th direct radiative decay}\label{THEO.DECEXC.GAMMADEC}
The transition from the $^{229\text{m}}$Th $3/2^+[631]$ nuclear excited state to the $5/2^+[633]$ nuclear ground state can proceed via two multipolarities: $M1$ and $E2$. It has been shown, however, that the $E2$ transition probability is significantly lower than the $M1$ probability (see, e.g., \cite{Reich3}). For this reason the radiative $^{229\text{m}}$Th lifetime is dominated by the $M1$ transition to the nuclear ground state.\\[0.2cm]
Generally, the radiative decay rate $\Gamma_\gamma=1/\tau_\gamma$ of magnetic multipole transitions can be expressed in terms of the energy-independent reduced transition probability $B_\downarrow(ML)$ as \cite{Blatt1952}
\begin{equation}
\Gamma_\gamma=\frac{2\mu_0}{\hbar}\frac{L+1}{L[(2L+1)!!]^2}\left(\frac{\omega_n}{c}\right)^{2L+1} B_\downarrow(ML).
\end{equation}
Here $L$ denotes the multipolarity and $\omega_n$ the angular frequency of the nuclear transition. For the magnetic dipole transition of $^{229\text{m}}$Th with $L=1$, the radiative decay rate scales with $\omega_n^3$. The problem of calculating the isomer's decay rate is therefore shifted to the problem of determining $B_\downarrow(ML)$. Various models of different complexity and accuracy exist to estimate the reduced transition probability. Usually, $B_\downarrow(ML)$ is given in Weisskopf units. One Weisskopf unit is defined for magnetic multipole transitions as \cite{Blatt1952}
\begin{equation}
1\ \text{W.u.}=\frac{10}{\pi}\left(\frac{3}{L+3}\right)^2R^{2L-2}\mu_N^2.
\end{equation}
Here $R=1.2 A_N^{1/3}$~fm is the nuclear radius (where $A_N$ denotes the mass number) and $\mu_{N}=5.051\cdot10^{-27}$~J/T denotes the nuclear magneton. For the case of $^{229\text{m}}$Th with $L=1$, one obtains $1\ \text{W.u.}=45/(8\pi)\mu_N^2$.\\[0.2cm]
A list of the different values for $B_\downarrow(M1)$ of $^{229\text{m}}$Th found in literature is provided in Tab.~\ref{halflives}. Realistic radiative lifetimes vary between $1.3\cdot10^3$~s and $1.7\cdot10^4$~s. Today, the values published in \cite{Minkov,Minkov2017b,Minkov2019} are the most accepted ones. Their calculation is based on a sophisticated nuclear structure model, which includes Coriolis coupling. Assuming the latest isomeric energy value of about 8.1~eV, and the latest value for $B_\downarrow(M1)$ of $5.0\cdot10^{-3}$~W.u. \cite{Minkov2019}, the radiative lifetime is obtained as $1.2\cdot10^4$~s.
\begin{table}[t]
\begin{center}
\caption{\footnotesize Different reported literature values for the reduced transition probability and corresponding radiative lifetimes.}
\begin{footnotesize}
\begin{tabular}{ccc}
\noalign{\smallskip}\hline\noalign{\smallskip}
 $B_\downarrow(M1)$ & \parbox[pt][2em][c]{2.5cm}{\centering{$\tau_\gamma$ for 8.1~eV}}  & \parbox[pt][2em][c]{2.5cm}{\centering{Reference}} \\
\noalign{\smallskip}\hline\noalign{\smallskip}
\parbox[0pt][1.5em][c]{2.5cm}{\centering{$0.93$ W.u.$^*$}} &  \parbox[pt][1em][c]{2cm}{\centering{$65$ s}} & \parbox[pt][1em][c]{2.5cm}{\cite{Reich_Helmer}} \\
\parbox[0pt][1.5em][c]{2.5cm}{\centering{$7.3\cdot10^{-3}$ W.u.}} &  \parbox[pt][1em][c]{2cm}{\centering{$8.2\cdot10^3$ s}} & \parbox[pt][1em][c]{2.5cm}{\cite{Reich3}} \\
\parbox[0pt][1.5em][c]{2.5cm}{\centering{$1.4\cdot10^{-2}$ W.u.}} &  \parbox[pt][1em][c]{2cm}{\centering{$4.3\cdot10^3$ s}} & \parbox[pt][1em][c]{2.5cm}{\cite{Strizhov}} \\
\parbox[0pt][1.5em][c]{2.5cm}{\centering{$3.9\cdot10^{-2}$ W.u.}} &  \parbox[pt][1em][c]{2cm}{\centering{$1.5\cdot10^3$ s}} & \parbox[pt][1em][c]{2.5cm}{\cite{Wycech}} \\
\parbox[0pt][1.5em][c]{2.5cm}{\centering{$3.4\cdot10^{-3}$ W.u.}} &  \parbox[pt][1em][c]{2cm}{\centering{$1.7\cdot10^4$ s}} & \parbox[pt][1em][c]{2.5cm}{\cite{Helmer_Reich}} \\
\parbox[0pt][1.5em][c]{2.5cm}{\centering{$1$ W.u.$^*$}} &  \parbox[pt][1em][c]{2cm}{\centering{$60$ s}} & \parbox[pt][1em][c]{2.7cm}{\cite{Karpeshin2}} \\
\parbox[0pt][1.5em][c]{2.5cm}{\centering{$1.2\cdot10^{-2}$ W.u.}} &  \parbox[pt][1em][c]{2cm}{\centering{$4.9\cdot10^3$ s}} & \parbox[pt][1em][c]{2.5cm}{\cite{Typel}} \\
\parbox[0pt][1.5em][c]{2.5cm}{\centering{$4.8\cdot10^{-2}$ W.u.}} &  \parbox[pt][1em][c]{2cm}{\centering{$1.3\cdot10^3$ s}} & \parbox[pt][1em][c]{2.5cm}{\cite{Dykhne3}} \\
\parbox[0pt][1.5em][c]{2.5cm}{\centering{$2.4\cdot10^{-2}$ W.u.}} &  \parbox[pt][1em][c]{2cm}{\centering{$2.5\cdot10^3$ s}} & \parbox[pt][1em][c]{2.5cm}{\cite{Barci}}\\
\parbox[0pt][1.5em][c]{2.5cm}{\centering{$1.4\cdot10^{-2}$ W.u.}} &  \parbox[pt][1em][c]{2cm}{\centering{$4.3\cdot10^3$ s}} & \parbox[pt][1em][c]{2.7cm}{\cite{Ruchowska}}\\
\parbox[0pt][1.5em][c]{2.5cm}{\centering{$3.2\cdot10^{-2}$ W.u.}} &  \parbox[pt][1em][c]{2cm}{\centering{$1.8\cdot10^3$ s}} & \parbox[pt][1em][c]{2.5cm}{\cite{Tkalya3}}\\
\parbox[0pt][1.5em][c]{2.5cm}{\centering{$\le4.8\cdot10^{-2}$ W.u.}} &  \parbox[pt][1em][c]{2cm}{\centering{$\ge1.3\cdot10^3$ s}} & \parbox[pt][1em][c]{2.5cm}{\cite{Tkalya99}}\\
\parbox[0pt][1.5em][c]{2.5cm}{\centering{$\ge1.4\cdot10^{-2}$ W.u.}} &  \parbox[pt][1em][c]{2cm}{\centering{$\le4.1\cdot10^3$ s}} & \parbox[pt][1em][c]{2.5cm}{\cite{Tkalya99}}\\
\parbox[0pt][1.5em][c]{2.5cm}{\centering{$3.0\cdot10^{-2}$ W.u.}} &  \parbox[pt][1em][c]{2cm}{\centering{$2.0\cdot10^3$ s}} & \parbox[pt][1em][c]{2.5cm}{\cite{Tkalya10}}\\
\parbox[0pt][1.5em][c]{2.5cm}{\centering{$7.0\cdot10^{-3}$ W.u.}} &  \parbox[pt][1em][c]{2cm}{\centering{$8.5\cdot10^3$ s}} & \parbox[pt][1em][c]{2.5cm}{\cite{Minkov}}\\
\parbox[0pt][1.5em][c]{2.5cm}{\centering{$1.0\cdot10^{-2}$ W.u.}} &  \parbox[pt][1em][c]{2cm}{\centering{$5.9\cdot10^3$ s}} & \parbox[pt][1em][c]{2.7cm}{\cite{Karpeshin2018}}\\
\parbox[0pt][1.5em][c]{2.5cm}{\centering{$5.0\cdot10^{-3}$ W.u.}} &  \parbox[pt][1em][c]{2cm}{\centering{$1.2\cdot10^4$ s}} & \parbox[pt][1em][c]{2.5cm}{\cite{Minkov2019}}\\
\noalign{\smallskip}\hline\noalign{\smallskip}
\end{tabular}
\end{footnotesize}
\begin{tablenotes}
\footnotesize $^{*}$ These numbers are based on the Moszkowski and the Weisskopf approximation, respectively, which are known to overestimate the reduced transition probability.
\end{tablenotes}
\label{halflives}
\end{center}
\end{table}
\subsubsection{Direct photon excitation}\label{THEO.DECEXC.GAMMAEXC}
The reverse process of the isomer's radiative decay, the direct excitation of the nuclear isomer with photons, was considered by E.V. Tkalya in 1992 \cite{Tkalya2}. Later discussions can also be found in \cite{Tkalya1,Varlamov,Karpeshin3,Tkalya2b,Tkalya9}. In these early publications the excitation with broad-band (incoherent) light sources was in the focus. The density matrix formalism, as required to describe narrow-band excitation with coherent light sources, was applied to $^{229\text{m}}$Th in \cite{Kazakov2,Kazakov1}. Also nuclear two-photon excitation with a frequency comb was discussed \cite{Romanenko}. Coherent population transfer in $^{229}$Th was further considered in \cite{Liao2,Liao1,Das}. It was proposed that the $^{229}$Th isomeric decay could be identified via electromagnetically modified nuclear forward scattering. An optomechanical microlever was proposed to bridge the gap between an optical laser and X-rays in \cite{Liao3}. It was emphasized that such a device could have significant implications for $^{229\text{m}}$Th and, in particular, could be applied for the isomer's energy determination. Collective effects in $^{229}$Th doped VUV transparent crystals during coherent laser excitation were studied in \cite{Nickerson2018} (see also \cite{Nickerson2019}). The density matrix formalism for direct laser excitation of nuclear transitions was recently reviewed \cite{Wense2019d}. In the following a short discussion of the most important equations is provided.\\[0.2cm]
The cross section for the direct excitation process under resonant irradiation can be expressed as
\begin{equation}
\label{sigmagamma}
\sigma_\gamma=\frac{\lambda_n^2}{4\pi}\frac{\Gamma_\gamma}{\tilde{\Gamma}_n},
\end{equation}
where $\lambda_n$ is the nuclear transition wavelength. Here the decay rate of the coherences was introduced as 
\begin{equation}
\label{decoherence}
\tilde{\Gamma}_n=\frac{\Gamma_n+\Gamma_\ell}{2},
\end{equation}
with $\Gamma_n=\Gamma_\gamma+\Gamma_\text{nr}$ being the total width of the nuclear transition including $\gamma$ decay ($\Gamma_\gamma$) as well as other non-radiative decay channels ($\Gamma_\text{nr}$). $\Gamma_\ell$ denotes the bandwidth of the light used for excitation. A potential extra term, $\tilde{\Gamma}_\text{add}$, that accounts for additional decoherence expected in a solid-state environment due to fluctuations of the nuclear spin states was neglected. For $^{229}$Th, $\tilde{\Gamma}_\text{add}$ was estimated to be between about $2\pi\cdot150$~Hz \cite{Kazakov1} and $2\pi\cdot1$ to $2\pi\cdot10$~kHz \cite{Rellergert}.\\[0.2cm]
By defining the cross section according to Eq.~(\ref{sigmagamma}) the two different scenarios $\Gamma_\ell\gg\Gamma_n$ as well as $\Gamma_n\gg\Gamma_\ell$ are automatically correctly described. If the bandwidth of the laser light is significantly broader than the nuclear transition linewidth one obtains $\sigma_\gamma=\lambda_n^2 \Gamma_\gamma/(2\pi\Gamma_\ell)$. For ordinary laser parameters (e.g., a bandwidth of about 1~GHz) a cross section of $\approx10^{-25}$~cm$^2$ was obtained \cite{Tkalya2}. It is obvious that a narrow laser bandwidth is favorable for a large cross section. The excitation rate per nucleus can then be calculated from the cross section via \cite{Steck}
\begin{equation}
\label{Gammaexc}
\Gamma_\text{exc}^\gamma=\frac{\sigma_\gamma I_\ell}{\hbar\omega_n}=\frac{\pi c^2I_\ell}{\hbar\omega_n^3\tilde{\Gamma}_n}\Gamma_\gamma,
\end{equation}
with $I_\ell$ being the intensity of the light used for excitation and $\omega_n$ the angular frequency corresponding to the nuclear transition. Due to the small radiative decay rate of $^{229\text{m}}$Th, $\Gamma_\gamma=1/\tau_\gamma\approx10^{-4}$~Hz, the direct excitation rate is small unless many nuclei are irradiated in parallel, e.g., in a solid-state environment, or a narrow-band laser is used for excitation.\\[0.2cm]
For the following we define the population density $\rho_\text{exc}(t)$ of the nuclear excited state as the number of excited nuclei divided by the number of irradiated nuclei $\rho_\text{exc}(t)=N_\text{exc}(t)/N_0$. Further, the Rabi frequency $\Omega_\gamma$ for the nuclear transition is introduced as \cite{Steck}
\begin{equation}
\label{Rabi}
\Omega_\gamma^2=2\Gamma_\text{exc}^\gamma\tilde{\Gamma}_n=\frac{2\pi c^2 I_\ell \Gamma_\gamma}{\hbar\omega_n^3}.
\end{equation} 
Approximating the nuclear ground and excited state as a two-level system, the population density $\rho_\text{exc}(t)$ under resonant laser irradiation can be modeled with Torrey's solution of the optical Bloch equations \cite{Wense2019d},
\begin{equation}
\label{Torrey}
\begin{aligned}
\rho_\text{exc}(t)=&\frac{\Omega_\gamma^2}{2\left(\Gamma_n\tilde{\Gamma}_n+\Omega_\gamma^2\right)}\times\\ &\left[1-e^{-\frac{1}{2}(\Gamma_n+\tilde{\Gamma}_n)t}\left(\cos(\lambda t)+\frac{\Gamma_n+\tilde{\Gamma}_n}{2\lambda}\sin(\lambda t)\right)\right],
\end{aligned}
\end{equation}
where $\lambda=|\Omega_\gamma^2-(\tilde{\Gamma}_n-\Gamma_n)^2/4|^{1/2}$. Eq.~(\ref{Torrey}) is only valid for $\lvert\tilde{\Gamma}_n-\Gamma_n\rvert/2<\Omega_\gamma$. If $\lvert\tilde{\Gamma}_n-\Gamma_n\rvert/2>\Omega_\gamma$, the $\sin$ and $\cos$ functions have to be exchanged by $\sinh$ and $\cosh$, respectively.\\[0.2cm]
In the low saturation limit, defined by the condition $\rho_\text{exc}\ll 1$, the population density simplifies to \cite{Wense2019d}
\begin{equation}
\rho_\text{exc}(t)=\frac{\Omega_\gamma^2\tilde{\Gamma}_n/(2\Gamma_n)}{\Delta_n^2+\tilde{\Gamma}_n^2}\left(1-e^{-\Gamma_n t}\right),
\end{equation}
where the detuning of the laser frequency from the nuclear resonance $\Delta_n=\omega_\ell-\omega_n$ was introduced. In the steady state case, in which excitation and decay of the excited state are in equilibrium, one has
\begin{equation}
\label{Dieter}
\rho_\text{exc}=\frac{\Omega_\gamma^2\tilde{\Gamma}_n/(2\Gamma_n)}{\Delta_n^2+\tilde{\Gamma}_n^2}=\Gamma_\text{exc}^\gamma\frac{\tilde{\Gamma}_n^2/\Gamma_n}{\Delta_n^2+\tilde{\Gamma}_n^2}.
\end{equation}
\subsection{Higher order processes of isomer deexcitation: internal conversion, bound-internal conversion and electronic bridge decay}\label{THEO.HIGHDEC}
\subsubsection{Internal conversion}\label{THEO.HIGHDEC.IC}
In the internal conversion (IC) process the nucleus transfers its excitation energy to the electronic shell, leading to the emission of an electron as shown in Fig. \ref{excdecay}~b. The IC process was experimentally described by Meitner in 1924 \cite{Meitner1924,Hahn1924} and the theory was developed in the following decades (see \cite{Tralli1951} and references therein). A detailed theoretical discussion can be found in \cite{Rose1955a,Rose1955b}. The IC decay rate is usually expressed in terms of the IC coefficient, defined as the ratio between the IC- and the direct radiative-decay rate
\begin{equation}
\alpha_{ic}=\frac{\Gamma_{ic}}{\Gamma_\gamma}.
\end{equation}
Calculation of $\alpha_{ic}$ requires numerical computation. In case of a non-vanishing IC decay channel, the total decay rate of the nuclear excited state is found to be
\begin{equation}
\Gamma_n=\Gamma_\gamma+\Gamma_{ic}=\left(1+\alpha_{ic}\right)\Gamma_\gamma.
\end{equation}
Internal conversion for the special case of $^{229\text{m}}$Th was considered in 1991 \cite{Strizhov}. Here the IC coefficient was predicted to be on the order of $10^{9}$ if energetically allowed. However, in the following years the IC process was assumed to be energetically forbidden in the thorium atom, as the $3.5\pm1$~eV energy estimate from 1994 \cite{Helmer_Reich} was below the thorium ionization threshold of 6.3~eV (see, e.g., \cite{Karpeshin2,Typel,Tkalya7,Tkalya9,Karpeshin5}). Instead, the EB mechanism was considered to be the main decay channel of $^{229\text{m}}$Th (see App.~\ref{THEO.HIGHDEC.EB}).\\[0.2cm]
It was not before 2007, when an improved measurement changed the isomer's energy to 7.6~eV \cite{Beck1}, that IC as a potential decay channel of $^{229\text{m}}$Th in the neutral atom in its electronic ground-state was reconsidered \cite{Karpeshin6,Tkalya99}. In both studies, the IC coefficient was found to be about $10^9$, resulting in an IC shortened isomeric lifetime of about 10~$\mu$s, as was later experimentally confirmed \cite{Seiferle3}. Interestingly, it was found that for the IC decay channel both the $M1$ and $E2$ multipole orders play an important role \cite{Bilous2018a,Karpeshin2020}.\\[0.2cm]
Since 1999, also the isomeric decay in a solid-state environment was theoretically investigated \cite{Tkalya7}. It was emphasized that, even if the internal conversion process were energetically forbidden in the free thorium atom, the isomer could still transfer its energy to the conduction electrons of a metal surface if the thorium atom is attached to it. The electron could leave the metal surface if the surface work function is below the energy of the isomeric state. Such effects are expected to shorten the isomer's lifetime and could be used for the energy determination of the isomeric state \cite{Tkalya7}. Comprehensive investigations of the influence of the chemical environment on the isomeric decay were published in 2000 \cite{Tkalya2b,Tkalya8,Koltsov}. These publications also contain the deexcitation of the isomeric state in dielectric media, in which case the radiative decay is expected to dominate as long as the material's band gap is larger than the isomeric energy. The work provided the basis for the later proposal of a solid-state nuclear frequency standard \cite{Peik}. A compact overview of the theoretical investigations at that time can be found in \cite{Tkalya9}. This paper also contains a detailed list of potential applications. The isomeric decay in a dielectric sphere, thin film or metal cavity was analyzed in \cite{Tkalya2018}. It was found that the isomer's lifetime will be influenced by the geometry of the solid-state environment. IC from excited electronic levels was considered in \cite{Strizhov,Dykhne3} and investigated in detail in \cite{Bilous2} (see also \cite{Bilous2018c}). The internal conversion decay of $^{229\text{m}}$Th from excited atomic Rydberg states was discussed in \cite{Tkalya2019} and for the $^{229}$Th anion in \cite{Tkalya2020}.
\subsubsection{Bound-internal conversion and electronic bridge decay}\label{THEO.HIGHDEC.EB}
\begin{figure}[t]
 \begin{center}
 \includegraphics[width=6cm]{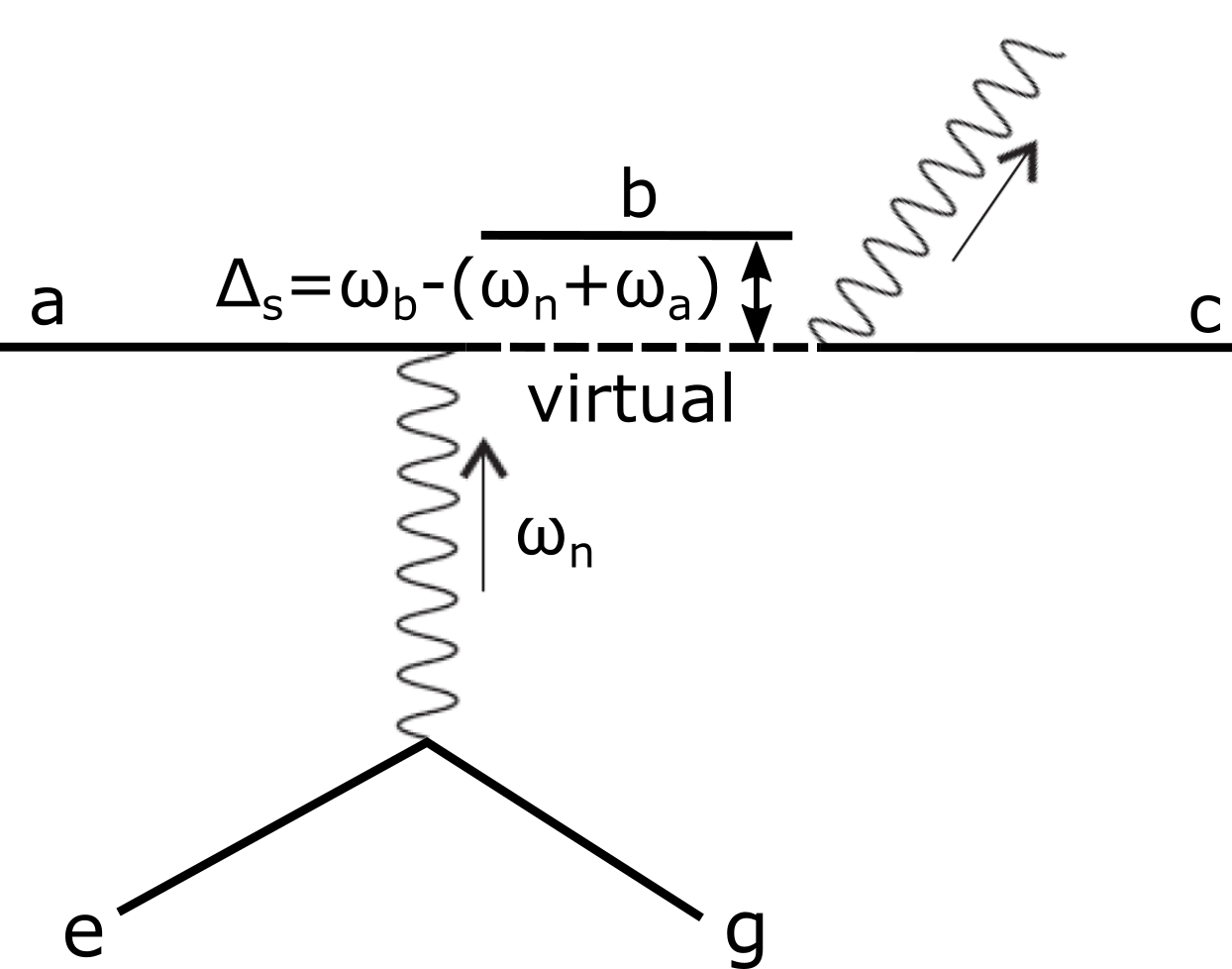}
  \caption{\footnotesize Electronic bridge decay as introduced in \cite{Krutov1958}. During deexcitation of the nucleus from its excited state $(e)$ to its ground state $(g)$ an electron in state $(a)$ is excited to a virtual level close to state $(b)$ and subsequently decays via emission of a photon to state $(c)$. EB and BIC processes can be considered as equivalent \cite{Karpeshin2018b}.}
 \label{EBdecay}
 \end{center}
\end{figure}

If the nuclear energy is not sufficiently large to lead to the ejection of an electron from the atomic shell and thus IC is energetically forbidden, nuclear decay can still occur by exciting an electron to a higher lying bound-state. This process is known as bound internal conversion (BIC). Sometimes the same process is also referred to as ``discrete", ``sub-threshold" or ``resonance internal conversion" \cite{Band}. As the excited electronic state will soon decay to the ground state, BIC will usually be followed by photon emission. In contrast to IC, the BIC process has a relatively short history. Experimentally, BIC was observed in 1995 for highly charged $^{125}$Te \cite{Attallah1995} (see also \cite{Carreyre2000}), followed by theoretical calculations \cite{Karpeshin1996}.\\[0.2cm]
In a simplistic picture one would introduce BIC as a second-order process in analogy to the Feynman diagram for IC shown in Fig. \ref{excdecay}~b. However, for discrete atomic levels populated in the decay, energy conservation would only be fulfilled if the energy of the nuclear transition exactly matched the energy difference of excited electronic shell states. As this will in general not be the case, the corresponding Feynman diagram does not satisfy energy conservation and has vanishing probability. Instead, processes of higher order have to be considered. The dominant process in this context is the diagram shown in Fig.~\ref{EBdecay}. Note, that the depicted diagram is not a strict Feynman diagram, as it spatially distinguishes the real and the virtual intermediate energy levels. In fact, this process was already proposed long before BIC by Krutov in 1958 under the name electronic bridge (EB) \cite{Krutov1958,Krutov}. In the EB decay a virtual electronic shell state gets excited during the nuclear decay and subsequently a photon is emitted in the deexcitation of the shell state. Experimental observation of the EB decay was achieved for $^{93}$Nb \cite{Kekez1985} and for $^{193}$Ir \cite{Zheltonozhskii1988}. Today, BIC and EB are considered as equivalent processes \cite{Karpeshin2018b}.\\[0.2cm]
The deexcitation of $^{229\text{m}}$Th via an electronic bridge channel was discussed by Strizhov and Tkalya in 1991 \cite{Strizhov}, following a discussion of the same decay channel for $^{235\text{m}}$U by Zon and Karpeshin \cite{Zon}. Later papers that discussed the isomer's deexcitation via an EB mechanism, however under the name ``resonance conversion", are \cite{Karpeshin4,Karpeshin5,Karpeshin6}. The potential for an EB decay in $^{229}$Th$^{+}$ ions was analyzed in \cite{Porsev2} for isomeric energies of 3.5~eV and 5.5~eV, respectively. The case of 7.6~eV was later investigated in \cite{Porsev3}. The obtained result, however, is in disagreement with an experimentally obtained upper limit of the isomeric lifetime in $^{229}$Th$^{+}$ \cite{Seiferle3,Seiferle2019c}. A detailed discussion of this discrepancy was carried out in \cite{Karpeshin2018}. EB decay in the $3+$ charge state was considered in \cite{Porsev1} and more recently re-investigated in \cite{Mueller}. In the following, a short mathematical discussion of the EB decay will be provided. A complete treatment requires numerical computation.\\[0.2cm]
As for IC decay, one can define an EB-coefficient $\alpha_{eb}$ as the ratio between the EB- and the $\gamma$-decay rate (the same coefficient was denoted as $R$ in \cite{Karpeshin6} and as $\beta$ in \cite{Porsev1})
\begin{equation}
\label{alphaeb}
\alpha_{eb}=\frac{\Gamma_{eb}}{\Gamma_\gamma}.
\end{equation}
In order to derive an expression for $\alpha_{eb}$ we will first assume that an individual electronic state ($b$) dominates the deexcitation as indicated in Fig.~\ref{EBdecay} to establish an expression for the partial EB decay rate from $(a)$ to $(c)$. This is obtained as the rate of excitation of state $(b)$ during the IC decay with detuning $\Delta_s=\omega_b-(\omega_n+\omega_a)$, multiplied by the probability of $(b)$ decaying to $(c)$. We therefore have
\begin{equation}
\label{Heidrun}
\Gamma_{eb}^{(a\rightarrow c)}=\Gamma_{ic}^{(a\rightarrow b)}(\Delta_s)P^{(b\rightarrow c)}=\Gamma_{ic}^{(a\rightarrow b)}(\Delta_s)\frac{\Gamma^{(b\rightarrow c)}}{\Gamma_b}.
\end{equation} 
Here $\Gamma^{(b\rightarrow c)}$ denotes the partial natural decay rate from state $(b)$ to $(c)$, whereas $\Gamma_b$ is the total natural decay rate (linewidth) of state $(b)$. The rate of IC decay of the nucleus that leads to the electronic transition from state $(a)$ to $(b)$ ($\Gamma_{ic}^{(a\rightarrow b)})$ equals the atomic excitation rate $\Gamma_\text{exc}^{(a\rightarrow b)}$ that is caused by the virtual photons emitted from the nucleus. This is similar to the laser excitation of a two-level system discussed in App.~\ref{THEO.DECEXC.GAMMAEXC}. If the population of state $(b)$ is in equilibrium, the excitation rate will equal the total decay rate, which can be expressed as the product of the population density and the natural decay rate. We can therefore make use of Eq.~(\ref{Dieter}) for the population density to obtain
\begin{equation}
\label{Thomas}
\begin{aligned}
\Gamma_{ic}^{(a\rightarrow b)}(\Delta_s)&=\Gamma_\text{exc}^{(a\rightarrow b)}(\Delta_s)=\rho_\text{exc}^{(b)}(\Delta_s)\Gamma_b\\
&=\Gamma_\text{exc}^{(a\rightarrow b)}(0)\frac{\tilde{\Gamma}_s^2}{\Delta_s^2+\tilde{\Gamma}_s^2}\\
&=\Gamma_{ic}^{(a\rightarrow b)}(0)\frac{\tilde{\Gamma}_s^2}{\Delta_s^2+\tilde{\Gamma}_s^2}.
\end{aligned}
\end{equation}
In complete equivalence to Eq.~(\ref{decoherence}), $\tilde{\Gamma}_s=(\Gamma_s+\Gamma_n)/2$ denotes the decay rate of the coherences of the transition, with $\Gamma_s=\Gamma_b+\Gamma_a$ being the total width of the shell resonance. Inserting Eq.~(\ref{Thomas}) into Eq.~(\ref{Heidrun}), the resulting expression is
\begin{equation}
\label{Jens}
\Gamma_{eb}^{(a\rightarrow c)}=\Gamma_{ic}^{(a\rightarrow b)}\frac{\tilde{\Gamma}_s^2}{\Delta_s^2+\tilde{\Gamma}_s^2}\frac{\Gamma^{(b\rightarrow c)}}{\Gamma_b}.
\end{equation} 
It is convenient to express $\Gamma_{ic}^{(a\rightarrow b)}$ by means of so-called discrete conversion coefficients $\alpha_d$ that were introduced in \cite{Zon} and carry the dimension of angular frequency or energy:
\begin{equation}
\label{Rudi}
\Gamma_{ic}^{(a\rightarrow b)}=\frac{\alpha_d^{(a\rightarrow b)}\Gamma_\gamma}{\Gamma_b}.
\end{equation}
Calculation of the discrete conversion coefficient requires numerical computation. For $^{229}$Th tabulated values can be found in \cite{Karpeshin3,Karpeshin5,Karpeshin6,Karpeshin2018}. To obtain the total EB decay rate, summation over all possible intermediate and final states has to be carried out, leading to the final expression for $\Gamma_{eb}$ as
\begin{equation}
\begin{aligned}
\Gamma_{eb}&=\sum_{b,c}\alpha_d^{(a\rightarrow b)}\frac{\tilde{\Gamma}_s^2}{\Delta_s^2+\tilde{\Gamma}_s^2}\frac{\Gamma^{(b\rightarrow c)}\Gamma_\gamma}{\Gamma_b^2}\\
&=\sum_{b}\alpha_d^{(a\rightarrow b)}\frac{\tilde{\Gamma}_s^2}{\Delta_s^2+\tilde{\Gamma}_s^2}\frac{\Gamma_\gamma}{\Gamma_b}.
\end{aligned}
\end{equation}
Accordingly, the EB coefficient is obtained as (see, e.g., \cite{Karpeshin6,Karpeshin2018})
\begin{equation}
\alpha_{eb}=\sum_{b}\frac{\alpha_d^{(a\rightarrow b)}}{\Gamma_b}\frac{\tilde{\Gamma}_s^2}{\Delta_s^2+\tilde{\Gamma}_s^2}.
\end{equation}
Note that usually it is assumed that $\Gamma_b\gg(\Gamma_n,\Gamma_a)$, resulting in $\tilde{\Gamma}_s=\Gamma_b/2$. The obtained values for $\alpha_{eb}$ heavily depend on the electronic environment and a coincidental overlap of the electronic and nuclear transition. For $^{229}$Th$^+$ typical values vary between $10$ and $10^4$. For a non-vanishing EB decay channel, the decay rate of the nuclear excited state is obtained as
\begin{equation}
\Gamma_n=\Gamma_\gamma+\Gamma_{eb}=\left(1+\alpha_{eb}\right)\Gamma_\gamma.
\end{equation}
\subsubsection{Laser-assisted bound internal conversion}\label{THEO.HIGHDEC.LABIC}
\begin{figure}[t]
 \begin{center}
 \includegraphics[width=6cm]{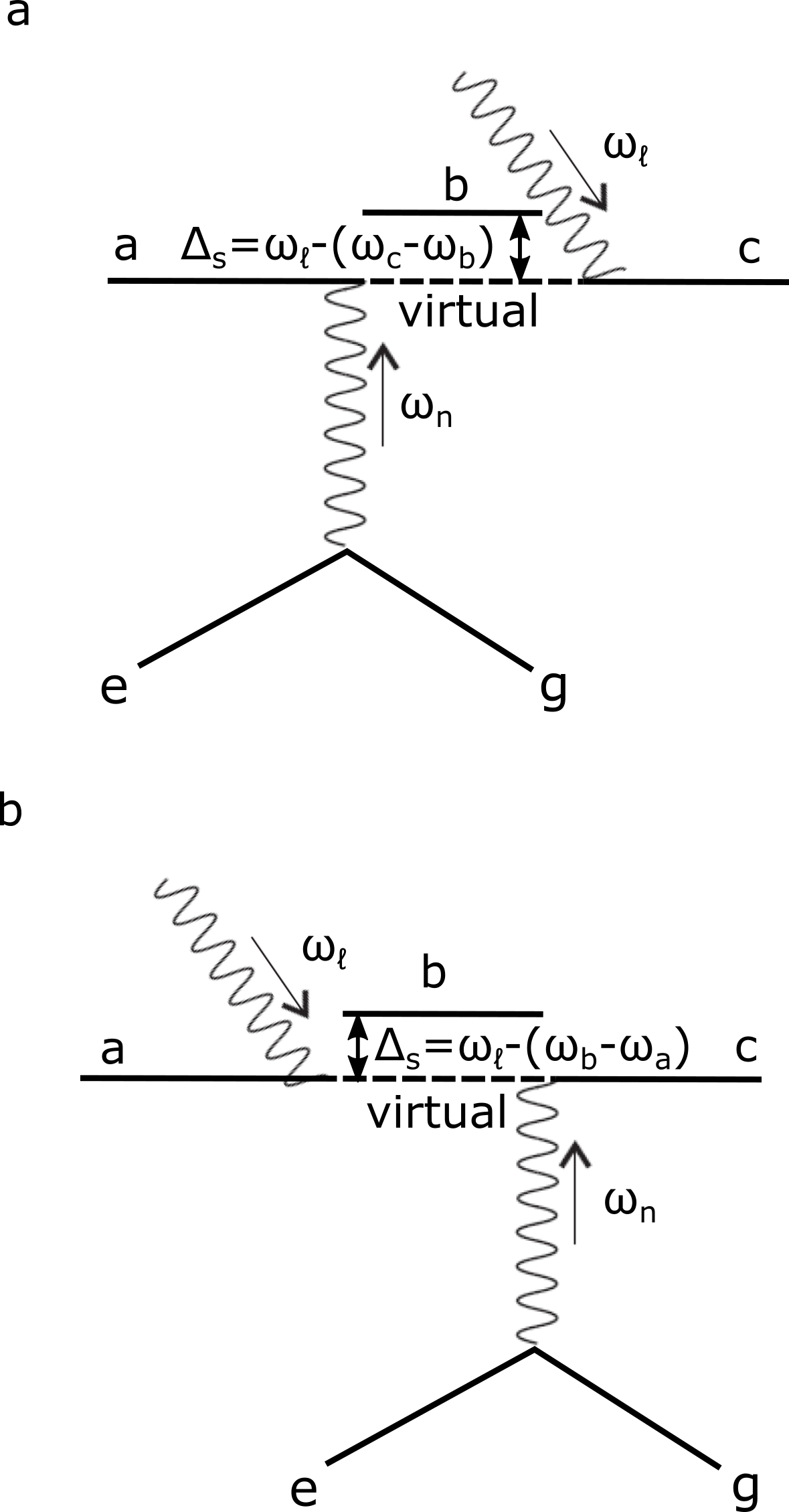}
  \caption{\footnotesize Diagrams for laser-assisted bound internal conversion as described in \cite{Karpeshin1}.}
 \label{LABIC}
 \end{center}
\end{figure}

Laser-assisted BIC was discussed by Karpeshin et al. in \cite{Karpeshin1}. The basic idea is to use laser light to introduce the missing energy into the shell-nucleus system, which is required to fulfill the resonance condition between the nuclear transition and a given excited electronic state. It was inferred that the deexcitation probability could be enhanced by a factor of $10^3$ or more in this way. The same effect was later also discussed by Typel and Leclercq-Willain \cite{Typel} and reviewed in \cite{Matinyan}. A recent publication considering the same process for thorium ions under the name ``laser-induced electronic bridge" (LIEB) is found in \cite{Bilous3} (see also \cite{Bilous2018c}). The process has not yet been experimentally observed.\\[0.2cm]
Laser-assisted BIC can be considered as an excitation process of the electronic shell. Diagrammatic visualizations corresponding to the process are shown in Fig.~\ref{LABIC}. The excitation rate for Fig.~\ref{LABIC} a can be calculated as 
\begin{equation}
\label{Moni}
\Gamma_\text{exc}^{(a\rightarrow c)}=\sum_b\rho_\text{exc}^{(b)}(\Delta_s)\Gamma_\text{exc}^{(b\rightarrow c)}.
\end{equation}
From Eq.~(\ref{Thomas}) we know that the population density for $(b)$ is obtained as
\begin{equation}
\rho_\text{exc}^{(b)}(\Delta_s)= \frac{\Gamma_{ic}^{(a\rightarrow b)}}{\Gamma_b}\frac{\tilde{\Gamma}_s^2}{\Delta_s^2+\tilde{\Gamma}_s^2},
\end{equation}
with $\Delta_s=\omega_\ell-(\omega_c-\omega_b)$, $\tilde{\Gamma}_s=(\Gamma_s+\Gamma_\ell)/2$ and $\Gamma_s=\Gamma_b+\Gamma_c$. One can use Eq.~(\ref{Gammaexc}) to express $\Gamma_\text{exc}^{(b\rightarrow c)}$ as
\begin{equation}
\label{Bene}
\Gamma_\text{exc}^{(b\rightarrow c)}=\frac{\pi c^2 I_\ell}{\hbar \omega_\text{res}^3 \tilde{\Gamma}_\text{res}}\Gamma^{(c\rightarrow b)}.
\end{equation}
Here the resonance frequency follows from the resonance condition of the diagram as $\omega_\text{res}=\omega_c-\omega_a-\omega_n$ and the decoherence rate is introduced as $\tilde{\Gamma}_\text{res}=(\Gamma_\text{res}+\Gamma_\ell)/2$ with $\Gamma_\text{res}=\Gamma_c+\Gamma_a+\Gamma_n$. Inserting Eq.~(\ref{Bene}) into Eq.~(\ref{Moni}) and using Eq.~(\ref{Rudi}) for $\Gamma_{ic}^{(a\rightarrow b)}$, one arrives at the final expression for the excitation rate:
\begin{equation}
\label{Felix}
\Gamma_\text{exc}^{(a\rightarrow c)}=\frac{\pi c^2 I_\ell}{\hbar \omega_\text{res}^3 \tilde{\Gamma}_\text{res}}\sum_b\frac{\alpha_d^{(a\rightarrow b)}\Gamma_\gamma}{\Gamma_b^2}\frac{\tilde{\Gamma}_s^2\Gamma^{(c\rightarrow b)}}{\Delta_s^2+\tilde{\Gamma}_s^2}.
\end{equation}
A similar discussion leads to the excitation rate of the second diagram Fig.~\ref{LABIC}~b, which is identical, however with $\alpha_d^{(a\rightarrow b)}$ replaced by $\alpha_d^{(b\rightarrow c)}$ and $\Gamma^{(c\rightarrow b)}$ replaced by $\Gamma^{(b\rightarrow a)}$. Here the following definitions hold: $\Delta_s=\omega_\ell-(\omega_b-\omega_a)$, $\tilde{\Gamma}_s=(\Gamma_s+\Gamma_\ell)/2$ and $\Gamma_s=\Gamma_a+\Gamma_b$. It is usually assumed that $\Gamma_b\gg(\Gamma_\ell,\Gamma_a,\Gamma_c,\Gamma_n)$, in which case one obtains $\tilde{\Gamma}_s\approx\Gamma_b/2$.  
\subsection{Higher-order processes of isomer excitation: electronic bridge, NEET and NEEC}\label{THEO.HIGHEXC}
\subsubsection{Electronic bridge excitation}\label{THEO.HIGHEXC.IEB}
\begin{figure}[t]
 \begin{center}
 \includegraphics[width=6cm]{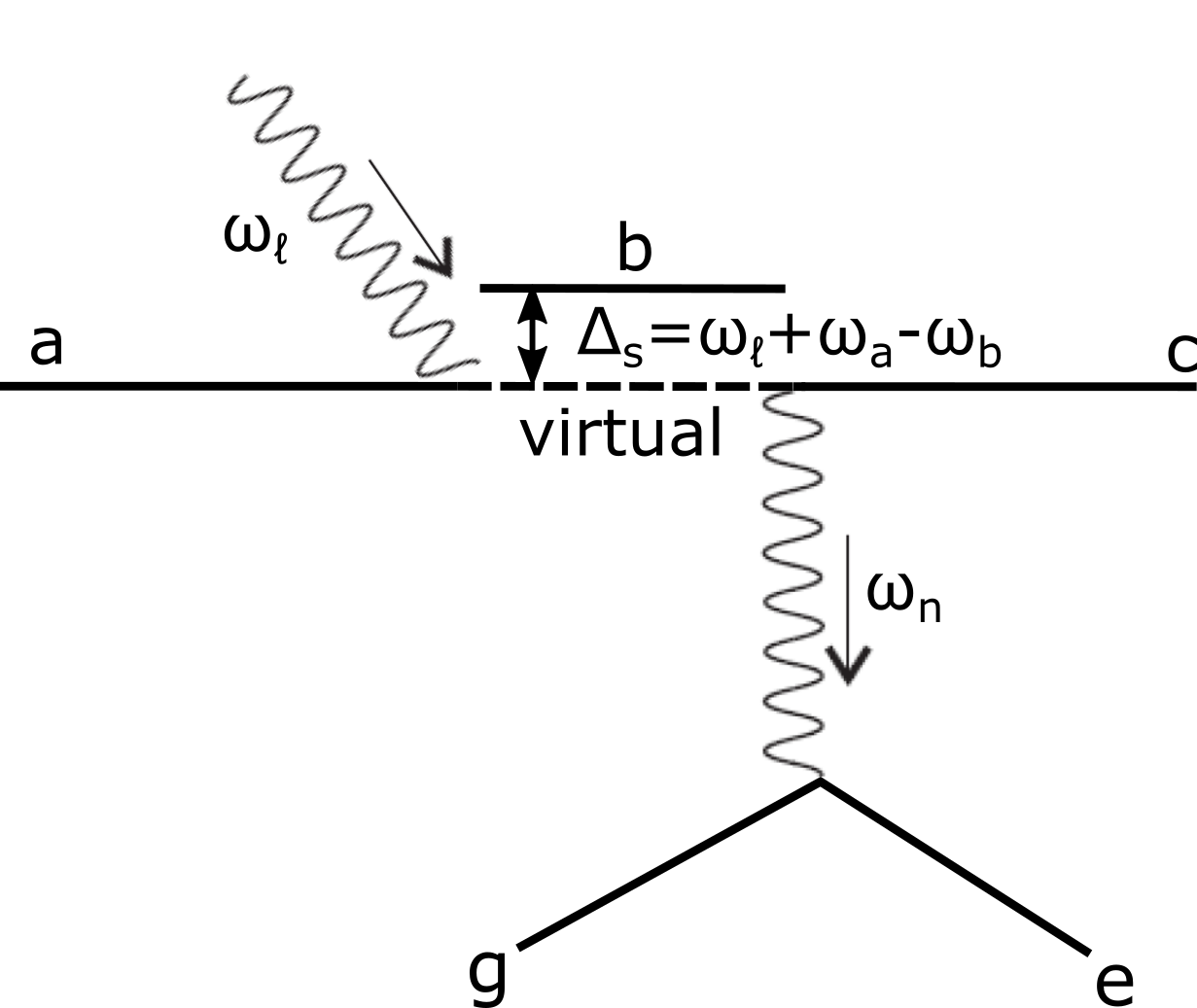}
  \caption{\footnotesize Diagrammatic visualization of nuclear excitation via the EB mechanism. Starting from the initial atomic state $(a)$, a virtual intermediate state $(b)$ is excited, which transfers part of its energy to the nucleus during deexcitation to the final state $(c)$. The expressions ``reverse bound internal conversion", ``reverse resonance conversion" (RRC) and ``inverse electronic bridge" (IEB) are frequently used synonyms for EB excitation. See text for details.}
 \label{EBexcitation}
 \end{center}
\end{figure}

\begin{figure}[t]
 \begin{center}
 \includegraphics[width=7cm]{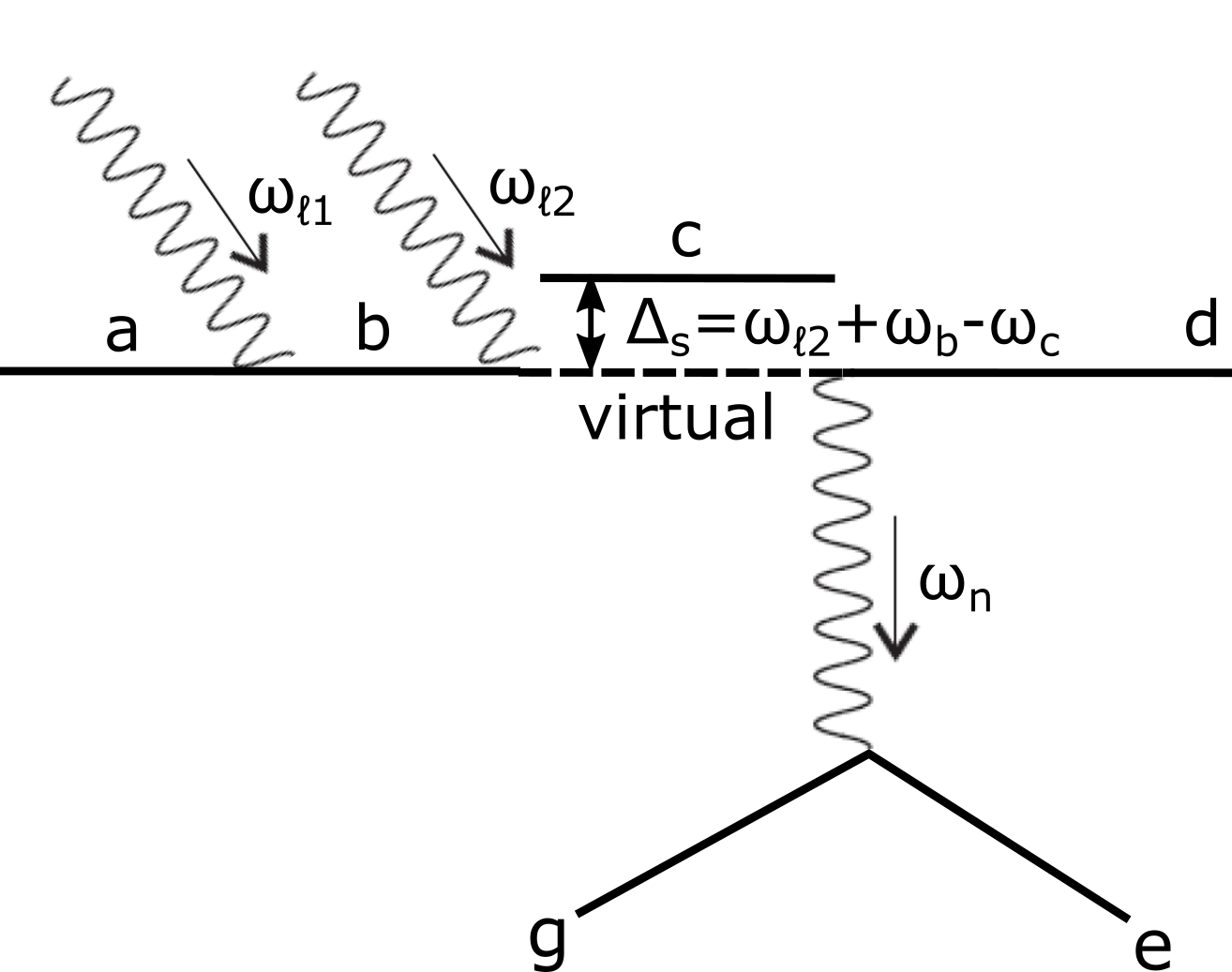}
  \caption{\footnotesize Two-step EB excitation mechanism proposed in \cite{Campbell2,Porsev3} for laser excitation of $^{229\text{m}}$Th. Starting at ground state $(a)$, the atom is excited into state $(b)$ by means of Laser 1. A second laser is used to drive the EB excitation starting from state $(b)$. This scheme has the advantage that longer laser wavelengths can be used, where large laser powers are easily available.}
 \label{twostep}
 \end{center}
\end{figure}

The electronic bridge (EB) excitation shown in Fig.~\ref{EBexcitation} is the reverse of the EB decay given in the diagram in Fig.~\ref{EBdecay}. The concept of the EB mechanism is to excite a virtual electronic level, which subsequently transfers its energy to the nucleus. Importantly, the incident photons have to fulfill the resonance condition, e.g., the energy of the virtual excited shell state minus the energy of the final electronic state has to equal the nuclear transition energy. This poses a central difference compared to the NEET excitation scheme discussed in App.~\ref{THEO.HIGHEXC.NEET}, where a real electronic state is excited, e.g., by laser light, and the nucleus is populated during the free decay of the excited electronic state. EB excitation has not yet been experimentally observed.\\[0.2cm]
Historically, EB was introduced under the name ``Compton excitation of the nucleus" by Batkin in 1979 \cite{Batkin1979}. Later the name ``inverse electronic bridge" (IEB) was introduced for the process \cite{Tkalya1990}. Considering the similarity between EB exc. and NEET, credit is also given to Morita \cite{Morita1973}.\\[0.2cm]
Early papers applying the EB excitation mechanism to $^{229\text{m}}$Th were published by E.V. Tkalya in 1992 \cite{Tkalya5,Tkalya2}. In these publications it was assumed that the laser was detuned from the nuclear resonance, which was later called ``non-resonant IEB" \cite{Tkalya1}. In particular this would permit tuning the laser to a given shell transition, resulting in the excitation of a real instead of a virtual electronic level. In this sense, the proposed process resembles the NEET excitation discussed in App.~\ref{THEO.HIGHEXC.NEET} (NEET and EB exc. were used as synonyms at that time \cite{Tkalya2}). However, it was pointed out in \cite{Kalman} that no emission of additional photons was taken into account as would be required for energy conservation in the NEET process. Therefore the process described in \cite{Tkalya2} requires close proximity between the nuclear and the atomic transitions.\\[0.2cm]
Two different EB excitation mechanisms were distinguished \cite{Tkalya2}: an {\it elastic EB}, in which the final electronic level equals the initial one, and an {\it inelastic EB}, where the electronic levels of the initial and the final states differ. For $^{229\text{m}}$Th, the inelastic EB process was found to be more efficient, as it allows for shell excitation with an $E1$ transition and a decay via an $M1$ transition, which most efficiently excites the nucleus.\\[0.2cm]
Populating $^{229\text{m}}$Th by tuning a laser frequency to the nuclear resonance instead of the atomic shell resonance, which corresponds to the modern definition of the EB excitation shown in Fig.~\ref{EBexcitation}, was proposed by F.F. Karpeshin et al.: ``{\it We propose three different schemes of radiative pumping [...]: (i) By a laser tuned at the frequency of the nuclear transition [...].}" \cite{Karpeshin1994} and ``{\it Now we conclude that one must tune the laser just in resonance with the state $M$ [nuclear transition] in order to really excite the isomeric state [...]}" \cite{Karpeshin2}. The process was then discussed in more detail under the name ``resonant IEB" in \cite{Tkalya1,Tkalya6}.\\[0.2cm]
Various different processes for the isomer's excitation, including EB, were discussed in \cite{Karpeshin3,Karpeshin2002} (EB was considered as one process out of the larger class of NEET processes). A short discussion can also be found in \cite{Tkalya9}. A review of different nuclear laser excitation processes that use the electronic shell, including EB, was provided in \cite{Matinyan}.\\[0.2cm]
After the change of the isomeric energy to a value above the IC threshold for thorium atoms of 6.3~eV in 2007, the idea of a two-step EB excitation scheme (shown in Fig.~\ref{twostep}) was developed. In the two-step EB excitation, an atomic shell state is excited from the atomic ground state prior to the application of the EB mechanism for nuclear excitation. This has the advantage that light of longer wavelength can be applied, which is easier to produce with larger intensity. Two-step EB excitation of the nuclear isomer was proposed in \cite{Campbell2} and later discussed in more detail in \cite{Porsev3,Karpeshin7,Karpeshin11}. The main difference between the excitation schemes proposed in \cite{Porsev3} and \cite{Karpeshin7} is that in \cite{Porsev3} the electronic shell is considered to be in its ground-state after nuclear excitation, whereas in \cite{Karpeshin7} it was argued that larger nuclear excitation rates could be obtained if a particularly chosen electronic excited state is populated after nuclear excitation. The highest calculated EB excitation rates occur for an 8s to 7s shell transition. The same process was also discussed in \cite{Karpeshin12} in comparison to NEET as defined in terms of the diagram shown in Fig. \ref{NEET}~b (note that the expressions ``reverse resonance conversion" and ``reverse bound internal conversion" are frequently used as synonyms for EB excitation). In 2019 the EB decay and excitation processes were discussed for the case of $^{229}$Th-doped VUV transparent crystals \cite{Nickerson2019,Nickerson2020}. The possibility to perform EB excitation in highly charged $^{229}$Th$^{35+}$ ions was also considered \cite{Bilous2020} (see also \cite{Bilous2018c}). In the following, the central equations for EB excitation will be derived.
\\[0.2cm]
\begin{figure}[t]
 \begin{center}
 \includegraphics[width=6cm]{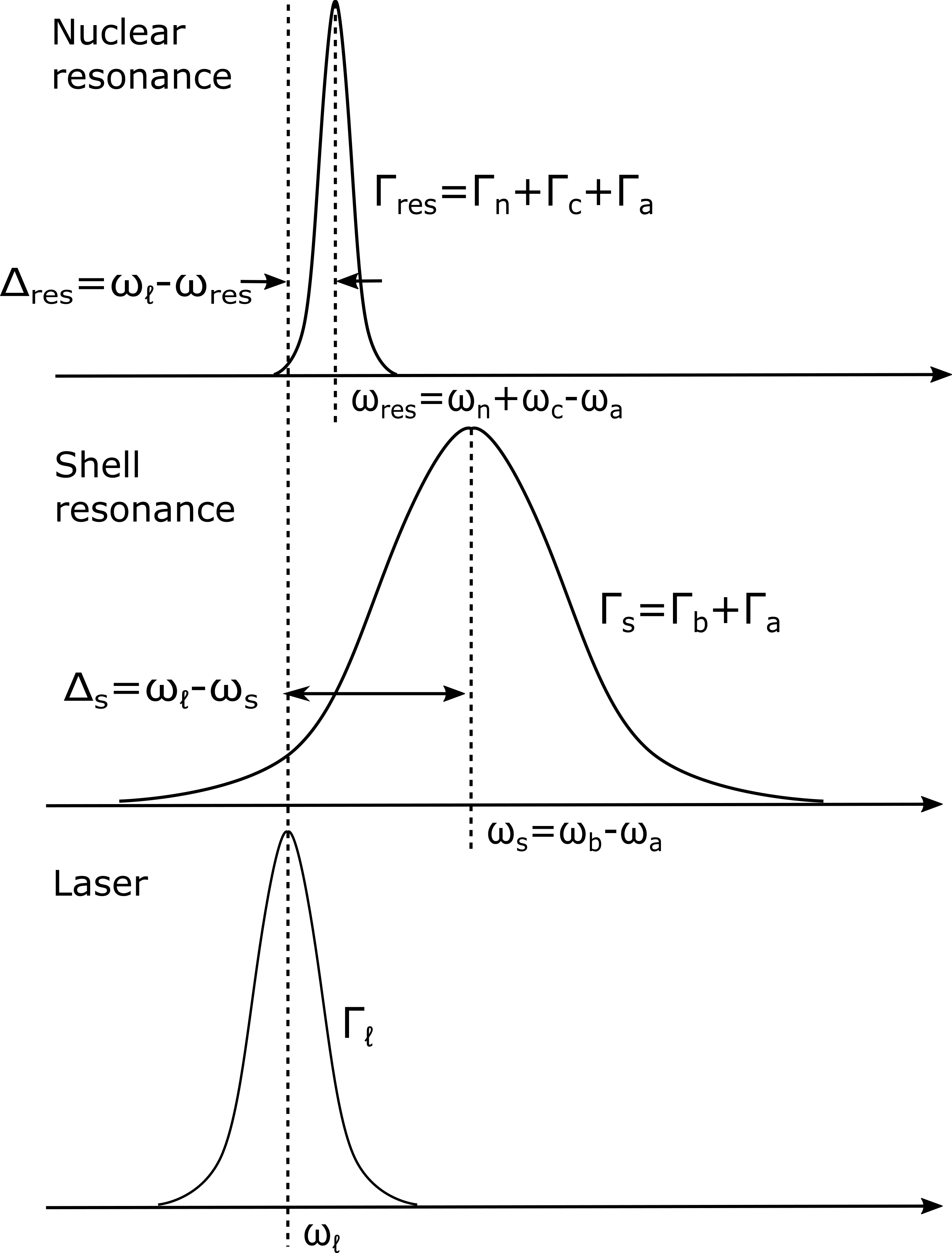}
  \caption{\footnotesize Resonance curves and parameters used in the derivation of EB excitation rate. The laser couples to the nuclear resonance via a virtual layer of electronic shell resonances. See text for explanation.}
 \label{EBresonances}
 \end{center}
\end{figure}
Generally, nuclear excitation via the EB process is found to be more efficient than direct excitation. Therefore it is convenient to introduce an enhancement factor $R_{eb}$ as the ratio between the EB cross section $\sigma_{eb}$ and the cross section for direct photon excitation $\sigma_\gamma$. According to the EB diagram shown in Fig.~\ref{EBexcitation}, an EB resonance occurs if the condition $\omega_\ell+\omega_a=\omega_n+\omega_c$ is fulfilled. This means that there is one resonance for each final state $(c)$ that can be populated during the decay of the virtual excited state $(b)$. Usually the final states $(c)$ are assumed to be sufficiently narrow and distant such that the laser frequency $\omega_\ell$ can be tuned to an individual EB resonance, thereby fixing a final state $(c)$. For the following, we introduce the EB resonance frequency for final state $(c)$ as $\omega_\text{res}^{(c)}=\omega_n+\omega_c-\omega_a$. Based on this consideration and by making use of Eq.~(\ref{Gammaexc}) for the cross sections, one can define an EB enhancement factor for each final state $(c)$ as 
\begin{equation}
R_{eb}^{(c)}=\frac{\sigma_{eb}^{(c)}}{\sigma_\gamma}\simeq\frac{\Gamma_\text{exc}^{eb\ (c)}}{\Gamma_\text{exc}^\gamma}.
\end{equation}
Here $\Gamma_\text{exc}^{eb\ (c)}$ denotes the EB-excitation rate at resonance and $\Gamma_\text{exc}^\gamma$ is the excitation rate for direct photon excitation.\\[0.2cm]
The most straightforward way to obtain an explicit expression for the EB excitation rate is by making use of Eq.~(\ref{Gammaexc}) for the relation between excitation and natural decay rate. This relation is known to be of general validity and leads to \cite{Porsev3}\footnote{Note that an additional factor of $4\pi$ occurs in \cite{Porsev3}, which originates from the definition of the laser spectral intensity as $I_\omega=c\rho_\omega/(4\pi)$. A further factor of $\pi/2$ difference originates from a Lorentzian line-shape that we assume.}
\begin{equation}
\label{Peter}
\Gamma_\text{exc}^{eb\ (c)}=\frac{\pi c^2 I_\ell}{\hbar(\omega_\text{res}^{(c)})^3\tilde{\Gamma}_\text{res}^{(c)}}\Gamma_{eb}^{(c\rightarrow a)}.
\end{equation}
Here $\tilde{\Gamma}_\text{res}^{(c)}=(\Gamma_\text{res}^{(c)}+\Gamma_\ell)/2$ was introduced as the decoherence rate of the EB resonance. As the center frequency of the resonance is influenced by the frequencies of the atomic initial and final states, the total linewidth $\Gamma_\text{res}^{(c)}$ of the EB resonance is obtained as the convolution of the individual linewidths of the atomic and nuclear transitions:
\begin{equation}
\Gamma_\text{res}^{(c)}=\Gamma_n+\Gamma_a+\Gamma_c.
\end{equation}
As before, $\Gamma_n$ denotes the total linewidth of the nuclear transition, including potential non-radiative decay channels. The partial natural EB decay rate $\Gamma_{eb}^{(c\rightarrow a)}$ can be obtained from Eq.~(\ref{Jens}) after summation over $(b)$ and exchanging the initial and final states $(a)$ and $(c)$:
\begin{equation}
\label{Johanna}
\Gamma_{eb}^{(c\rightarrow a)}=\sum_b\Gamma_{ic}^{(c\rightarrow b)}\frac{\tilde{\Gamma}_s^2}{\Delta_s^2+\tilde{\Gamma}_s^2}\frac{\Gamma^{(b\rightarrow a)}}{\Gamma_b}.
\end{equation}
According to the diagram Fig.~\ref{EBexcitation}, the detuning with respect to the intermediate transition also has to be re-defined as $\Delta_s=\omega_\ell-\omega_s$, with $\omega_s$ as the frequency of the atomic transition from $(a)$ to $(b)$: $\omega_s=\omega_b-\omega_a$. The decoherence rate is defined as $\tilde{\Gamma}_s=(\Gamma_s+\Gamma_\ell)/2$, with $\Gamma_s=\Gamma_b+\Gamma_a$ denoting the transition bandwidth. The meaning of the different variables is visualized in Fig.~\ref{EBresonances}. By inserting Eq.~(\ref{Johanna}) into Eq.~(\ref{Peter}) one obtains
\begin{equation}
\label{Rosi}
\Gamma_\text{exc}^{eb\ (c)}=\frac{\pi c^2 I_\ell}{\hbar(\omega_\text{res}^{(c)})^3\tilde{\Gamma}_\text{res}^{(c)}}\sum_b\frac{\tilde{\Gamma}_s^2 \Gamma_{ic}^{(c\rightarrow b)}}{\Delta_s^2+\tilde{\Gamma}_s^2}\frac{\Gamma^{(b\rightarrow a)}}{\Gamma_b}.
\end{equation}
Note that this expression equals the nuclear excitation rate at exact EB resonance $\Delta_\text{res}^{(c)}=\omega_\ell-\omega_\text{res}^{(c)}=0$. It allows us to define the EB Rabi frequency in equivalence to Eq.~(\ref{Rabi}) as $\Omega_{eb}^{(c)\ 2}=2\Gamma_\text{exc}^{eb\ (c)}\tilde{\Gamma}_\text{res}^{(c)}$. Accordingly, Torrey's solution can be used to model the on-resonance nuclear population density as a function of time for EB excitation when substituting $\Omega_\gamma$ by $\Omega_{eb}^{(c)}$, and $\tilde{\Gamma}_n$ by $\tilde{\Gamma}_\text{res}^{(c)}$ in Eq.~(\ref{Torrey}). In the low-saturation limit, and assuming equilibrium and non-zero detuning $\Delta_\text{res}^{(c)}$ of the laser light with respect to the EB resonance, one can use Eq.~(\ref{Dieter}) for the population density yielding
\begin{equation}
\label{Mandy}
\begin{aligned}
\Gamma_\text{exc}^{eb}(\Delta_\text{res})&=\rho_\text{exc}^{eb}(\Delta_\text{res})\Gamma_n\\
&=\Gamma_\text{exc}^{eb}(0)\frac{\tilde{\Gamma}_\text{res}^{2}}{\Delta_\text{res}^{2}+\tilde{\Gamma}_\text{res}^{2}}\\
&=\frac{\pi c^2 I_\ell}{\hbar\omega_\text{res}^{3}}\frac{\tilde{\Gamma}_\text{res}}{\Delta_\text{res}^{2}+\tilde{\Gamma}_\text{res}^{2}}\sum_b\frac{\tilde{\Gamma}_s^2 \Gamma_{ic}^{(c\rightarrow b)}}{\Delta_s^2+\tilde{\Gamma}_s^2}\frac{\Gamma^{(b\rightarrow a)}}{\Gamma_b}.
\end{aligned}
\end{equation}
Here the $(c)$ index was dropped for easier notation. It is convenient to relate the partial IC decay rates $\Gamma_{ic}^{(c\rightarrow b)}$ to the discrete conversion coefficients $\alpha_d$ by means of Eq.~(\ref{Rudi}). By doing so and applying Eq.~(\ref{Gammaexc}) to relate $\Gamma_\gamma$ to $\Gamma_\text{exc}^\gamma$, the following expression for the on-resonance ($\Delta_\text{res}=0$) EB-enhancement factor is obtained
\begin{equation}
\label{Carsten}
R_{eb}^{(c)}\simeq\frac{\tilde{\Gamma}_n}{\tilde{\Gamma}_\text{res}}\sum_b\frac{\tilde{\Gamma}_s^2\alpha_d^{(c\rightarrow b)}}{\Delta_s^2+\tilde{\Gamma}_s^2}\frac{\Gamma^{(b\rightarrow a)}}{\Gamma_b^2}.
\end{equation}
Different derivations of the EB excitation rate can be found in the literature (see, e.g., \cite{Tkalya2,Tkalya1,Karpeshin3}) leading to comparable results. Care has to be taken with the underlying assumptions. Eq.~(\ref{Carsten}) accounts for all possible widths of the atomic, nuclear and laser resonances. It allows for a rough estimate of the enhancement factor: first let us assume that $R_{eb}^{(c)}$ is dominated by a single intermediate state $(b)$ and drop the sum. Further, consider a laser bandwidth that dominates over all widths of atomic and nuclear transitions and takes a value of $\Gamma_\ell=1\cdot10^{-5}$~eV. Then it is reasonable that $\Delta_s\gg\tilde{\Gamma}_s$ holds. Finally, assume that the partial decay rate $\Gamma^{(b\rightarrow a)}$ can be approximated by the total natural decay rate of state $(b)$. In this case Eq.~(\ref{Carsten}) simplifies to 
\begin{equation}
R_{eb}^{(c)}\approx\frac{\Gamma_\ell^2 }{\Delta_s^2}\frac{\alpha_d^{(c\rightarrow b)}}{\Gamma_b}.
\end{equation}
Listed values for $\alpha_d$ vary between $10^6$ and $10^{11}$~eV \cite{Karpeshin3}. By further assuming $\Delta_s\approx 1$~eV and $\Gamma_b\approx\Gamma_\ell$, a characteristic range of values for $R_{eb}^{(c)}$ between $10$ and $10^6$ is obtained.\\[0.2cm]
For the two-step EB process, shown in Fig.~\ref{twostep}, the excitation rate for fixed final state $(c)$ is \cite{Karpeshin7,Karpeshin12}
\begin{equation}
\Gamma_\text{2-step}^{eb\ (c)}=\rho_\text{exc}^{(b)}\Gamma_\text{exc}^{eb\ (c)}.
\end{equation}
Here $\rho_\text{exc}^{(b)}$ denotes the population density of the first excited electronic state $(b)$. In the most general case it can be obtained from Eq.~(\ref{Torrey}). In \cite{Karpeshin7, Karpeshin12}, the low-saturation limit defined by the condition $\rho_\text{exc}^{(b)}\ll 1$ was discussed. In this limit and considering the population of state $(b)$ to be in equilibrium, Eq.~(\ref{Dieter}) can be used for $\rho_\text{exc}^{(b)}$. Under the assumption of resonant excitation one obtains
\begin{equation}
\label{Julia}
\rho_\text{exc}^{(b)}=\frac{\Gamma_\text{exc}^{(a\rightarrow b)}}{\Gamma_b}=\frac{\pi c^2 I_{\ell 1} }{\hbar\omega_{\ell 1}^3\tilde{\Gamma}_b}\frac{\Gamma^{(b\rightarrow a)}}{\Gamma_b},
\end{equation}
where the atomic transition frequency was denoted as $\omega_{\ell 1}$. In \cite{Porsev3} the other extreme was considered, assuming a complete population of state $(b)$ resulting in $\rho_\text{exc}^{(b)}=1$ and correspondingly $\Gamma_\text{2-step}^{eb\ (c)}=\Gamma_\text{exc}^{eb\ (c)}$. In this case Eq.~(\ref{Carsten}) can also be used to calculate a two-step enhancement factor (denoted as $\beta$ in \cite{Porsev3}). Values ranging from $10$ to $10^4$ were predicted, depending on the overlap with electronic transitions.
\subsubsection{Nuclear excitation by electron transition}\label{THEO.HIGHEXC.NEET}
\begin{figure}[t]
 \begin{center}
 \includegraphics[width=6cm]{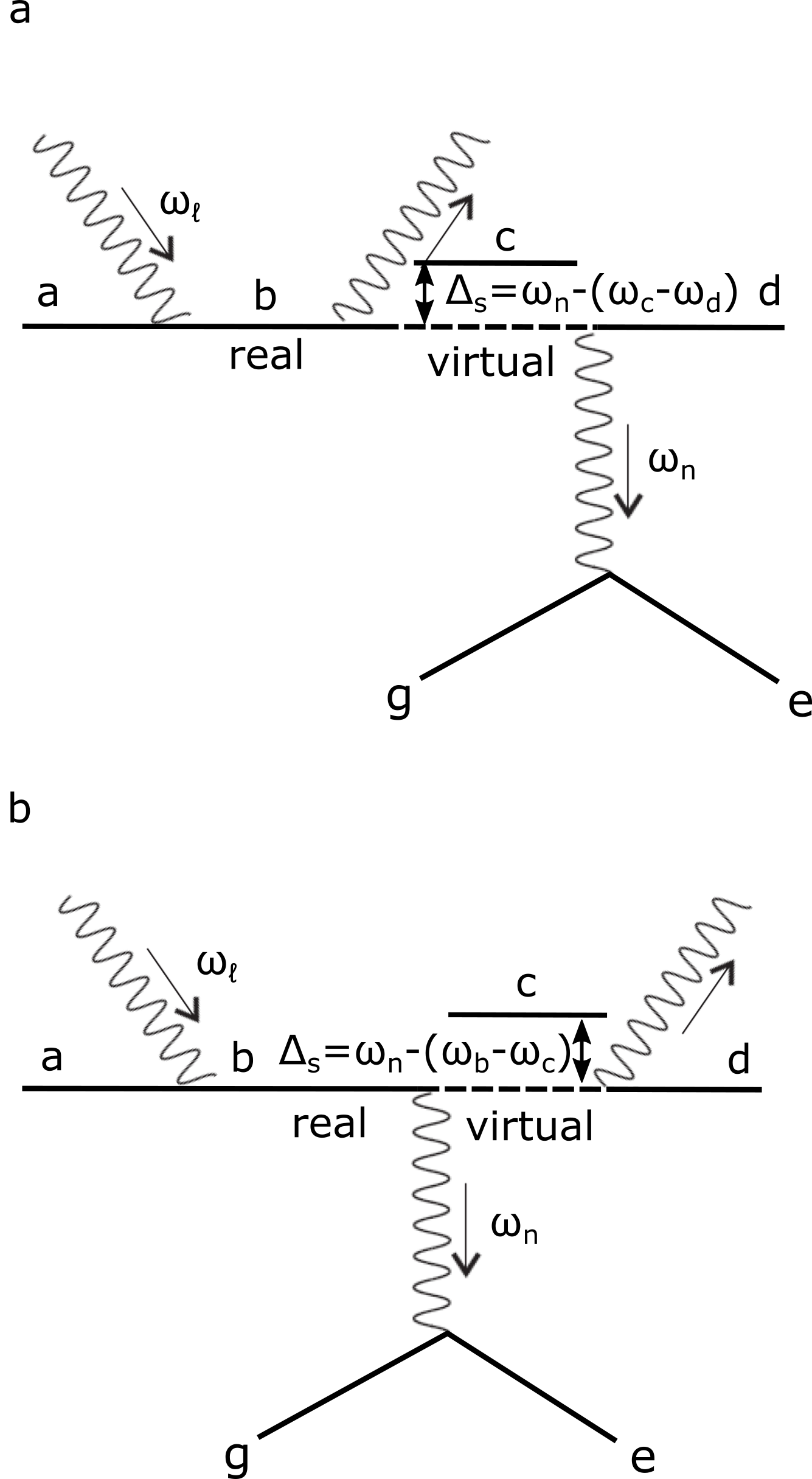}
  \caption{\footnotesize Diagrams contributing to nuclear excitation by electron transition (NEET). Importantly, during NEET, nuclear excitation occurs from a real electronic shell state. This is the main difference compared to EB excitation, where nuclear population occurs directly during laser excitation of a virtual electronic level. While in early publications NEET and EB exc. was used as synonyms (see, e.g., \cite{Tkalya2}), more recently it was proposed to clearly distinguish the two processes \cite{Karpeshin12}.}
\label{NEET}
\end{center}
\end{figure}

There is no consistent definition of nuclear excitation by electron transition (NEET) in the literature. Historically, NEET was proposed in 1973 by Morita \cite{Morita1973} as a process in which an electron from a bound excited state decays, thereby exciting the nucleus. It can therefore be considered as the reverse of BIC. It was assumed that an electron is removed from a deep shell level by, e.g., synchrotron radiation and subsequently an electron from a higher lying shell state fills the vacancy. The energy that is set free in this process can couple to the nucleus if a nuclear level matches the resonance condition. Experimentally the process was studied for $^{189}$Os \cite{Otozai1978}, $^{237}$Np \cite{Saito1980}, $^{197}$Au \cite{Fujioka1984} (see also \cite{Kishimoto2000}) and $^{193}$Ir \cite{Kishimoto2005}. Excitation via NEET was also reported for $^{235}$U \cite{Izawa1979}. The interpretation of the earlier experiments in terms of NEET was critically discussed in \cite{Tkalya5b}. In particular, the result for $^{235}$U was considered doubtful as it could not be reproduced despite significant efforts (see \cite{Chodash2016} and references therein).\\[0.2cm]
Originally, NEET was introduced as a second-order process shown in the Feynman diagram Fig. \ref{excdecay}~g. However, as for the BIC process, in the discrete case this Feynman diagram does not obey energy conservation. Instead, in a modern interpretation of NEET, higher-order processes have to be taken into consideration. The next higher-order process in this context is that shown in Fig.~\ref{EBexcitation} and already described as the EB excitation process. For this reason, NEET and EB were for a long time used interchangeably for the same process; although, the general term ``NEET" encompasses an entire class of even higher-order processes (see, e.g., \cite{Tkalya2,Kalman,Karpeshin3,Karpeshin2002}).\\[0.2cm]
It was then proposed in \cite{Karpeshin12} to clearly distinguish between NEET and EB excitation (BIC was used as a synonym for EB exc.): EB processes were considered as those in which a virtual electronic level is populated before nuclear excitation, whereas in NEET processes the virtual level is populated after nuclear excitation.\\[0.2cm]
Here we follow a slightly different path of defining NEET by assuming that these are processes in which nuclear excitation occurs by the free decay of a real electronic shell state. Note the difference to EB excitation, where a virtual electronic level is continuously populated. This definition is motivated by the fact, that the two most important NEET diagrams, shown in Fig.~\ref{NEET}, cannot be experimentally distinguished. According to this definition, NEET is the time reverse of the laser-assisted BIC process described in App.~\ref{THEO.HIGHDEC.LABIC}.\\[0.2cm]
Laser excitation of $^{229\text{m}}$Th via the NEET process shown in Fig. \ref{NEET}~a was proposed in 1992 by Kálmán and Keszthelyi under the name IEB \cite{Kalman} (note that it took 2 years for the paper to pass the publication process). The same paper also proposed stimulating the emission of the second photon with laser irradiation. Today, one would consider this as ``laser-assisted NEET" (LANEET). The NEET process was further considered in 1994 by Karpeshin et al. \cite{Karpeshin1994} and later discussed in \cite{Karpeshin2,Karpeshin3,Karpeshin2002,Karpeshin12}. In \cite{Mueller2019} the combination of the two processes shown in Fig.~\ref{NEET} is considered for $^{229}$Th$^{2+}$ under the name ``nuclear excitation by two-photon electron transition" (NETP). The obvious advantage of the NEET process compared to EB excitation is that the laser does not have to be tuned to the unknown nuclear transition energy. This may, however, come at the cost of smaller excitation efficiency. In 2019 the process of NEET was discussed for the case of $^{229}$Th-doped VUV transparent crystals in \cite{Nickerson2019} under the name EB excitation.\\[0.2cm] 
Similar to the discussion of the EB excitation process, it is easiest to obtain the natural NEET excitation rate during deexcitation of the electronic shell state $(b)$, $\Gamma_{neet}^{(b)}$, by considering the reverse process. The reverse process, the laser-assisted BIC discussed in App.~\ref{THEO.HIGHDEC.LABIC}, excites the shell with rate $\Gamma_\text{exc}^{(d\rightarrow b)}$ during nuclear decay. Applying Eq.~(\ref{Gammaexc}) to the NEET process, one has
\begin{equation}
\Gamma_\text{exc}^{(d\rightarrow b)}=\frac{\pi c^2 I_\ell}{\hbar \omega_\text{res}^3 \tilde{\Gamma}_\text{res}}\Gamma_{neet}^{(b)}.
\end{equation}
Now using Eq.~(\ref{Felix}) for the atomic excitation rate one obtains for Fig.~\ref{NEET}~a after summation over the intermediate and final states $(c)$ and $(d)$:
\begin{equation}
\Gamma_{neet}^{(b)}=\sum_{c,d}\frac{\alpha_d^{(d\rightarrow c)}\Gamma_\gamma}{\Gamma_c^2}\frac{\tilde{\Gamma}_s^2\Gamma^{(b\rightarrow c)}}{\Delta_s^2+\tilde{\Gamma}_s^2},
\end{equation}
with $\Delta_s=\omega_n-(\omega_c-\omega_d)$, $\tilde{\Gamma_s}=(\Gamma_s+\Gamma_n)/2$ and $\Gamma_s=\Gamma_c+\Gamma_d$. Similarly, the result for Fig.~\ref{NEET}~b is identical, however with $\alpha_d^{(d\rightarrow c)}$ replaced by $\alpha_d^{(c\rightarrow b)}$ and $\Gamma^{(b\rightarrow c)}$ replaced by $\Gamma^{(c\rightarrow d)}$, with $\Delta_s=\omega_n-(\omega_b-\omega_c)$, $\tilde{\Gamma}_s=(\Gamma_s+\Gamma_n)/2$ and $\Gamma_s=\Gamma_b+\Gamma_c$.\\[0.2cm]
Note that the way $\Gamma_{neet}^{(b)}$ was defined above corresponds to the rate of nuclear excitation in the natural decay of the electronic shell state $(b)$. One can also define the probability of nuclear excitation per decay of $(b)$ as
\begin{equation}
P_{neet}^{(b)}=\frac{\Gamma_{neet}^{(b)}}{\Gamma_b}.
\end{equation}
The actual rate of nuclear excitation will depend on the population of state $(b)$. More explicitly, if one assumes equilibrium, then
\begin{equation}
\begin{aligned}
\Gamma_\text{exc}^{neet}&=\Gamma_\text{exc}^{(a\rightarrow b)}P_{neet}^{(b)}=\frac{\Gamma_\text{exc}^{(a\rightarrow b)}}{\Gamma_b}\Gamma_{neet}^{(b)}\\
&=\rho_\text{exc}^{(b)}\Gamma_{neet}^{(b)},
\end{aligned}
\end{equation} 
where $\rho_\text{exc}^{(b)}$ denotes the population density of state $(b)$, which can be obtained for resonant laser irradiation from Torrey's solution Eq.~(\ref{Torrey}). It is also possible to define a NEET enhancement factor as
\begin{equation}
R_{neet}=\frac{\sigma_{neet}}{\sigma_{\gamma}}\simeq\frac{\Gamma_\text{exc}^{neet}}{\Gamma_\text{exc}^\gamma}.
\end{equation}
Using again Eq.~(\ref{Gammaexc}), the enhancement factor is explicitly obtained as
\begin{equation}
R_{neet}\simeq\frac{\tilde{\Gamma_n}}{\tilde{\Gamma}_b}\frac{\Gamma^{(b\rightarrow a)}}{\Gamma_b}\sum_{c,d}\frac{\alpha_d^{(d\rightarrow c)}}{\Gamma_c^2}\frac{\tilde{\Gamma}_s^2\Gamma^{(b\rightarrow c)}}{\Delta_s^2+\tilde{\Gamma}_s^2}.
\end{equation}
A similar expression is obtained for the diagram shown in Fig.~\ref{NEET}~b. Typical predictions for the NEET enhancement factor vary between 1 and $10^3$.\\[0.2cm]
As for EB, two-step excitation processes were also proposed for NEET (see \cite{Karpeshin12,Mueller2019}). The excitation rate for the two-step NEET process is
\begin{equation}
\Gamma_\text{2-step}^{neet}=\rho_\text{exc}^{(c)}\Gamma_{neet}^{(c)}=\rho_\text{exc}^{(b)}\frac{\Gamma_\text{exc}^{(b\rightarrow c)}}{\Gamma_c}\Gamma_{neet}^{(c)}.
\end{equation}
Note that $\rho_\text{exc}^{(c)}=\rho_\text{exc}^{(b)}\Gamma_\text{exc}^{(b\rightarrow c)}/\Gamma_c$ holds. Here $\rho_\text{exc}^{(b)}$ denotes the population density of the first atomic state $(b)$, given in Eq.~(\ref{Julia}). 
\subsubsection{Nuclear excitation by electron capture}\label{THEO.HIGHEXC.NEEC}
The reverse process of internal conversion is called nuclear excitation by electron capture (NEEC, see Fig.~\ref{excdecay}~g). In the NEEC process, a free electron is captured by an ion and the released binding energy excites the nucleus.\\[0.2cm]
The NEEC process was theoretically introduced in 1976 \cite{Godanskii} under the name reverse internal electron conversion (RIEC) and experimentally observed in 2018 for $^{93}$Mo \cite{Chiara}. A recent discussion of the NEEC process can be found in \cite{Gunst2018}.\\[0.2cm]
The excitation of $^{229\text{m}}$Th via NEEC was proposed in \cite{Strizhov} under the original name RIEC. Importantly, any electron capture process requires the production of an electron vacancy in the atomic shell. Two different excitation schemes are usually considered: Excitation in a laser-generated plasma (see, e.g., \cite{Strizhov,Harston1999,Tkalya2004,Gunst2018} and references therein) and electron capture by ions in a storage ring \cite{Palffy2,Palffy3}. Due to the large expected enhancement factor for the NEEC process of up to 6 orders of magnitude compared to direct photon excitation \cite{Gunst2014}, both concepts are experimentally investigated. An experiment aimed at plasma excitation is discussed in \cite{Borisyuk2018b,Lebedinskii2019} and references therein. Experiments at storage rings are described in \cite{Brandau1,Brandau2,Ma}.\\[0.2cm]
The process of laser ionization followed by NEEC was introduced as laser-assisted nuclear excitation by electron capture (LANEEC) in \cite{Bilous2018c}. The same process was theoretically investigated for the excitation of $^{229\text{m}}$Th under the name ``electronic bridge via the continuum" in Refs. \cite{Borisyuk2019,Dzyublik}. Large enhancement factors for the nuclear excitation were predicted. This process will be investigated in more detail in the following.\\[0.2cm]
The EB excitation rate given in Eq.~(\ref{Mandy}) is used as a starting point to estimate the LANEEC excitation rate for $^{229\text{m}}$Th. As electron excitation to the continuum is always resonant, $\Delta_s=0$ is assumed. Further, only the case of zero detuning with respect to the nuclear resonance is considered leading to $\Delta_\text{res}=0$. In this case Eq.~(\ref{Mandy}) transforms to
\begin{equation}
\label{Sarah}
\begin{aligned}
\Gamma_\text{exc}^{neec}&=\frac{\pi c^2 I_\ell}{\hbar\omega_\text{res}^{3}\tilde{\Gamma}_\text{res}}\sum_b \frac{\Gamma^{(b\rightarrow a)}}{\Gamma_b} \Gamma_{ic}^{(c\rightarrow b)}\\
&\simeq\frac{1}{\tilde{\Gamma}_\text{res}}\sum_b \frac{\tilde{\Gamma}_b}{\Gamma_b}\Gamma_\text{exc}^{(a\rightarrow b)} \Gamma_{ic}^{(c\rightarrow b)}.
\end{aligned}
\end{equation} 
For the last approximation Eq.~(\ref{Gammaexc}) was used. For the continuum we have $\Gamma_b\gg\Gamma_\ell$ and therefore $\tilde{\Gamma}_b=\Gamma_b/2$. Further, one can define the ionization cross section as
\begin{equation}
\sigma_\text{ion}^{(a\rightarrow b)}=\frac{\hbar\omega_\text{res}\Gamma_\text{exc}^{(a\rightarrow b)}}{I_\ell}.
\end{equation}
Inserting this into Eq.~(\ref{Sarah}) and by assuming that the ionization cross section takes the same value $\sigma_\text{ion}$ for each transition (which is in general not valid, but which later allows us to obtain an upper limit for the excitation rate) one obtains
\begin{equation}
\Gamma_\text{exc}^{neec}\approx \frac{I_\ell\sigma_\text{ion}}{2\hbar\omega_\text{res}\tilde{\Gamma}_\text{res}} \sum_b \Gamma_{ic}^{(c\rightarrow b)}.
\end{equation}
The summation over the partial IC decay rates $\Gamma_{ic}^{(c\rightarrow b)}$ corresponds to the total IC decay rate from state $(c)$:
\begin{equation}
\sum_b\Gamma_{ic}^{(c\rightarrow b)}=\Gamma^{(c)}_{ic}=\alpha^{(c)}_{ic}\Gamma_\gamma.
\end{equation}
Here $\Gamma^{(c)}_{ic}$ was introduced as the total IC decay rate of state (c). Therefore we obtain
\begin{equation}
\Gamma_\text{exc}^{neec}\approx \frac{I_\ell\sigma_\text{ion}\alpha^{(c)}_{ic}\Gamma_\gamma}{2\hbar\omega_\text{res}\tilde{\Gamma}_\text{res}}.
\end{equation}
In a last step the NEEC enhancement factor is defined as
\begin{equation}
R_{neec}=\frac{\sigma_{neec}}{\sigma_\gamma}\simeq\frac{\Gamma_\text{exc}^{neec}}{\Gamma_\text{exc}^{\gamma}}
\end{equation}
and by again applying Eq.~(\ref{Gammaexc}) for $\Gamma_\text{exc}^\gamma$ one obtains
\begin{equation}
R_{neec}\simeq\frac{\omega_n^2\tilde{\Gamma}_n}{2\pi c^2\tilde{\Gamma}_\text{res}}\sigma_\text{ion}\alpha_{ic}^{(c)}.
\end{equation}
For the following we approximate $\alpha_{ic}^{(c)}$ by $\alpha_{ic}\approx10^{9}$. The ionization cross section will be transition dependent. In order to obtain an upper limit for the enhancement factor, a large value of $\sigma_\text{ion}\approx10^{-17}$~cm$^2$ is assumed \cite{Borisyuk2019}. The remaining factor is calculated to be $\omega_n^2/(2\pi c^2)\approx 3\cdot10^{10}$~cm$^{-2}$. Based on these considerations the NEEC enhancement ratio is estimated to be
\begin{equation}
R_{neec}\simeq 300\frac{\tilde{\Gamma}_n}{\tilde{\Gamma}_\text{res}}.
\end{equation}
This is in agreement with the result from \cite{Borisyuk2019} for the case that the bandwidth of the laser light dominates both decoherences, e.g., $\Gamma_\ell\gg (\Gamma_n,\Gamma_a,\Gamma_c)$.\\[0.2cm]
In 2019 a process of ``nonlinear laser nuclear excitation" was proposed by Andreev et al. \cite{Andreev2019} (see also \cite{Andreev2019b}). In this process the atomic shell is used to generate high harmonics from short laser pulses. These high harmonics couple to the nucleus via different mechanisms. High harmonic generation can be described as the radiative recombination of an electron after tunnel ionization. Nuclear excitation can then occur by NEEC during the recombination process, via direct excitation from the high harmonics, inelastic scattering at the nuclei or higher-order processes in the atomic shell. High excitation efficiency is expected from this mechanism.\\[0.2cm]
Several alternative ways to excite the nucleus have been proposed in the literature. These include $^{229\text{m}}$Th excitation via surface plasmons \cite{Varlamov}, via Coulomb excitation \cite{Alexandrov,Isakov}, low-energy ion scattering spectroscopy \cite{Borisyuk1} and electron-beam surface irradiation \cite{Troyan4,Borisyuk2016,Tkalya2020b}. Also several theoretical proposals were put forward that make use of a storage ring for the investigation $^{229\text{m}}$Th: the isomeric decay in hydrogen-like $^{229}$Th (Th$^{89+}$) was considered by Karpeshin et al. in 1998 \cite{Karpeshin8}. An experiment to probe the special predictions, making use of the ESR storage ring at GSI, has been proposed. Hydrogen-like $^{229}$Th had already been discussed earlier in the context of nuclear spin mixing \cite{Wycech,Pachucki}. Here also a storage ring experiment was proposed and the case of muonic thorium was discussed. The isomeric state in $^{229}$Th$^{89+}$ and $^{229}$Th$^{87+}$, as well as in muonic thorium, has been re-investigated by E.V. Tkalya in several independent publications \cite{Tkalya10,Tkalya11,Tkalya12}.\\[0.2cm]
\end{appendices}
\noindent\textbf{Acknowledgements}\\
During our work on the $^{229}$Th nuclear clock project we have been in contact with many people working in the field and the comprehensive review presented here is the outcome of years of discussions and thoughts on the topic. In particular, we would like to acknowledge discussions with (in alphabetic order) K. Beeks, J. Berengut, P.V. Bilous, P.V. Borisyuk, S. Brewer, J.T. Burke, B. Changala, J. Crespo, T. Dickel, S. Ding, S. Dörscher, C.E. Düllmann, C. Enss, A. Fleischmann, J. Geist, S. Geldorf, P. Glowacki, F. Greiner, D. Habs, J. Hall, T.W. Hänsch, C. Heyl, C. Hornung, E.R. Hudson, J. Jeet, F.F. Karpeshin, G.A. Kazakov, M. Laatiaoui, D.R. Leibrandt, M. Mallweger, D.M. Meier, P. Micke, I. Moore, R.A. Müller, S.W. Nam, J.B. Neumayr, B. Nickerson, M. Okhapkin, G.C. O'Neil, A. Ozawa, A. Pálffy, E. Peik, W. Plass, I. Pohjalainen, F. Ponce, G. Porat, O. Pronin, I. Pupeza, A. Radnaev, M.S. Safronova, C. Sanner, P.O. Schmidt, L. Schmöger, C. Schneider, S. Schoun, T. Schumm, Y. Shigekawa, S. Stellmer, C. Tamm, J. Thielking, P.G. Thirolf, E.V. Tkalya, T. Udem, P. van Duppen, J. Weitenberg, D. Wineland, J. Ye, K. Yoshimura, C. Zhang and G.A. Zitzer. We thank B. Changala and S. Dörscher for proof-reading the manuscript. This work was supported by the European Union's Horizon 2020 research and innovation
programme under grant agreement 664732 ``nuClock", by DFG (Th956/3-2) and by the Humboldt Foundation.


\begin{thebibliography}{99}
\bibitem[Ahmad\ 2015]{Ahmad2015} Ahmad, I. et al., {\it Electron capture decay of 58-min $^{229}_{92}$U and levels in $^{229}_{91}$Pa}, Phys. Rev. C \textbf{92} (2015) 024313.
\bibitem[Akavoli\ 1989]{Akavoli} Akavoli, Y.A., {\it Nuclear Data Sheets for $A=229$}, Nuclear Data Sheets \textbf{58} (1989) 555.
\bibitem[Alexandrov\ 2011]{Alexandrov} Alexandrov, V.S. et al., {\it Nuclear transitions and new standards of length and time}, Acta Physica Polonica B \textbf{42} (2011) 853-858 .
\bibitem[Andreev\ 2019a]{Andreev2019} Andreev, A.V. et al., {\it Nuclear isomer excitation in $^{229}$Th atoms by superintense laser fields}, Phys. Rev. A \textbf{99} (2019) 013422.
\bibitem[Andreev\ 2019b]{Andreev2019b} Andreev, A.V. et al., {\it Toward the possibility of $^{229}$Th isomeric nuclear state excitation by two-color laser field}, EPJ Web of Conferences \textbf{220} (2019) 01001.
\bibitem[Attallah\ 1995]{Attallah1995} Attallah, F. et al., {\it Charge state blocking of K-shell internal conversion in $^{125}$Te}, Phys. Rev. Lett. \textbf{75} (1995) 1715-1718.
\bibitem[Band\ 2001]{Band} Band, I.M., {\it Discrete conversion of gamma rays in $^{229}$Th and in highly ionized $^{125}$Te$^Q$ ions}, Jour. Exp. Theor. Phys. \textbf{93} (2001) 948-956.
\bibitem[Barci\ 2003]{Barci} Barci, V. et al., {\it Nuclear structure of $^{229}$Th from $\gamma$-ray spectroscopy study of $^{233}$U $\alpha$-particle decay}, Phys. Rev. C \textbf{68} (2003) 034329.
\bibitem[Barker\ 2018]{Barker2018} Barker, B.J. et al., {\it Oxidation state of $^{229}$Th recoils implanted into MgF$_2$ crystals}, Science Journal of Chemistry \textbf{6} (2018) 66-76.
\bibitem[Batkin\ 1979]{Batkin1979} Batkin, I.S., {\it Compton excitation of the nuclear levels}, Yad. Fiz \textbf{29} (1979) 903 [Sov. J. Nucl. Phys. \textbf{29} (1979)  464].
\bibitem[Beck\ 2007]{Beck1} Beck, B.R. et al., {\it Energy splitting of the ground-state doublet in the nucleus $^{229}$Th}, Phys. Rev. Lett. \textbf{109} (2007) 142501.
\bibitem[Beck\ 2009]{Beck2} Beck, B.R. et al., {\it Improved value for the energy splitting of the ground-state doublet in the nucleus $^{229m}$Th}, LLNL-PROC-415170 (2009).
\bibitem[Beloy\ 2014]{Beloy} Beloy, K., {\it Hyperfine structure in $^{229g}$Th$^{3+}$ as a probe of the $^{229g}$Th $\rightarrow$ $^{229m}$Th nuclear excitation energy}, Phys. Rev. Lett. \textbf{112} (2014) 062503.
\bibitem[Bemis\ 1988]{Bemis} Bemis, C.E., et al., {\it Coulomb excitation of states in $^{229}$Th}, Physica Scripta \textbf{38} (1988) 657-663.
\bibitem[Benko\ 2014]{Benko} Benko, C. et al., {\it Extreme ultraviolet radiation with coherence time greater than 1~s}, Nature Photonics \textbf{8} (2014) 530-536.
\bibitem[Benko\ 2016]{Benko2016} Benko, C., {\it Extreme ultraviolet frequency combs for precision measurement and strong-field physics}, Ph.D. Thesis, University of Colorado, USA (2016).
\bibitem[Bennet\ 2002]{Bennet} Bennet, M. et al., {\it Huygens'clocks}, Proceedings of the Royal Society of Lodon A \textbf{458} (2002) 563-579.
\bibitem[Berengut\ 2009]{Berengut} Berengut, J.C. et al., {\it Proposed experimental method to determine $\alpha$ sensitivity of splitting between ground and 7.6 eV isomeric states in $^{229}$Th}, Phys. Rev. Lett. \textbf{102} (2009) 210801.
\bibitem[Berengut\ 2010]{Berengut2} Berengut, J.C., Flambaum, V.V., {\it Testing time-variation of fundamental constants using a $^{229}$Th nuclear clock}, Nucl. Phys. News \textbf{20} (2010) 19-22.
\bibitem[Berengut\ 2012]{Berengut3} Berengut, J.C., Flambaum, V.V., {\it Manifestation of a spatial variation of fundamental constants in atomic and nuclear clocks, Oklo, meteorites, and cosmological phenomena}, EPL \textbf{97} (2012) 20006.
\bibitem[Berengut\ 2018]{Berengut2018} Berengut, J.C., {\it Resonant electronic-bridge excitation of the $^{235}$U nuclear transition in ions with chaotic spectra}, Phys. Rev. Lett. \textbf{121} (2018) 253002.
\bibitem[Bernhardt\ 2012]{Bernhardt2012} Bernhardt, B. et al., {\it Vacuum ultraviolet frequency combs generated by a femtosecond enhancement cavity in the visible}, Optics Letters \textbf{37} (2012) 503-505.
\bibitem[Bilous\ 2015]{Bilous} Bilous, P.V., Yatsenko, L.P., {\it Analysis of parasitic signals in the method of recoil nuclei applied to direct observation of the $^{229\mathrm{m}}$Th isomeric state}, Ukr. J. Phys. \textbf{60} (2015) 376-381.
\bibitem[Bilous\ 2017]{Bilous2} Bilous, P.V. et al., {\it Internal conversion from excited electronic states of $^{229}$Th ions}, Phys. Rev. A \textbf{95} (2017) 032503.
\bibitem[Bilous\ 2018a]{Bilous2018a} Bilous, P.V., Minkov, N., Pálffy, A., {\it Electric quadrupole channel of the 7.8~eV $^{229}$Th transition}, Phys. Rev. C \textbf{97} (2018) 044320.
\bibitem[Bilous\ 2018b]{Bilous3} Bilous, P.V., Peik, E., Pálffy, A., {\it Laser induced electronic bridge for characterization of the $^{229\text{m}}$Th $\rightarrow$ $^{229\text{g}}$Th nuclear transition with a tunable optical laser}, New Journal of Physics \textbf{20} (2018) 013016.
\bibitem[Bilous\ 2018c]{Bilous2018c} Bilous, P.V., {\it Towards a nuclear clock with the $^{229}$Th isomeric transition}, Ph.D. Thesis, University of Heidelberg, Germany (2018).
\bibitem[Bilous\ 2020]{Bilous2020} Bilous, P.V. et al., {\it Electronic bridge excitation in highly charged $^{229}$Th ions}, Phys. Rev. Lett. \textbf{124} (2020) 192502.
\bibitem[Blatt\ 1952]{Blatt1952} Blatt, J.M., Weisskopf, V.F., {\it Theoretical Nuclear Physics}, John Wiley and Sons, New York (1952).
\bibitem[Bohnet\ 2012]{Bohnet2012} Bohnet, J.G. et al., {\it A steady-state superradiant laser with less than one intracavity photon}, Nature \textbf{484} (2012) 78-81.
\bibitem[Borisyuk\ 2014a]{Borisyuk2} Borisyuk, P.V. et al., {\it The band structure of submonolayer thorium coatings on a silicon oxide surface}, Colloid Journal \textbf{76} (2014) 645-650.
\bibitem[Borisyuk\ 2014b]{Borisyuk3} Borisyuk, P.V. et al., {\it The formation of submonolayer thorium coatings on a silicon oxide surface by electrochemical deposition}, Colloid Journal \textbf{76} (2014) 514-421.
\bibitem[Borisyuk\ 2015a]{Borisyuk} Borisyuk, P.V. et al., {\it Band structure and decay channels of thorium-229 low lying isomeric state for ensemble of thorium atoms adsorbed on calcium fluoride}, Phys. Status Solidi C \textbf{12} (2015) 1333-1337.
\bibitem[Borisyuk\ 2015b]{Borisyuk1} Borisyuk, P.V. et al., {\it Preparation technique of thorium films by electrochemical deposition for nuclear optical frequency standard based on thorium-229}, Journal of Sol-Gel Science and Technology \textbf{73} (2015) 580-585.
\bibitem[Borisyuk\ 2016]{Borisyuk2016} Borisyuk, P.V. et al., {\it Thorium silicate compound as a solid-state target for production of isomeric thorium-229 nuclei by electron beam irradiation}, AIP Advances \textbf{6} (2016) 095304.
\bibitem[Borisyuk\ 2017a]{Borisyuk4} Borisyuk, P.V. et al., {\it Trapping, retention and laser cooling of Th$^{3+}$ ions in a multisection linear quadrupole trap}, Quantum Electron \textbf{47} (2017) 406-411.
\bibitem[Borisyuk\ 2017b]{Borisyuk5} Borisyuk, P.V. et al., {\it Mass selective laser cooling of $^{229}$Th$^{3+}$ in a multisectional linear Paul trap loaded with a mixture of thorium isotopes}, Europ. Jour. Mass. Spectr. \textbf{23} (2017) 136-139.
\bibitem[Borisyuk\ 2017c]{Borisyuk2017c} Borisyuk, P.V. et al., {\it Doppler cooling of thorium ions in a multisectional linear Paul trap}, IOP Conf. Series: Journal of Physics: Conf. Series \textbf{941} (2017) 012111.
\bibitem[Borisyuk\ 2017d]{Borisyuk2017d} Borisyuk, P.V. et al., {\it Formation of local thorium silicate compound by electrochemical deposition from an acetone solution of thorium nitrate}, J. Sol-Gel Sci. Technol. \textbf{81} (2017) 313-320.
\bibitem[Borisyuk\ 2017e]{Borisyuk2017e} Borisyuk, P.V. et al., {\it Method of the production and trapping of thorium ions for nuclear transition investigation}, IOP Conf. Series: Journal of Physics: Conf. Series \textbf{941} (2017) 012107.
\bibitem[Borisyuk\ 2018a]{Borisyuk2018a} Borisyuk, P.V. et al., {\it Experimental studies of thorium ion implantation from pulse laser plasma into thin silicon oxide layers}, Laser Phys. Lett. \textbf{15} (2018) 056101.
\bibitem[Borisyuk\ 2018b]{Borisyuk2018b} Borisyuk, P.V. et al., {\it Excitation of $^{229}$Th nuclei in laser plasma: the energy and half-life of the low-lying isomeric state}, arXiv:1804.00299 (2018).
\bibitem[Borisyuk\ 2018c]{Borisyuk2018c} Borisyuk, P.V. et al., {\it Surface physicochemical properties and decay of the low-lying isomer in the $^{229}$Th nucleus}, Quantum Electronics \textbf{48} (2018) 460-463.
\bibitem[Borisyuk\ 2019]{Borisyuk2019} Borisyuk, P.V. et al., {\it Excitation of the low-energy $^{229\text{m}}$Th isomer in the electron bridge process via the continuum}, Phys. Rev. C \textbf{100} (2019) 044306.
\bibitem[Bothwell\ 2019]{Bothwell2019} Bothwell, T. et al., {\it JILA SrI optical lattice clock with uncertainty of $2.0\times10^{-18}$}, Metrologia \textbf{56} (2019) 065004.
\bibitem[Brandau\ 2010]{Brandau1} Brandau, C. et al., {\it Resonant recombination at ion storage rings: a conceptual alternative for isotope shift and hyperfine studies}, Hyperfine Interact. \textbf{196} (2010) 115-127.
\bibitem[Brandau\ 2013]{Brandau2} Brandau, C. et al., {\it Probing nuclear properties by resonant atomic collisions between electrons and ions}, Phys. Scr. \textbf{T156} (2013) 014050.
\bibitem[Brewer\ 2019]{Brewer} Brewer, S.M. et al., {\it An $^{27}$Al$^{+}$ quantum-logic clock with systematic uncertainty below $10^{-18}$}, Phys. Rev. Lett. \textbf{123} (2019) 033201.
\bibitem[Browne\ 2001]{Browne} Browne, E. et al., {\it Search for decay of the 3.5-eV level in $^{229}$Th}, Phys. Rev. C \textbf{64} (2001), 014311.
\bibitem[BurkeD\ 1990]{Burke1} Burke, D.G. et al., {\it Additional evidence for the proposed excited state at $\le$ 5 eV in $^{229}$Th}, Phys. Rev. C \textbf{42} (1990) 499.
\bibitem[BurkeD\ 2008]{Burke2} Burke, D.G. et al., {\it Nuclear structure of $^{229,231}$Th studied with the $^{230,232}$Th(d,t) reactions}, Nucl. Phys. A \textbf{809} (2008) 129-170.
\bibitem[BurkeJ\ 2010]{JBurke} Burke, J.T. et al., {\it $^{229}$Th the bridge between nuclear and atomic interactions}, Law\ rence Livermore National Laboratory Technical Report LLNL-TR-463538 (2010).
\bibitem[BurkeJ\ 2019]{Burke2019} Burke, J.T., {\it One tick closer to a nuclear clock}, Nature \textbf{573} (2019) 202-203.
\bibitem[Bussmann\ 2007]{Bussmann} Bussmann, M., {\it Laser-cooled ion beams and strongly coupled plasmas for precision experiments}, Ph.D. Thesis, Ludwig-Maximilians-Universität München, Germany (2007). 

\bibitem[Campbell\ 2009]{Campbell2} Campbell, C.J. et al., {\it Multiply charged thorium crystals for nuclear laser spectroscopy}, Phys. Rev. Lett. \textbf{102} (2009) 233004.
\bibitem[Campbell\ 2011a]{Campbell3} Campbell, C.J. et al., {\it Wigner Crystals of $^{229}$Th for optical excitation of the nuclear isomer}, Phys. Rev. Lett. \textbf{106} (2011) 223001.
\bibitem[Campbell\ 2011b]{Campbell4} Campbell, C.J., {\it Trapping, laser cooling, and spectroscopy of thorium IV}, Ph.D. Thesis, Georgia Institute of Technology, USA (2011).
\bibitem[Campbell\ 2012]{Campbell1} Campbell, C.J. et al., {\it Single-Ion nuclear clock for metrology at the 19th decimal place}, Phys. Rev. Lett. \textbf{108} (2012) 120802.
\bibitem[Canty\ 1977]{Canty} Canty, M.J. et al., {\it The decay of $^{233}$U}, J. Phys. G: Nucl. Phys. \textbf{3} (1977) 421.
\bibitem[Carreyre2000]{Carreyre2000} Carreyre, T. et al., {\it First direct proof of internal conversion between bound states}, Phys. Rev. C \textbf{62} (2000) 024311. 
\bibitem[Carstens\ 2016]{Carstens} Carstens, H. et al., {\it High-harmonic generation at 250~MHz with photon energies exceeding 100~eV}, Optica \textbf{3}, (2016) 366-369.
\bibitem[CGPM\ 1967]{Cesium} Comptes Rendus de la 13$^{e}$ Conférence Générale des Poids et Mesures (CGPM) Paris 1967/68, (1969) 103. Online: https://www.bipm.org/utils/common/pdf /CGPM/CGPM13.pdf
\bibitem[CGPM\ 2018]{CGPM2018}  Comptes Rendus de la 26$^{e}$ Conférence Générale des Poids et Mesures (CGPM) Paris 2018, (2019) 472. Online: https://www.bipm.org/utils/common/pdf /CGPM/CGPM26.pdf
\bibitem[Chayawattanangkur1973]{Chaya} Chayawattanangkur, K., Herrmann, G., Trautmann, N., {\it Heavy isotopes of actinium: $^{229}$Ac, $^{230}$Ac, $^{231}$Ac, $^{232}$Ac}, J. Inorg. Nucl. Chem. \textbf{35} (1973) 3061.
\bibitem[Chen\ 2009]{Chen2009} Chen, C.T. et al., {\it Deep UV nonlinear optical crystal KBe$_2$BO$_3$F$_2$ - discovery, growth, optical properties and applications}, Appl. Phys. B \textbf{97} (2009) 9-25.
\bibitem[Chiara\ 2018]{Chiara} Chiara, C.J. et al., {\it Isomer depletion as experimental evidence of nuclear excitation by electron capture}, Nature \textbf{554} (2018) 216-218.
\bibitem[Chodash\ 2016]{Chodash2016} Chodash, P.A. et al., {\it Nuclear excitation by electronic transition of $^{235}$U}. Phys. Rev. C \textbf{93}, 034610 (2016).
\bibitem[Churchill\ 2010]{Churchill2010} Churchill, L.R., {\it Trapping triply ionized thorium isotopes}, Ph.D. Thesis, Georgia Institute of Technology, Georgia (2010).
\bibitem[Churchill\ 2011]{Churchill2011} Churchill, L.R., DePalatis, M.V., Chapman, M.S., {\it Charge exchange and chemical reactions with trapped Th$^{3+}$}, Phys. Rev. A \textbf{83} (2011) 012710.
\bibitem[Cingöz\ 2012]{Cingoez} Cingöz, A. et al., {\it Direct frequency comb spectroscopy in the extreme ultraviolet}, Nature \textbf{482}, (2012) 68-71.
\bibitem[Clairon\ 1991]{Clairon1991} Clairon, A. et al., {\it Ramsey resonance in a Zacharias fountain}, Europhys. Lett. \textbf{16} (1991) 165-170.
\bibitem[Cohen\ 1971]{Cohen1971} Cohen, B.L., {\it Concepts of nuclear physics}, McGraw-Hill, New York (1971) 298.

\bibitem[Das\ 2013]{Das} Das, S. et al., {\it Quantum interference effects in an ensemble of $^{229}$Th nuclei interacting with coherent light}, Phys. Rev. C \textbf{88} (2013) 024601.
\bibitem[Delva\ 2017]{Delva} Delva, P. et al., {\it Test of special relativity using a fiber network of optical clocks}, Phys. Rev. Lett. \textbf{118} (2017) 221102.
\bibitem[Delva\ 2019]{Delva2019} Delva, P. et al., {\it Chronometric Geodesy: Methods and Applications}, in: Relativistic Geodesy, Foundations and Applications, Ed. D. Puetzfeld, C. Lämmerzahl, Fundamental Theories of Physics \textbf{196}, Springer (2019).  
\bibitem[Dembczynski\ 2015]{Dembczynski} Dembczynski, J., Elantkowska, M., Ruczkowski, J., {\it Method for detecting the isomeric state $I=(3/2)^+$ in $^{229}$Th with laser-induced fluorescence}, Phys. Rev. A \textbf{92} (2015) 012519.
\bibitem[Derevianko\ 2014]{Derevianko} Derevianko, A., Pospelov, M., {\it Hunting for topological dark matter with atomic clocks}, Nature Physics \textbf{10} (2014) 933-936.
\bibitem[Dessovic\ 2014]{Dessovic} Dessovic P. et al., {\it $^{229}$Thorium-doped calcium fluoride for nuclear laser spectroscopy}, J. Phys. Condens. Matter \textbf{26} (2014) 105402.
\bibitem[vanDuppen\ 2017]{Duppen} van Duppen, P., et al., {\it Characterization of the low-energy $^{229\text{m}}$Th isomer}, Letter of intent to the ISOLDE and Neutron Time-of-flight Committee (2017). Online available at: {\it cds.cern.ch/record/2266840}.
\bibitem[Dykhne\ 1996]{Dykhne2} Dykhne, A.M. et al., {\it Alpha decay of the first excited state of the Th-229 nucleus}, JETP Lett. \textbf{64} (1996) 345-349.
\bibitem[Dykhne\ 1998a]{Dykhne3} Dykhne, A.M., Tkalya, E.V., {\it Matrix element of the anomalously low-energy (3.5$\pm$0.5 eV) transition in $^{229}$Th and the isomer lifetime}, JETP Lett. \textbf{67} (1998) 251.
\bibitem[Dykhne\ 1998b]{Dykhne} Dykhne, A.M., Tkalya, E.V., {\it $^{229m}$Th(3/2$^+$, 3.5 eV) and a check of the exponentiality of the decay law}, JETP Lett. \textbf{67} (1998) 549-552.
\bibitem[Dzuba\ 2010]{Dzuba} Dzuba, V.A., Flambaum, V.V., {\it Exponential increase of energy level density in atoms: Th and Th II}, Phys. Rev. Lett. \textbf{104} (2010) 213002.
\bibitem[Dzyublik\ 2020]{Dzyublik} Dzyublik, A.Y., {\it Excitation of $^{229\text{m}}$Th in the electron bridge via continuum, as a scattering process}, Phys. Rev. C \textbf{102} (2020) 024604.

\bibitem[Ellis\ 2014]{Ellis} Ellis, J.K. et al., {\it Investigation of thorium salts as candidate materials for direct observation of the $^{229m}$Th nuclear transition}, Inorg. Chem. \textbf{53} (2014) 6769-6774.
\bibitem[Essen\ 1955]{Essen} Essen, L., Parry, J.V.L., {\it An atomic standard of frequency and time interval: A caesium resonator}, Nature \textbf{176} (1955) 280-282.
\bibitem[Fadeev\ 2020]{Fadeev} Fadeev, P. et al., {\it Sensitivity of $^{229}$Th nuclear clock transition to variation of the fine-structure constant}, arXiv:2007.00408 (2020).
\bibitem[Faul\ 1966]{radiocarbon} Faul, H., {\it Nuclear clocks}, United States Atomic Energy Commission Division of Technical Information. Library of Congress Catalog Card Number 67-60195 (1966). Online: https://www.osti.gov/includes/opennet/includes/ Understanding\%20the\%20Atom/Nuclear\%20Clocks.pdf
\bibitem[Feldmeier\ 2017]{Feldmeier} Feldmeier, H. et al., {\it Variation of fundamental constants and $^{229}$Th}, Proceedings of the 14th Marcel Grossmann Meeting (2017) 3670-3675.
\bibitem[Flambaum\ 2006]{Flambaum1} Flambaum, V.V., {\it Enhanced effect of temporal variation of the fine structure constant and the strong interaction in $^{229}$Th}, Phys. Rev. Lett. \textbf{97} (2006) 092502.
\bibitem[Flambaum\ 2007]{Flambaum2} Flambaum, V.V., {\it Variation of fundamental constants: theory and observations}, Int. J. Mod. Phys. A \textbf{22} (2007) 4937-4950.
\bibitem[Flambaum\ 2009a]{Flambaum3} Flambaum, V.V., Wiringa, R.B., {\it Enhanced effect of quark mass variation in $^{229}$Th and limits from Oklo data}, Phys. Rev. C \textbf{79} (2009) 034302.
\bibitem[Flambaum\ 2009b]{Flambaum4} Flambaum, V.V. et al., {\it Coulomb energy contribution to the excitation energy in $^{229}$Th and enhanced effect of $\alpha$ variation}, Europhys. Lett. \textbf{85} (2009) 50005.
\bibitem[Flambaum\ 2016]{Flambaum6} Flambaum, V.V., {\it Enhancing the effect of Lorentz invariance and Einstein's equivalence principle violation in nuclei and atoms}, Phys. Rev. Lett. \textbf{117} (2016) 072501.
\bibitem[Flambaum\ 2019]{Flambaum2019} Flambaum, V.V., {\it Enhanced nuclear Schiff moment and time-reversal violation in $^{229}$Th-containing molecules}, Phys. Rev. C \textbf{99} (2019) 035501.
\bibitem[Fleischmann\ 2009]{Fleischmann2009} Fleischmann, L. et al., {\it Metallic magnetic calorimeters for X-ray spectroscopy}, IEEE/CSC \& ESAS European Superconductivity news forum (ENSF) 7 (2009). AIP Conference Proceedings \textbf{1185} (2009) 571. Online: https://snf.ieeecsc.org/sites/ieeecsc.org/files/CR11 \_MMC\_final\_110408.pdf
\bibitem[Friedrich\ 2015]{Friedrich} Friedrich, S., {\it The world's lowest nuclear state in thorium-229m}, LDRL Annual Report 2015, Law\-rence Livermore National Laboratory. Online: https://ldrd-annual.llnl.gov/ldrd-annual-2015/nuclear/lowest
\bibitem[Fujioka\ 1984]{Fujioka1984} Fujioka, H. et al., {\it Observation of Nuclear Excitation by Electron Transition (NEET) in $^{197}$Au} Zeitschrift für Physik A - Atoms and Nuclei \textbf{315} (1984)  121-122.

\bibitem[Gangrsky\ 2005]{Gangrsky} Gangrsky, Yu.P. et al., {\it Search for light radiation in decay of $^{229}$Th isomer with anomalously low excitation energy}, Bull. Rus. Acad. Sci. Phys. \textbf{69} (2005) 1857.
\bibitem[Geist\ 2020]{Geist2020} Geist, J., {\it Bestimmung der Isomerenergie von $^{229}$Th mit dem hochauflösenden Mikrokalorimeter-Array maXs30}, Ph.D. Thesis, University of Heidelberg, Germany (2020).
\bibitem[Gill\ 2008]{Gill2008} P. Gill et al., {\it Optical atomic clocks for space}, National Physical Laboratory (NPL) Technical Supporting Document (2008). Online: http://www.researchgate.net/ publication/241531359\_Optical\_Atomic\_Clocks\_for\_Space
\bibitem[Gill\ 2011]{Gill2011} Gill, P., {\it When should we change the definition of the second?}, Phil. Trans. R. Soc. A \textbf{369} (2011) 4109-4130.
\bibitem[Gill\ 2016]{Gill2016} Gill, P., {\it Is the time right for a redifinition of the second by optical atomic clocks?}, Jour. of Physics: Conference Series \textbf{723} (2016) 012053.
\bibitem[Godanskii\ 1976]{Godanskii} Goldanskii, V. I., Namiot, V. A., {\it On the excitation of isomeric nuclear levels by laser radiation through inverse internal electron conversion} Phys. Lett. B \textbf{62} (1976) 393-394.
\bibitem[Godun\ 2014]{Godun} R.M. Godun et al., {\it Frequency ratio of two optical clock transitions in $^{171}$Yb$^+$ and costraints on the time variation of fundamental constants}, Phys. Rev. Lett. \textbf{113} (2014) 210801.
\bibitem[Gohle\ 2005]{Gohle2005} Gohle, C. et al., {\it A frequency comb in the extreme ultraviolet}, Nature \textbf{436} (2005) 234-237.
\bibitem[Gouder\ 2019]{Gouder2019} Gouder, T. et al., {\it Measurements of the band gap of ThF$_4$ by electron spectroscopy techniques}, Physical Review Research \textbf{1} (2019) 033005.
\bibitem[Gould\ 1923]{Gould} Gould, R.T., {\it The marine chronometer: its history and development} (1923) J.D. Potter.
\bibitem[Groot-Ber\-ning\ 2019]{Groot-Berning2019} Groot-Berning, K. et al., {\it Trapping and sympathetic cooling of single thorium ions for spectroscopy}, Phys. Rev. A \textbf{99} (2019) 023420.
\bibitem[Guimaraes\ 2005]{Filho} Guimaraes-Filho, Z.O., Helene, O., {\it Energy of the 3/2$^{+}$ state of $^{229}$Th reexamined}, Phys. Rev. C \textbf{71} (2005) 044303.
\bibitem[Gulda\ 2002]{Gulda} Gulda, K. et al., {\it The nuclear structure of $^{229}$Th}, Nucl. Phys. A \textbf{703} (2002) 45.
\bibitem[Gusev\ 2016]{Gusev} Gusev, Yu.I., et al., {\it Studying the decay of thorium-229 isomer by means of conversion spectroscopy}, Bull. Rus. Acad. Sci. Phys. \textbf{80} (2016) 875-879.
\bibitem[Gusev\ 2019]{Gusev2019} Gusev, Yu.I. et al., {\it Measuring the energy of $^{229}$Th isomer decay}, Bull. Rus. Acad. Sci. Phys. \textbf{83} (2019) 1179-1182.
\bibitem[Gunst\ 2014]{Gunst2014} Gunst, J. et al., {\it Dominant Secondary Nuclear Photoexcitation with the X-Ray Free-Electron Laser}, Phys. Rev. Lett. \textbf{112}, 082501 (2014).
\bibitem[Gunst\ 2018]{Gunst2018} Gunst, J. et al., {\it Nuclear excitation by electron capture in optical-laser-generated plasma}, Phys. Rev. E \textbf{97}, 063205 (2018).

\bibitem[Haas\ 2020]{Haas2019} Haas, R. et al., {\it Development of a recoil ion source providing slow Th ions including $^{229(\text{m})}$Th in a broad charge state distribution}, Hyperfine Interactions \textbf{241} (2020) 25.
\bibitem[Hahn\ 1924]{Hahn1924} Hahn, O., Meitner, L., {\it Das $\beta$-Strahlenspektrum von Radium und seine Deutung}, Zeitschrift für Physik \textbf{26} (1924) 161-168.
\bibitem[Halasyamani\ 2017]{Halasyamani2017} Halasyamani, P.S., Zhang, W., {\it Viewpoint: Inorganic materials for UV and Deep-UV nonlinear optical applications}, Inorg. Chem. \textbf{56} (2017) 12077-12085.
\bibitem[Hall\ 2006]{Hall2006} Hall, J.L., {\it Nobel Lecture: Defining and measuring optical frequencies}, Rev. Mod. Phys. \textbf{78} (2006) 1279.
\bibitem[Hänsch\ 2006]{Haensch2006} Hänsch, T.W., {\it Nobel Lecture: Passion for precision}, Rev. Mod. Phys. \textbf{78} (2006) 1297.
\bibitem[Harston\ 1999]{Harston1999} Harston, M.R., Chemin, J.F., {\it Mechanisms of nuclear excitation in plasmas}, Phys. Rev. C \textbf{59}, 2462-2473 (1999).
\bibitem[Hayes\ 2007]{Hayes1} Hayes, A.C., Friar, J.L., {\it Sensitivity of nuclear transition frequencies to temporal variation of the fine structure constant or the strong interaction}, Phys. Lett. B \textbf{650} (2007) 229-232.
\bibitem[Hayes\ 2008]{Hayes2} Hayes, A.C. et al., {\it Splitting sensitivity of the ground and 7.6 eV isomeric states of $^{229}$Th}, Phys. Rev. C \textbf{78} (2008) 024311.
\bibitem[He\ 2007]{He2} He, X.T., Ren, Z.Z., {\it Enhanced sensitivity to variation of fundamental constants in the transition of Th-229 and Bk-249}, Journal of Physics G, Nuclear and Patricle Physics \textbf{34} (2007) 1611-1619.
\bibitem[He\ 2008a]{Xiao} He, X.T., Ren, Z.Z., {\it Temporal variation of the fine structure constant and the strong interaction parameter in the $^{229}$Th transition}, Nucl. Phys. A \textbf{806} (2008) 117-123.
\bibitem[He\ 2008b]{He3} He, X.T., Ren, Z.Z., {\it Enhanced sensitivity to variation of fundamental constants in the narrow nuclear transitions (II)}, Journal of Physics G, Nuclear and Patricle physics \textbf{35} (2008) 035106. 
\bibitem[Hehlen\ 2013]{Hehlen} Hehlen, M.P. et al., {\it Optical spectroscopy of an atomic nucleus: Progress toward direct observation of the $^{229}$Th isomer transition}, J. Lumin. \textbf{133} (2013) 91-95.
\bibitem[Helmer\ 1993]{Helmer1993} Helmer, R.G., Reich, C.W., {\it An improvement in the value of the energy of the first excited state in $^{229}$Th}, Report of the Idaho National Engineering Laboratory (1993). Online: https://digital.library.unt.edu/ark: /67531/metadc1212960/m2/1/high\_res\_d/6326722.pdf
\bibitem[Helmer\ 1994]{Helmer_Reich} Helmer, R., Reich, C.W., {\it An excited state of $^{229}$Th at 3.5 eV}, Phys. Rev. C \textbf{49} (1994) 1845-1858.
\bibitem[Herrera\ 2012a]{Herrera1} Herrera-Sancho, O.A. et al., {\it Two-photon laser excitation of trapped $^{232}$Th$^+$ ions via the 402-nm resonance line}, Phys. Rev. A \textbf{85} (2012) 033402.
\bibitem[Herrera\ 2012b]{Herrera3} Herrera-Sancho, O.A., {\it Laser excitation of 8-eV electronic states in Th$^+$: A first pillar of the electronic bridge toward excitation of the Th-229 nucleus}, Ph.D. Thesis, University of Hannover, Germany (2012).
\bibitem[Herrera\ 2013]{Herrera2} Herrera-Sancho, O.A. et al., {\it Energy levels of Th$^+$ between 7.3 and 8.3 eV}, Phys. Rev. A \textbf{88} (2013) 012512.
\bibitem[Herschbach\ 2012]{Herschbach2012} Herschbach, N. et al., {\it Linear Paul trap design for an optical clock with Coulomb crystals}, Applied Physics B \textbf{107} (2012) 891-906.
\bibitem[Herten\ 2018]{Herten2018} Herten, F., Waldmann, G., {\it Functional principles of early time measurement at Stonehenge and Nebra}, Archäologische Informationen \textbf{41} (2018) 275-288.
\bibitem[Higgins\ 2004]{Higgins2004} Higgins, K. et al., {\it A Walk Through Time} (version 1.2.1) (2004) National Institute of Standards and Technology, Gaithersburg, MD. Online: http://physics.nist.gov/time
\bibitem[Hofmann\ 2008]{Hofmann} Hofmann-Wellenhof, B., Lichtenegger, H., Wasle, E., {\it GNSS Global Navigation Satellite Systems}, Springer Wien New York (2008).
\bibitem[Holzberger\ 2015]{Holzberger2015} Holzberger, S. et al., {\it Femtosecond Enhancement Cavities in the Nonlinear Regime}, Phys. Rev. Lett. \textbf{115} (2015)023902.
\bibitem[Horton\ 1928]{Horton1928} Horton, J.W., Marrison, W.A., {\it Precision determination of frequency}, I.R.E. Proc. \textbf{16} (1928) 137-154. 
\bibitem[Huntemann\ 2016]{Huntemann} Huntemann, N. et al., {\it Single-ion atomic clock with $3\cdot10^{-18}$ systematic uncertainty}, Phys. Rev. Lett. \textbf{116} (2016) 063001.

\bibitem[Inamura\ 2003]{Inamura1} Inamura, T.T., Karpeshin, F.F., Trzhaskovskaya, M.B., {\it Bridging atomic and nuclear states in $^{229}$Th}, Czech. J. Phys. Suppl. B \textbf{53} (2003) 349.
\bibitem[Inamura\ 2005a]{Inamura2} Inamura, T.T., {\it Laser methods in the study of nuclei, atoms and molecules}, Phys. Scr. \textbf{71} (2005) C1-C4. 
\bibitem[Inamura\ 2005b]{Inamura3} Inamura, T.T., Mitsugashira, T., {\it Pumping $^{229m}$Th by hollow-cathode discharge}, Hyperfine Interactions \textbf{162} (2005) 115-123. 
\bibitem[Inamura\ 2009]{Inamura4} Inamura, T.T., Haba, H., {\it Search for a ``3.5-eV isomer" in $^{229}$Th in a hollow-cathode electric discharge}, Phys. Rev. C \textbf{79} (2009) 034313. 
\bibitem[Irwin\ 1997]{Irwin} Irwin, G.M., Kim, K.H., {\it Observation of electromagnetic radiation from deexcitation of the $^{229}$Th isomer.}, Phys. Rev. Lett. \textbf{79} (1997) 990-993.
\bibitem[Irwin\ 1999]{Irwin2} Irwin, G.M., Kim, K.H., {\it Irwin and Kim reply}, Phys. Rev. Lett. \textbf{83} (1999) 1073.
\bibitem[Isakov\ 2017]{Isakov} Isakov, V.I., {\it Isomeric level of the $^{229}$Th nucleus and its population in the Coulomb excitation reaction}, Physics of Atomic Nuclei \textbf{80} (2017) 1080-1087.
\bibitem[Itano\ 1993]{Itano} Itano, W.M. et al., {\it Quantum projection noise: Population fluctuations in two-level systems}, Phys. Rev. A \textbf{47} (1993), 3554.
\bibitem[Izawa\ 1979]{Izawa1979} Izawa, Y., Yamanaka, C., {\it Production of $^{235}$U$^{\text{m}}$ by nuclear excitation by electron transition in a laser produced uranium plasma}, Phys. Lett. \textbf{88} B (1979) 59-61.

\bibitem[JacksonJ\ 1928]{Jackson1928} Jackson, J., Bowyer, B., {The accuracy of Schortt free pendulum clocks}, Nature \textbf{121} (1928) 868-870.
\bibitem[JacksonRA\ 2009]{Jackson} Jackson, R.A. et al., {\it Computer modelling of thorium doping in LiCaAlF$_6$ and LiSrAlF$_6$: application to the development of solid state optical frequency devices}, J. Phys.: Condens. Matter \textbf{21} (2009) 325403.
\bibitem[Jeet\ 2015]{Jeet} Jeet, J. et al., {\it Results of a direct search using synchrotron radiation for the low-energy $^{229}$Th nuclear isomeric transition}, Phys. Rev. Lett. \textbf{114} (2015) 253001.
\bibitem[Jeet\ 2018]{Jeet2018} Jeet, J., {\it Search for the low lying transition in the $^{229}$Th nucleus}, Ph.D. Thesis, University of California, UCLA, USA (2018).
\bibitem[Johnsen\ 1974]{Johnsen} Johnsen, R. et al., {\it Rate coefficients for oxidation of Ti$^+$ and Th$^+$ by O$_2$ and NO at low energies}, J. Phys. Chem. \textbf{61} (1974) 5404.
\bibitem[Jones\ 2005]{Jones2005} Jones, R.J. et al., {\it Phase-coherent frequency combs in the vacuum ultraviolet via high-harmonic generation inside a femtosecond enhancement cavity}, Phys. Rev. Lett. \textbf{94} (2005) 193201.

\bibitem[Kálmán\ 1994]{Kalman} Kálmán P., Keszthelyi, T., {\it Laser-driven inverse electronic bridge process: An optical tool for determining low-energy separation of nearby nuclear states}, Phys. Rev. C \textbf{49} (1994) 324-328.
\bibitem[Kálmán\ 2001]{Kalman2} Kálmán, P., Bükki, T., {\it Deexcitation of $^{229}$Th$^{m}$: Direct $\gamma$ decay and electronic-bridge process}, Phys. Rev. C \textbf{63} (2001) 027601.
\bibitem[Kang\ 2018]{Kang2018} Kang, L. et al., {\it Removal of A-site alkali and alkaline earth metal cations in KBe$_2$BO$_3$F$_2$-type layered structures to enhance the deep-ultraviolet nonlinear optical capability}, Inorg. Chem. \textbf{57} (2018) 11146-11156.
\bibitem[Karpeshin\ 1992]{Karpeshin1} Karpeshin, F.F. et al., {\it Study of $^{229}$Th through laser-induced resonance internal conversion}, Phys. Lett. B \textbf{282} (1992) 267-270.
\bibitem[Karpeshin\ 1994]{Karpeshin1994} Karpeshin, F.F., Band, I.M., Trzhaskows\-kaya, M.B., {\it NEET revisited in connection with the resonance radiative pumping of $^{229m}$Th}, International conference on nuclear shapes and nuclear structure at low excitation energies, Editions Frontiers, Antibes (France) June 20-25 (1994), Conference abstract page 50. Online available at {\it https://inis.iaea.org/collection/ NCLCollectionStore/\_Public/26/022/26022866.pdf}. See also: AIP conference proceedings \textbf{329} (1995) 468-471.
\bibitem[Karpeshin\ 1996a]{Karpeshin2} Karpeshin, F.F. et al., {\it Optical pumping $^{229m}$Th through NEET as a new effective way of producing nuclear isomers}, Phys. Lett. B \textbf{372} (1996) 1-7.
\bibitem[Karpeshin\ 1996b]{Karpeshin1996} Karpeshin, F.F. et al., {\it Subthreshold internal conversion to bound states in highly ionized $^{125}$Te ions}, Phys. Rev C \textbf{53} (1996) 1640-1645.
\bibitem[Karpeshin\ 1998]{Karpeshin8} Karpeshin, F.F. et al., {\it Rates of transitions between the hyperfine-splitting components of the ground-state and the 3.5 eV isomer in $^{229}$Th$^{89+}$}, Phys. Rev. C \textbf{57} (1998) 3085.
\bibitem[Karpe\-shin\ 1999a]{Karpeshin3} Karpeshin, F.F., Band, I.M., Trzhaskows\-kaya, M.B., {\it 3.5 eV isomer of $^{229m}$Th: How it can be produced}, Nucl. Phys. A \textbf{654} (1999) 579-596.
\bibitem[Karpeshin\ 1999b]{Karpeshin9} Karpeshin, F.F. et al., {\it On the question of electron bridge for the 3.5 eV isomer of $^{229}$Th}, Phys. Rev. Lett. \textbf{83} (1999) 1072. 
\bibitem[Karpeshin\ 1999c]{Karpeshin10} Karpeshin, F.F. et al., {\it Role of the electron bridge in de-excitation of the 3.5 eV $^{229}$Th isomer}, Bull. Rus. Acad. Sci. Phys. \textbf{63} (1999) 30. 
\bibitem[Karpeshin\ 2002]{Karpeshin2002} Karpeshin, F.F., {\it Electron Shell as a Resonator}, Hyperfine Interactions \textbf{143}, 79-96 (2002).
\bibitem[Karpeshin\ 2005]{Karpeshin4} Karpeshin, F.F., Trzhaskovskaya, M.B., {\it Resonance conversion as the predominant decay mode of $^{229m}$Th}, Hyperfine Interact. \textbf{162} (2005) 125-132.
\bibitem[Karpeshin\ 2006]{Karpeshin5} Karpeshin, F.F., Trzhaskovskaya, M.B., {\it Resonance conversion as a dominant decay mode for the 3.5-eV isomer in $^{229m}$Th}, Phys. At. Nucl. \textbf{69} (2006) 571. 
\bibitem[Karpeshin\ 2007]{Karpeshin6} Karpeshin, F.F., Trzhaskovskaya, M.B., {\it Impact of the electron environment on the lifetime of the $^{229}$Th$^m$ low-lying isomer}, Phys. Rev. C \textbf{76} (2007) 054313.
\bibitem[Karpeshin\ 2015]{Karpeshin7} Karpeshin, F.F., Trzhaskovskaya, M.B., {\it Excitation of the $^{229m}$Th nuclear isomer via resonance conversion in ionized atoms}, Phys. Atom. Nucl. \textbf{78} (2015) 715-719.
\bibitem[Karpeshin\ 2016]{Karpeshin11} Karpeshin, F.F., Trzhaskovskaya, M.B., {\it The $^{229}$Th isomer line as a reference for a high-precision frequency standard}, Measurement Techniques \textbf{59} (2016) 722-727.
\bibitem[Kar\-pe\-shin\ 2017]{Karpeshin12} Karpeshin, F.F., Trzhaskovskaya, M.B., {\it Bound internal conversion versus nuclear excitation by electron transition: Revision of the theory of optical pumping of the $^{229\text{m}}$Th isomer}, Phys. Rev. C \textbf{95} (2017) 034310.
\bibitem[Karpeshin\ 2018a]{Karpeshin2018} Karpeshin, F.F., Trzhaskovskaya, M.B., {\it Impact of the ionization of the atomic shell on the lifetime of the $^{229\text{m}}$Th isomer}, Nucl. Phys. A \textbf{969} (2018) 173-183.
\bibitem[Karpeshin\ 2018b]{Karpeshin2018b} Karpeshin, F.F., Trzhaskovskaya, M.B., Vitushkin, L.F., {\it Physics of laser-assisted nuclear processes as the base for creation of the nuclear clock}, Proceedings of the 676. WE-Heraeus-Seminar ``Novel optical clocks in atoms and nuclei" (NOCAN), Bad Honnef, Germany (2018).
\bibitem[Karpeshin\ 2020]{Karpeshin2020} Karpeshin, F.F., Trzhaskovskaya, M.B., Vitushkin, L.F., {\it On the E2 admixture in the deexcitation of the $^{229\text{m}}$Th isomer}, arXiv:2003.08655 (2020).
\bibitem[Kasamatsu\ 2005]{Kasamatsu} Kasamatsu, Y. et al., {\it Search for the decay of $^{229m}$Th by photon detection}, Radiochim. Acta \textbf{93} (2005) 511-514. 
\bibitem[Kasamatsu\ 2018]{Kasamatsu2018} Kasamatsu, Y., Kikunaga, H., {\it Variation of the nuclear decay constants of $^{229\text{m}}$Th and $^{235\text{m}}$U affected by their chemical environments}, Radioisotopes \textbf{67} (2018) 471-482. (in Japanese)
\bibitem[Kazakov\ 2011]{Kazakov2} G.A. Kazakov et al., {\it Atomic clock with a nuclear transition: current status in TU Wien} Presented at the conference ``Isomers in Nuclear and Interdisciplinary Research", 4-10 July 2011, Sankt-Petersburg, Russia. arXiv:1110.0741v2 (2013).
\bibitem[Kazakov\ 2012]{Kazakov1} Kazakov, G.A., et al., {\it Performance of a $^{229}$Thorium solid-state nuclear clock}, New J. Phys. \textbf{14} (2012) 083019.
\bibitem[Kazakov\ 2014]{Kazakov3} Kazakov, G.A. et al., {\it Prospects for measuring the $^{229}$Th isomer energy using a metallic magnetic micro-calorimeter}, Nucl. Instrum. Methods A \textbf{735} (2014) 229-239.
\bibitem[Kazakov\ 2017] {Kazakov4} Kazakov, G.A. et al., {\it Re-evaluation of the Beck et al. data to constrain the energy of the $^{229}$Th isomer}, arXiv:1702.00749 (2017). 
\bibitem[Kekez\ 1985]{Kekez1985} Kekez, D. et al., {\it Nuclear deexcitation via the electronic-bridge mechanism}, Phys. Rev. Lett. \textbf{55} (1985) 1366-1368. 
\bibitem[Keller\ 2019a]{Keller2019} Keller, J. et al., {\it Probing time dilation in Coulomb crystals in a high-precision ion trap}, Phys. Rev. Applied \textbf{11} (2019) 011002.
\bibitem[Keller\ 2019b]{Keller2019b} Keller, J. et al., {\it Controlling systematic frequency uncertainties at the 10$^{-19}$ level in linear Coulomb crystals}, Phys. Rev. A \textbf{99} (2019) 013405.
\bibitem[Kessler\ 2012]{Kessler} Kessler, T. et al., {\it A sub-40-mHz-linewidth laser based on a silicon single-crystal optical cavity}, Nat. Phot. \textbf{6} (2012) 687-692.
\bibitem[Kikunaga\ 2005]{Kikunaga1} Kikunaga, H. et al., {\it Search for $\alpha$-decay of $^{229m}$Th produced from $^{229}$Ac $\beta$-decay following $^{232}$Th$(\gamma,p2n)$ reaction}, Radiochim. Acta \textbf{93} (2005) 507-510. 
\bibitem[Kikunaga\ 2007]{Kikunaga3} Kikunaga, H., {\it Search for an isomer state of $^{229}$Th with extremely low energy using alpha-spectrometry}, Ph.D. Thesis, Kanazawa University, Japan (2007).
\bibitem[Kikunaga\ 2009]{Kikunaga2} Kikunaga, H. et al., {\it Half-life estimation of the first excited state of $^{229}$Th by using $\alpha$-particle spectrometry}, Phys. Rev. C \textbf{80} (2009) 034315.
\bibitem[Kim\ 2011]{Kim} Kim, K.H., Irwin, G.M., {\it Excitation of the low-lying isomeric level in $^{229}$Th}, Fizika B (Zagreb) \textbf{20} (2011) 319-330.
\bibitem[Kishimoto\ 2000]{Kishimoto2000} Kishimoto, S. et al., {\it Observation of nuclear excitation by electron transition in $^{197}$Au with synchrotron X rays and an avalanche photodiode}, Phys. Rev. Lett. \textbf{85} (2000) 1831-1834.
\bibitem[Kishimoto\ 2005]{Kishimoto2005} Kishimoto, S. et al., {\it Evidence for nuclear excitation by electron transition on $^{193}$Ir and its probability}, Nuclear Physics A \textbf{748} (2005) 3-11.
\bibitem[Knize\ 2020]{Knize2019} Knize, R.J. et al., {\it Search for direct detection of thorium-229m nuclear VUV photons}, Journal of Physics B: Atomic, Molecular and Optical Physics \textbf{53} (2020) 015002. 
\bibitem[Kolkowitz\ 2016]{Kolkowitz} Kolkowitz, S. et al., {\it Gravitational wave detection with optical lattice atomic clocks}, Phys. Rev. D \textbf{94} (2016) 124043.
\bibitem[Koltsov\ 2000]{Koltsov} Koltsov, V.V., {\it Effect of dielectric properties of a medium on the probability of the isomeric transiton with energy 3.5 eV in the $^{229\text{m}}$Th nucleus}, Bull. Rus. Acad. Sci. Phys. \textbf{64} (2000) 447. 
\bibitem[Kotthaus\ 2012]{Kotthaus} Kotthaus, T. et al., {\it Energy splitting between the first two quasi-particle states in $^{230}$Pa in comparison with the isotone $^{229}$Th}, Phys. Lett. B \textbf{718} (2012) 460-464.
\bibitem[Kozlov\ 2018]{Kozlov2018} Kozlov, M.G. et al., {\it Highly charged ions: optical clocks and applications in fundamental physics}, Rev. Mod. Phys. \textbf{90} (2018) 045005.
\bibitem[Kraemer\ 2020]{Kraemer} Kraemer, S. et al., {\it Study of the radiative decay of the low-energy isomer in $^{229}$Th}, Proposal to the ISOLDE and Neutron Time-of-flight Committee (2020). Online available at: {\it cds.cern.ch/record/2717784}.
\bibitem[Kroger\ 1972]{Kroger} Kroger, L.A., {\it Level-scheme studies in $^{229}$Th and $^{231}$Th from the alpha-decay of $^{233}$U and $^{235}$U}, Ph.D. Thesis, University of Wyoming, USA (1972).
\bibitem[Kroger\ 1976]{Kroger_Reich} Kroger, L.A., Reich, C.W., {\it Features of the low energy level scheme of $^{229}$Th as observed in the $\alpha$ decay of $^{233}$U}, Nucl. Phys. A  \textbf{259} (1976) 29.
\bibitem[Krutov\ 1958]{Krutov1958} Krutov, V.A., {\it K teorii vnutrennei konversii (To the theory of internal conversion)}, Izv. Akad. Nauk SSSR, Ser. Fiz. \textbf{22} (1958) 162.
\bibitem[Krutov\ 1990]{Krutov} Krutov, V.A., {\it Internal conversion in the field of an ``electronic bridge"}, JETP Lett. \textbf{52} (1990) 584-588.
\bibitem[Kudryav\-tsev\ 2001]{Kudryavtsev} Kudryavtsev, Y. et al., {\it A gas cell for thermalizing, storing and transporting radioactive ions and atoms. Part I: Off-line studies with a laser ion source}, Nucl. Instrum. Methods Phys. Res. B \textbf{179} (2001) 412.

\bibitem[Lebedinskii\ 2019]{Lebedinskii2019} Lebedinskii, Y.Y. et al., {\it A unique system for registering one-photon signals in the ultraviolet range from an isomeric $^{229\text{m}}$Th nucleus implanted on thin SiO$_2$/Si films}, Phys. Status Solidi A (2019) 1900551.
\bibitem[Lezius\ 2016]{Lezius} Lezius M. et al., {\it Space-born frequency comb metrology}, Optica \textbf{3} (2016) 1381-1387.
\bibitem[Liao\ 2012]{Liao2} Liao, W.T. et al., {\it Coherence-enhanced optical determination of the $^{229}$Th isomeric transition}, Phys. Rev. Lett. \textbf{109} (2012) 262502.
\bibitem[Liao\ 2013]{Liao1} Liao, W.T., {\it Coherent control of nuclei and X-rays}, Ph.D. Thesis, University of Heidelberg, Germany (2013). Published as Springer Thesis (Springer, Berlin, 2014).
\bibitem[Liao\ 2017]{Liao3} Liao, W.T., Pálffy, A., {\it Optomechanically induced transparency of x-rays via optical control}, Sci. Rep. \textbf{7} (2017) 321.
\bibitem[Litvinova\ 2009]{Litvinova} Litvinova, E. et al., {\it Nuclear structure of lowest $^{229}$Th states and time-dependent fundamental constants}, Phys. Rev. C \textbf{79} (2009) 064303.
\bibitem[Ludlow\ 2015]{Ludlow} Ludlow, A.D. et al., {\it Optical Atomic clocks}, Rev. Mod. Phys. \textbf{87} (2015) 637-699.
\bibitem[Lyons\ 1949]{Lyons} Lyons, H., {\it The atomic clock}, Instruments Vol. \textbf{22} (1949) 133-135.

\bibitem[Ma\ 2015]{Ma} Ma, X. et al., {\it Proposal for precision determination of 7.8 eV isomeric state in $^{229}$Th at heavy ion storage ring}, Phys. Scr. \textbf{T166} (2015) 014012.
\bibitem[Madjarov\ 2019]{Madjarov2019} Madjarov, I.S. et al., {\it An Atomic-Array Optical Clock with Single-Atom Readout}, Phys. Rev. X \textbf{9} (2019) 041052.
\bibitem[Marrison\ 1948]{Marrison} Marrison, W.A., {\it The evolution of the quartz crystal clock}, The Bell System Technical Journal \textbf{27} (1948) 510-588.
\bibitem[Masuda\ 2019]{Masuda2019} Masuda, T. et al., {\it X-ray pumping of the $^{229}$Th nuclear clock isomer}, Nature \textbf{573} (2019) 238-242.
\bibitem[Matei\ 2017]{Matei2017} Matei, D.G. et al., {\it 1.5 $\mu$m lasers with sub-10 mHz linewidth}, Phys. Rev. Lett. \textbf{118} (2017) 263202.
\bibitem[Matinyan\ 1998]{Matinyan} Matinyan, S., {\it Lasers as a bridge between atomic and nuclear physics}, Phys. Rep. \textbf{298} (1998) 199.
\bibitem[Meitner\ 1924]{Meitner1924} Meitner, L., {\it Über die Rolle der $\gamma$-Strahlen beim Atomzerfall}, Zeitschrift für Physik \textbf{26} (1924) 169-177.
\bibitem[McCarthy\ 2018]{McCarthy2018} McCarthy, D.D., Seidelmann, P.K., {\it Time: from earth rotation to atomic physics}, Cambridge University Press, 2nd Edition (2018).
\bibitem[McGrew\ 2018]{McGrew} McGrew, W.F. et al., {\it Atomic clock performance enabling geodesy below the centimetre level}, Nature \textbf{564}, (2018) 87-90.
\bibitem[Mehlstäubler\ 2018]{Mehlstaeubler} Mehlstäubler, T. et al., {\it Atomic clocks for geodesy}, Rep. Prog. Phys. \textbf{81} (2018) 064401.
\bibitem[Meier\ 2019]{Meier2019} Meier, D.M. et al., {\it Electronic level structure of Th$^+$ in the range of the $^{229\text{m}}$Th isomer energy}, Phys. Rev. A \textbf{99} (2019) 052514.
\bibitem[Meyer\ 2018]{Meyer2018} Meyer, E.R. et al., {\it Thorium-doped CsI: implications for the thorium nuclear clock transition}, Phys. Rev. A \textbf{97} (2018) 060503(R).
\bibitem[Micke\ 2020]{Micke2020} Micke, P. et al., {\it Coherent laser spectroscopy of highly charged ions using quantum logic}, Nature \textbf{578} (2020) 60-65.
\bibitem[Minkov\ 2017a]{Minkov} Minkov, N., Pálffy, A., {\it Reduced transition probabilities for the $\gamma$-decay of the 7.8~eV isomer in $^{229}$Th}, Phys. Rev. Lett. \textbf{118} (2017) 212501.
\bibitem[Minkov\ 2017b]{Minkov2017b} Minkov, N., Pálffy, A., {\it Model mechanism for radiative decay of the 7.8~eV isomer in $^{229}$Th}, Nuclear Theory \textbf{36} (2017) 205-214.
\bibitem[Minkov2018]{Minkov2018} Minkov, N., Pálffy, A., {\it Electromagnetic properties of the $^{229\text{m}}$Th isomer}, Nuclear Theory \textbf{37} (2018) 33-40.
\bibitem[Minkov\ 2019a]{Minkov2019} Minkov, N., Pálffy, A., {\it Theoretical predictions for the magnetic dipole moment of $^{229\text{m}}$Th}, Phys. Rev. Lett. \textbf{122} (2019) 162502.
\bibitem[Minkov\ 2019b]{Minkov2019b} Minkov, N., Pálffy, A., {\it The magnetic moment as a constraint in determining the $^{229\text{m}}$Th isomer decay rates}, Acta Physica Polonica B Proceedings Supplement \textbf{12} (2019) 629-636.
\bibitem[Mitsu\-gashira\ 2003]{Mitsugashira} Mitsugashira, T. et al., {\it Alpha-decay from the 3.5 eV isomer of $^{229}$Th}, J. Radioanal. Nucl. Chem. \textbf{255} (2003) 63-66.
\bibitem[Moore\ 2004]{Moore} Moore, I.D. et al., {\it Search for a low-lying 3.5-eV isomeric state in $^{229}$Th}, ANL Phys. Dev. Rep. PHY-10990-ME-2004 (2004). 
\bibitem[Morita\ 1973]{Morita1973} Morita, M., {\it Nuclear excitation by electron transition and its application to uranium 235 separation} Progress of Theoretical Physics \textbf{49} (1973) 1574-1586.
\bibitem[Müller\ 2017]{Mueller} Müller, R.A. et al., {\it Theoretical analysis of the electron bridge process in $^{229}$Th$^{3+}$}, Nucl. Instrum. Meth. B \textbf{408} (2017) 84-88.
\bibitem[Müller\ 2018]{Mueller2018} Müller, R.A. et al., {\it Hyperfine interaction with the $^{229}$Th nucleus and its low-lying isomeric state}, Phys. Rev. A \textbf{98} (2018) 020503(R).
\bibitem[Müller\ 2019]{Mueller2019} Müller, R.A., Volotka, A.V., Surzhykov, A., {\it Excitation of the $^{229}$Th nucleus via a two-photon electronic transition}, Phys. Rev. A \textbf{99} (2019) 042517.
\bibitem[Muramatsu\ 2020]{Muramatsu} Muramatsu, H. et al., {\it Optimized TES microcalorimeters with 14 eV energy resolution at 30 keV for $\gamma$-ray measurements of the $^{229}$Th isomer}, Journal of Low Temperature Physics (2020).

\bibitem[Nagle\ 1960]{Nagle1960} Nagle, D.E. et al., {\it Ultra-high resolution $\gamma$-ray resonance in Zinc-67}, Nature \textbf{186} (1960) 707-708.
\bibitem[Nakazato\ 2016]{Nakazato2016} Nakazato, T. et al., {\it Phase-matched frequency conversion below 150~nm in KBe$_2$BO$_3$F$_2$}, Optics Express \textbf{24} (2016) 17149.
\bibitem[Natarajan\ 2012]{Natarajan2012} Natarajan, C.M. et al., {\it Superconducting nanowire single-photon detectors: physics and applications}, Superconductor Science and Technology \textbf{25} (2012) 063001.
\bibitem[Neumayr\ 2006]{Neumayr2006} Neumayr, J.B. et al., {\it  Performance of the MLL-Ion catcher}, Rev. Sci. Instrum. \textbf{77} (2006) 065109.
\bibitem[Nicholson\ 2015]{Nicholson} Nicholson, T.L. et al., {\it Systematic evaluation of an atomic clock at $2\cdot10^{-18}$ total uncertainty}, Nature Communications \textbf{6} (2015) 6896.
\bibitem[Nickerson\ 2018]{Nickerson2018} Nickerson, B.S., Liao, W.T., Pálffy, A., {\it Collective effects in $^{229}$Th-doped crystals}, Phys. Rev. A \textbf{98} (2018) 062520.
\bibitem[Nickerson\ 2019]{Nickerson2019} Nickerson, B.S., {\it Towards coherent control of the $^{229}$Th isomeric transition in VUV-transparent crystals}, Ph.D. Thesis, University of Heidelberg, Germany (2019).
\bibitem[Nickerson\ 2020]{Nickerson2020} Nickerson, B.S. et al., {\it Nuclear excitation of the $^{229}$Th isomer via defect states in doped crystals}, Phys. Rev. Lett. \textbf{125} (2020) 032501.
\bibitem[Nilsson\ 1955]{Nilsson} Nilsson, S.G., {\it Binding states of individual nucleons in strongly deformed nuclei}, Dan. Mat. Fys. Medd. \textbf{29} (1955) 1-69.
\bibitem[Norcia\ 2016]{Norcia2016} Norcia, M.A. et al., {\it Superradiance on the millihertz linewidth strontium clock transition}, Science Advances \textbf{2} (2016) 1601231.
\bibitem[Norcia\ 2019]{Norcia2019} Norcia, M.A. et al., {\it Seconds-scale coherence on an optical clock transition in a tweezer array}, Science eaay0644 (2019).

\bibitem[Oelker\ 2019]{Oelker2019} Oelker, E. et al., {\it Demonstration of $4.8\times10^{-17}$ stability at 1~s for two independent optical clocks}, Nature Photonics \textbf{13} (2019) 714-719.
\bibitem[Oganessian\ 1995]{Oganessian} Oganessian, Y.T., Karamian, S.A., {\it On the question of a gamma-ray laser on nuclear levels}, Laser Physics \textbf{5} (1995) 336-342.
\bibitem[Okhapkin\ 2015]{Okhapkin} Okhapkin, M.V. et al., {\it Observation of an unexpected negative isotope shift in $^{229}$Th$^+$ and its theoretical explanation}, Phys. Rev. A \textbf{92} (2015) 020503.
\bibitem[Otozai\ 1978]{Otozai1978} Otozai, K., Arakawa, R., Saito, T., {\it Nuclear excitation by electron transition in $^{189}$Os}, Nucl. Phys. A \textbf{297} (1978) 97-104.
\bibitem[Ozawa\ 2008]{Ozawa2008} Ozawa, A. et al., {\it High harmonic frequency combs for high resolution spectroscopy}, Phys. Rev. Lett. \textbf{100} (2008) 253901.
\bibitem[Ozawa\ 2013]{Ozawa2013} Ozawa, A. et al., {\it VUV frequency-comb spectroscopy of atomic xenon}, Phys. Rev. A \textbf{87} (2013) 022507.
\bibitem[Ozawa\ 2015]{Ozawa2015} Ozawa, A. et al., {\it High average power coherent VUV generation at 10~MHz repetition frequency by intracavity high harmonic generation}, Optics Express \textbf{23} (2015) 15107.
\bibitem[Ozawa\ 2017]{Ozawa2017} Ozawa, A. et al., {\it Single ion fluorescence excited with a single mode of an UV frequency comb}, Nature Communications \textbf{8} (2017) 44.

\bibitem[Pachucki\ 2001]{Pachucki} Pachucki, K., et al., {\it Nuclear spin mixing oscillations in $^{229}$Th$^{89+}$}, Phys. Rev. C \textbf{64} (2001) 064301.
\bibitem[Pálffy\ 2007]{Palffy2} Pálffy, A., Evers, J., Keitel, C.H., {\it Isomer triggering via nuclear excitation by electron capture}, Phys. Rev. Lett. \textbf{99} (2007) 172502.
\bibitem[Pálffy\ 2008]{Palffy3} Pálffy, A. et al., {\it Nuclear excitation by electron capture followed by fast X-ray emission}, Phys Lett. B \textbf{661} (2008) 330-334. 
\bibitem[Peik\ 2003]{Peik} Peik, E., Tamm, C., {\it Nuclear laser spectroscopy of the 3.5 eV transition in $^{229}$Th}, Eur. Phys. Lett. \textbf{61} (2003) 181.
\bibitem[Peik\ 2009]{Peik2} Peik, E. et al., {\it Prospects for a nuclear optical frequency standard based on thorium-229}, proceedings of 7th symposium on frequency standards and metrology (5-11 October 2008) (2009) 532-538.
\bibitem[Peik\ 2013]{Peik3} Peik, E., Zimmermann, K., {\it Comment on ``Observation of the deexcitation of the $^{229m}$Th nuclear isomer"}, Phys. Rev. Lett. \textbf{111} (2013) 018901.
\bibitem[Peik\ 2015]{Peik4} Peik, E., Okhapkin, M., {\it Nuclear clocks based on resonant excitation of $\gamma$-transitions}, C. R. Physique \textbf{16} (2015) 516-523.
\bibitem[Pimon\ 2020]{Pimon} Pimon, M. et al., {\it DFT calculation of $^{229}$thorium-doped magnesium fluoride for nuclear laser spectroscopy}, Journal of Physics: Condensed Matter \textbf{32} (2020) 255503.
\bibitem[Piotrowski\ 2016]{Piotrowski} Piotrowski, M., Gensemer, S., {\it A network of nuclear atomic clocks for precision gravity measurement}, Joint 13th asia pacific physics conference and 22nd australian institute of physics congress (APPC-AIP) abstracts (2016) Brisbane Convention and Exhibition Centre.
\bibitem[Piotrowski\ 2020]{Piotrowski2020} Piotrowski, M. et al., {\it Studies of thorium and ytterbium ion trap loading from laser ablation for gravity monitoring with nuclear clocks}, OSA Continuum \textbf{3} (2020) 2210-2221.
\bibitem[Pohjalainen\ 2018]{Pohjalainen2018} Pohjalainen, I., {\it Gas-phase chemistry, recoil source characterization and in-gas-cell resonance laser ionization of actinides at IGISOL}, Ph.D. Thesis, University of Jyväskylä, Finland (2018).
\bibitem[Pohjalainen\ 2020]{Pohjalainen2019} Pohjalainen, I., Moore, I.D., Sajavaara, T., {\it Characterization of $^{233}$U alpha recoil sources for $^{229(\text{m})}$Th beam production}, Nuclear Inst. and Methods in Physics Research B \textbf{463} (2020) 441-448.
\bibitem[Poli\ 2013]{Poli} Poli, N. et al., {\it Optical atomic clocks}, La Rivista del Nuovo Cimento \textbf{36} (2013) 555-624.
\bibitem[Ponce\ 2017]{Ponce} Ponce, F., {\it High accuracy measurements of the nuclear decay of U-235m and search for the nuclear decay of Th-229m}, Ph.D. Thesis, University of California, Davis, USA (2017).
\bibitem[Ponce\ 2018a]{Ponce2018a} Ponce, F. et al., {\it Accurate measurement of the first excited nuclear state in $^{235}$U}, Phys. Rev. C \textbf{97} (2018) 054310.
\bibitem[Ponce\ 2018b]{Ponce2018b} Ponce, F. et al., {\it A search for the decay of metastable $^{229\text{m}}$Th with superconducting tunnel junctions} Jour. of Low Temp. Phys. \textbf{193} (2018) 1214-1221.
\bibitem[Porat\ 2018]{Porat2018} Porat, G., Heyl, C.M. et al., {\it Phase-matched extreme-ultraviolet frequency-comb generation}, Nature Photonics \textbf{12} (2018) 387-391.
\bibitem[Porsev\ 2010a]{Porsev1} Porsev, S.G., Flambaum, V.V., {\it Effect of atomic electrons on the 7.6 eV nuclear transition in $^{229m}$Th$^{3+}$}, Phys. Rev. A \textbf{81} (2010) 032504.
\bibitem[Porsev\ 2010b]{Porsev2} Porsev, S.G., Flambaum, V.V., {\it Electronic bridge process in $^{229}$Th$^+$}, Phys. Rev. A \textbf{81} (2010) 042516.
\bibitem[Porsev\ 2010c]{Porsev3} Porsev, S.G. et al., {\it Excitation of the isomeric $^{229m}$Th nuclear state via an electronic bridge process in $^{229}$Th$^+$}, Phys. Rev. Lett. \textbf{105} (2010) 182501.
\bibitem[Pupeza\ 2013]{Pupeza} I. Pupeza et al., {\it Compact high-repetition-rate source of coherent 100 eV radiation}, Nature Photonics \textbf{7} (2013) 608-612.
\bibitem[Pyka\ 2014]{Pyka2014} Pyka, K. et al., {\it A high-precision segmented Paul trap with minimized micromotion for an optical multiple-ion clock}, Applied Physics B \textbf{114} (2014) 231-241.
\bibitem[Radnaev\ 2011]{Radnaev2011} Radnaev, A.G. et al., {\it Towards an optical nuclear clock with thorium-229}, Proceedings of the 43rd annual precise time and time interval (PTTI) systems and applications meeting, Long Beach, California (2011) 469-476. 
\bibitem[Rad\-na\-ev\ 2012a]{Radnaev} Radnaev, A.G., Campbell, C.J., Kuzmich, A., {\it Observation of the 717-nm electric quadrupole transition in triply charged thorium}, Phys. Rev. A \textbf{86} (2012) 060501.
\bibitem[Radnaev\ 2012b]{Radnaev2} Radnaev, A.G., {\it Towards quantum telecommunication and a thorium nuclear clock}, Ph.D. Thesis, Georgia Institute of Technology, USA (2012).
\bibitem[Raeder\ 2011]{Raeder} Raeder, S. et al., {\it Resonance ionization spectroscopy of thorium isotopes - towards a laser spectroscopic identification of the low-lying 7.6 eV isomer of $^{229}$Th}, J. Phys. \textbf{B 44} (2011) 165005.
\bibitem[Ramsey\ 1983]{Ramsey} Ramsey, N.F., {\it History of Atomic clocks}, Journal of Research of the National Bureau of Standards \textbf{88} (1983) 301-318.
\bibitem[Reich\ 1984]{Reich_Helmer1} Reich, C.W. et al., {\it Emission probabilities and energies of $\gamma$-ray transitions from the decay of $^{233}$U}, Int. J. Appl. Radiat. Isot. \textbf{35} (1984) 185-188.
\bibitem[Reich\ 1990]{Reich_Helmer} Reich, C.W., Helmer, R., {\it Energy separation of the doublet of intrinsic states at the ground state of $^{229}$Th}, Phys. Rev. Lett. \textbf{64} (1990) 271.
\bibitem[Reich\ 1991]{Reich3} Reich, C.W., {\it An interesting finding in $^{229}$Th}, Izv. Akad. Nauk SSSR, Ser. Fiz. \textbf{55}, 878
(1991) (in Russian). Invited paper presented at the 40th all-union conference on nuclear spectroscopy and  nuclear structure, Leningrad, USSR, 10-13 April 1990. English version online: https://www.osti.gov/scitech/servlets/purl/6159406
\bibitem[Reich\ 1993]{Reich_Helmer2} Reich, C.W., Helmer, R.G., Proc. Int. Symp. Nucl. Phys. of our Times (Singapore: World Scientific) (1993) 474.
\bibitem[Rellergert\ 2010a]{Rellergert} Rellergert, W.G., et al., {\it Constraining the evolution of the fundamental constants with a solid-state optical frequency reference based on the $^{229}$Th nucleus}, Phys. Rev. Lett. \textbf{104} (2010) 200802.
\bibitem[Rellergert\ 2010b]{Rellergert2} Rellergert, W.G. et al., {\it Progress towards fabrication of $^{229}$Th-doped high energy band-gap crystals for use as a solid-state optical frequency reference}, IOP Conf. Ser.: Mater. Sci. Eng. \textbf{15} (2010) 012005.
\bibitem[Richardson\ 1998]{Richardson} Richardson, D.S. et al., {\it Ultraviolet photon emission observed in the search for the decay of the $^{229}$Th isomer}, Phys. Rev. Lett. \textbf{80} (1998) 3206-3208.
\bibitem[Riehle\ 2006]{Riehle2006} Riehle, F., {\it Frequency standards: Basics and Applications}, John Wiley and Sons (2006).
\bibitem[Riehle\ 2015]{Riehle2015} Riehle, F., {\it Towards a redefinition of the second based on optical atomic clocks}, Comptes Rendus Physique \textbf{16} (2015) 506-515.
\bibitem[Riehle\ 2017]{Riehle2017} Riehle, F., {\it Optical clock networks}, Nature Photonics \textbf{11} (2017) 25-31.
\bibitem[Riehle\ 2018]{Riehle2018} Riehle, F. et al., {\it The CIPM list of recommended frequency standard values: guidelines and procedures}, Metrologia \textbf{55} (2018) 188.
\bibitem[Rivlin\ 2007]{Rivlin} Rivlin, L.A., {\it Nuclear gamma-ray laser: the evolution of the idea}, Quantum Electronics \textbf{37} (2007) 723-744.
\bibitem[Romanenko2012]{Romanenko} Romanenko, V.I. et al., {\it Direct two-photon excitation of isomeric transition in thorium-229 nucleus}, Ukr. J. Phys. \textbf{57} (2012) 1119-1131.
\bibitem[Roberts\ 2017]{Roberts2017} Roberts, B.M. et al., {\it Search for domain wall dark matter with atomic clocks on board global positioning system satellites}, Nature Communications \textbf{8} (2017) 1195.
\bibitem[Rose\ 1955a]{Rose1955a} Rose, M.E., {\it Multipole fields}, Wiley, New York (1955).
\bibitem[Rose\ 1955b]{Rose1955b} Rose, M.E., {\it Theory of internal conversion}, in Alpha-, beta- and gamma-ray spectroscopy, K. Siegbahn (Ed.), North-Holland Publishing Company Amsterdam (1955).
\bibitem[Rosenband\ 2008]{Rosenband} Rosenband, T. et al., {\it Frequency ratio of Al$^+$ and Hg$^+$ single-ion optical clocks; metrology at the 17th decimal place}, Science \textbf{319} (2008) 1808-1812.
\bibitem[Ruchowska\ 2006]{Ruchowska} Ruchowska, E. et al., {\it Nuclear structure of $^{229}$Th}, Phys. Rev. C \textbf{73} (2006) 044326.
\bibitem[Safronova\ 2013]{Safronova2} Safronova, M.S. et al., {\it Magnetic dipole and electric quadrupole moments of the $^{229}$Th nucleus}, Phys. Rev. A \textbf{88} (2013) 060501.
\bibitem[Safronova\ 2014]{Safronova3} Safronova, M.S., Safronova, U.I., Clark, W., {\it Relativistic all-order calculations of Th, Th$^+$, and Th$^{2+}$ atomic properties}, Phys. Rev. A \textbf{90} (2014) 032512.
\bibitem[Safronova\ 2016]{Safronova} Safronova, M.S., {\it Elusive transition spotted in thorium}, Nature \textbf{533} (2016) 44-45.
\bibitem[Safronova\ 2018a]{Safronova2018a} Safronova, M.S., {\it In search of the nuclear clock}, Nature Physics \textbf{14} (2018) 198.
\bibitem[Safronova\ 2018b]{Safronova2018b} Safronova, M.S. et al., {\it Nucelar charge radii of $^{229}$Th from isotope and isomer shifts}, Phys. Rev. Lett. \textbf{121} (2018) 213001.
\bibitem[Safronova\ 2018c]{Safronovareview} Safronova, M.S., et al., {\it Search for new physics with atoms and molecules}, Reviews of Modern Physics \textbf{90} (2018) 025008.
\bibitem[Safronova\ 2019]{Safronova2019} Safronova, M.S., {\it Search for variation of fundamental constants with clocks}, Ann. Phys. (Berlin) \textbf{531} (2019) 1800364.
\bibitem[Saito\ 1980]{Saito1980} Saito, T., Shinohara, A., Otozai, K., {\it Nuclear excitation by electron transition (NEET) in $^{237}$Np following K-shell photoionization}, Phys. Lett. \textbf{92} B (1980) 293-296.
\bibitem[Sakharov\ 2010]{Sakharov} Sakharov, S.L., {\it On the energy of the 3.5 eV level in $^{229}$Th}, Phys. At. Nucl. \textbf{73} (2010) 1.
\bibitem[Sanner\ 2019]{Sanner2019} Sanner, C. et al., {\it Optical clock comparison for Lorentz symmetry testing}, Nature \textbf{567} (2019) 204-208.
\bibitem[Saule\ 2019]{Saule2019} Saule, T. et al., {\it High-flux ultrafast extreme-ultraviolet photoemission spectroscopy at 18.4 MHz pulse repetition rate}, Nature Communications \textbf{10} (2019) 458.
\bibitem[Scheibe\ 1936]{Scheibe1936} Scheibe, A., Adelsberger, U., {\it Nachweis von Schwankungen der astronomischen Tageslänge im Jahre 1935 mittels Quarzuhren}, Phys. Z. \textbf{37} (1936) 38.
\bibitem[Schioppo\ 2017]{Schioppo} Schioppo, M. et al., {\it Ultrastable optical clock with two cold-atom ensembles}, Nat. Phot. \textbf{11} (2017) 48-52.
\bibitem[Schmöger\ 2015]{Schmoeger} Schmöger, L. et al., {\it Coulomb crystallization of highly charged ions}, Science \textbf{347} (2015) 1233.
\bibitem[Schnatz\ 1996]{Schnatz1996} Schnatz, H. et al., {\it First phase-coherent frequency measurement of visible radiation}, Phys. Rev. Lett. \textbf{76} (1996) 18. 
\bibitem[Schneider\ 2016]{Schneider} Schneider, P., {\it Spektroskopische Messungen an Thorium-229 mit einem Detektor-Array aus metallischen magnetischen Kalorimetern}, Master Thesis, Ruprecht-Karls-University, Heidelberg (2016) in German.
\bibitem[Schreitl\ 2016]{Schreitl2016} Schreitl, M., {\it Growth and characterization of (doped) calcium fluoride crystals for the nuclear spectroscopy of Th-229}, Ph.D. Thesis, Technical University of Vienna, Austria (2016).
\bibitem[Seiferle\ 2015]{Seiferle2} Seiferle, B., {\it Setup of a VUV detection system for the direct identification of the fluorescence radiation of $^{229\mathrm{m}}$Th}, Master Thesis, Ludwig-Maximilians-Universität München, Germany (2015).
\bibitem[Seiferle\ 2016]{Seiferle} Seiferle, B. et al., {\it A VUV detection system for the direct photonic identification of the first excited isomeric state of $^{229}$Th}, Eur. Phys. J. D \textbf{70} (2016) 58.
\bibitem[Seiferle\ 2017a]{Seiferle3} Seiferle, B., von der Wense, L., Thirolf, P.G., {\it Lifetime measurement of the $^{229}$Th nuclear isomer}, Phys. Rev. Lett. \textbf{118} (2017) 042501.
\bibitem[Seiferle\ 2017b]{Seiferle4} Seiferle, B. et al., {\it Feasibility study of internal conversion electron spectroscopy of $^{229\text{m}}$Th}, Eur. Phys. J. A \textbf{53} (2017) 108.
\bibitem[Seiferle\ 2019a]{Seiferle2019b} Seiferle, B. et al., {\it Energy of the $^{229}$Th nuclear clock transition}, Nature \textbf{573} (2019) 243-246.
\bibitem[Seiferle\ 2019b]{Seiferle2019c} Seiferle, B., {\it Characterization of the $^{229}$Th nuclear clock transition}, Ph.D. Thesis, Ludwig-Maximilians-Universität München, Germany (2019).
\bibitem[Seiferle\ 2020]{Seiferle2019a} Seiferle, B. et al., {\it Towards a precise determination of the excitation energy of the thorium nuclear isomer using a magnetic bottle spectrometer}, Nucl. Inst. and Meth. Phys. Res. B \textbf{463} (2020) 499-503.
\bibitem[Seres\ 2019]{Seres2019} Seres, J. et al., {\it All-solid-state VUV frequency comb at 160 nm using high-harmonic generation in nonlinear femtosecond enhancement cavity}, Optics Express \textbf{27} (2019) 6618-6628.
\bibitem[Shaw\ 1999]{Shaw} Shaw, R.W. et al., {\it Spontaneous ultraviolet emission from $^{233}$Uranium/$^{229}$Thorium samples}, Phys. Rev. Lett. \textbf{82} (1999) 1109-1111.
\bibitem[Shigekawa\ 2019]{Shigekawa2019} Shigekawa, Y. et al., {\it Observation of internal-conversion electrons emitted from $^{229\text{m}}$Th produced by $\beta$ decay of $^{229}$Ac}, Phys. Rev. C \textbf{100} (2019) 044304.
\bibitem[Shimamura\ 2005]{Shimamura2005} Shimamura, K. et al., {\it Advantageous growth characteristics and properties of SrAlF$_5$ compared with BaMgF$_4$ for UV/VUV nonlinear optical applications}, Journal of Crystal Growth \textbf{275} (2005) 128-134.
\bibitem[Sikorsky\ 2020]{Sikorsky2020} Sikorsky, T. et al., {\it Measurement of the $^{229}$Th isomer energy with a magnetic micro-calorimeter}, Phys. Rev. Lett. \textbf{125}, 142503 (2020).
\bibitem[Sonnenschein\ 2012a]{Sonnenschein1} Sonnenschein, V. et al., {\it The search for the existence of $^{229m}$Th at IGISOL}, Eur. Phys. J. A \textbf{48} (2012) 52.
\bibitem[Sonnenschein\ 2012b]{Sonnenschein2} Sonnenschein, V. et al., {\it Determination of the ground-state hyperfine structure in neutral $^{229}$Th}, J. Phys. B: At. Mol. Opt. Phys. \textbf{45} (2012) 165005.
\bibitem[Sonnenschein\ 2014]{Sonnenschein3} V. Sonnenschein, {\it Laser developments and high resolution resonance ionization spectroscopy of actinide elements}, Ph.D. Thesis, University of Jyväskylä, Finland (2014).
\bibitem[Sorge\ 2016]{Sorge} Sorge, F., Cammalleri, M., Genchi, G., {\it On the birth and growth of pendulum clocks in the early modern era}, in: Essays on the history of mechanical engineering, 273-290 Springer (2016).
\bibitem[Steck2007]{Steck} Steck, D.A., {\it Quantum and Atom Optics}. Online: http://atomoptics-nas.uoregon.edu/$\sim$dsteck /teaching/quantum-optics/quantum-optics-notes.pdf
\bibitem[Steele\ 2008]{Steele} Steele, A.V., {\it Barium ion cavity QED and triply ionized thorium ion trapping}, Ph.D. Thesis, Georgia Institute of Technology, USA (2008).
\bibitem[Stellmer\ 2015]{Stellmer2} Stellmer, S. et al., {\it Radioluminescence and photoluminescence of Th:CaF$_2$ crystals}, Sci. Rep. \textbf{5} (2015) 15580.
\bibitem[Stellmer\ 2016a]{Stellmer4} Stellmer, S. et al., {\it Feasibility study of measuring the $^{229}$Th nuclear isomer transition with $^{233}$U-doped crystals}, Phys. Rev. C \textbf{94} (2016) 014302.
\bibitem[Stellmer\ 2016b]{Stellmer3} Stellmer, S. et al., {\it Towards a measurement of the nuclear clock transition in $^{229}$Th}, Journal of Physics: Conference Series \textbf{723} (2016) 012059.
\bibitem[Stellmer\ 2018a]{Stellmer2018a} Stellmer S. et al., {\it Attempt to optically excite the nuclear isomer in $^{229}$Th}, Phys. Rev. A \textbf{97} (2018) 062506.
\bibitem[Stellmer\ 2018b]{Stellmer2018b} Stellmer, S. et al., {\it Toward an energy measurement of the internal conversion electron in the deexcitation of the $^{229}$Th isomer}, Phys. Rev. C \textbf{98} (2018) 014317.
\bibitem[Stenholm\ 1985]{Stenholm1985} Stenholm, S., {\it Dynamics of trapped particle cooling in the Lamb-Dicke limit}, Journal of the Optical Society of America B \textbf{2} (1985) 1743.
\bibitem[Stopp\ 2019]{Stopp2019} Stopp, F. et al., {\it Catching, trapping and in-situ-identification of thorium ions inside Coulomb crystals of $^{40}$Ca$^+$ ions}, Hyperfine Interact. \textbf{240} (2019) 33.
\bibitem[Strizhov\ 1991]{Strizhov} Strizhov, V.F., Tkalya, E.V., {\it Decay channel of low-lying isomer state of the $^{229}$Th nucleus. Possibilities of experimental investigation}, Sov. Phys. JETP \textbf{72} (1991) 387.
\bibitem[Swan\-berg\ 2012]{Swanberg} Swanberg, E., {\it Searching for the decay of $^{229m}$Th}, Ph.D. Thesis, University of California, Berkeley, USA (2012).
\bibitem[Thielking\ 2018]{Thielking2018} Thielking, J. et al., {\it Laser spectroscopic characterization of the nuclear clock isomer $^{229\text{m}}$Th}, Nature \textbf{556} (2018) 321-325.
\bibitem[Thirolf\ 2007]{Thirolf1} Thirolf, P.G. et al., {\it Optical access to the lowest nuclear transition in $^{229\mathrm{m}}$Th}, Annual Report of the Maier-Leibnitz Laboratory, Garching (2007) 18.
\bibitem[Thirolf2008]{Thirolf2008} Thirolf, P.G. et al., {\it Towards optical control over the lowest nuclear excited state in $^{229}$Th}, Annual Report of the Maier-Leibnitz-Laboratory, Garching (2008).
\bibitem[Thirolf\ 2013]{Thirolf2} Thirolf, P.G. et al., {\it Towards an all-optical access to the lowest nuclear excitation in $^{229\mathrm{m}}$Th}, Acta Physica Polonica B \textbf{44} (2013) 391-394.
\bibitem[Thirolf\ 2017]{Thirolf2017} Thirolf, P.G. et al., {\it Direct detection of the elusive $^{229}$thorium isomer: Milestone towards a nuclear clock}, 2017 Joint Conference of the European Frequency and Time Forum and IEEE International Frequency Control Symposium (EFTF/IFCS) (2017). 
\bibitem[Thirolf\ 2019a]{Thirolf2019} Thirolf, P.G. et al., {\it Improving our knowledge on the $^{229\text{m}}$Thorium isomer: Toward a test bench for time variations of fundamental constants}, Ann. Phys. (Berlin) (2019) 1800381.
\bibitem[Thirolf\ 2019b]{Thirolf2019b} Thirolf, P.G. et al., {\it The 229-thorium isomer: doorway to the road from the atomic clock to the nuclear clock}, J. Phys. B: At. Mol. Opt. Phys. \textbf{52} (2019) 203001.
\bibitem[Thirolf\ 2020]{Thirolf2020} Thirolf, P.G. et al., {\it Phase transition in the thorium isomer story}, Acta Physica Polonica B \textbf{51} (2020) 561.
\bibitem[Tkal\-ya\ 1990]{Tkalya1990} Tkayla, E.V., {\it Excitation of atomic nuclei in a plasma via the mechanism of an inverse electron bridge}, Sov. Phys. Dokl. \textbf{35}, 1069 (1990).
\bibitem[Tkalya\ 1992a]{Tkalya5} Tkalya, E.V., {\it Excitation of low-lying isomer level of the nucleus $^{229}$Th by optical photons}, JETP Lett. (USSR) \textbf{55} (1992) 211.
\bibitem[Tkalya\ 1992b]{Tkalya5b} Tkalya, E.V., {\it Probability of nonradiative excitation of nuclei in transitions of an electron in an atomic shell}, Sov. Phys. JETP \textbf{75} (1992) 200-209.
\bibitem[Tkalya\ 1992c]{Tkalya2} Tkalya, E.V., {\it Cross section for excitation of the low-lying ($\le$ 5 eV) $^{229}$Th isomer with laser radiation by the inverse electron bridge}, Sov. J. Nucl. Phys. \textbf{55} (1992) 1611-1617.
\bibitem[Tkalya\ 1995]{Tkalya1995} Tkalya, E.V. et al., {\it Investigation of the channels of excitation and decay of the anomalously low-energy isomeric level $3/2+(3.5\pm1.0)$~eV in the $^{229}$Th nucleus}, research proposal submitted to the European Union's INTAS programme (1995), funded within the fourth framework programme FP4 in 1996 under grant-agreement 95-IN-RU-828. Online: http://en.ibrae.ac.ru/pubtext/287/
\bibitem[Tkalya\ 1996a]{Tkalya1} Tkalya, E.V. et al., {\it Processes of the nuclear isomer $^{229m}$Th(3/2$^+$, 3.5$\pm$1.0 eV) resonant excitation by optical photons}, Phys. Scripta \textbf{53} (1996) 296-299.
\bibitem[Tkalya\ 1996b]{Tkalya6} Tkalya, E.V. et al., {\it Various mechanisms of the resonant excitation of the isomeric level $^{229m}$Th ((3/2)$^+$, 3.5 eV) by photons}, Phys. Atomic Nuclei \textbf{59} (1996) 779.
\bibitem[Tkalya\ 1999]{Tkalya7} Tkalya, E.V., {\it Nonradiative decay of the low-lying nuclear isomer $^{229m}$Th (3.5eV) in a metal}, JETP Letters \textbf{70} (1999) 371-374.
\bibitem[Tkalya\ 2000a]{Tkalya2b} Tkalya, E.V. et al., {\it Decay of the low-energy nuclear isomer $^{229}$Th$^m$(3/2$^+$, 3.5$\pm$1.0 eV) in solids (dielectrics and metals): A new scheme of experimental research}, Phys. Rev. C \textbf{61} (2000) 064308.
\bibitem[Tkalya\ 2000b]{Tkalya8} Tkalya, E.V., {\it Spontaneous emission probability for M1 transition in a dielectric medium: $^{229m}$Th(3/2$^+$, 3.5$\pm$1.0 eV) decay}, JETP Lett. \textbf{71} (2000) 311.
\bibitem[Tkalya\ 2003]{Tkalya9} Tkalya, E.V., {\it Properties of the optical transition in the $^{229}$Th nucleus}, Phys. Usp. \textbf{46} (2003) 315-324.
\bibitem[Tkalya\ 2004]{Tkalya2004} Tkayla, E.V., {\it Mechanisms for the excitation of atomic nuclei in hot dense plasma}, Laser Physics \textbf{14} (2004) 360-377.
\bibitem[Tkalya\ 2011]{Tkalya3} Tkalya, E.V., {\it Proposal for a nuclear gamma-ray laser of optical range}, Phys. Rev. Lett. \textbf{106} (2011) 162501.
\bibitem[Tkalya\ 2013]{Tkalya3b} Tkalya, E.V., Yatsenko, L.P., {\it Creation of inverse population in the $^{229}$Th ground-state doublet by means of a narrow-band laser}, Laser Phys. Lett. \textbf{10} (2013) 105808.
\bibitem[Tkalya\ 2015]{Tkalya99} Tkalya, E.V. et al., {\it Radiative lifetime and energy of the low-energy isomeric level in $^{229}$Th}, Phys. Rev. C \textbf{92} (2015) 054324.
\bibitem[Tkalya\ 2016a]{Tkalya10} Tkalya, E.V., {\it Anomalous magnetic hyperfine structure of the $^{229}$Th ground-state doublet in muonic atoms}, Phys. Rev. A \textbf{94} (2016) 012510.
\bibitem[Tkalya\ 2016b]{Tkalya11} Tkalya, E.V., Nikolaev, A.V., {\it Magnetic hyperfine structure of the ground-state doublet in highly charged ions $^{229}$Th$^{89+,87+}$ and the Bohr-Weisskopf effect}, Phys. Rev. C \textbf{94} (2016) 014323.
\bibitem[Tkalya\ 2017a]{Tkalya100} Tkalya, E.V. et al., {\it Erratum: Radiative lifetime and energy of the low-energy isomeric level in $^{229}$Th [Phys. Rev. C \textbf{92}, 054324 (2015)]}, Phys. Rev. C \textbf{95} (2017) 039902(E).
\bibitem[Tkalya\ 2017b]{Tkalya12} Tkalya, E.V., {\it Low-energy E0 transition between the components of the ground-state doublet in the muonic atom $^{229}$Th}, Phys. Rev. A \textbf{95} (2017) 042512.
\bibitem[Tkalya\ 2018]{Tkalya2018} Tkalya, E.V., {\it Decay rate of the nuclear isomer $^{229}$Th(3/2$^+$, 7.8 eV) in a dielectric sphere, thin film and metal cavity}, Phys. Rev. Lett. \textbf{120} (2018) 122501.
\bibitem[Tkalya\ 2019]{Tkalya2019} Tkalya, E.V., {\it Decay of the low-energy nuclear $^{229\text{m}}$Th isomer via atomic Rydberg states}, Phys. Rev. C \textbf{100} (2019) 054316.
\bibitem[Tkalya\ 2020a]{Tkalya2020} Tkalya, E.V., Si, R., {\it Internal conversion of the low-energy $^{229\text{m}}$Th isomer in the thorium anion}, Phys. Rev. C \textbf{101} (2020) 054602.
\bibitem[Tkalya\ 2020b]{Tkalya2020b} Tkalya, E.V., {\it Excitation of $^{229\text{m}}$Th at inelastic scattering of low energy electrons}, Phys. Rev. Lett. \textbf{124} (2020) 242501.
\bibitem[Tordoff\ 2006a]{Tordoff} Tordoff, B. et al., {\it Investigation of the low-lying isomer in $^{229}$Th by collinear laser spectroscopy}, Hyperfine Interact. \textbf{171} (2006) 197-201.
\bibitem[Tordoff\ 2006b]{Tordoff2} Tordoff, B. et al., {\it An ion guide for the production of a low energy ion beam of daughter products of $\alpha$-emitters}, Nucl. Instrum. Methods B \textbf{252} (2006) 347.
\bibitem[Tralli\ 1951]{Tralli1951} Tralli, N., Goertzel, G., {\it The theory of internal conversion}, Phys. Rev. \textbf{83} (1951) 399-404.
\bibitem[Trautmann\ 1994]{Trautmann1994} Trautmann, N. et al., {\it Accurate determination of the first ionization potential of actinides by laser spectroscopy}, J. Alloys Compd. \textbf{28} (1994) 213-214.
\bibitem[Troyan\ 2013]{Troyan} Troyan, V.I. et al., {\it Generation of thorium ions by laser ablation and inductively coupled plasma techniques for optical nuclear spectroscopy}, Las. Phys. Lett. \textbf{10} (2013) 105301.
\bibitem[Troyan\ 2014a]{Troyan1} Troyan, V.I. et al., {\it Quadrupole Paul ion trap in complex for optical spectroscopy of multiply charged thorium ions for the development of a nuclear frequency standard}, Measurement Techniques \textbf{57} (2014) 777-782.
\bibitem[Troyan\ 2014b]{Troyan5} Troyan, V.I. et al., {\it Experimental treatments for the investigation of nuclear optical transition in thorium-229}, IEEE European Frequency and Time Forum (EFTF) (2014) 15615881.
\bibitem[Troyan\ 2015a]{Troyan2} Troyan, V.I. et al., {\it The development of nuclear frequency standard with the use of ion crystal manipulation system}, Physics Procedia \textbf{72} (2015) 245-248.
\bibitem[Troyan\ 2015b]{Troyan3} Troyan, V.I. et al., {\it Multisectional linear ion trap and novel loading method for optical spectroscopy of electron and nuclear transitions}, Eur. J. Mass Spectrom. \textbf{21} (2015) 1-12.
\bibitem[Troyan\ 2015c]{Troyan2015c} Troyan, V.I. et al., {\it Formation of thorium-disodium fluoride crystals by electron-beam evaporation}, Physics Procedia \textbf{72} (2015) 175-178.
\bibitem[Troyan\ 2015d]{Troyan4} Troyan, V.I. et al., {\it Local Electrochemical depostion of thorium on SiO$_2$/Si(111) Surface}, Physics Procedia \textbf{72} (2015) 179-183. 
\bibitem[Typel\ 1996]{Typel} Typel, S., Leclercq-Willain, C., {\it Nuclear excitation by laser-assisted electronic transitions}, Phys. Rev. A \textbf{53} (1996) 2547-2561.

\bibitem[Udem\ 2002]{Udem} Udem, Th., Holzwarth, R., Hänsch, T.W., {\it Optical frequency metrology}, Nature \textbf{416} (2002) 233-237.
\bibitem[Usher\ 1929]{Usher} Usher, A.P., {\it History of Mechanical Inventions}, Oxford University Press (1929).
\bibitem[Utter\ 1999]{Utter} Utter, S.B. et al., {\it Reexamination of the optical gamma ray decay in $^{229}$Th}, Phys. Rev. Lett. \textbf{82} (1999) 505-508.
\bibitem[Uzan\ 2003]{Uzan} Uzan, J.P., {\it The fundamental constants and their variation: observational and theoretical status}, Rev. Mod. Phys. \textbf{75} (2003) 403.
\bibitem[Varga\ 2014]{Varga2014} Varga, Z., Nicholl, A.,  Mayer, K., {\it Determination of the $^{229}$Th half-life}, Phys. Rev. C \textbf{89} (2014) 064310. 
\bibitem[Varlamov\ 1996]{Varlamov} Varlamov, V.O. et al., {\it Excitation of a $^{229m}$Th ((3/2)$^+$, 3.5 eV) isomer by surface plasmons}, Phys. Doklady \textbf{41} (1996) 47. 
\bibitem[Verlinde\ 2019]{Verlinde2019} Verlinde, M. et al., {\it An alternative approach to populate and study the $^{229}$Th nuclear clock isomer}, Phys. Rev. C \textbf{100} (2019) 024315.
\bibitem[Vorykhalov\ 1995]{Vorykhalov} Vorykhalov, O.V., Koltsov, V.V., {\it Search for an isomeric transition of energy below 5 eV in $^{229}$Th nucleus.}, Bull. Rus. Acad. Sci.: Physics \textbf{59} (1995) 20-24.

\bibitem[Wakeling\ 2014]{Wakeling} Wakeling, M.A., {\it Charge states of $^{229\text{m}}$Th: Path to finding the half-life}, Thesis performed at LLNL, Livermore (2014).
\bibitem[vdWense\ 2013]{Wense1} von der Wense, L. et al., {\it Towards a direct transition energy measurement of the lowest nuclear excitation in $^{229}$Th}, JINST \textbf{8} (2013) P03005.
\bibitem[vdWense\ 2015]{Wense4} von der Wense, L. et al., {\it Determination of the extraction efficiency for $^{233}$U source $\alpha$-recoil ions from the MLL buffer-gas stopping cell}, Eur. Phys. J. A \textbf{51} (2015) 29.
\bibitem[vdWense\ 2016a]{Wense2} von der Wense, L. et al., {\it Direct detection of the $^{229}$Th nuclear clock transition}, Nature \textbf{533} (2016) 47-51.
\bibitem[vdWense\ 2016b]{Wense5} von der Wense, L. et al., {\it The extraction of $^{229}$Th$^{3+}$ from a buffer-gas stopping cell}, Nucl. Instrum. Methods B \textbf{376} (2016) 260-264.
\bibitem[vdWense\ 2016c]{Wense3} von der Wense, L., {\it On the direct detection of $^{229\text{m}}$Th}, Ph.D. Thesis, Ludwig-Maximilians-Universität München, Germany (2016). Published as Springer Thesis (Springer, Berlin, 2018).
\bibitem[vdWense\ 2017]{Wense6} von der Wense, L. et al., {\it A laser excitation scheme for $^{229\text{m}}$Th}, Phys. Rev. Lett. \textbf{119} (2017) 132503.
\bibitem[vdWense\ 2018]{Wense2018} von der Wense, L., Seiferle, B., Thirolf, P.G., {\it Towards a $^{229}$Th-based nuclear clock}, Measurement Techniques \textbf{60} (2018) 1178-1192.
\bibitem[vdWense\ 2019a]{Wense2019a} von der Wense, L. et al., {\it Preparing an isotopically pure $^{229}$Th ion beam for studies of $^{229\text{m}}$Th}, Jove video article doi:10.3791/58516 (2019).
\bibitem[vdWense\ 2019b]{Wense2019b} von der Wense, L. et al., {\it The concept of laser-based conversion electron Mössbauer spectroscopy for a precise energy determination of $^{229\text{m}}$Th}, Hyperfine Interact. \textbf{240} (2019) 23.
\bibitem[vdWense\ 2020a]{Wense2019d} von der Wense, L. et al., {\it The theory of direct laser excitation of nuclear transitions}, Eur. Phys. J. A \textbf{56} (2020) 176.
\bibitem[vdWense\ 2020b]{Wense2019c} von der Wense, L., Zhang, C., {\it Concepts for direct frequency-comb spectroscopy of $^{229\text{m}}$Th and an internal-conversion-based solid-state nuclear clock}, Eur. Phys. J. D \textbf{74} (2020) 146.

\bibitem[Winkler2016]{Winkler} Winkler, G., et al., {\it Non-planar femtosecond enhancement cavity for VUV frequency comb applications}, Optics Express \textbf{24} (2016) 5253-5262. 
\bibitem[Wycech\ 1993]{Wycech} Wycech, S., Zylicz, J., {\it Predictions for nuclear spin mixing in magnetic fields}, Acta Phys. Pol. B \textbf{24} (1993) 637. 
\bibitem[Wynands\ 2005]{Wynands} Wynands, R., Weyers, S., {\it Atomic fountain clocks}, Metrologica \textbf{42} (2005) 64-79.
\bibitem[Yamaguchi\ 2015]{Yamaguchi} Yamaguchi, A. et al., {\it Experimental search for the low-energy nuclear transition in $^{229}$Th with undulator radiation}, New J. Phys. \textbf{17} (2015) 053053.
\bibitem[Yamaguchi\ 2019]{Yamaguchi2019} Yamaguchi, A. et al., {\it Energy of the $^{229}$Th nuclear clock isomer determined by absolute $\gamma$-ray
energy difference}, Phys. Rev. Lett. \textbf{123} (2019) 222501.
\bibitem[Yasuda\ 2017]{Yasuda2017} Yasuda, Y. et al., {\it Search for the ultraviolet photons emitted from $^{229\text{m}}$Th}, 6th Asia-Pacific Symposium on Radiochemistry (2017). Online: http://www.apsorc2017.org/upload/abs/1485936747 \_yasuday11@chem.sci.osaka-u.ac.jp2.pdf
\bibitem[Yoshimi\ 2018]{Yoshimi2018} Yoshimi, A. et al., {\it Nuclear resonant scattering experiment with fast time response: Photonuclear excitation of $^{201}$Hg}, Phys. Rev. C \textbf{97} (2018) 024607.
\bibitem[Yost\ 2009]{Yost2009} Yost, D.C. et al., {\it Vacuum-ultraviolet frequency combs from below-threshold harmonics}, Nature Physics \textbf{5} (2009) 815-820.
\bibitem[Young\ 1999]{Young} Young, J.P. et al., {\it Radioactive origin of emissions observed from uranium compounds and their silica cells}, Inorg. Chem. \textbf{38} (1999) 5192-5194.
\bibitem[ZhangX\ 2011]{Zhang2011} Zhang, X. et al., {\it Optical properties of the vacuum-ultraviolet nonlinear optical crystal BPO$_4$}, J. Opt. Soc. Am. B \textbf{28} (2011) 2236-2239.
\bibitem[ZhangC\ 2020]{Zhang2020} Zhang, C. et al., {\it Noncollinear Enhancement Cavity for Record-High Out-Coupling Efficiency of an Extreme-UV Frequency Comb}, Phys. Rev. Lett. \textbf{125} (2020) 093902.
\bibitem[Zhao\ 2012]{Zhao} Zhao, X. et al., {\it Observation of the deexcitation of the $^{229m}$Th nuclear isomer}, Phys. Rev. Lett. \textbf{109} (2012) 160801.
\bibitem[Zheltonozhskii\ 1988]{Zheltonozhskii1988} Zheltonozhskii, V.A. et al., {\it Decay of $^{193\text{m}}$Ir via an electron bridge}, Zh. Eksp. Teor. Fiz. \textbf{94} (1988) 32 [Sov. Phys. JETP \textbf{67} (1988) 16].
\bibitem[Zimmermann\ 2010]{Zimmermann} Zimmermann, K., {\it Experiments towards optical nuclear spectroscopy with thorium-229}, Ph.D. Thesis, University of Hannover, Germany (2010).
\bibitem[Zimmermann\ 2012]{Zimmermann1} Zimmermann, K. et al., {\it Laser ablation loading of a radiofrequency ion trap}, Appl. Phys. B \textbf{107} (2012) 883-889.
\bibitem[Zon\ 1990]{Zon} Zon, B.A., Karpeshin, F.F., {\it Acceleration of the decay of $^{235\text{m}}$U by laser-induced resonant internal conversion}, Sov. Phys. JETP \textbf{70} (1990) 224-227.
 
\end{thebibliography}
\end{document}